\documentclass[final,3p,times]{elsyearticle}

\usepackage{graphicx}

\usepackage{amssymb}

\usepackage[english,french]{babel}
\usepackage[utf8]{inputenc}
\usepackage[T1]{fontenc}

\def\og{\leavevmode\raise.3ex\hbox{$\scriptscriptstyle\langle\!\langle$~}}
\def\fg{\leavevmode\raise.3ex\hbox{~$\!\scriptscriptstyle\,\rangle\!\rangle$}}
\def\og{\lq\lq}
\def\fg{\rq\rq}

\usepackage[colorinlistoftodos]{todonotes}
\usepackage[colorlinks=true, allcolors=blue]{hyperref}
\usepackage{braket}
\usepackage{subfig}
\usepackage{bbold}
\usepackage{amsmath}
\usepackage{graphicx}
\usepackage{natbib}
\usepackage{wrapfig}
\usepackage{comment}

\newcommand{\dd}{\mathrm{d}}
\newcommand{\qq}{\mathbf{q}}
\newcommand{\ii}{\mathrm{i}}
\newcommand{\eee}{\mathrm{e}}
\newcommand{\bea}{\begin{eqnarray}}
\newcommand{\eea}{\end{eqnarray}}
\newcommand{\be}{\begin{equation}}
\newcommand{\ee}{\end{equation}}

\begin{document}

\selectlanguage{english}
\begin{frontmatter}



\title{Nuclear spin squeezing by continuous quantum non-demolition measurement: a theoretical study}


\author[ad1]{Alan Serafin}
\author[ad1]{Yvan Castin}
\author[ad2]{Matteo Fadel}
\author[ad2]{Philipp Treutlein}
\author[ad1]{Alice Sinatra}

\address[ad1]{Laboratoire Kastler Brossel, ENS-Universit\'e PSL, CNRS, Universit\'e de la Sorbonne and Coll\`ege de France, 24 rue Lhomond, 75231 Paris, France}
\address[ad2]{Department of physics, University of Basel, Klingelbergstrasse 82, 4056 Basel, Switzerland}

\begin{abstract}
\noindent
We propose to take advantage of the very weak coupling of the ground-state helium-3 nuclear spin to its environment to produce very long-lived macroscopic quantum states, here nuclear spin squeezed states, in a gas cell at room temperature. To perform a quantum non-demolition measurement of a transverse component of the previously polarized collective nuclear spin, a discharge is temporarily switched on in the gas, which populates helium-3 metastable state. The collective spin corresponding to the $F=1/2$ metastable level then hybridizes slightly with the one in the ground state by metastability exchange collisions. To access the nuclear spin fluctuations, one continuously measures the light field leaking out of an optical cavity, where it has interacted dispersively with the metastable state collective spin. In a model of three coupled collective spins (nuclear, metastable and Stokes for light) in the Primakoff approximation, and for two measurement schemes, we calculate the moments of the collective nuclear spin squeezed component $ I_z $ conditioned on the optical signal averaged over the observation time $ t $. In the photon counting scheme, we find that the squeezed observable is $ I_z^2 $ rather than $ I_z $. In the homodyne detection scheme, we analytically solve the stochastic equation for the state of the system conditioned on the measurement; the conditional expectation value of $ I_z $ depends linearly on the signal and the conditional variance of $ I_z $ does not depend on it. The conditional variance decreases as $ (\Gamma_{\rm sq}t)^{-1} $, where the squeezing rate $ \Gamma_{\rm sq}$, which we calculate explicitly, depends linearly on the light intensity in the cavity at weak atom-field coupling and saturates at strong coupling to the ground state metastability exchange effective rate, proportional to the metastable atom density. Finally, we take into account the de-excitation of metastable atoms at the walls, which induces nuclear spin decoherence with an effective rate $ \gamma_\alpha$. It imposes a limit $ \propto (\gamma_\alpha/\Gamma_{\rm sq})^{1/2} $ on the conditional variance reached in a time $ \propto(\gamma_\alpha\Gamma_{\rm sq})^{-1/2}$. A multilingual version is available on the open archive HAL at \url{https://hal.archives-ouvertes.fr/hal-03083577}.

\noindent 
\vskip 0.5\baselineskip

\end{abstract} 
\end{frontmatter}

\section{Introduction}
Helium-3 in its ground state enjoys the remarkable property of having a purely nuclear spin {$1/2$}, perfectly isolated from the outside world even in an environment as hostile to quantum coherences as a {gas} of helium in a centimetric cell at room temperature and a pressure of the order of a millibar. By well-mastered nuclear polarization techniques, reaching a polarization of 90 $ \% $, we can then routinely prepare (for example for lung imaging by nuclear magnetic resonance \cite{MacFall1996}) a giant collective nuclear spin with an extremely long {lifetime}. Recently, a coherence time $ T_2 $ {larger} than 60 hours was measured in ultra-precise magnetometry devices \cite{Heil2010}, that seems limited only by the longitudinal decay time $ T_1 $ due to collisions with the cell walls. \footnote{Times $ T_1 $ of several hundred hours can even be obtained \cite{Nacher2017}.} These numbers make the macroscopic nuclear spin in a {room temperature} {gas} an {\sl ideal} system for the production, the study and the use of entangled states, and therefore a competitor of cold atomic gases and Bose-Einstein condensates in metrology and quantum information processing \cite{PezzeRMP}. Already in 2005, we suggested that the nuclear spins of helium-3 could give rise to quantum memories \cite{DantanReinaudi2005} or to non-local quantum states \cite{Reinaudi2007} with very long lifetimes. Since then, experimental breakthroughs have been made in the field of spin {squeezing}, notably by means of non-demolition quantum measurements (QND) in atomic alkali gases interacting with the electromagnetic field {\cite{PezzeRMP,stroboscopic,Kasevich,Xiao}}, which recently made it possible to obtain a {squeezed} spin state with a lifetime of one second in the hyperfine ground state of rubidium under metrological conditions \cite{Myles2020}. 
{We recall that, in analogy with the squeezed states of light in quantum optics, a (here collective) spin $\vec{I}$ is said to be squeezed if it admits reduced fluctuations with respect to the standard quantum limit in a direction $z$ orthogonal to that $x$ of the mean spin \footnote{\label{note0}{The standard quantum noise corresponds to $\Delta I_y^{\rm st}=\Delta I_z^{\rm st}=(|\langle I_x\rangle|/2)^{1/2}$ (our spins being dimensionless here) i.e. to the equality in the case of cylindrical symmetry of the fluctuations around the mean spin in the Heisenberg inequality $\Delta I_y \Delta I_z \geq |\langle I_x\rangle|/2$. A more detailed discussion shows that, in the considered geometry, the noise-to-signal ratio in a precession frequency measurement is in fact proportional to $\xi=(2I)^{1/2}\Delta I_z/|\langle I_x\rangle|$ (this naturally brings in the uncertainty angle on the spin direction in the $xz$ plane) rather than to the naively expected $\Delta I_z/\Delta I_z^{\rm st}$ ratio. In our proposed scheme, we degrade $\xi^2$ by a factor $\approx 2$ using a partially polarized nuclear spin state $\langle I_x\rangle/I<1$, see figure \ref{fig3}, but we aim to gain much more with spin squeezing.}}; this leads to increased accuracy in pointing the mean spin in that direction, hence in its precession frequency around the $y$ axis, in any metrological device such as an atomic clock or a magnetometer \cite{Wineland,Ueda} that is sufficiently free from technical noise to reach the standard quantum limit  \cite{Santarelli}.}

Transposing to the nuclear spin of helium-3 the technique of squeezing by QND measurement used for the hyperfine spins of alkalis, represents a real challenge, however, due to the specificity of the nuclear spin: its weak coupling to the environment. The singlet ground state of helium-3, separated in energy by about 20 eV from {the first excited state}, is not directly accessible by laser. However, by means of an oscillating discharge, a small fraction of the {gas} atoms, on the order of $ 10^{-6} $, can be brought into the metastable triplet state, an excellent starting point for near infrared optical transitions. The orientation of the nuclear spins is then obtained through an indirect process, metastability exchange optical pumping \cite{Nacher2017}. Initially, the angular momentum is transferred by laser-matter interaction from photons to metastable atoms, a priori to their electron spin (the only one to be strongly coupled to the {laser} field) but a posteriori also to their nuclear spin thanks to hyperfine coupling. Secondly, we take advantage of the metastability exchange collisions between metastable and ground state atoms to orient the nuclear spins in the ground state, with a time scale of the order of a second, limited by the low density of the atoms in the metastable state. Even though the metastability exchange collision can transfer quantum correlations (see {references} \cite{DantanReinaudi2005, Reinaudi2007} and our section \ref{subsec:MEC}), we cannot expect that a single measurement on a small fraction of the atoms ($ 10^{-6} $) projects the whole system into a {squeezed} state. The solution we propose is to perform a continuous QND {measurement amplified} by a resonant optical cavity
{in which the cell is placed}.
Indeed, although the metastable atoms individually have a relatively short {lifetime} (they lose their quantum correlations and fall back into the ground state in each collision {with} the cell walls), a continuous destructive measurement of the light {leaking out of the cavity} after interaction with the metastable atoms amounts to performing a continuous QND measurement on the collective nuclear spin in the ground state, which {prepares} it into the desired {squeezed} state without affecting its lifetime. 
\footnote{{This type of measurement thus differs from the instantaneous measurement of quantum mechanics textbooks, which projects the state of the system into an eigenspace of the measured observable according to the von Neumann postulate.}} {As we will see, our procedure for reducing spin fluctuations is more precisely summarized by the diagram:}
{\small
\begin{center}
\begin{tabular}{ccccc}
\!\!\!\fbox{\parbox{2.8cm}{collective spin $\vec{I}$ in the ground state}} & \parbox{2.7cm}{metastability exchange $\leftarrow\!\!\!\xrightarrow{\hspace{2.2cm}}$ \hspace*{2mm}} & \fbox{\parbox{3cm}{collective spin in the metastable state}} & \parbox{1.8cm}{Faraday effet $\leftarrow\!\!\!\xrightarrow{\hspace{1.3cm}}$} & \fbox{\parbox{2.5cm}{cavity field polarized along $x$} } \\
$\Big\uparrow$ & & & & $\Big\downarrow$  \\
\!\!\!\fbox{\parbox{2.8cm}{reduction of fluctuations of $I_z^2$,  $I_z$}} & $\xleftarrow{\hspace{2cm}}$ & \fbox{\parbox{3cm}{continuous non-demolition measurement of $I_z^2$, $I_z$}} & \parbox{2.9cm}{photon counting, $\xleftarrow{\hspace{2.3cm}}$ homodyne detection } &  \fbox{\parbox{2.5cm}{continous measurement of the field polarized along $y$ leaking out from the cavity}}
\end{tabular}
\end{center}
}

This work gives a detailed theoretical description of the {squeezing} mechanism and its limits; a more detailed feasibility study taking into account the experimentally accessible values of the parameters is carried out in {reference} \cite{letter}. Very recently, similar ideas have been put forward in a different physical system, the alkali-rare gas mixture \cite{Firstenberg2020, katz2019quantum}. We are confident that quantum manipulation of long-lived nuclear spins is promised rapid development, opening up new perspectives for basic research and applications,  {in particular in magnetometry \cite{Xiao,magnetth1,magnetth2,magnetexp1,magnetth3}}.

\section{Overview and semi-classical description} 
\label{sec:view} 

\begin{figure} [t] 
\centerline{\includegraphics [scale = 0.42]{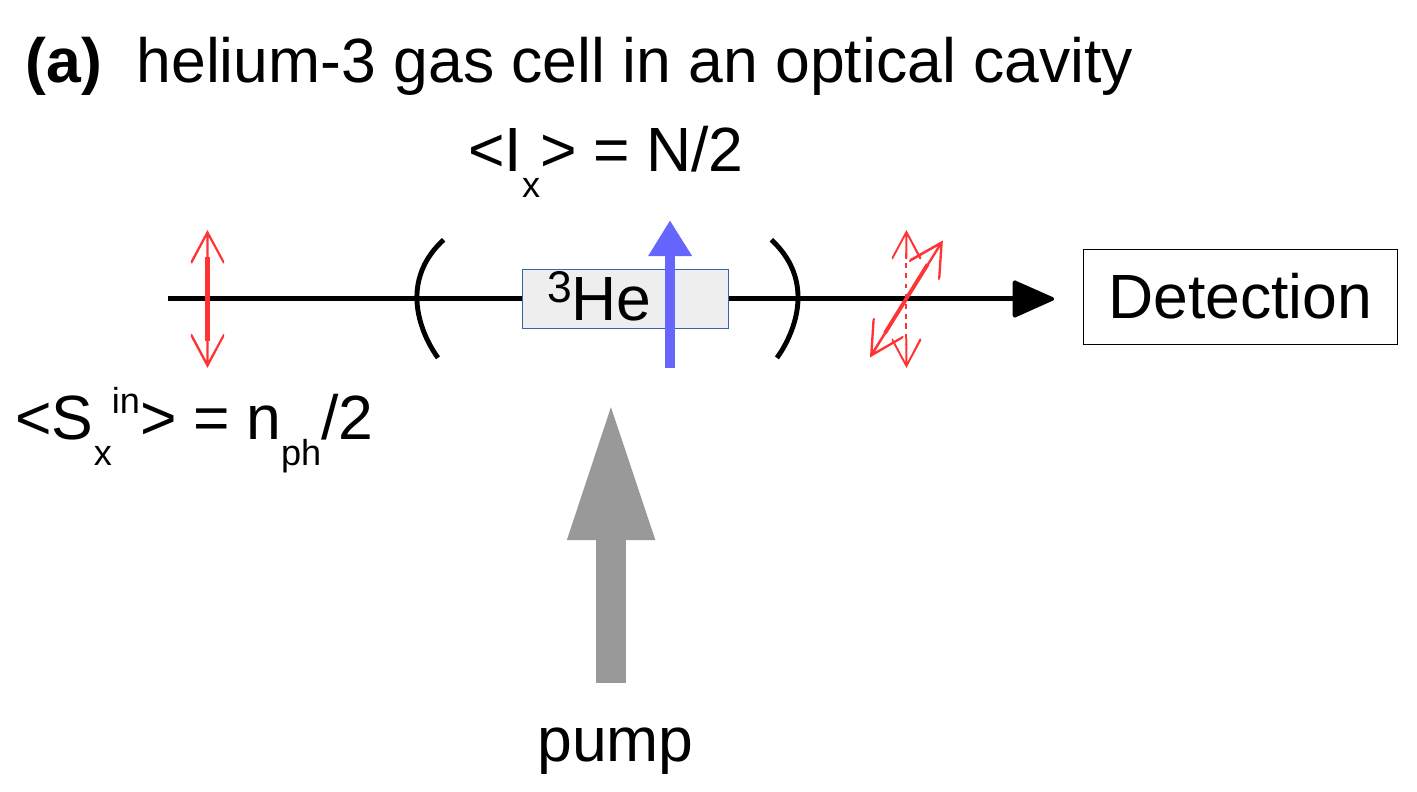} 
\includegraphics [scale = 0.4]{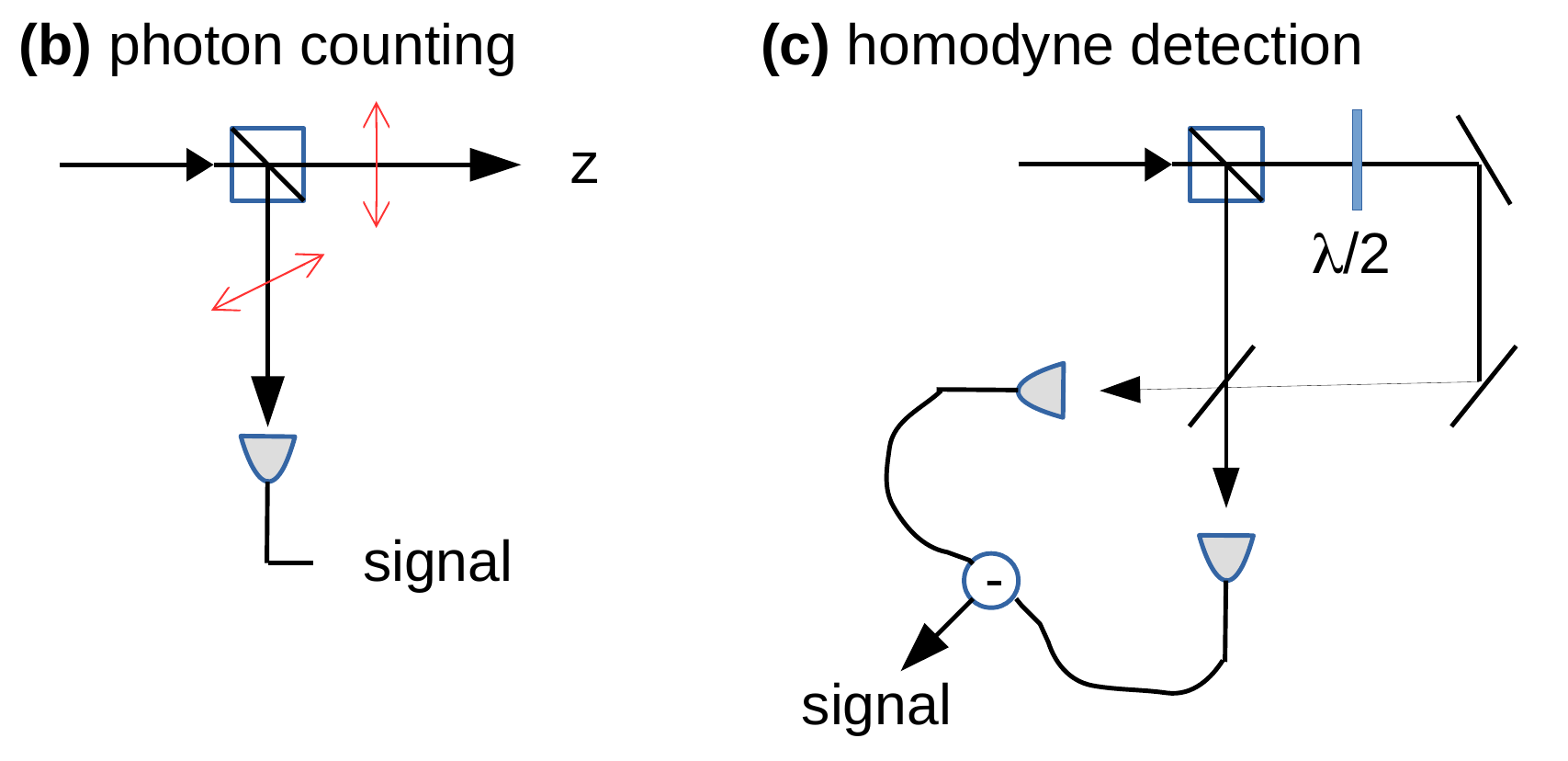}} 
\caption{Overview of the set-up. (a) A centimetric glass cell is filled with a helium-3 {gas} at room temperature and placed in an optical cavity with axis $ {z} $ (horizontal axis in the figure). The Stokes spin of light and the atomic spins (nuclear and metastable) are linearly polarized along $ {x} $ (vertical axis in the figure). The {cavity mode} polarized along $ {y} $, initially empty, is populated by the Faraday effect due to the quantum fluctuations of the {metastable spin}s along $ {z} $ during its propagation in the gas. 
{It is measured continuously outside the cavity by one of the following two methods:}
(b) the photons {leaking out of the cavity} polarized along $ {y} $ are separated from those polarized along $ {x} $ by a polarizing beam-splitter and then detected in the photon counting regime. (c) Outside the cavity, a homodyne detection of one quadrature of the outgoing field {polarized along} $ {y} $ is carried out, using as local oscillator the outgoing field {polarized along} $ {x} $ (whose polarization has been previously rotated with a half-wave plate to bring it along $ {y} $).} 
\label{fig1} 
\end{figure}

The considered physical system is shown in figure \ref{fig1}. A cell filled with a partially polarized {gas} of a few mbar of pure helium-3 atoms is placed inside an optical cavity. While the majority of atoms remain in the singlet ground state $ 1^{1} S $ of helium, a weak {discharge brings} a tiny fraction of the atoms, usually $ \simeq 10^{-6} $, {into} the metastable triplet state 2$^{3} S $. On the one hand, the cavity is {driven} by a laser beam propagating along the {cavity axis} $ {z} $ and linearly polarized in the direction $ {x} $, which is also the direction of polarization of the atomic sample, to excite the 2$^{3} S-2^{3} P $ transition with a large frequency detuning; on the other hand, atoms in the metastable state 2$^{3} S $ (of electronic and nuclear hyperfine spin) are coupled to atoms in the ground state (of purely nuclear spin) by {metastability exchange collisions};
{remarkably, although each exchange collision is individually incoherent, this leads to a well-defined macroscopic coupling between the corresponding collective spins \cite{DupontRoc,LaloeDupontLeduc}.}
As the Faraday interaction with the metastable atoms causes the polarization of the light initially directed {along} $ {x} $ to rotate slightly around the $ {z} $ axis, {by an angle proportional to} the component of the collective {metastable spin}s {along} $ {z} $ as we will see, a continuous destructive measurement of the polarization component {along} $ {y} $ of the field {leaking out of the cavity} (i) by counting photons as indicated in figure \ref{fig1}b or (ii) by homodyne detection as in figure figure \ref{fig1}c, ultimately performs a non-demolition continuous quantum measurement of the collective nuclear spin along $ {z} $ of helium-3 atoms in the ground state. 

In the rest of this section, by a semi-classical treatment of the {spin fluctuations} around {their expectation values in} the stationary state, we reduce our complex physical system to the simpler one of three coupled collective spins, of which {section} \ref{sec:qt} will give a quantum description. 

\begin{figure} [t] 
\vspace*{-15mm}
\centerline{\includegraphics [scale = 0.30,clip=]{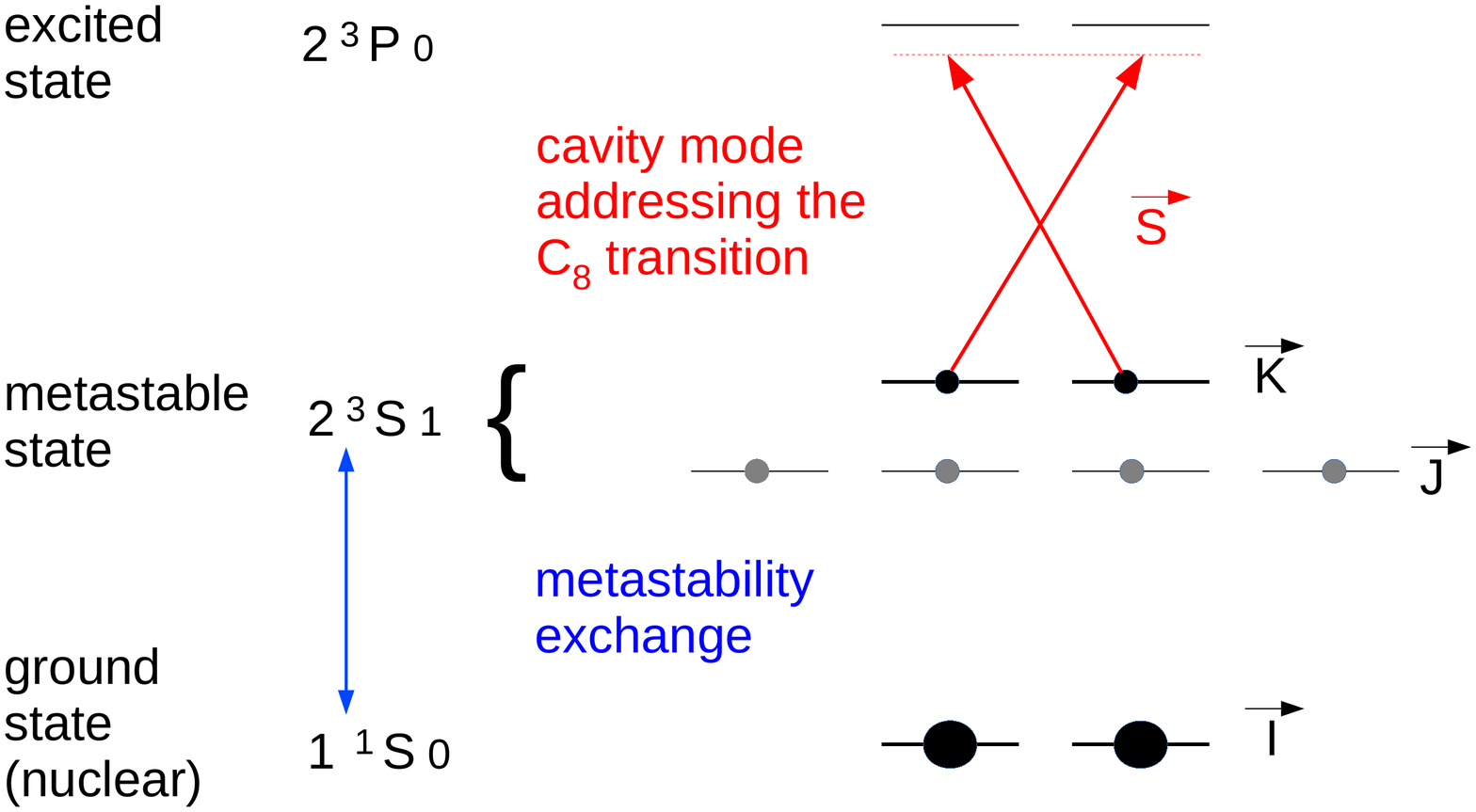}} 
\caption{Relevant energy levels of the ${}^3$He atoms (the Zeeman sublevels correspond to the choice of $ {z} $ as the quantization axis, the atoms being {polarized along} $ {x} $). The {cavity mode} polarized along $ {x} $ excites the transition $ C_8 $ between the level $ F = 1/2 $ of the metastable state 2$^{3} S_1 $ and the highest energy level $ F = 1/2 $ of the excited state 2${}^3P $, with a negative frequency detuning much {larger} in absolute value than the Doppler half-width of the excited state (of the order of 1 GHz), so that the {velocity class} resonant with the laser is almost empty, but {much smaller} than the 6.74 GHz hyperfine {splitting} in the metastable state (and a fortiori than the fine {splitting} 2${}^3P_1-2{}^3P_0 $ of 29.6 GHz in the excited state), so that the metastable level $ F = 3/2 $ is very weakly affected by the laser. (Note: The frequency separation does not make it possible to largely satisfy these two constraints, and we cannot exclude that the coupling of $ F = 3/2 $ to the field has a small effect on the {squeezing} dynamics; we neglect it here but we could take it into account with a more complete than our {minimal model} Hamiltonian (\ref{eq002}), like that of {reference} \cite{Dantan2007}). The six sublevels of the 2$^{3} S_1 $ metastable state are coupled to the two (purely nuclear) sublevels of the $ 1^{1} S_0 $ ground state by metastability exchange collisions.}
\label{fig2} 
\end{figure}
\begin{figure} [t] 
\centerline{\includegraphics [width = 0.40 \textwidth]{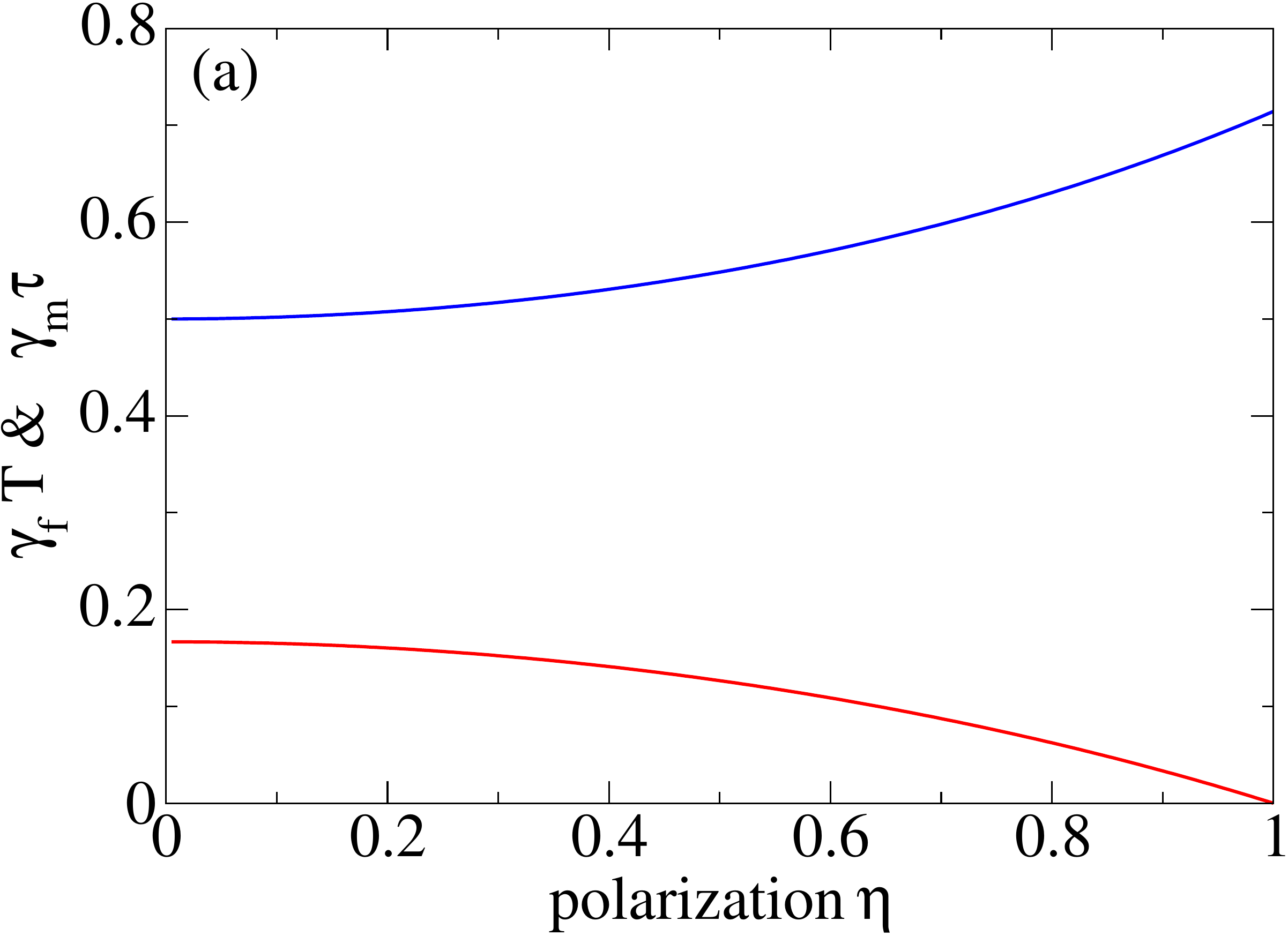} \quad \quad \includegraphics [width = 0.40 \textwidth]{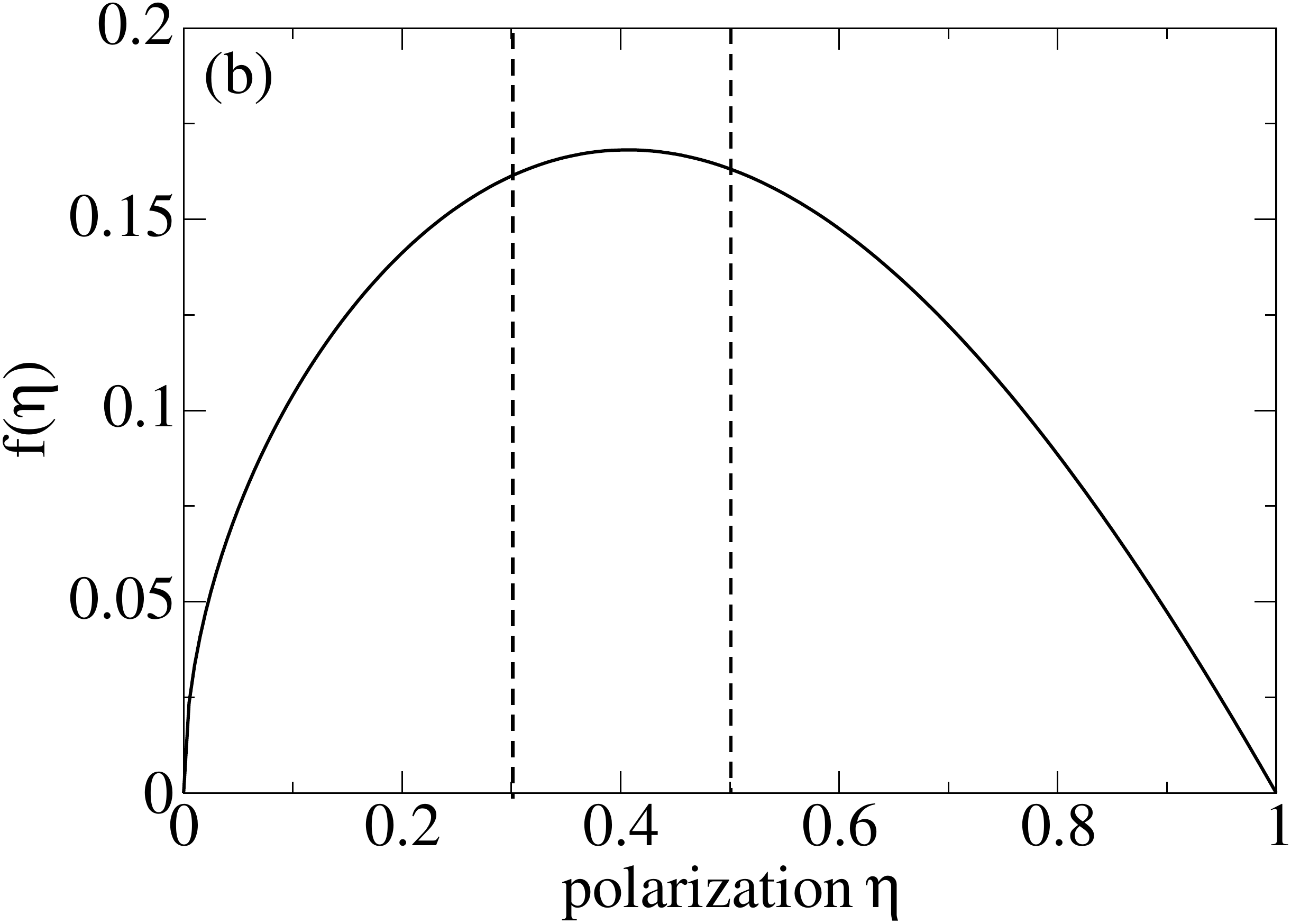}} 
\caption{(a) Effective metastability exchange rates $ \gamma_f $ {(lower red curve)} and $ \gamma_m $ {(upper blue curve)} as given by equation (\ref{eq006}) {in the ground and metastable states}, as functions of nuclear polarization $ \eta $, normalized by the metastability exchange collision rates $ 1/T $ and $ 1/\tau $ {experienced} by {a ground-state and a metastable} atom in the {gas}. (b) Nuclear polarization dependence of the Faraday {frequency} $ \Omega_\alpha $ 
{entering in the rate (\ref{eq054}) at which excitations are created by Faraday coupling in the hybridized nuclear bosonic mode, and in the spin squeezing rates (\ref{eq868}) and (\ref{eq135}),}
in the limit $ \gamma_f \ll \gamma_m $; more precisely, we plot the factor $ f (\eta) = \sqrt{\eta} \frac{1-\eta^2}{3+\eta^2} $ such that 
$\Omega_\alpha \simeq \Omega (\gamma_f/\gamma_m)^{1/2} = \chi \sqrt{n_{\rm ph}n_{\rm cell}}\sqrt{\frac{n_{\rm cell}}{N_{\rm cell}}}f(\eta)$.
When the polarization varies between 0.3 and 0.5 (vertical dashed lines), $ f (\eta) $ deviates by 4 \% from its maximum $ \simeq $ 0.17 reached in $ \eta = $ 0.42.
{It is therefore advantageous to place oneself close to this value of $\eta$ to reduce the temporal drift of $\Omega_\alpha$ due to a slight damping of the nuclear polarization during spin squeezing (indeed, the optical pumping process is then interrupted and the lifetime of the polarization is reduced by the presence of the discharge, it becomes of the order of $\gamma_\alpha^{-1}$, where $\gamma_\alpha$ is the induced decoherence rate of section \ref{sec:edld}).}
} 
\label{fig3} 
\end{figure}

 
The relevant atomic structure of the ${}^3$He atom and the transitions excited by the cavity field are shown in figure \ref{fig2}. We call $ \Vec{I} $ the collective nuclear spin in the ground state, $ \Vec{J} $ and $ \Vec{K} $ the collective spins associated with the hyperfine manifolds $ F = 3/2 $ and $ F = $ 1/2 in the metastable state. For light propagating along $ {z} $, we introduce the Stokes spin \cite{Dantan2007} built from the {creation and annihilation operators} of a photon in the linearly polarized {cavity} modes along $ {x} $ and $ {y} $: \footnote{\label{note2} Equivalently, we can build the Stokes spin $ \vec{S} $ using annihilation operators in circularly polarized modes, $ c_1 = \frac{1}{\sqrt{2}} (c_x-\ii c_y) $, $ c_2 = \frac{1}{\sqrt{2}} (c_x+\ii c_y) $ \cite{Mandel1999}, in which case $ S_z = \frac{1}{2} \left (c_1^\dagger c_1-c_2^\dagger c_2 \right) $.} 
\begin{equation}
S_x = \frac{1}{2}\left( c_x^\dagger c_x -  c_y^\dagger c_y \right) \quad ; \quad 
S_y = \frac{1}{2}\left( c_x^\dagger c_y +  c_y^\dagger c_x \right) \quad ; \quad 
S_z =  \frac{1}{2\ii}\left( c_x^\dagger c_y -  c_y^\dagger c_x \right) \label{eq001} \,.
\end{equation}
We assume for simplicity that the cell is uniformly illuminated by the cavity mode. Within the limit of {a large detuning} and a weak saturation of the atomic transition by the field, the excited state 2$^3 P $ can be eliminated adiabatically and the {interaction Hamiltonian} between the {metastable spin} $ \vec{K} $ and the Stokes spin $ \vec{S} $ takes {the} Faraday form \cite{Dantan2007}: 
\begin{equation}
H = \hbar \chi K_z S_z 
\label{eq002}
\end{equation}
which is none other than the {lightshift} operator of Zeeman sublevels in the metastable {level $F=1/2$}, as we can clearly see {from} the form of $ S_z $ in {footnote} \ref{note2}. The coupled nonlinear equations describing the evolution of the mean spins are given in \ref{app:semiclassical}, see {equations} (\ref{DerivativesNLTotSx})-(\ref{eq202c}). Besides the evolution due to the Faraday Hamiltonian (\ref{eq002}) and to the metastability exchange collisions, they include the contribution of the usual Liouvillian terms in the {quantum master equation} describing the injection of a polarized coherent field {along} $ {x} $ in the cavity and the losses due to the output mirror, whose combined effect leads to $ \langle S_x \rangle = n_{\rm ph}/2 $ in the stationary state in the absence of atoms, $ n_{\rm ph} $ being the average number of photons in the polarized mode {along} $ {x} $. These equations are then linearized around a partially polarized stationary solution (\ref{eq207})-(\ref{eq208}), and the fluctuations of the spin $ \vec{J} $ and the collective alignment tensor in $ F = 3/2 $ are eliminated adiabatically \footnote{We think that this non-mathematically controlled approximation is reasonable for the proposed experiment, because the spin $ \vec{J} $ is not directly coupled to light so {it} is not directly affected by {the} continuous field measurement. On the other hand, by eliminating in the same way the fluctuations of the spin $ \vec{K} $, directly coupled to the field, one would commit a non-negligible error on the {spin squeezing dynamics} in the case of the detection by {photon counting} (amounting to omitting the double jump $ C_d $ in the {quantum master equation} (\ref{eq030oneMode}) and the rate $ \Gamma_0 $ in the average number of photons counted (\ref{eq064})) therefore strongly underestimating the number of photodetections required to achieve a given {squeezing} level), but a negligible error in the case of homodyne detection, as we have verified on the one-mode model in section \ref{sec:ana_omm}.} to obtain coupled equations {for} the fluctuations of the three collective spins $ \vec{I} $, $ \vec{K} $ and $ \vec{S} $, whose stationary mean values are given by: 
\begin{equation}
\langle \vec{I}\rangle_s= \frac{N}{2} \, \vec{u}_x 
\quad ; \quad
\langle \vec{K} \rangle_s= \frac{n}{2} \, \vec{u}_x
\quad ; \quad
\langle \vec{S} \rangle_s=\frac{n_{\rm ph}}{2} \, \vec{u}_x \label{eq003}
\end{equation}
Here $ \vec{u}_x $ is the unit vector {along} $ {x} $, $ N $ and $ n $ are the effective numbers of {ground-state and metastable} atoms participating in the dynamics of {the} collective spins. As we show in \ref{app:semiclassical}, these effective numbers are renormalized with respect to the total true numbers $ N_{\rm cell} $ and $ n_{\rm cell} $ in the cell, by {polarization dependent factors}: 
\begin{equation}
N=\eta \, N_{\rm cell} 
\quad ; \quad
 n=\left( \frac{1-\eta^2}{3+\eta^2} \right) \eta \, n_{\rm cell} \label{eq004} \\
\end{equation}
where $ \eta \in [0,1] $ is the nuclear polarization, \footnote{Note that $ n = 0 $ in the fully polarized case $ \eta = 1 $. Indeed, the entire population of the metastable state is then in the extreme Zeeman sublevel $ m_x = 3/2 $ of the hyperfine state $ F = 3/2 $ and the manifold $ F = 1/2 $ is empty.} and the semi-classical equations {for} the fluctuations of the three collective spins are: 
\begin{align}    
        \frac{\dd}{\dd t}\delta S_z &=  -\frac{\kappa}{2} \delta S_z \quad \quad \quad \quad \quad \quad \quad  \quad \quad \:\:
        \frac{\dd}{\dd t}\delta S_y   =   -\frac{\kappa}{2} \delta S_y + \chi\langle S_x \rangle_s  \delta K_z \label{eq:redS} \\
                \frac{\dd}{\dd t}\delta I_z &=  - \gamma_f \delta I_z + \gamma_m \delta K_z \quad \quad \quad \quad \quad \quad 
                  \frac{\dd}{\dd t}\delta I_y =  - \gamma_f \delta I_y + \gamma_m \delta K_y \label{eq:redI} \\
    \frac{\dd}{\dd t}\delta K_z &= - \gamma_m \delta K_z + \gamma_f \delta I_z \quad \quad \quad \quad  \quad \quad
        \frac{\dd}{\dd t}\delta K_y =    - \gamma_m \delta K_y + \gamma_f \delta I_y+ \chi \langle K_x \rangle_s \delta S_z \label{eq005}
\end{align}
Here, $ \kappa $ is the {cavity loss rate}, $ \gamma_m $ and $ \gamma_f $ are the {effective metastability exchange rates} in the metastable state and in the ground state. The latter depend on the nuclear polarization as below and in figure \ref{fig3}a, and are in the same ratio as the effective atom numbers $ N $ and $ n $ (\ref{eq004}) forming the collective spins: 
\begin{equation}
{\gamma_f = \frac{4 + \eta^2}{8 - \eta^2} \, \frac{1-\eta^2}{3+\eta^2} \, \frac{1}{T} \quad ;\quad \gamma_m =  \frac{4 + \eta^2}{8 - \eta^2}  \, \frac{1}{\tau} \quad ;}\quad  \frac{\gamma_m}{\gamma_f}= \frac{N}{n} \gg 1
\label{eq006}
\end{equation}
the individual {metastability exchange} collisions rates $1/T$ and $1/\tau$ {experienced} by an atom in the ground state and in the {metastable} state being proportional to $ n_{\rm cell} $ and $ N_{\rm cell} $ {respectively,} {with the same proportionality constant}. In figure \ref{fig3}b, we also show the nuclear polarization dependence of the effective Faraday coupling $ \Omega_\alpha $ (\ref{eq036}) between light and the nuclear spin hybridized by the metastable, which controls the {spin squeezing rate} in (\ref{eq054}). 

\section{Quantum description} 
\label{sec:qt} 
In {section} \ref{sec:view}, we have seen that we can model our complex physical system {as} three coupled collective spins (\ref{eq003}): the nuclear spin $ \vec{I} $ in the ground state, the spin $ \vec{K} $ in the hyperfine level $ F = 1/2 $ of the metastable state and the Stokes spin $ \vec{S} $ of the {cavity field}. In this section, we present the full quantum {treatment of} this model. After having introduced the Primakoff approximation, we move on to the quantum description of the metastability exchange which couples the nuclear and metastable spins. 

\subsection{Primakoff approximation {and metrological gain due to squeezing}} 
\label{sub:primakov} 
Initially, the collective nuclear spin $ \vec{I} $, the collective {metastable spin} $ \vec{K} $ and the Stokes spin $ \vec{S} $ of light are polarized along $ {x} $, and will remain so throughout the experimental procedure. In the Holstein-Primakoff approximation, which {treats} the macroscopic spin components along $ {x} $ {as} classical variables, the remaining $ {y} $ and $ {z} $ components, orthogonal to the mean spins, behave like the quadrature operators (Hermitian and antihermitian parts of annihilation operators, therefore canonically conjugated, $ [X, P] = \ii/2 $) of three bosonic modes $ a $, $ b $, $ c $: \footnote{If we consider a large spin $ \vec{S} $ fully {polarized along} $ {x} $, we can approximate the spin component in this direction by a classical variable, by setting 
$\hat{S}_x \simeq \langle \hat{S}_x \rangle$ so that $[ \hat{S}_y/\!\sqrt{2\langle{\hat{S}_x\rangle}}\, , \, \hat{S}_z/\!\sqrt{2\langle{\hat{S}_x\rangle}}] \simeq \ii /2$.
} 
\begin{align}
\frac{I_y}{\sqrt{N}} \stackrel{\mbox{\scriptsize{Primakoff}}}{\simeq}  X_a = \frac{a+a^\dagger}{2} \quad &;  \quad
\frac{K_y}{\sqrt{n}} \stackrel{\mbox{\scriptsize{Primakoff}}}{\simeq}  X_b = \frac{b+b^\dagger}{2} \quad &;  \quad
\frac{S_y}{\sqrt{n_{\rm ph}}} \stackrel{\mbox{\scriptsize{Primakoff}}}{\simeq}  X_c = \frac{c+c^\dagger}{2} 
\label{eq014a} \\
\frac{I_z}{\sqrt{N}} \stackrel{\mbox{\scriptsize{Primakoff}}}{\simeq}  P_a = \frac{a-a^\dagger}{2\ii} \quad &; \quad 
\frac{K_z}{\sqrt{n}} \stackrel{\mbox{\scriptsize{Primakoff}}}{\simeq}  P_b = \frac{b-b^\dagger}{2\ii} \quad &; \quad
\frac{S_z}{\sqrt{n_{\rm ph}}} \stackrel{\mbox{\scriptsize{Primakoff}}}{\simeq}  P_c = \frac{c-c^\dagger}{2\ii}
\label{eq014b} 
\end{align}
{We used the mean values (\ref{eq003}) in the normalization.}
Let us make the link with the exact bosonic representation (\ref{eq001}) of {Stokes' spin}, writing: 
\begin{align}
\frac{S_y}{\sqrt{n_{\rm ph}}} - \ii \frac{S_z}{\sqrt{n_{\rm ph}}} =\frac{1}{\sqrt{n_{\rm ph}}} c^\dagger_y c_x \stackrel{\mbox{\scriptsize{Primakoff}}}{\simeq} c_y^\dagger \quad \quad ; \quad \quad \frac{S_y}{\sqrt{n_{\rm ph}}} + \ii\frac{S_z}{\sqrt{n_{\rm ph}}} =\frac{1}{\sqrt{n_{\rm ph}}}  c^\dagger_x c_y\stackrel{\mbox{\scriptsize{Primakoff}}}{\simeq} c_y
\end{align}
This shows that the creation operator $ c^\dagger $ in (\ref{eq014a})-(\ref{eq014b}), identified with $ c_y^\dagger $ in Primakoff's approximation, transfers a photon from the highly populated coherent state cavity mode {polarized along} $ {x} $ into the {initially empty cavity mode} {polarized along} $ {y} $. In Primakoff's approximation, the {atom-field Faraday coupling Hamiltonian} (\ref{eq002}) is written: 
\begin{align}
H = \hbar \Omega P_b P_c \quad \quad \mbox{with}  \quad \quad \Omega = \chi \sqrt{nn_{\rm ph}} \label{eq016} \,.
\end{align}
As $ \chi $ does not depend on the field strength in the cavity, $ \Omega^2$ is proportional to its intensity. 
{Finally, let's write in terms of Primakoff variables the parameter $\xi$ of reference \cite{Wineland} quantifying the level of spin squeezing usable in a magnetometer (the metrological gain is the higher as $\xi$ is the lower, see our footnote \ref{note0}), $z$ being the maximal squeezing transverse direction to the average spin {and the collective nuclear spin being of quantum number $I=N_{\rm cell}/2$:}
\begin{equation}
\label{eq017}
\boxed{\xi^2 \equiv \frac{2I\,\mbox{Var}_{\rm signal}(I_z)}{\langle I_x\rangle^2} = \frac{4\,\mbox{Var}_{\rm signal}(P_a)}{\eta}}
\end{equation}
The nuclear polarization $\eta$ being fixed, and a non-demolition quantum measurement being carried out continuously on the nuclear spin, one must try to minimize the variance of $P_a$ conditioned on the measurement signal to be defined, by making it decrease as much as possible below its initial value $1/4$.}

\subsection{{Quantum master equation} for metastability exchange} 
\label{subsec:MEC} 
Let us consider in this subsection the evolution of the system due to {metastability exchange} only ($ \chi = 0 $). In a {quantum treatment}, the classical equations (\ref{eq:redI})-(\ref{eq005}) become stochastic equations including quantum fluctuations. In Primakoff's approximation, this gives for the quadratures $ X $ in the metastable and fundamental state: 
\begin{equation}
\dd  X_a = -\gamma_f X_a  \dd t  + \sqrt{\gamma_m \gamma_f} X_b  \dd t + \dd X_a^{\rm stoch}
\quad ; \quad 
\dd X_b = -\gamma_m X_b  \dd t + \sqrt{\gamma_m \gamma_f} X_a  \dd t + \dd X_b^{\rm stoch} 
\label{eq018}
\end{equation} 
where we used the third equality of {equation} (\ref{eq006}). Langevin noises $ \dd X_i^{\rm stoch} $, with $ i \in \{a, b \} $, have zero mean, are independent random variables at different times, and have variances and equal-time covariances calculated in {reference} \cite{DantanReinaudi2005}: 
\begin{equation}
\langle  \dd X_i^{\rm stoch} \dd X_j^{\rm stoch} \rangle = D_{ij} \dd t \quad \mbox{with} \quad  
{\underline{\underline{D}}}=\frac{1}{2} \begin{pmatrix} \gamma_f & -\sqrt{\gamma_m \gamma_f} \\  -\sqrt{\gamma_m \gamma_f} & \gamma_m \end{pmatrix}
\label{eq019}
\end{equation}
We have equations of the same form as (\ref{eq018}) for the quadratures $ P_i $, with other Langevin noises $ \dd P_i^{\rm stoch} $, with the same covariance matrix as {equation} (\ref{eq019}) between them but with a covariance matrix with the noises $ \dd X_i^{\rm stoch} $ given by 
\begin{equation}
\langle  \dd X_i^{\rm stoch} \dd P_j^{\rm stoch} \rangle = {\cal D}_{ij} \dd t \quad \mbox{with} \quad  
{\underline{\underline{\cal D}}}= \ii {\underline{\underline{D}}}
\end{equation}
For calculating the mean values and variances of atomic observables, this stochastic formulation is equivalent to a {quantum master equation} {for} the atomic density operator $ \rho_{\rm at} $ of the two bosonic modes $ a $ and $ b $: 
\begin{equation}
\frac{\dd \rho_{\rm at}}{\dd t} =  C\rho_{\rm at} C^\dagger -\frac{1}{2}\{ C^\dagger C,\rho_{\rm at} \} \quad \mbox{with} \quad  C=\sqrt{2\gamma_f}a-\sqrt{2\gamma_m}b
\label{eq021}
\end{equation}
Indeed, the Langevin stochastic representation of the {quantum master equation} (\ref{eq021}) for any operator $ A $ is written 
\begin{equation}
\dd A = \frac{\dd t}{2} \left\{ C^\dagger [A,C ] - [A,C^\dagger] C \right\} + \dd A^{\rm stoch} \quad \mbox{where} \quad  
 \dd A^{\rm stoch}  = [C^\dagger,A] \dd B + \dd B^\dagger [A,C] 
\end{equation}
and $ \dd B $ is a Markovian stochastic operator with zero mean, with an equal-time covariance matrix 
\begin{equation}
\langle \dd B \, \dd B^\dagger \rangle = \dd t \quad;\quad \langle \dd B \, \dd B \rangle =  \langle \dd B^\dagger \dd B^\dagger \rangle= \langle \dd B^\dagger \dd B \rangle =0
\end{equation}
To be complete, let us sketch another reasoning, which avoids quantum Langevin noises. It suffices to admit that the equations of evolution on the means $ \langle X_i \rangle $ and $ \langle P_i \rangle $ taken from (\ref{eq:redI})-(\ref{eq005}) derive from a {quantum master equation} of the {Lindblad form} (\ref{eq030Arbitrary}). Since these equations are linear, the jump operators $ C_m $ surrounding $ \rho_{\rm at} $ in the {quantum master equation} are linear combinations of $ a $ and $ b $, and {we recover} (\ref{eq021}). 

\subsection{Three-mode {quantum master equation}} 
The complete evolution, including the {atom-field coupling of} Hermitian Hamiltonian (\ref{eq016}), metastability exchange and cavity losses, is described by the {quantum master equation} \footnote{We neglect here the internal evolution of the atomic modes (spin precession) by supposing that the Zeeman sublevels are {degenerate} in the ground state and in the metastable {level}  $ F = 1/2 $, {that is that} the external magnetic field is zero, $ \vec{B} = \vec{0} $.
{This simplifying assumption calls for the following comments. (i) In the experiment, we plan to impose a
guiding field along $x$ of the order of $\mu T$ to avoid a “wild” precession of the mean spin around a residual magnetic field of unknown direction \cite{letter}. From a theoretical point of view, we then place ourselves in the frame rotating around $x$ at the corresponding nuclear spin Larmor frequency $\omega_L^{(x)}$ to eliminate this guiding field [in principle it would be necessary to compensate for the difference between the metastable and ground-state Larmor frequencies, for example by means of a fictitious magnetic field created by a light shift, but this precaution seems superfluous because the metastable Larmor frequency remains small compared to the effective metastability exchange rate $\gamma_m$]; in principle, the cavity must also be rotated at the angular speed $\omega_L^{(x)}$ so that it is stationary in the rotating frame; if the cavity remains fixed in the laboratory frame, we can switch on the Faraday coupling and perform a measurement of the field leaving the cavity no longer continuously but stroboscopically (each time the spin component to be squeezed merges with optical axis)  \cite{stroboscopic}. (ii) If there exists during the squeezing phase a small parasitic static magnetic field (in addition to the guiding field) in the $yz$ plane, at an angle $\phi$ with $y$, this field rotates at the angular frequency $-\omega_{\rm L}^{(x)}$ in the rotating frame; in the stochastic equation of the one-mode model (\ref{eq078}) for homodyne detection, this adds a contribution of Hermitian Hamiltonian 
$H_\alpha = \sqrt{N} \hbar\omega_L^{(y)} (\Omega_\beta/\Omega) [\cos(\phi-\omega_L^{(x)}t) X_\alpha+\sin(\phi-\omega_L^{(x)}t) P_\alpha]$ 
where $\omega_L^{(y)}\ll\omega_L^{(x)}$ is the Larmor angular frequency of the nuclear spin in the parasitic field and $\Omega_\beta/\Omega$ is given by equation  (\ref{eq036}).  The $P_\alpha$ term is absorbed in
a phase change of ansatz (\ref{eq080}). The $X_\alpha$ term does not change the parameter $u$ of ansatz (\ref{eq080}), and hence the variance of $P_\alpha$ in a realization, but adds an oscillating deterministic part to $\bar{P}_\alpha$, namely
$(\sqrt{N}\omega_L^{(y)}/2\omega_L^{(x)})(\Omega_\beta/\Omega)[\sin(\phi-\omega_L^{(x)}t)-\sin\phi]$
and, according to (\ref{eq102}), a deterministic part
$\sqrt{\Gamma_{\rm ex}} (\Omega_\beta/\Omega) (\sqrt{N} \omega_L^{(y)}/2\omega_L^{(x)})\{[\cos(\phi-\omega_L^{(x)}t)-\cos\phi]/(\omega_L^{(x)}t)-\sin\phi\}$
to the integrated signal $\sigma(t)$. The $\sin\phi$ term in the signal can be canceled by choosing the initial time ``$t=0$'' of the squeezing procedure; the other contributions cancel out for a duration $t$ of the whole experiment multiple of  $2\pi/\omega_L^{(x)}$. (iii) If we want to use the spin squeezed state to measure a magnetic field, we turn off the discharge and the guiding field and we arrange that the field to be measured $\vec{B}_{\rm mes}$ is oriented along $y$. The collective nuclear spin then precesses in the $xz$ plane by an angle which must be measured to access $\vec{B}_{\rm mes}$, and the initial direction of squeezing $z$ is precisely that which is necessary for reducing the angular uncertainty on the spin. Another strategy consists in starting from an ordinary polarized state, unsqueezed, of the nuclear spin and in carrying out, discharge on but guiding field off, the continuous measurement of the light leaking out of the cavity in the presence of $B_{\rm mes} \vec{u}_y$; in the one-mode model with homodyne detection of section \ref{sec:3modesTo1Homo}, we then recover the magnetometry proposals of references \cite{magnetth1,magnetth2}.}
}
\begin{equation}
\frac{\dd \rho}{\dd t} = \frac{1}{\ii\hbar} \left[H ,\rho \right] + \kappa \left( c\rho c^\dagger -\frac{1}{2}\{ c^\dagger c,\rho\}\right) + C\rho C^\dagger -\frac{1}{2}\{ C^\dagger C,\rho \}
\label{eq030}
\end{equation}
where $ C $ is the jump operator for metastability exchange (\ref{eq021}), $ \kappa $ is the {cavity loss rate}, $ \gamma_m $ and $ \gamma_f $ are the {effective} metastability exchange {rates} {in the ground state and in the metastable state}. 

Initially, the three modes are in {vacuum state} corresponding to a polarized state for the three spins. For this initial state, the first moments of the quadratures remain zero, and one can obtain a closed system of equations {for} the second moments. We find that the quadratures $ P $ maintain constant variances and zero covariances in the three modes, 
\begin{equation}
\langle P_a^2 \rangle(t) =  \langle P_b^2 \rangle(t) =  \langle P_c^2 \rangle(t) =  \frac{1}{4}\quad ; \quad \langle P_a P_b\rangle(t)=\langle P_a P_c\rangle(t)=\langle P_b P_c\rangle(t)=0
\label{eq032} 
\end{equation}
that the variance $ \langle X_c^2 \rangle $ remains bounded and the covariances $ \langle X_a X_c \rangle $ and $ \langle X_b X_c \rangle $ remain zero, while the variances and covariance of the quadratures $ X_a $ and $ X_b $, and therefore the number of excitations in the atomic modes, \footnote{For the initial state considered, we have at all times $ \langle X_a \rangle = 0 $ and $ \langle X_a^2 \rangle-\frac{1}{4} = \langle a^\dagger a \rangle $, where $ \langle a^\dagger a \rangle $ is the average number of excitations in the nuclear spin mode, so that Var $ X_a = \langle a^\dagger a \rangle+\frac{1}{4} $; {one has indeed $a^\dagger a+1/2=X_a^2+P_a^2$}. The same relations hold for the other two modes.} diverge linearly in time, at least as long as the Primakoff approximation is applicable. We give here explicitly only long-time behaviors: 
\be
\begin{array}{ll}
\displaystyle\langle X_a^2 \rangle (t)\underset{t \to +\infty}{=} \frac{\gamma_m \gamma_f}{(\gamma_m+\gamma_f)^2}\frac{\Omega^2 t}{4\kappa} +O(1) & \quad\displaystyle \langle X_b^2 \rangle (t) \underset{t \to +\infty}{=}  \frac{\gamma_f^2}{(\gamma_m+\gamma_f)^2}\frac{\Omega^2 t}{4\kappa} +O(1)\\
&\\
\displaystyle \langle X_a X_b\rangle(t) \underset{t \to +\infty}{=} \frac{\gamma_m^{1/2}\gamma_f^{3/2}}{(\gamma_m+\gamma_f)^2}\frac{\Omega^2 t}{4\kappa}+O(1)  & \quad\displaystyle \langle X_c^2\rangle (t) -  \frac{1}{4} \underset{t \to +\infty}{\to} \left(\frac{\Omega}{2\kappa}\right)^2 \left( 1 - \frac{2\gamma_m}{\kappa + 2(\gamma_m+\gamma_f)} \right)
\end{array}
\label{eq031}
\ee

\subsection{One-mode model} 
\label{sec:ana_omm} 

In this subsection, we derive a one-mode {quantum master equation} describing the slow evolution of the nuclear spin within the limit 
\begin{equation}
\Gamma_{\rm ex} \ll \gamma_f < \gamma_m \quad\mbox{and}\quad \Gamma_{\rm ex} \ll \kappa
\label{eq033}
\end{equation}
{where $\Gamma_{\rm ex}$ is the rate at which excitations are created under the effect of Faraday coupling in the hybridized nuclear bosonic mode $\alpha$ defined below}
(it suffices to know here that $ \Gamma_{\rm ex} \propto \Omega^2 $ so that (\ref{eq033}) is a weak Faraday coupling limit $ \Omega \to 0 $). To this end, it is {convenient} to introduce the bosonic annihilation operators into a cleverly rotated basis, by means of the following linear combinations of the operators $ a $ and $ b $: 
\begin{equation}
\alpha=\sqrt{\frac{\gamma_m}{\gamma_m+\gamma_f}}a+\sqrt{\frac{\gamma_f}{\gamma_m+\gamma_f}}b 
\quad \quad ; \quad \quad
\beta=\sqrt{\frac{\gamma_m}{\gamma_m+\gamma_f}}b-\sqrt{\frac{\gamma_f}{\gamma_m+\gamma_f}}a
\label{eq034}
\end{equation}
$ \alpha $ and $ \beta $ indeed correspond to the eigenmodes of the metastability exchange part of the three-mode {quantum master equation} (\ref{eq030}) (in practice, we have $ \gamma_m \gg \gamma_f $, see {equation} (\ref{eq006}), so that the mode $ \beta $ corresponds to the {metastable spin} slightly hybridized with the spin of the ground state, and $ \alpha $ to the nuclear spin slightly hybridized with the {metastable spin}). While the $ \alpha $ mode undergoes a time divergence of its average number of excitations {(hence the possibility of defining a rate $\Gamma_{\rm ex}$)}, the $ \beta $ mode is strongly damped and tends towards a stationary value (see the results (\ref{eq032}) and (\ref{eq031}), which show that {$\langle P_\beta^2\rangle=1/4$ and} $ \langle X_\beta^2 \rangle = O (1) $ where $ X_\beta = (\beta+\beta^\dagger)/2 $), which will allow to eliminate it adiabatically, just like the cavity field. In this new basis, the {three-mode master equation} (\ref{eq030}) takes the form 
\begin{equation}
\boxed{
\frac{\dd \rho}{\dd t} = \frac{1}{\ii\hbar} \left[ H,\rho \right] + \kappa \left( c\rho c^\dagger -\frac{1}{2}\{ c^\dagger c,\rho\}\right) 
+ \gamma_\beta\left( \beta \rho \beta^\dagger -\frac{1}{2}\{ \beta^\dagger \beta,\rho\}\right)}
\label{eq030t}
\end{equation}
where $ \gamma_\beta \equiv 2 (\gamma_m+\gamma_f) $ and, noting $ P_\alpha = (\alpha-\alpha^\dagger)/2 \ii $ and $ P_\beta = (\beta-\beta^\dagger)/2 \ii $ the $ P $ quadratures of the new modes, 
\begin{equation}
H=\hbar(\Omega_\alpha P_\alpha + \Omega_\beta P_\beta) P_c \quad \mbox{with}\quad \Omega_\alpha\equiv \Omega \sqrt{\frac{\gamma_f}{\gamma_m+\gamma_f}} \quad\mbox{and}\quad \Omega_\beta{\equiv}\Omega  \sqrt{\frac{\gamma_m}{\gamma_m+\gamma_f}}
\label{eq036}
\end{equation}
{Reference \cite{Sorensen} explains in all generality how to perform an adiabatic elimination at the level of the master equation.}
Here we prefer to carry it out, as in {reference} \cite{CastinMolmer_adel}, in the weak Faraday coupling limit $ \Omega \to 0 $ in the Monte Carlo wave function formalism \cite{JOSAB, CastinDalibard}, where the density operator solution of the {quantum master equation} (\ref{eq030t}) is obtained by averaging {pure state}s over independent stochastic realizations, each realization corresponding to the deterministic evolution of {an} {unnormalized} state vector $ | \psi (t) \rangle $ under the action of the effective non-Hermitian Hamiltonian 
\begin{equation}
H_{\rm eff}=H-\frac{\ii\hbar}{2} \left( \kappa c^\dagger c + \gamma_\beta \beta^\dagger \beta \right)
\label{eq010_MC}
\end{equation}
interrupted randomly by quantum jumps (discontinuous evolutions $ | \psi \rangle \to C | \psi \rangle $) of jump operators 
\begin{equation}
C_c=\sqrt{\kappa}c \quad \mbox{and} \quad C_\beta=\sqrt{\gamma_\beta}\beta \,.
\label{eq038}
\end{equation}
In the absence of the coherent coupling $ \Omega $ in (\ref{eq036}) the hybridized metastable mode and the cavity mode remain in the initial empty state. To first order in $ \Omega $, this state is coupled to states with an excitation in the cavity (by the action of $ P_c $) and with zero or one excitation in the mode of the hybridized metastable (by the action of $ P_\alpha $ or $ P_\beta $). We can then truncate the Monte Carlo state vector $ | \psi \rangle $ in the {Fock basis} $ \{| n_\alpha \rangle_{\rm background} | n_\beta \rangle_{\scriptsize \mbox{meta}} | n_c \rangle_{\rm cav} \} $ as follows 
\begin{equation}
|\psi\rangle = |\psi_\alpha^{00} \rangle |0\rangle |0\rangle +
|\psi_\alpha^{01} \rangle |0\rangle |1\rangle +  |\psi_\alpha^{11} \rangle |1\rangle |1\rangle
\label{eq039}
\end{equation}
committing an error of norm $ O (\Omega^2) $. 
Under the effect of the effective Hamiltonian (\ref{eq010_MC}), the fast components $ | \psi_\alpha^{01} \rangle $ and $ | \psi_\alpha^{11} \rangle $ exponentially join an adiabatic {following regime} of the slow component $ | \psi_\alpha^{00} \rangle $ with rates $ \kappa/2$ or $ (\kappa+\gamma_\beta)/2$. Hence their adiabatic elimination within the limit (\ref{eq033}) \footnote{In adiabatic following, the {occupation} probabilities of the excited components are 
$\langle\psi_\alpha^{11}|\psi_\alpha^{11}\rangle_{\rm adiab}/\langle\psi|\psi\rangle = [\Omega_\beta^2/4(\kappa+\gamma_\beta)^2] \langle\psi_\alpha^{00}|\psi_\alpha^{00}\rangle/\langle\psi|\psi\rangle $
and 
$ \langle \psi_\alpha^{01}|\psi_\alpha^{01}\rangle_{\rm adiab}/\langle\psi|\psi\rangle =(\Gamma_{\rm ex}/\kappa) \langle \psi_\alpha^{00}|P_\alpha^2|\psi_\alpha^{00}\rangle/\langle\psi|\psi\rangle$
where we used (\ref{eq054}). Within the limit (\ref{eq033}), we can easily verify that they are $ \ll 1 $, so that almost all the population is in the component $ | \psi_\alpha^{00} \rangle | 0 \rangle | 0 \rangle $ as it should be, which will allow us to replace $ \langle \psi | \psi \rangle $ by $ \langle \psi_\alpha^{00} | \psi_\alpha^{00} \rangle $. We also verify that another condition for the validity of adiabatic elimination, namely the slowness of the evolution of the hybridized nuclear spin $ \alpha $ with respect to the fast variables, which reads here $ \Gamma_{\rm ex}, \Gamma_0 \ll \kappa, \kappa+\gamma_\beta $, is satisfied. However, these considerations do not allow us to show that the condition $ \Gamma_{\rm ex} \ll \gamma_f $ is necessary (unless $ \kappa \ll \gamma_\beta $). To see it in general terms, we push to the order $ \Omega^4 $ the computation of the effective Hamiltonian 
$H_{\rm eff}^{00}=P H_{\rm eff}P + P H Q (zQ-Q H_{\rm eff} Q)^{-1} QHP$
in the subspace $ n_\beta = n_c = 0 $ onto which $ P $ projects (here $ Q = \mathbb{1}-P $ and $ z = O (\Omega^2) $). Qualitatively, at this order, by action of $ H_\alpha $ then of $ H_\beta $ on $ | \psi_\alpha^{00} \rangle | 0 \rangle | 0 \rangle $ (with the obvious notation $ H = H_\alpha+H_\beta $), we virtually create an excitation $ \beta $ alone, relaxing at the rate $ \gamma_\beta/2 $, hence the additional adiabaticity condition $ \Gamma_0 \ll \gamma_\beta $; joined to $ \Gamma_0 \ll \kappa $ and $ \gamma_f <\gamma_m $, it implies $ \Gamma_{\rm ex} \ll \gamma_f $ since 
$\Gamma_{\rm ex}/\gamma_f = (\Gamma_0/\kappa+\Gamma_0/\gamma_\beta)(4\gamma_\beta/\gamma_m) < 16 (\Gamma_0/\kappa+\Gamma_0/\gamma_\beta)$.
Quantitatively, we find a correction to the coefficient of $ P_\alpha^2 $ in $ H_{\rm eff}^{\rm 00} $ of type $ H_\alpha G_0 H_\beta G_0 H_\beta G_0 H_\alpha $ ($ G_0 $ is the resolvent of $ H_{\rm eff} $ for $ \Omega = 0 $) of the form $ \hbar \Gamma_{\rm ex} \Omega_\beta^2/\gamma_\beta \kappa $, which must be negligible, which imposes $ \Omega_\beta^2/\gamma_\beta \kappa \ll 1 $, i.e. $ \Gamma_{\rm ex} \ll \gamma_f $ taking into account $ \gamma_f <\gamma_m $. The corrections to the scalar term are negligible as soon as $ \Gamma_0 \ll \gamma_\beta, \kappa $, and the new term in $ P_\alpha^4 $ which appears is {negligible compared to} $ \hbar \Gamma_{\rm ex} P_\alpha^2 $ for $ P_\alpha = O (1) $ if $ \Gamma_{\rm ex} \ll \kappa $.} 
\begin{equation}
|\psi_\alpha^{11}\rangle_{\rm adiab} \simeq  \frac{\ii\Omega_\beta}{2(\kappa+\gamma_\beta)} |\psi_\alpha^{00}\rangle \quad \mbox{and} \quad
|\psi_\alpha^{01}\rangle_{\rm adiab} \simeq \frac{\Omega_\alpha}{\kappa} \,P_\alpha |\psi_\alpha^{00}\rangle \label{eq043}
\end{equation}
We put the expressions of 
$|\psi_\alpha^{11}\rangle_{\rm adiab}$, $|\psi_\alpha^{01}\rangle_{\rm adiab}$ 
in the Hamiltonian evolution equation of $ | \psi_\alpha^{00} \rangle $ to obtain 
\begin{equation}
\ii\hbar \frac{\dd }{\dd t} |\psi_\alpha^{00}\rangle = -\frac{\ii\hbar}{2} \left( \Gamma_{\rm ex} P_\alpha^2  + \Gamma_0 \right) |\psi_\alpha^{00}\rangle \equiv H_{\rm eff}^{00} \: |\psi_\alpha^{00}\rangle 
\label{eq045}
\end{equation} 
where we have introduced the rates 
\begin{equation}
\boxed{
\Gamma_{\rm ex}=\frac{\Omega_\alpha^2}{\kappa} \quad\mbox{and}\quad \Gamma_0=\frac{\Omega_\beta^2}{4(\kappa+\gamma_\beta)}}
\label{eq054}
\end{equation}
By studying the effect of the cavity jump operator $ C_c $ and the metastability exchange jump operator $ C_\beta $ on the state vector (\ref{eq039}), we can interpret the effective Hamiltonian of {equation} (\ref{eq045}). (i) Let us first consider the effect of a cavity jump, which occurs at time $ t $ with a rate 
$\kappa (\langle \psi_\alpha^{11}|\psi_\alpha^{11}\rangle+\langle \psi_\alpha^{01}|\psi_\alpha^{01}\rangle)_{\rm adiab}/\langle\psi_\alpha^{00}|\psi_\alpha^{00}\rangle$.
Just after the jump, the state vector, initially in {the} adiabatic {following regime}, becomes 
\begin{align}
|\psi(t^+)\rangle= C_c|\psi(t^-)\rangle_{\rm adiab}  \propto  |\psi_\alpha^{01}(t^-)\rangle_{\rm adiab} |0\rangle|0\rangle + |\psi_\alpha^{11}(t^-)\rangle_{\rm adiab} |1\rangle|0\rangle
\end{align}
 It is the superposition of an unstable component $ | 1 \rangle | 0 \rangle $ and of a stable component $ | 0 \rangle | 0 \rangle $. 
With a probability 
$\langle\psi_\alpha^{11}|\psi_\alpha^{11}\rangle_{\rm adiab}$ $/(\langle\psi_\alpha^{01}|\psi_\alpha^{01}\rangle+\langle\psi_\alpha^{11}|\psi_\alpha^{11}\rangle)_{\rm adiab}$
the cavity jump is then followed by a metastability exchange jump before the system {state vector} has time to reach its adiabatic value. In this case, we have a \og double jump \fg, which ultimately does not affect the component $ | \psi_\alpha^{00} (t^-) \rangle $ since 
\begin{equation}
C_\beta C_c |\psi(t^-)\rangle_{\rm adiab} \propto  |\psi_\alpha^{00}(t^-)\rangle |0\rangle |0\rangle
\end{equation}
This process contributes to the scalar term (proportional to the identity) in the effective Hamiltonian of {equation} (\ref{eq045}). With the complementary probability 
$\langle\psi_\alpha^{01}|\psi_\alpha^{01}\rangle_{\rm adiab}/(\langle\psi_\alpha^{01}|\psi_\alpha^{01}\rangle+\langle\psi_\alpha^{11}|\psi_\alpha^{11}\rangle)_{\rm adiab}$
the {state vector} returns to its adiabatic value before other jumps occur, and is slaved to ${|\psi_\alpha^{(01)}(t^-)\rangle_{\rm adiab}\propto} P_\alpha | \psi_\alpha^{00} (t^-) \rangle $, that is, the slow component $ | \psi_\alpha^{00} (t^-) \rangle $ has effectively undergone a single quantum jump with a jump operator proportional to $ P_\alpha $. This process corresponds to the first term, proportional to $ P_\alpha^2 $, in the effective Hamiltonian of equation (\ref{eq045}). 
(ii) Suppose next that the jump at time $ t $ is a metastability exchange jump, which occurs with a rate 
$\gamma_\beta \langle \psi_\alpha^{11}|\psi_\alpha^{11}\rangle_{\rm adiab}/\langle\psi_\alpha^{00}|\psi_\alpha^{00}\rangle$.
We verify in this case that the state vector after the jump, $ C_\beta | \psi (t^-) \rangle $, is entirely unstable and almost immediately undergoes a second jump, a cavity jump. The total effect corresponds here again to a double jump and to the action of a scalar operator on the slow component. We derive from this discussion the following single jump and double jump 
rates : 
\bea
\Gamma_s &=& \frac{\kappa (\langle \psi_\alpha^{11}|\psi_\alpha^{11}\rangle+\langle \psi_\alpha^{01}|\psi_\alpha^{01}\rangle)_{\rm adiab}}{\langle\psi_\alpha^{00}|\psi_\alpha^{00}\rangle} \frac{\langle\psi_\alpha^{01}|\psi_\alpha^{01}\rangle_{\rm adiab}}{(\langle\psi_\alpha^{01}|\psi_\alpha^{01}\rangle+\langle\psi_\alpha^{11}|\psi_\alpha^{11}\rangle)_{\rm adiab}}=\Gamma_{\rm ex} \frac{\langle\psi_\alpha^{00}|P_\alpha^2|\psi_\alpha^{00}\rangle}{\langle\psi_\alpha^{00}|\psi_\alpha^{00}\rangle}\equiv\Gamma_{\rm ex} \langle P_\alpha^2\rangle \\
\Gamma_d &=& \frac{\kappa (\langle \psi_\alpha^{11}|\psi_\alpha^{11}\rangle+\langle \psi_\alpha^{01}|\psi_\alpha^{01}\rangle)_{\rm adiab}}{\langle\psi_\alpha^{00}|\psi_\alpha^{00}\rangle} \frac{\langle\psi_\alpha^{11}|\psi_\alpha^{11}\rangle_{\rm adiab}}{(\langle\psi_\alpha^{01}|\psi_\alpha^{01}\rangle+\langle\psi_\alpha^{11}|\psi_\alpha^{11}\rangle)_{\rm adiab}} + \frac{\gamma_\beta \langle \psi_\alpha^{11}|\psi_\alpha^{11}\rangle_{\rm adiab}}{\langle\psi_\alpha^{00}|\psi_\alpha^{00}\rangle} = \Gamma_0
\eea
We finally obtain the one-mode {quantum master equation} describing the slow evolution of the density operator $ \rho_\alpha $ of the bosonic mode $ \alpha $ (hybridized but almost purely nuclear spin): 
\begin{equation}
\boxed{
\frac{\dd \rho_{\alpha}}{\dd t} =  C_s\rho_{\alpha} C_s^\dagger -\frac{1}{2}\{ C_s^\dagger C_s,\rho_{\alpha}\} +  C_d\rho_{\alpha} C_d^\dagger -\frac{1}{2}\{ C_d^\dagger C_d,\rho_{\alpha}\}}
\label{eq030oneMode}
\end{equation}
in terms of two quantum jumps, the single jump (cavity only) $ C_s $ and the double jump (of cavity and metastability exchange in that order or in the other) $ C_d $: 
\begin{equation}
C_s= \sqrt{\Gamma_{\rm ex}} P_\alpha \quad ; \quad C_d=\sqrt{\Gamma_0} \mathbb{1}
\end{equation}
By solving equation (\ref{eq030oneMode}) for the empty initial state of $ \alpha $, we get: 
\begin{equation}
\langle X_\alpha^2 \rangle = \frac{1}{4} (1+\Gamma_{\rm ex}t) \quad ; \quad \langle P_\alpha^2 \rangle = \frac{1}{4} 
\end{equation}
{which effectively designates $\Gamma_{\rm ex}$ as a rate of creation of excitations in the $\alpha$ mode.}
Going back to the initial atomic basis (unrotated) and by limiting the state vector (\ref{eq039}) to its first term, we recover {equation} (\ref{eq032}) and the first three results of {equation} (\ref{eq031}) of the {three-mode model},
{yet valid at any Faraday coupling $\Omega$, not necessarily infinitesimal}.
Finally, the average number of photons polarized along $ {y} $ {leaking out of the cavity} per unit of time, given in the one-mode model by $ \Gamma_0+\Gamma_{\rm ex}/4 $ as shown by equation (\ref{eq064}), agrees with the exact value $ \kappa \langle c^\dagger c \rangle_{s} $ where the mean stationary number of $y$-polarized  photons in the cavity $ \langle c^\dagger c \rangle_{s} = \langle X_c^2 \rangle_{s}-1/4 $ is the last result of (\ref{eq031}). \footnote{On the other hand, the value of $ \langle c^\dagger c \rangle_{\rm adiab} $ in the adiabatic form (\ref{eq043}) of the state vector does not represent this number. The solution of the paradox is due to the existence of the de-excitation path (ii), that of the annihilation in the first jump of the excitation $ n_\beta = 1 $ in the metastable mode immediately followed by the loss of a {cavity photon}. The true output rate of $y$-polarized photons is therefore $ \kappa \langle c^\dagger c \rangle_{\rm adiab}+\gamma_\beta \langle \beta^\dagger \beta \rangle_{\rm adiab} $.} 

\section{{Continuous} non-demolition quantum measurement of nuclear spin} 
\label{sec:QND}

The {quantum averages} calculated in {section} \ref{sec:qt} correspond to the {ensemble averages} over an infinite number of {realizations of the experiment}. In this section we study what really interests us, the evolution of the system, in one or more given realizations, conditioned on the results of a continuous measurement on the $y$-polarized light {leaking out of the cavity}. For this, we return to the formulation in terms of Monte Carlo wave functions, as in section \ref{sec:qt}, where stochastic trajectories $ | \psi (t) \rangle $ corresponding to a particular {sequence of quantum jumps} reconstruct the density operator of the system conditioned on measurement results \cite{CastinDalibard}. The precise form of the Monte Carlo jump operators, which is not unique in the stochastic reformulation of a {quantum master equation}, is then determined by the particular measurements made.

\subsection{Squeezing by photon counting} 
\label{sec:dpcdp} 
Suppose that we continuously and directly count (by photodetection) the number of $y$-polarized photons leaking out of the cavity (see figure \ref{fig1}b), as proposed in {reference} \cite{Milburn1993}. The jump operator associated with this measurement is $ \sqrt{\kappa} c $, so the three-mode {quantum master equation} (\ref{eq030}) is already in the right form to analyze the evolution of the state vector $ | \psi (t) \rangle $ conditioned on the measurement. 

 The same is true within the limit of a weak Faraday coupling, $ \Omega \to 0 $, which leads to the one-mode model {of section \ref{sec:ana_omm}}. As the jump operators $ C_d $ and $ C_s $ of the {quantum master equation} (\ref{eq030oneMode}) both correspond to the cavity loss of a $y$-polarized photon (remember, $ C_d $ results from a cavity jump immediately followed or preceded by a metastability exchange jump, and $ C_s $ from a single cavity jump), the measurement cannot distinguish between the two, and the density operator conditioned on a given number $ n $ of detected photons is obtained by averaging over realizations having this same {\it total} number $ n $ of jumps. An {unnormalized} Monte Carlo state vector having undergone such $ n $ jumps during $ t $ is written 
\begin{equation}
|\psi(t)\rangle = \eee^{-\frac{\ii}{\hbar}{H}^{00}_{\rm eff}(t-t_n)} \, C_{\epsilon_{n}} \, \eee^{-\frac{\ii}{\hbar}{H}^{00}_{\rm eff}(t_n-t_{n-1})}
\, C_{\epsilon_{n-1}} \ldots \, C_{\epsilon_{1}} \,  \, \eee^{-\frac{\ii}{\hbar}{H}_{\rm eff}^{00} t_1} |\psi(0)\rangle \label{eq057}
\end{equation}
where $ \epsilon_{k} \in \{s, d \} $ and $ t_k $ are the type and time of the $ k $ th jump, $ H_{\rm eff}^{00} $ is the effective Hamiltonian (\ref{eq045})
{and we denote the state vector of the one-mode model $|\psi\rangle$ rather than $|\psi_\alpha^{00}\rangle$ to simplify}.
The quantum average of an observable $ O $ is obtained by averaging over all possible trajectories, therefore by summing over the number and type of jumps and by integrating over their times: 
\begin{equation}
\langle O \rangle(t) = \sum_n \int_{0<t_1<t_2 \ldots <t_n<t} \dd t_1 \, \dd t_2 \ldots \dd t_n \, \sum_{(\epsilon_k)_{1 \leq k \leq n} \in \{ s,d \}^n }
\langle \psi(t) |O| \psi(t) \rangle
\label{eq058}
\end{equation}
where the squared norm of each {unnormalized} state vector $ | \psi (t) \rangle $ automatically gives its probability density \cite{LiYun}. By taking $ O = \mathbb{1} $, we deduce the probability that $ n $ jumps occurred in the time interval $ [0, t] $:
\begin{equation}
\Pi_n(t) =  \int_{0<t_1<t_2 \ldots <t_n<t} \dd t_1 \, \dd t_2 \ldots \dd t_n \, \sum_{(\epsilon_k)_{1 \leq k \leq n} \in \{ s,d \}^n }
\langle \psi(t) | \psi(t) \rangle
\label{eq059}
\end{equation}
To evaluate (\ref{eq059}), we take advantage of the fact that all the jump operators in (\ref{eq057}) and their Hermitian conjugates commute with each other and with $ H_{\rm eff}^{00} $.
\footnote{{For this reason, keeping the information on the jump times does not allow to increase the efficiency of spin squeezing by post-selection. Indeed, the density operator $\rho_\alpha(t)|_{t_1,\ldots, t_n}$ knowing that $n$ jumps occurred at times $t_1,\ldots,t_n$ leads to the same probability distribution of $P_\alpha$ as the density operator $\rho_\alpha(t)|_n$ knowing only that $n$ jumps occurred during $[0,t]$.}}
By using the identities 
\begin{equation}
 \sum_{\epsilon_n=s,d}\ldots \sum_{\epsilon_1=s,d} \left( C_{\epsilon_{n}}^\dagger C_{\epsilon_{n}} \ldots   C_{\epsilon_{1}}^\dagger C_{\epsilon_{1}} \right) =
 \left( \sum_{\epsilon_n=s,d} C_{\epsilon_{n}}^\dagger C_{\epsilon_{n}} \right) \ldots  \left( \sum_{\epsilon_n=s,d}  C_{\epsilon_{1}}^\dagger C_{\epsilon_{1}} \right) = \left( \Gamma_{\rm ex} P_\alpha^2 + \Gamma_0\mathbb{1} \right)^n
\end{equation}
and by injecting a {closure relation} in the {eigenbasis} of $ P_\alpha $ such as 
$P_\alpha |p_\alpha\rangle = p_\alpha |p_\alpha\rangle$, 
after having integrated over the times $ t_k $ as allowed by the telescopic product of the evolution operators, we obtain 
\begin{equation}
\Pi_n(t)=\frac{t^n}{n!} \int_{-\infty}^{+\infty} \, \dd p_\alpha \, \left( \Gamma_{\rm ex} p_\alpha^2 +  \Gamma_0 \right)^n 
\eee^{-\Gamma_{\rm ex}p_\alpha^2t} \eee^{-\Gamma_0t} \Pi (p_\alpha,0)=
\binom{2n}{n}\frac{(\Gamma_{\rm ex}t/8)^n\, \eee^{-\Gamma_0 t}}{(1+\Gamma_{\rm ex}t/2)^{n+1/2}}  \,
\Phi \left(-n , \frac{1}{2} -n ; {\Gamma_0} t+  \frac{2\Gamma_0}{\Gamma_{\rm ex}}\right)
\label{eq062}
\end{equation}
where $ \Pi (p_\alpha, 0) $ is the initial probability distribution of $ p_\alpha $ (a Gaussian with zero mean and variance $ 1/4 $) and $ \Phi $ is {Kummer's hypergeometric confluent function} ${}_1F_1 $. We notice that (\ref{eq062}) is in fact a Gaussian average on $ p_\alpha $ of a Poisson distribution with parameter 
$\lambda=( \Gamma_{\rm ex} p_\alpha^2 + \Gamma_0)t$.
We deduce the mean and the variance of the number of photodetections during $ t $ : 
\begin{equation}
\langle n \rangle = \left(\Gamma_0 +\frac{1}{4} \Gamma_{\rm ex}\right) t \quad ; \quad 
\mbox{Var} \, n =  \langle n \rangle + \frac{(\Gamma_{\rm ex}t)^2 }{8} \label{eq064}
\end{equation}
Still using {equation} (\ref{eq062}), we access the probability distribution of $ p_\alpha $ knowing that $ n $ photons were detected in the time interval $ [0, t] $, an even function of $ p_\alpha $: 
\begin{equation}
\Pi_t(p_\alpha|n)=\frac{1}{\Pi_n(t)} \frac{t^n}{n!}  \left( \Gamma_{\rm ex} p_\alpha^2 +\Gamma_0\right)^n 
\eee^{-\Gamma_{\rm ex}p_\alpha^2t} \eee^{-\Gamma_0 t} \Pi (p_\alpha,0)
\label{eq065}
\end{equation} 
{As expected, it is an even function of $p_\alpha$, as photodetection only gives access to the outgoing $y$-polarized field intensity and cannot distinguish between opposite $\pm p_\alpha$ values of the $P_\alpha$ quadrature of the hybridized nuclear spin along $z$. This results in the squeezing of $P_\alpha^2$ fluctuations rather than $P_\alpha$, which we characterize by}
the conditional mean and variance of $ P_\alpha^2 $ knowing that $ n $ photons were detected during $ t $, {deduced from (\ref{eq065}):} 
\begin{equation}
\langle P_\alpha^2 \rangle_n = \frac{(n+1)}{\Gamma_{\rm ex} t} \frac{\Pi_{n+1}(t)}{\Pi_n(t)} - \frac{\Gamma_0}{\Gamma_{\rm ex}} \quad ; \quad {\mbox{Var}_n (P_\alpha^2)\equiv \langle P_\alpha^4\rangle_n -\langle P_\alpha^2\rangle_n^2 = \frac{(n+1)^2}{(\Gamma_{\rm ex} t)^2} \left[ \frac{(n+2)\Pi_{n+2}(t)}{(n+1)\Pi_n(t)} -\frac{\Pi^2_{n+1}(t)}{\Pi_n^2(t)}\right]}
\label{eq066}
\end{equation}
Finally, using equation~(\ref{eq065}), we find for $ \Gamma_{\rm ex} t \to+\infty $ that the probability distribution of $ p_\alpha^2 $ conditioned on {the} number $ n $ of photodetections is {peaked} around {a value $ p_0^2 $ with a conditional variance tending to zero} \footnote{According to equation (\ref{eq064}), the {right-hand side of the first equation in} (\ref{eq068}) is asymptotically of the order of unity for a typical photodetection sequence. {This equation} in fact only makes sense for $ p_0^2 $ positive therefore $ n> \Gamma_0 t $; then, the equivalents {(\ref{eq068})} apply when the gap between the two peaks in $ \Pi_t (p_\alpha | n) $ is much {larger} than their width, which imposes 
$2 p_0^2\gg n^{1/2}/\Gamma_{\rm ex }t = (\Gamma_0+ \Gamma_{\rm ex} p_0^2)^{1/2}/\Gamma_{\rm ex} t^{1/2}$.
{To obtain them, we pose $n=\gamma t$ with $\gamma>\Gamma_0$, then we write (\ref{eq065}) in the form $\exp[-t S(p_\alpha)]\Pi(p_\alpha,0)$ and we quadratize $S(p_\alpha)$ around its minima.}
}:
{
\begin{equation}
\boxed{
p_0^2-\frac{1}{4} = \frac{n-\langle n \rangle}{\Gamma_{\rm ex} t} \quad \mbox{hence}\quad 
\langle P_\alpha^2\rangle_n \underset{\Gamma_{\rm ex}t\to +\infty}{\sim} p_0^2} \quad ; \quad
\boxed{\mbox{Var}_n (P_\alpha^2)\underset{\Gamma_{\rm ex}t {\to+\infty}}{ \sim } \frac{n}{(\Gamma_{\rm ex}t)^2} \to 0 \label{eq068}}
\end{equation}
}
{By replacing in this expression $n$ by its mean value and taking into account the value $1/8$ of the variance of $P_\alpha^2$ in the initial state, we end up with the nuclear spin squeezing rate by photon counting:}
\be
\label{eq868}
{\boxed{\Gamma_{\rm sq} = \frac{\Gamma_{\rm ex}^2}{8(\Gamma_0+\frac{1}{4}\Gamma_{\rm ex})}}}
\ee
Correspondingly, the conditional probability distribution of $ p_\alpha $ has two peaks at $ \pm p_0 $ as visible on the Wigner function in figure \ref{fig7}b, obtained by numerical simulation of the conditional evolution of the system over long times in the one-mode model (\ref{eq030oneMode}). 
\footnote{\label{note600}{The absence of fringes shows that we have prepared a statistical mixture rather than a coherent superposition of two squeezed states of the $P_\alpha$ quadrature.  We find indeed from equation (\ref{eq058}) that
$\langle p_0|\rho_n(t)|-p_0\rangle/\langle p_0|\rho_n(t)|p_0\rangle=[(\Gamma_0-\Gamma_{\rm ex} p_0^2)/(\Gamma_0+\Gamma_{\rm ex} p_0^2)]^n\simeq \exp(-2\Gamma_{\rm ex} t p_0^2)\lll 1$
on figure \ref{fig7}b, $\rho_n(t)$ being the conditional density operator. The Laplace method gives at long times the conditional Wigner distribution
$W_t(x_\alpha,p_\alpha|n)\sim\pi\Pi_t(p_\alpha|n)\exp[-2 x_\alpha^2/\Gamma_{\rm ex}t s(p_\alpha)]/\sqrt{\pi\Gamma_{\rm ex}t s(p_\alpha)/2}$ with $s(p_\alpha)=[1+(\Gamma_0+\Gamma_{\rm ex}p_0^2)/(\Gamma_0+\Gamma_{\rm ex}p_\alpha^2)]/2$.
We notice that $s(\pm p_0)=1$ and $s(p_\alpha)\simeq 1$ for $\Gamma_0\gg\Gamma_{\rm ex}$.  While $W_t(0,0|n)$ is damped exponentially in time (in the limit $\Gamma_{\rm ex}\ll\Gamma_0$, it comes
$W_t(0,0|n)\sim(\Gamma_{\rm ex}/\Gamma_0)^{1/2}p_0\exp(2 p_0^2)\exp(-4 p_0^4 \Gamma_{\rm sq}t)$),
we also have $W_t(0,0|n)=\langle 2\varepsilon\rangle_n$ where $\varepsilon=\pm 1$ is the parity of the Monte Carlo wave function $\psi(p_\alpha,t)$. In a numerical simulation, we therefore have only a slow decay $\langle\varepsilon\rangle_n\approx 1/\sqrt{\mathcal{N}_n}$ where $\mathcal{N}_n$ is the number of trajectories that have undergone $n$ jumps during $t$ ; this leads to unphysical fringes with negative values in the Wigner distribution near $x_\alpha$ axis. To minimize this effect and make it imperceptible at a not too high $p_\alpha$ resolution ($\dd p_\alpha=0.044$ on figure \ref{fig7}b), we stop the Monte Carlo simulation at a stage where there is exactly the same number $\mathcal{N}_n/2$ of even and odd wave functions. {To obtain a coherent superposition of squeezed states, one would have to perform an additional post-selection, restricting oneself to Monte Carlo realizations of wavefunction $\psi(p_\alpha,t)$ of fixed parity $\varepsilon$ (having undergone an even number of single jumps if $\varepsilon=1$, an odd number otherwise, the jump operator $C_s$ changing the parity).  In the corresponding conditional density operator, one then has
$\langle p_0| \rho_{n,\varepsilon}(t)|-p_0\rangle/\langle p_0|\rho_{n,\varepsilon}(t)|p_0\rangle = \overline{\psi(-p_0,t)\psi(p_0,t)}^{n,\varepsilon}/\overline{\psi(p_0,t)\psi(p_0,t)}^{n,\varepsilon}=\varepsilon$
without the two-peak structure of the $P_\alpha$ distribution being affected at long times because
$\Pi_t(p_\alpha|n,\varepsilon)/\Pi_t(p_\alpha|n) \propto 1+\varepsilon[(\Gamma_0-\Gamma_{\rm ex} p_\alpha^2)/(\Gamma_0+\Gamma_{\rm ex}p_\alpha^2)]^n \to 1$
when $n\to +\infty$ at fixed nonzero $p_\alpha$. The Wigner distribution $W_t(x_\alpha,p_\alpha |n,\varepsilon)$ now exhibits positive and negative fringes of maximum amplitude $2$ on the $p_\alpha=0$ axis. This filtering technique also overcomes the decoherence mechanisms of section \ref{sec:edld}, because the Monte Carlo wavefunctions remain of well-defined parity after the action of the corresponding jump operator 
$\gamma_\alpha^{1/2}\alpha$. This idea of controlling decoherence through parity measurements is well known in cavity quantum electrodynamics  \cite{Raimond}.}}}
{This clearly shows that,} in a single realization of the experiment, the continuous {photodetection of the} $y$-polarized photons leaking out of the cavity makes more and more certain the value of $ P_\alpha^2 $, and therefore {to a large extent} of $ I_z^2 $, the square of the component {along} $ {z} $ of the collective nuclear spin, {as we clearly see by relating, in the} limit $ \Omega \to 0 $, the conditional moments of $ P_a^2 $, that is of $ I_z^2 $ to those of $ P_\alpha^2 $ : 
\begin{equation}
\label{eq968}
\langle P_a^2\rangle_n =  \frac{\gamma_m}{\gamma_f+\gamma_m} \langle P_\alpha^2\rangle_n + \frac{\gamma_f/4}{\gamma_f+\gamma_m}\quad ;\quad \mbox{Var}_n( P_a^2)= \frac{\gamma_m^2}{(\gamma_f+\gamma_m)^2}\mbox{Var}_n(P_\alpha^2) + 
\frac{\gamma_f \gamma_m}{(\gamma_f+\gamma_m)^2} \langle P_\alpha^2\rangle_n + \frac{\gamma_f^2/8}{(\gamma_f+\gamma_m)^2}
\end{equation}
{Since the squeezing is on $P_a^2$ rather than $P_a$, the conditional angular distribution of the collective nuclear spin is bimodal (it has, like that of $P_\alpha$, two well-separated peaks provided $\gamma_f/\gamma_m\ll (2 p_0)^2$); these structures, which are narrower than the standard quantum limit, still allow for a more accurate angular pointing than in the unsqueezed state. {We therefore redefine metrological gain (\ref{eq017}) by replacing in the right-hand side of this equation the conditional variance of $P_a$ by the square of the halfwidth $\delta P_a$ of the peaks centered at $\pm P_{a, 0}$ of the conditional probability distribution of $P_a$, then equating the center $P_{a,0}^2$ and the width $\delta(P_a^2)$ of the distribution of $P_a^2$ with the conditional mean and standard deviation of $P_a^2$:
\be
\label{eq067}
\xi^2 = \frac{4(\delta P_a)^2}{\eta} = \frac{(2 P_{a,0} \delta P_a)^2}{\eta P_{a,0}^2} = \frac{[\delta(P_a^2)]^2}{\eta P_{a,0}^2}=\frac{\mbox{Var}_n(P_a^2)}{\eta \langle P_a^2\rangle_n}
\ee
Note that this expression cannot be deduced from the method of moments explained in section II.B.6 of reference \cite{PezzeRMP} taking $P_a^2$ as estimator, because the effect of the unitary transformation $\exp(2 \ii\theta X_a)$ (in practice, a precession of the nuclear spin around a magnetic field along $y$) is not to shift the peak in the distribution of $P_a^2$ but to split it into two peaks centered at $(P_{a,0}\pm\theta)^2$.}}

\begin{figure} [t] 
\centerline{\includegraphics [width = 0.4 \textwidth, height = 0.23 \textwidth]{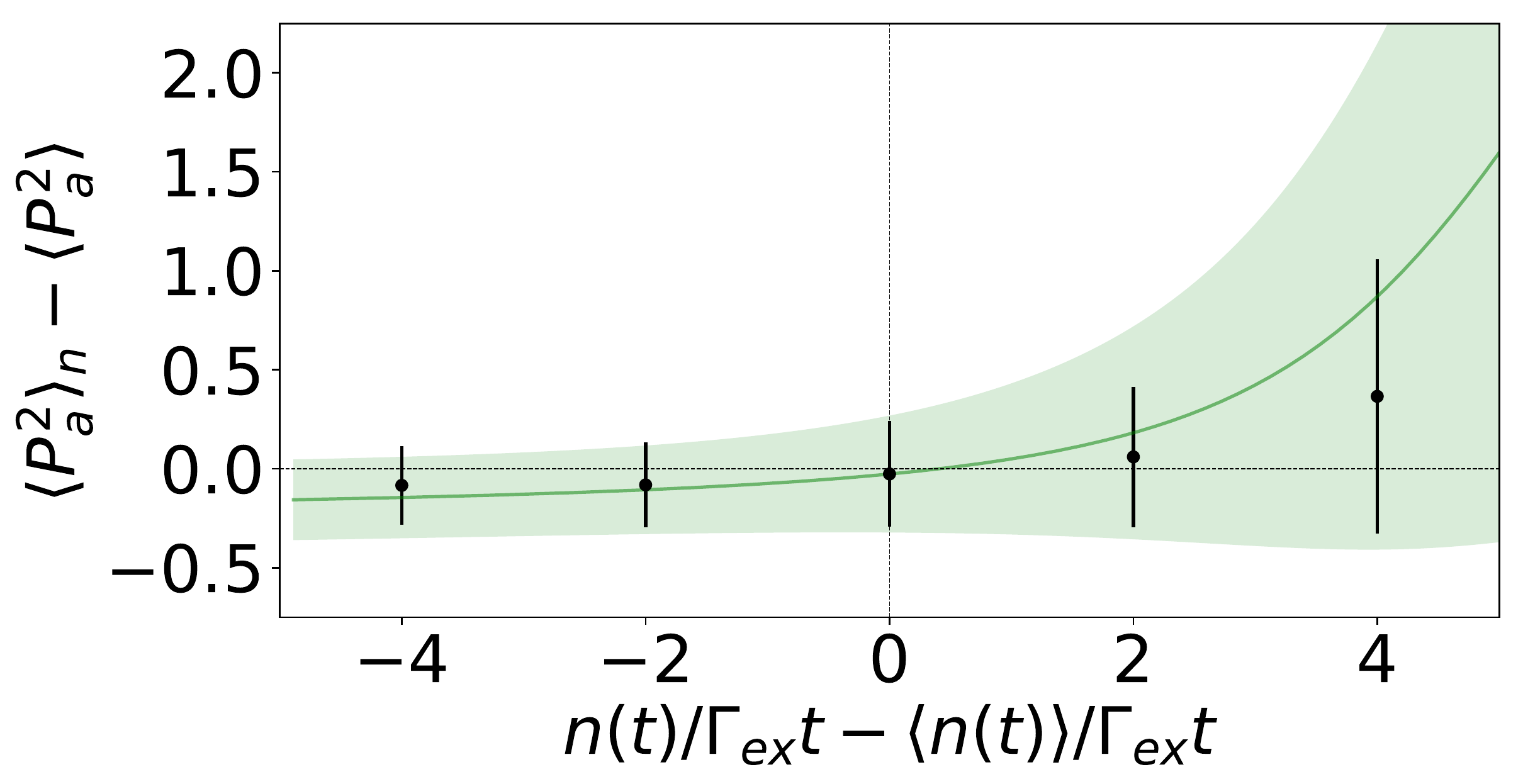} (a) \quad \includegraphics [width = 0.4 \textwidth, height = 0.23 \textwidth]{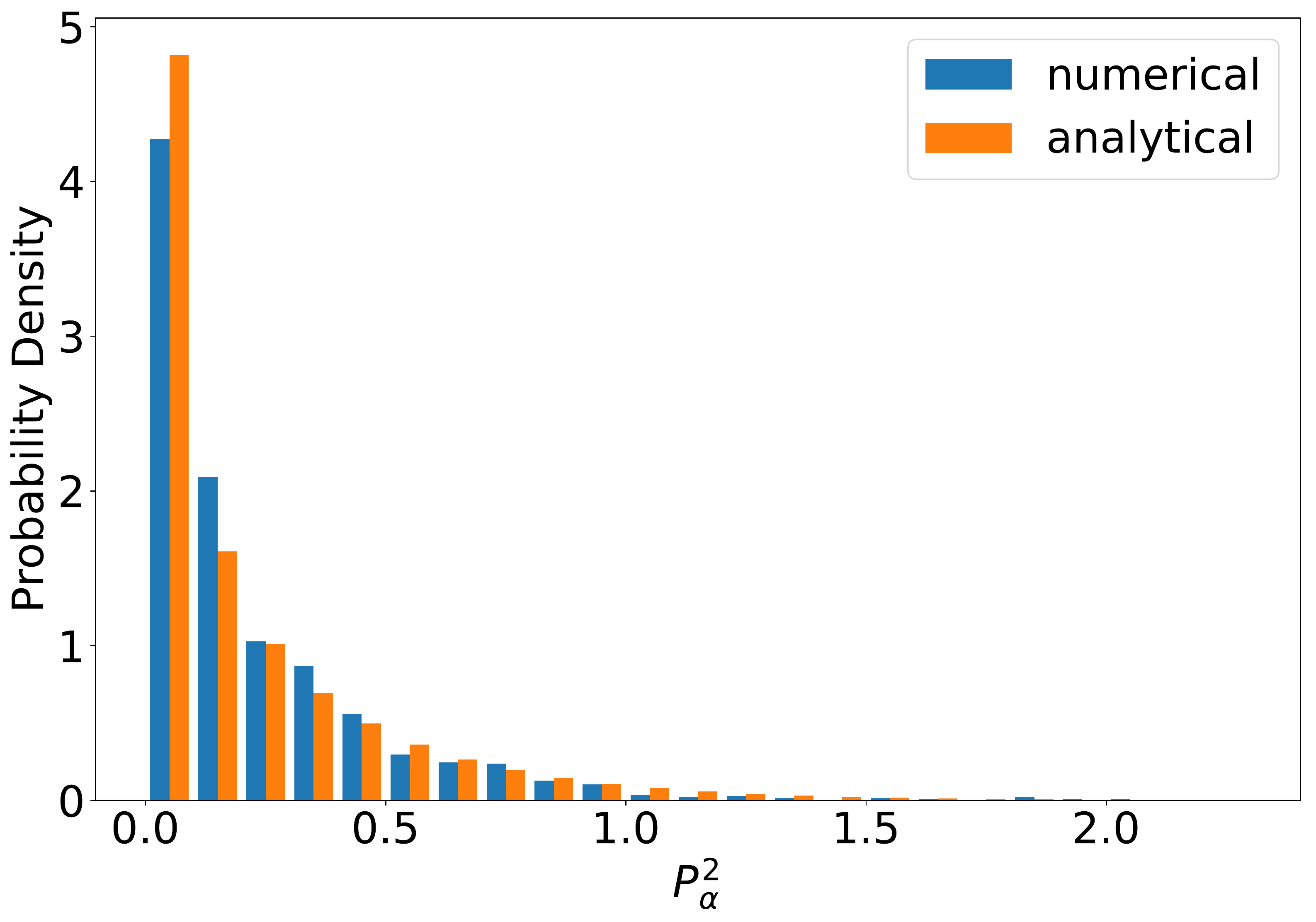} (b)} 
\caption{Squeezing of $ P_a^2 $ by photon counting at short times: $ \Gamma_{\rm ex} t = 15 $,
{where $\Gamma_{\rm ex}$ is the rate (\ref{eq054}) of creation of excitations in the hybridized nuclear bosonic mode $\alpha$}.
(a) Conditional mean and standard deviation of the {squared nuclear spin quadrature} $ P_a^2 $ knowing that $ n $ photons were detected in the time interval $ [0, t] $, as functions of this number $ n $. The standard deviation is plotted as a confidence interval. The unconditional mean $ \langle P_a^2 \rangle = 1/4 $ is independent of time, see {equation} (\ref{eq032}). Black dots and error bars: {numerical} simulation of the 3-mode model with 3000 realizations; green line and colored area: analytical predictions taken from {equations} (\ref{eq062}), (\ref{eq066}) and (\ref{eq968}) of the one-mode model. In practice, the black points are obtained after averaging over classes of values of $ n $ centered on these points (in a given class, the trajectories have close photodetection numbers $n$ but independent histories for the metastability exchange jumps which the experimenter cannot access). 3-mode model parameters: 
$\Omega/\kappa = 1/3$, $\gamma_m/\kappa = 1/10$,  $\gamma_f/\kappa = 1/1000$ (so that $\Gamma_{\rm ex}/\kappa=1/909$), $n_a^{\rm max} = 64, n_b^{\rm max} =n_c^{\rm max}=8$.  
This corresponds to 
$\Gamma_0/\Gamma_{\rm ex}=12\,500/601\simeq 20.8$ {and $\Gamma_{\rm sq}/\Gamma_{\rm ex}=601/101\, 202\simeq 1/168$}
{where $\Gamma_{\rm sq}$ is the squeezing rate (\ref{eq868}).} (b) For the class centered on $ n = \langle n (t) \rangle $, histogram of the conditional values of $ P_\alpha^2 $. Blue bars: {numerical} simulation of the three-mode model; orange bars: analytical predictions taken from {equation} (\ref{eq065}) of the one-mode model. 
} 
\label{fig6} 
\end{figure} 
Finally, we carry out a numerical verification of these analytical predictions in the three-mode model. In {figure} {\ref{fig6}a}, we plot the conditional mean of the square $ P_a^2 $ of the nuclear spin quadrature knowing that $ n $ photodetections occurred in the time interval $ [0, t] $, with $ \Gamma_{\rm ex} t = 15 $ (black dots), depending on this number $ n $. The ensemble of {realizations} is divided into 5 classes corresponding to a number of photodetections falling within a given interval, and the black dots are obtained by averaging over the {realizations} in the same class. The numerical results are close to the analytical predictions taken from (\ref{eq066}) and (\ref{eq968}) and plotted in green, except in the extreme classes which include a too low number of realizations. On the other hand, the asymptotic analytical predictions {(\ref{eq068})}, not shown, would be in disagreement with the simulations of the two models because the time 
{$t=15/\Gamma_{\rm ex}$ is not long enough, it is much less than the squeezing time $1/\Gamma_{\rm sq}$.} {In figure \ref{fig6}b, we plot the conditional probability distribution of $P_\alpha^2$ corresponding to the central class of figure \ref{fig6}a; there is also good agreement between one-mode analytical calculation and three-mode numerical simulation.}
\begin{figure} [t] 
\begin{center} 
\subfloat [] 
{\includegraphics [width = 0.45 \textwidth, height = 50mm]{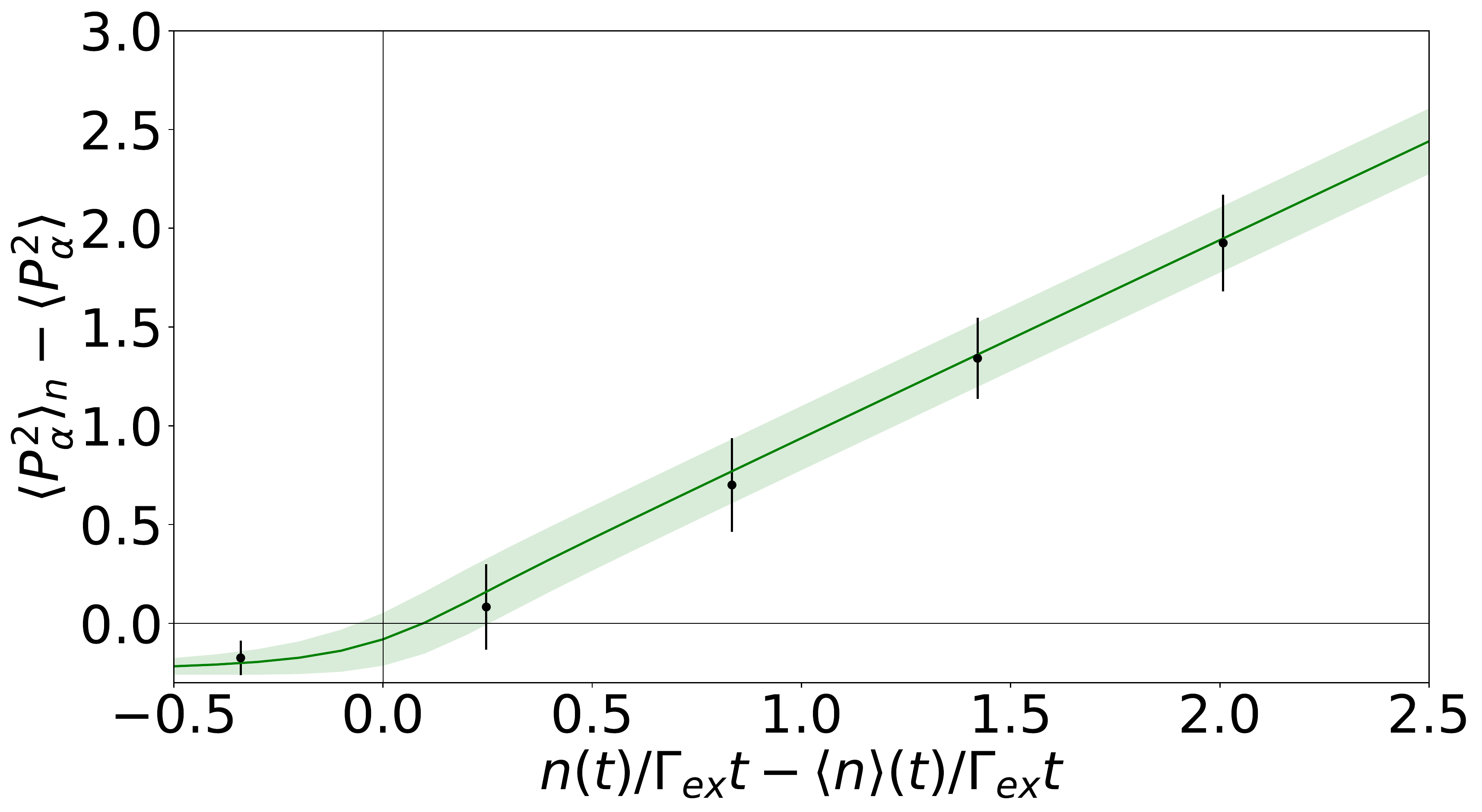}} \quad \quad 
\subfloat [] 
{\includegraphics [width = 0.4 \textwidth, height = 50mm]{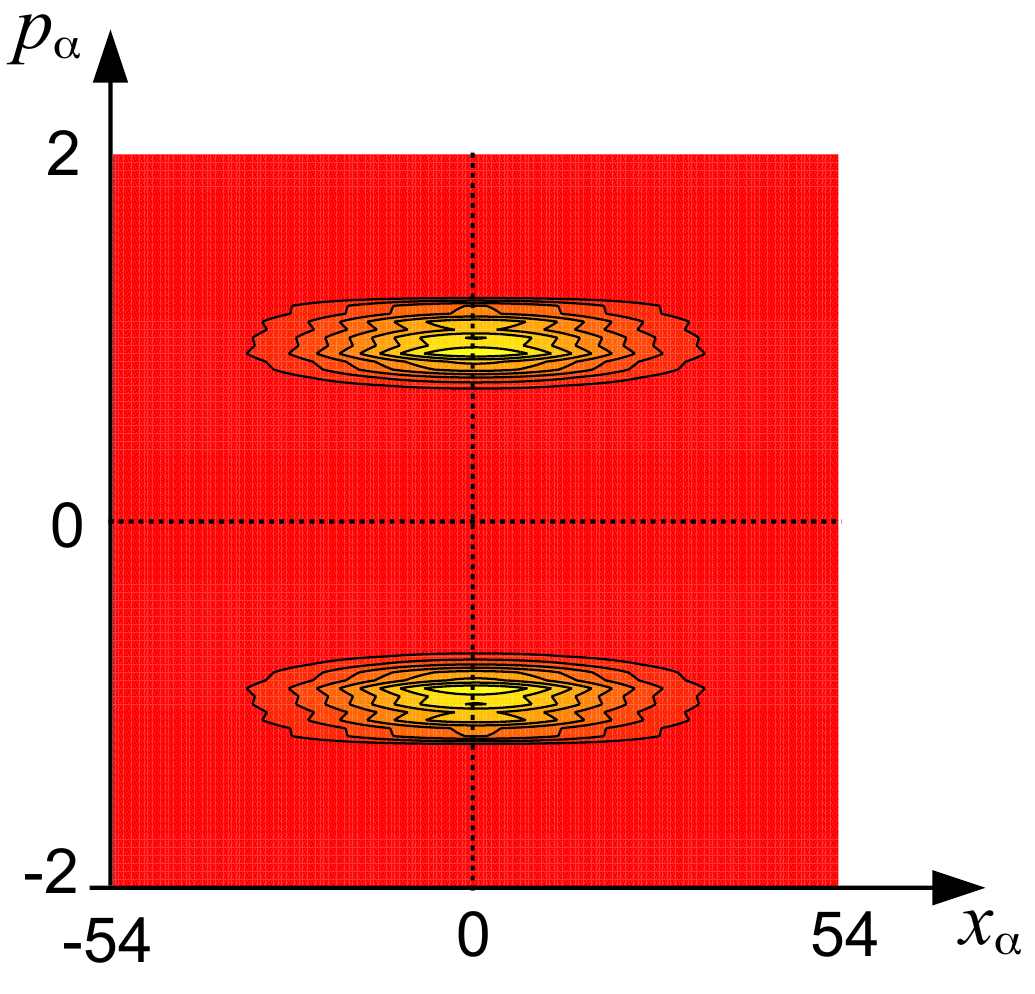}} 
\end{center} 
\caption{
Squeezing of $ P_\alpha^2 $ by photon counting at long times, $ \Gamma_{\rm ex} t = 1000 $, in the one-mode model (\ref{eq030oneMode}) (this long time {would make} a simulation in the three-mode model more difficult). 
(a) Conditional mean and standard deviation of $ P_\alpha^2 $ knowing that the number of photodetections $ n $ falls in a given class of values, similarly to figure \ref{fig6}a, for 2000 realizations. (b) Wigner distribution of the hybridized nuclear bosonic mode in the quadrature space $ (X_\alpha, P_\alpha) $ at $ \Gamma_{\rm ex} t = 1000 $, obtained by averaging the dyads $ | \psi (t) \rangle \langle \psi (t) | $ 
{whose number of photodetections falls in the 3rd class of (a) (302 trajectories out of 5000 realizations). It has two peak lines but no interference fringes (see note \ref{note600}).}
} 
\label{fig7} 
\end{figure} 
In figure \ref{fig7}, we are precisely exploring long times in the one-mode model, with $ \Gamma_{\rm ex} t = $ 1000 {that is $\Gamma_{\rm sq}t\simeq 5.94$}. Figure \ref{fig7}a, which is the equivalent of figure \ref{fig6}a, shows that $ \langle P_\alpha^2 \rangle_n $ is then related to the number of photodetections $ n $ as in the analytical prediction {(\ref{eq068})}, i.e. according to the {internal bisector} in the units of the figure, with a conditional standard deviation (\ref{eq068}) roughly constant {$\simeq (\Gamma_0 t)^{1/2}/\Gamma_{\rm ex}t\simeq 1/(\Gamma_{\rm sq} t)^{1/2}$} because $ \Gamma_0 $ is here $ \gg \Gamma_{\rm ex} $. 

\subsection{{Squeezing} by homodyne detection} 
We now assume that the $y$-polarized photons {leaking out of the cavity} are continuously measured by homodyne detection \cite{Wiseman2002}, as in figure \ref{fig1}c. We must first find the stochastic equations giving the evolution of the {system state vector} conditioned on homodyne detection, since the jump operators appearing naturally in (\ref{eq030t}) or (\ref{eq030oneMode}) of the three-mode or one-mode {quantum master equation} are unsuitable. We then present some analytical results obtained in the one-mode model and then in the three-mode model, before briefly discussing the effect of the finite coherence time of metastable atoms. 
{In order to obtain these results, we used the fact that, for the vacuum initial state considered here, the conditional state vector is given exactly at all times by a Gaussian ansatz \cite{KlausGauss}, whatever the number of modes of the model, in presence or absence of decoherence.}

\subsubsection{Suitable stochastic formulation of the {quantum master equation} } 
A general {quantum master equation} of the Lindblad form {\cite{Lindblad}}
\begin{equation}
\frac{\dd \rho}{\dd t} = \frac{1}{\ii\hbar} \left[H,\rho \right] + \sum_m C_m\rho C_m^\dagger - \frac{1}{2} \{ C_m^\dagger C_m, \rho \}
\label{eq030Arbitrary}
\end{equation}
with $ H $ the Hermitian part of the Hamiltonian and $ C_m $ the jump operators, can be rewritten in an equivalent way by adding {an arbitrary} constant to the jump operators and/or by mixing them by any linear unitary combination. In order to take into account a homodyne detection on the outgoing field, we form, from a jump operator $ C_m $ corresponding to a photodetection, the two \og homodyne \fg \, jump operators $ D_{m, \pm} $ \cite{CastinDalibard} 
\begin{equation}
D_{m,+}=\frac{\mu \mathbb{1} + C_m}{\sqrt{2}} \quad;\quad  D_{m,-}=\frac{\mu \mathbb{1} - C_m}{\sqrt{2}}
\label{eq072}
\end{equation}
where $ \mu^2 $ has the dimensions of a {frequency}. The measurement of the difference in the jump rates $ D^\dagger_+D_+-D^\dagger_-D_-$ then gives access to a quadrature of $ C_m $. Thus, for $ \mu $ real and $ C_m $ corresponding to the cavity jump operator $ C_c $, see {equation} (\ref{eq038}), the difference between the numbers of photons $ N_\pm $ detected during the short time interval $ \Delta t $ in the two output channels of figure \ref{fig1}c, which by definition constitutes the homodyne signal, 
\begin{equation}
N_+=(D_{c,+}^\dagger D_{c,+}^{\phantom{\dagger}}) \, \Delta t \quad ;\quad N_-=(D_{c,-}^\dagger D_{c,-}^{\phantom{\dagger}}) \, \Delta t  \quad ;\quad 
\frac{N_+-N_-}{2\, \mu}= \frac{c+c^\dagger}{2} \sqrt{\kappa} \Delta t
\label{eq075}
\end{equation}
gives access to $ X_c $ ~; it is indeed this quadrature of the field, conjugated to $ P_c $ therefore translated by a quantity proportional to $ P_b $ and to the time under the action of the Hamiltonian $ H $ (\ref{eq016}), which provides information on $ P_a $ through metastability exchange collisions. In the case of the {quantum master equation} with 3 modes (\ref{eq030t}), one has to apply the doubling procedure (\ref{eq072})  {\it a priori} only to the jump operator of {the} cavity. In practice, we will apply {it} also to the jump operator $ C_\beta $, that is we will double by homodyning {\it all} the jump operators $ C_m $, in order to avoid the discomfort of a hybrid representation mixing {discrete} quantum jumps and continuous stochastic evolution, see {equation} (\ref{eq076}) to come. In the case of the one-mode {quantum master equation} (\ref{eq030oneMode}), we need to \og homodyne \fg \, the two jump operators $ C_s $ and $ C_d $ anyway, since each of them comes with the loss of a photon in a cavity, as explained in section \ref{sec:ana_omm}. 

Within the limit of a large amplitude of the local oscillator $ \mu $, we can act as if $ \Delta t $ were infinitesimal \footnote{This approximation is valid for a time resolution, or a {time step} $ \Delta t $, such that $ \mu^{-2} \ll \Delta t \ll \kappa^{-1} $, where $ \kappa $ is in practice the fastest evolution rate in the system in the experiment.} and represent the evolution of the Monte Carlo wave function, {now normalized to unity}, by a continuous nonlinear stochastic equation without quantum jumps \cite{CastinDalibard, Helvetica, Percival} in Ito point of view: 
\be
\fbox{\parbox{0.85\textwidth}{
\vspace{-3mm}
\begin{align}
\dd|\phi(t)\rangle & = -\frac{\ii}{\hbar}H|\phi(t)\rangle \dd t 
    - \frac{1}{2} \sum_m \left(C_m^\dagger C_m - \langle \phi(t) | C_m + C_m^\dagger |\phi(t) \rangle C_m + \frac{1}{4}\langle \phi(t) | C_m + C_m^\dagger|\phi(t) \rangle^2\right)|\phi(t)\rangle \dd t\nonumber \\
   & + \sum_m \left( C_m -\frac{1}{2}\langle \phi(t) | C_m + C_m^\dagger |\phi(t) \rangle \right)|\phi(t)\rangle\,\dd\zeta_m(t) \nonumber
\end{align}
\vspace{-3mm}
}}
\label{eq076}
\ee
where, to each jump operator $ C_m $ in the initial {quantum master equation}, we associate a continuous-time stochastic process $ \dd \zeta_m (t) $, with real values, Gaussian, of zero mean, of variance $ \dd t $, statistically independent of other processes and without memory. At the same level of approximation, the homodyne signal operator (\ref{eq075}) is replaced by the sum of its average and a classical noise representing its fluctuations, which is none other than the corresponding $ \dd \zeta_m $ \cite{CastinDalibard}: 
\begin{equation}
\frac{N_+-N_-}{2 \mu } = \frac{\sqrt{\kappa}\langle \phi | c+c^\dagger |\phi\rangle}{2} \dd t + {\frac{1}{2}}\dd \zeta_c \label{eq077}
\end{equation}
 In practice, more than the {homodyning} history, that is the detailed time dependence of the homodyne detection signal, it is its time average over an interval of time $ [0, t] $ which is easily accessible in an experiment
{and allows a post-selection of states (i.e. experimental realizations) in a set of non negligible statistical weight}.
We thus introduce the integrated signal having the dimension of the {root of a frequency}, 
\be
\boxed{
\sigma(t) \equiv \frac{N_+^{\rm tot}-N_-^{\rm tot}}{2\mu t}= \frac{1}{t} \int_0^t  \dd t' \left[\sqrt{\kappa} \langle\phi(t')|X_c |\phi(t')\rangle + {\frac{1}{2}\frac{\dd\zeta_c(t')}{\dd t'}}\right]}
\label{eq977}
\ee
and we will calculate {in the following} the mean and the variance of the quadrature $ P_a $ of the nuclear spin conditioned on $ \sigma $. 
\footnote{{\label{note977} Recall that the mean and the variance of an observable (here $P_a$) in a single realization $|\phi(t)\rangle$ of the stochastic equation, thus of the experiment, have in general no physical meaning, because there is no possible measurement giving the mean value of an observable in a single realization, on the contrary it is necessary to average over a large number of realizations of the experiment having evolved during $t$ starting from the same initial pure case or density operator. A counter-example corresponds to the case where one can, by continuous measurements, trace back the time dependence of all the stochastic processes $\dd\zeta_m$ ; this is the case of the one-mode model subjected to a homodyne detection, the time dependence of the signal (\ref{eq077}) fixing that of the single stochastic process $\dd\zeta_c$. This would then allow, in principle, to select, among a large number of realizations of the experiment, those leading to the chosen $|\phi(t)\rangle$ state, and to deduce the expectation values of observables in $|\phi(t)\rangle$; in practice, this would be unrealistic, given the infinitesimal statistical weight of the realizations to keep.}}

\subsubsection{Analytical results in the one-mode model} \label{sec:3modesTo1Homo} 
\begin{figure} [t] 
\begin{center} 
{\includegraphics [width = 0.44 \textwidth]{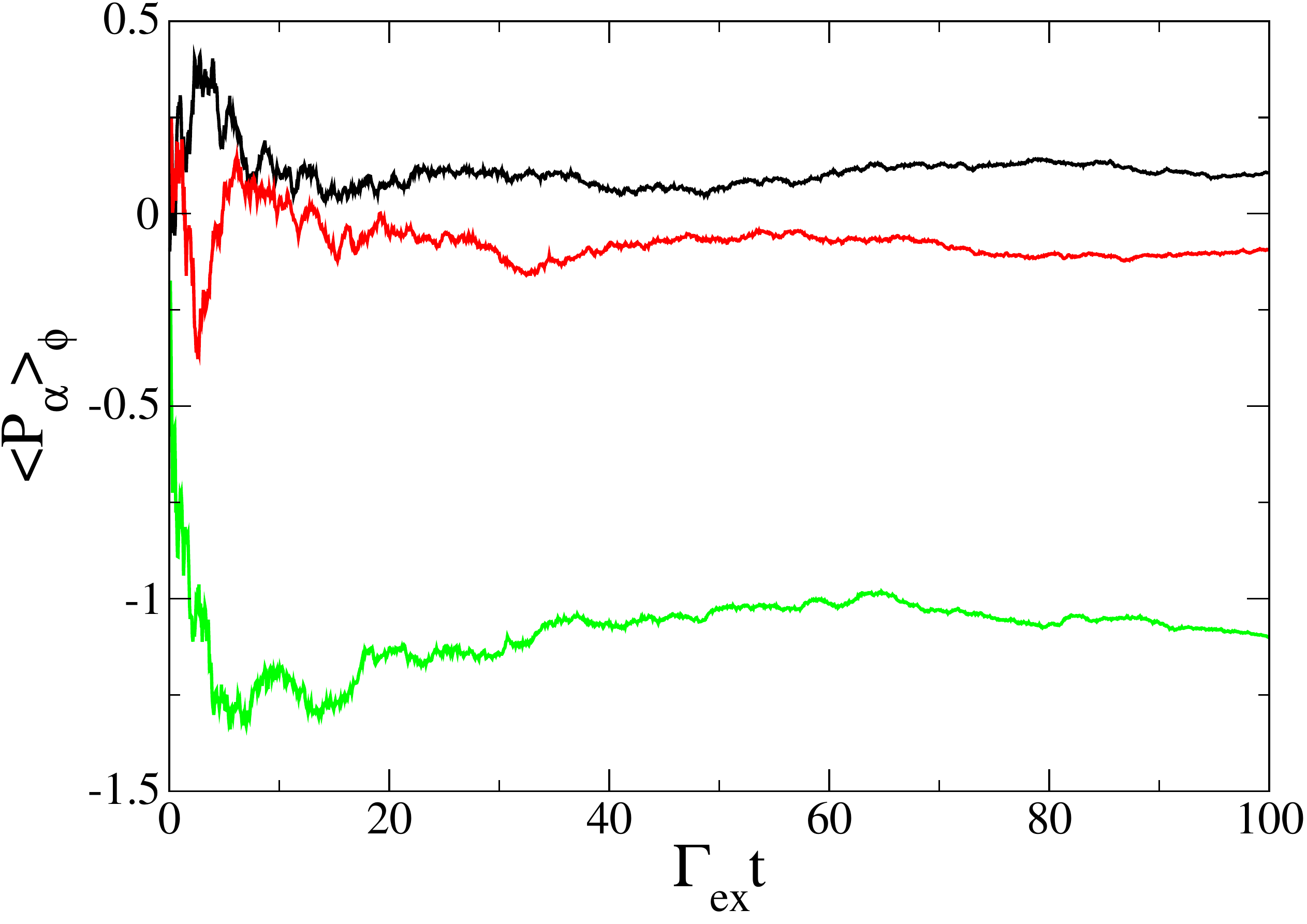} (a)} \quad 
{\includegraphics [width = 0.43 \textwidth]{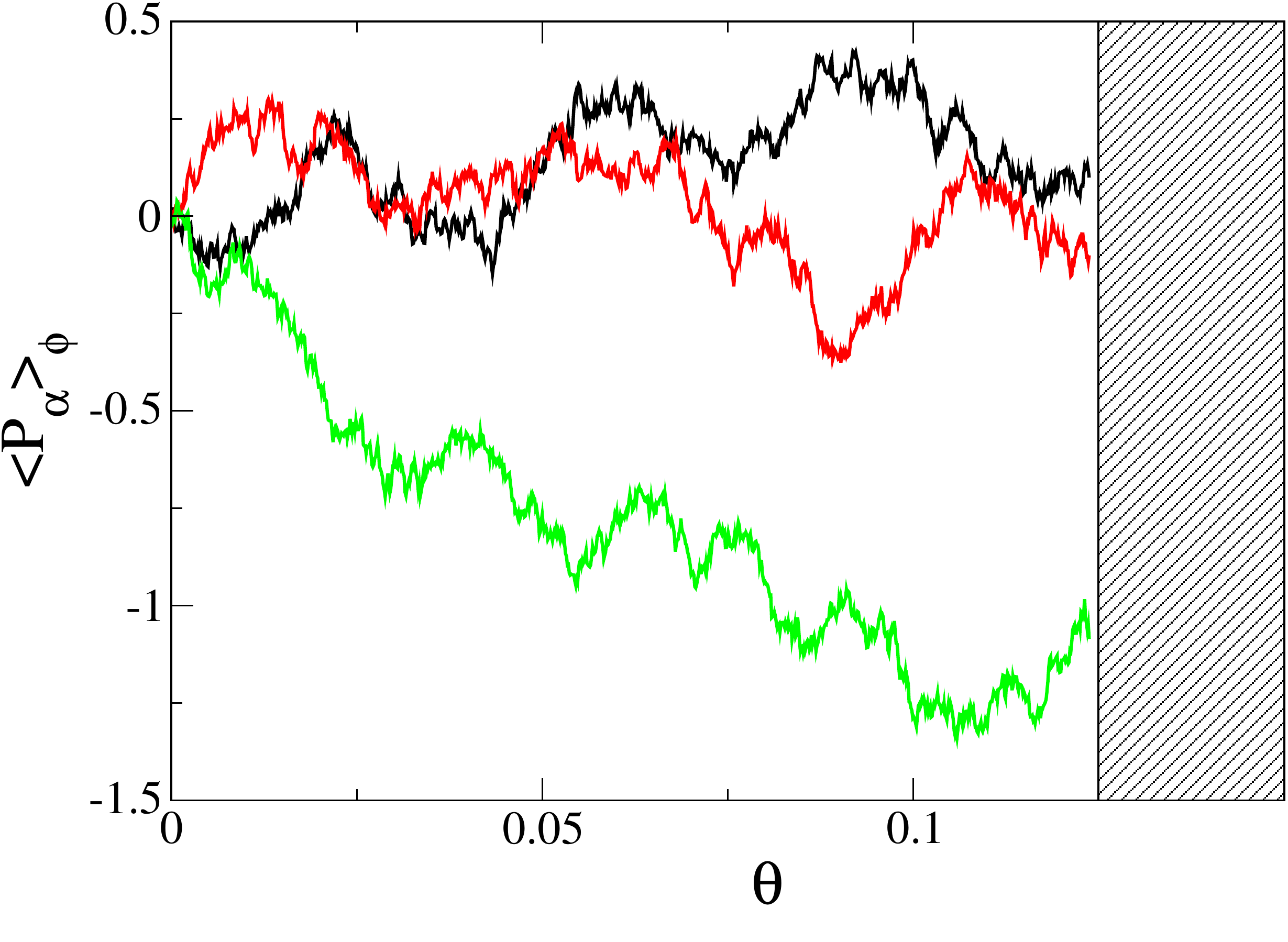} (b)} 
\end{center} 
\caption{In the case of {squeezing by} continuous homodyne detection: random walk (\ref{eq082}) performed in the one-mode model by the quantum average of the quadrature $ P_\alpha $ of the nuclear spin in a given realization of the experiment. (a) Quantum average as a function of true time $ t $ {reduced by the rate (\ref{eq054}) of creation of excitations} for three realizations of the experiment; it is a stretched Brownian motion converging at long times towards a fixed but unpredictable value. (b) Idem as a function of the compact renormalized time $ \theta $ (\ref{eq083}); this time it is an ordinary Brownian motion but limited to $ \theta \leq 1/8 $.} 
\label{fig8} 
\end{figure} 
Let us explicitly write the stochastic equation (\ref{eq076}) for the one-mode model (\ref{eq030oneMode}): 
\begin{equation}
\dd|\phi(t) \rangle = -\frac{\dd t}{2} \Gamma_{\rm ex} [P_\alpha-\bar{P}_\alpha(t)]^2  |\phi(t) \rangle + \sqrt{\Gamma_{\rm ex}} \dd\zeta_s(t) [P_\alpha-\bar{P}_\alpha(t)] |\phi(t) \rangle 
\label{eq078}
\end{equation}
with 
$\bar{P}_\alpha(t)\equiv \langle \phi(t) | P_\alpha |\phi(t)  \rangle$.
The highlight is that the jump operator $C_d$ proportional to the identity, which added noise in the photon counting detection scheme of {section} \ref{sec:dpcdp}, gives no contribution and completely disappears in the homodyne case. Indeed, the photons emitted during these jumps come from the component $ | 1 \rangle | 1 \rangle $ of the state vector (\ref{eq039}) containing one excitation $ \beta $, which makes them {optically incoherent} with the light field injected into the cavity, i.e. with the component $ | 0 \rangle | 0 \rangle $ of (\ref{eq039}), in the sense that $ | 1 \rangle | 1 \rangle $ contributes to $ \langle c^\dagger c \rangle $ but not to $ \langle c+c^\dagger \rangle $. So only the stochastic process $ \dd \zeta_s $ associated with the jump operator $ C_s $ remains. This process coincides with {the one} $ \dd \zeta_c $ appearing in the homodyne detection signal (\ref{eq077}), $ \dd \zeta_s \equiv \dd \zeta_c $, a fact admitted here but which will be derived in section \ref{sec:sdmatm}. 

 The stochastic equation (\ref{eq078}) {exhibits a} linear noise term and a quadratic deterministic term in the operator $ P_\alpha $, real in Fourier space. For the initial state considered here, it is thus solved exactly by a Gaussian ansatz on the wave function in {momentum representation}, real and correctly normalized for the commutation relation $ [X_\alpha, P_\alpha] = \ii/2 $: 
\begin{equation}
\langle p_\alpha | \phi(t)\rangle = [2\pi u(t)]^{1/4} \exp\{-u(t)[p_\alpha-\bar{P}_\alpha(t)]^2\}
\label{eq080}
\end{equation}
{ On the other hand, the Gaussianity is lost in the {squeezing} by photodetection protocol of {section} \ref{sec:dpcdp}.} 
Using the Ito calculation, \footnote{We only keep the linear terms in $ \dd t $ or in noise, and we systematically replace the quadratic terms $ \dd \zeta_s^2 $ by their mean $ \dd t $.} we find that $ u $ follows {a deterministic evolution equation}, to be integrated with the initial condition $ u (0) = 1 $: 
\begin{equation}
\dd u(t)=\Gamma_{\rm ex}\dd t \quad \mbox{so that}\quad u(t)=1+\Gamma_{\rm ex} t\quad\mbox{and}\quad\mbox{Var}_{\phi} P_\alpha(t)\equiv \frac{1}{4 u(t)} = \frac{1}{4} \frac{1}{1+\Gamma_{\rm ex}t}
\label{eq081}
\end{equation}
where we have also given the variance of $ P_\alpha $ in the state $ | \phi \rangle $. On the contrary, the equation for the mean value of $ P_\alpha $ in $ | \phi \rangle $ is purely stochastic, with a diffusion coefficient $ D (t) $ depending on time and the initial condition $ \bar{P}_\alpha (0) = 0 $: 
\begin{equation}
\dd\bar{P}_\alpha(t)=[2 D(t)]^{1/2} \dd\zeta_s(t) \quad\mbox{with}\quad D(t)= \frac{\Gamma_{\rm ex}}{8 u(t)^2} = \frac{\Gamma_{\rm ex}}{8(1+\Gamma_{\rm ex}t)^2}
\label{eq082}
\end{equation}
As $ D (t) $ is of finite integral, $ \bar{P}_\alpha (t) $ stabilizes asymptotically (at long times) at a fixed value on a single realization, as seen in figure \ref{fig8}, with a variance in the quantum state $ \mbox{Var}_\phi P_\alpha $ tending to $ 0 $. This phenomenon of \og stochastic convergence \fg \, towards an eigenstate of the measured observable (in this case $ P_\alpha $) is expected in the description of a quantum measurement by a diffusion equation of the {state vector} \cite{Helvetica, Percival, Gisin}. To show it here, we introduce a renormalized time $\theta$ in terms of which $ \bar{P}_\alpha $ performs an ordinary Brownian motion with a unity diffusion coefficient, and we notice that this time is bounded: 
\begin{equation}
\theta=\int_0^t\dd t' D(t') = \frac{{\Gamma_{\rm ex}} t}{8(1+\Gamma_{\rm ex} t)}\underset{t\to +\infty}{\to} \theta_\infty=\frac{1}{8}
\label{eq083}
\end{equation}
At the renormalized instant $ \theta_\infty $, $ \bar{P}_\alpha $ follows a Gaussian law with zero mean and variance $ 1/4 $: $ \bar{P}_\alpha $ has therefore the same asymptotic probability distribution ($ t \to+\infty $) as that of the observable $ P_\alpha $ in the initial quantum state of the nuclear spin. 

\begin{figure} [t] 
\centerline{\includegraphics [width = 0.33 \textwidth]{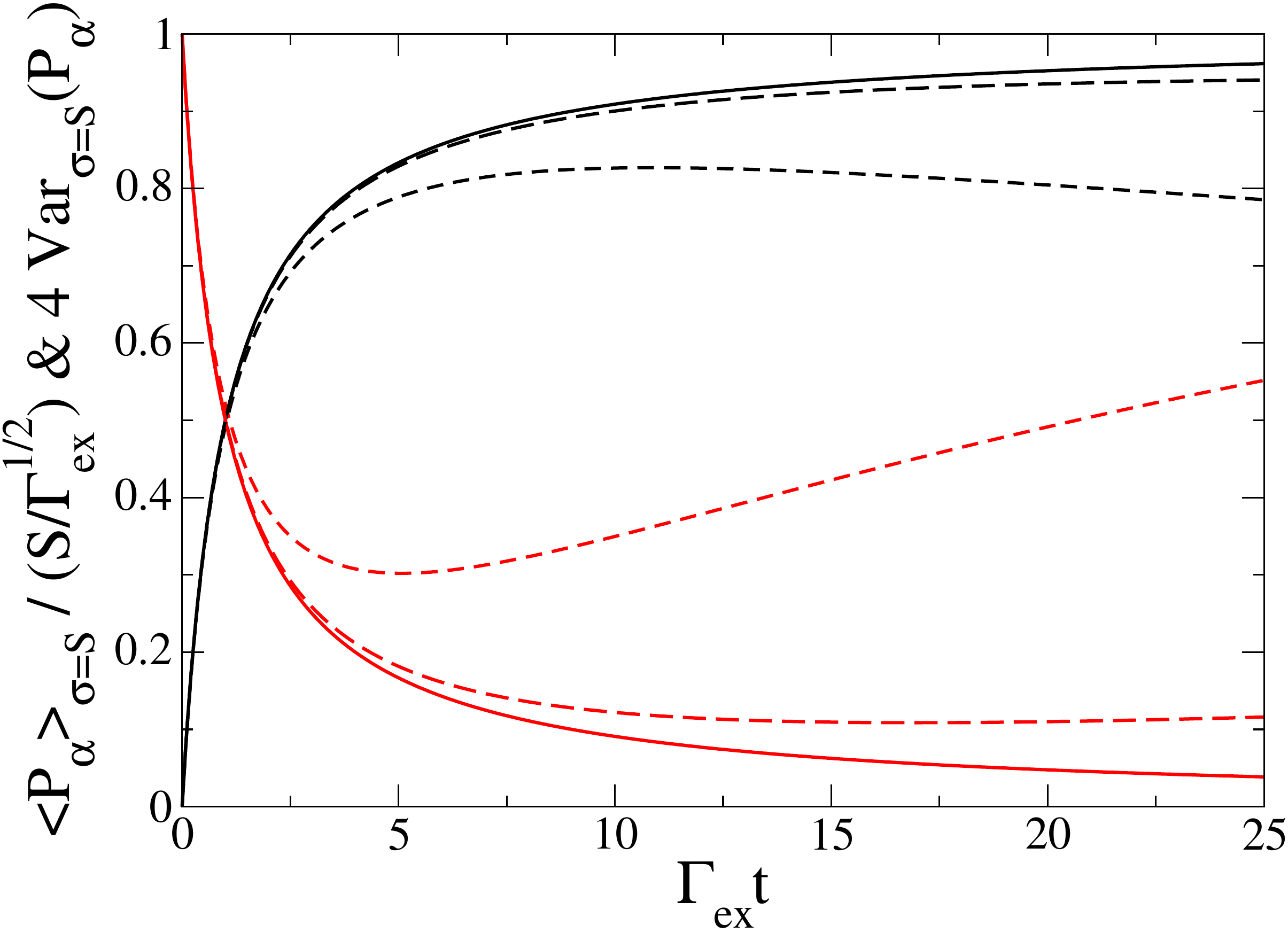} (a) \includegraphics [width = 0.33 \textwidth, height = 0.23 \textwidth]{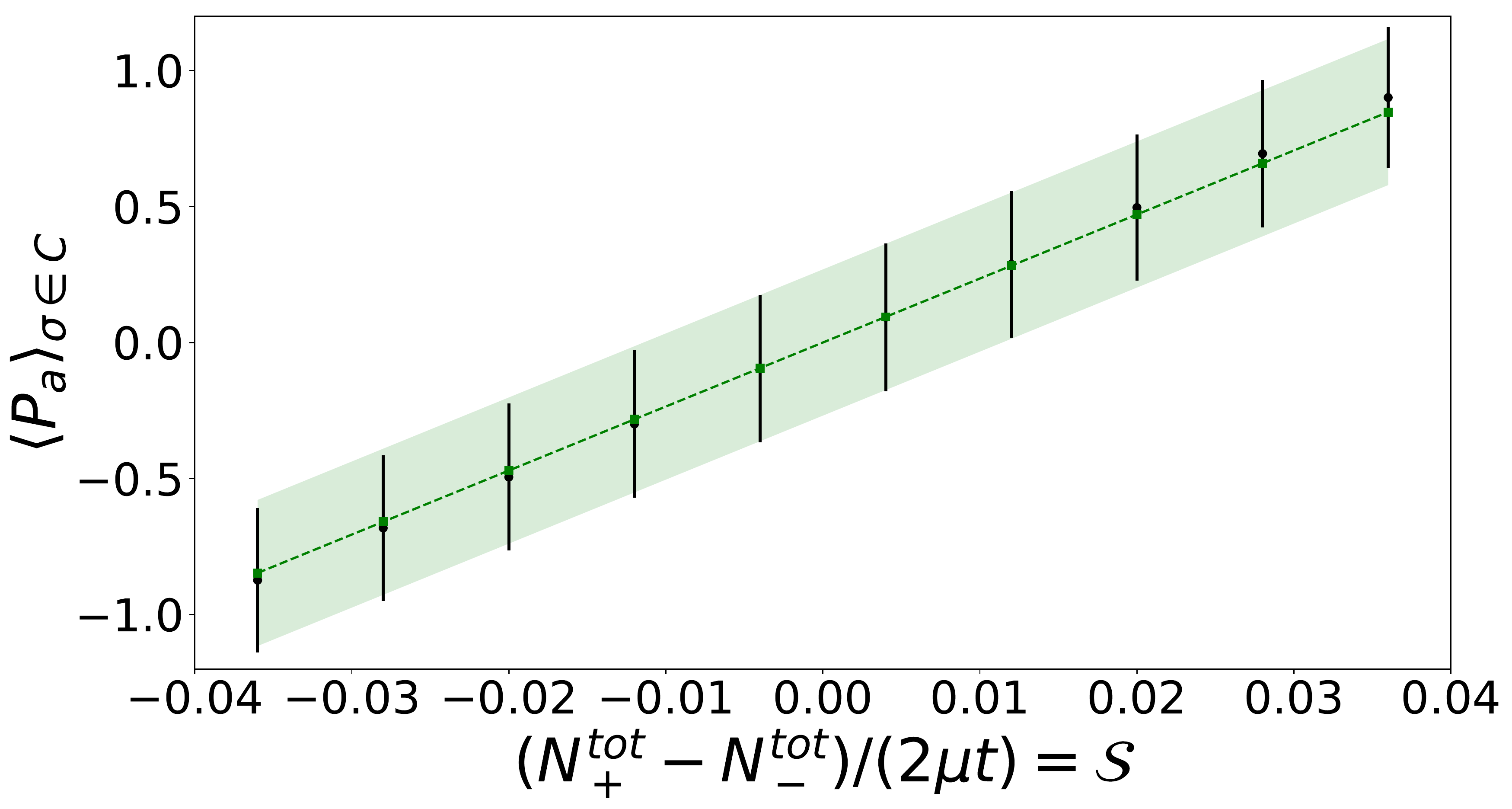} (b) \includegraphics [width = 0.33 \textwidth]{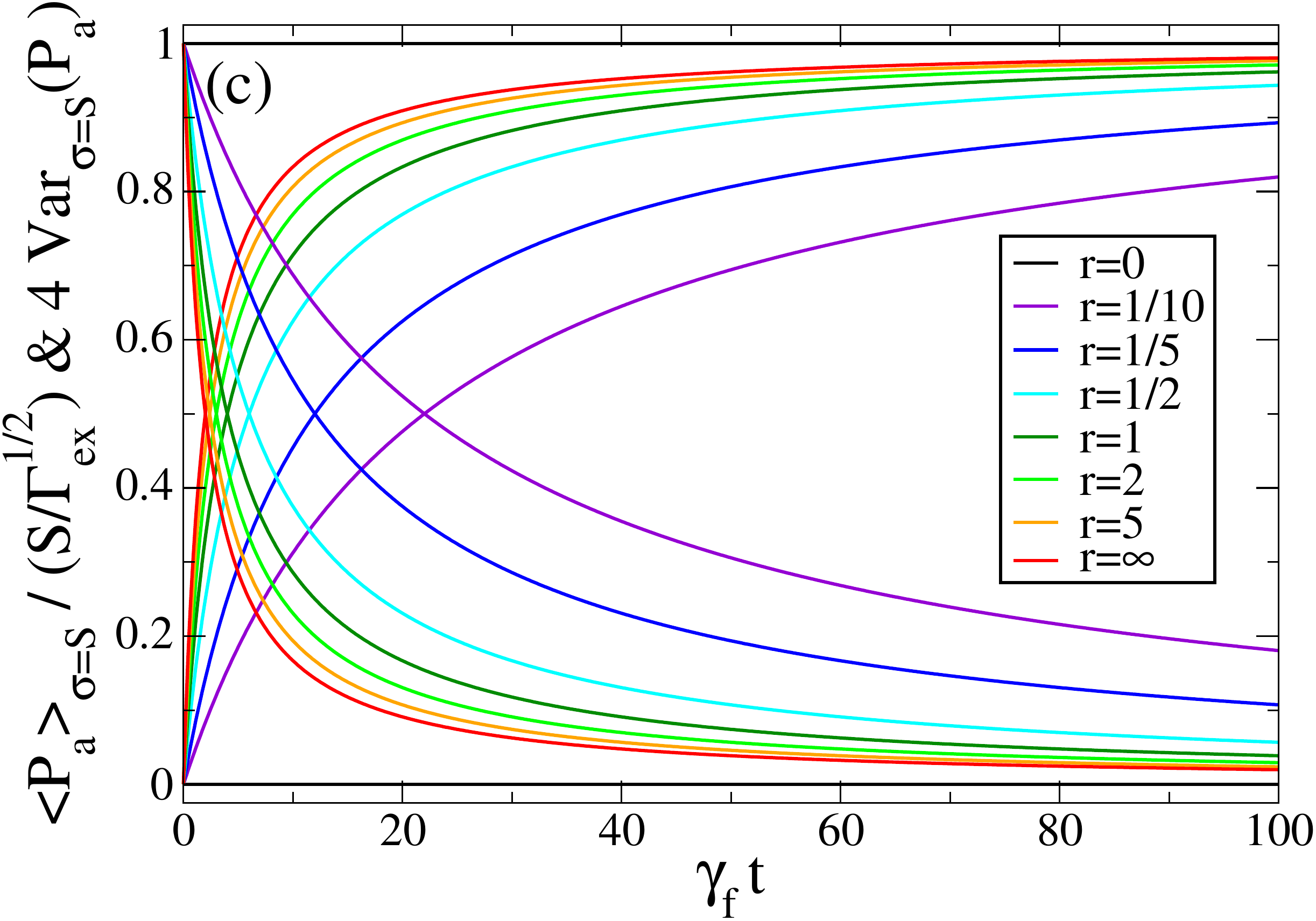} (c)} 
\caption{{Nuclear} spin squeezing by continuous homodyne measurement. (a) In the one-mode model, mean (in black) and variance (multiplied by 4, in red) of the quadrature $ P_\alpha $ of the hybridized nuclear spin conditioned on the integrated homodyne signal $ \sigma $, as functions of the integration time $ t $. Solid lines: analytical expressions (\ref{eq101}). Dashed lines: expressions (\ref{eq155a}) and (\ref{eq155b}) in the presence of decoherence {by de-excitation of metastable atoms at the cell walls} (long dashed line: $ \epsilon = 1/100 $, short dashed line: $ \epsilon = 1/10 $, with $ \epsilon = \gamma_\alpha/\Gamma_{\rm ex} $ and $ \gamma_\alpha $ the {effective decoherence} rate (\ref{eq152})). (b) In the three-mode model, for $ \Gamma_{\rm ex} t = 5 $, mean and standard deviation of the quadrature $ P_a $ of the nuclear spin conditioned on the signal $ \sigma $ belonging to a class $ C $, the range of values of $ \sigma/\sqrt{\Gamma_{\rm ex}} $ having been divided into 10 classes of the same width {(the values $\mathcal{S}$ of $\sigma$ are in units of $\Gamma_{\rm ex}^{1/2}$ on the horizontal axis)}. The standard deviation is plotted as a confidence interval. In black: numerical simulation of the stochastic equation (\ref{eq076}) with $ 1079 $ realizations. Dashed green and colored area: exact results taken from relations (\ref{eq127a}), (\ref{eq127b}) and the analytical expression of the conditional probability distribution of $ \bar{P}_a $ in terms of the variances and covariance (\ref{eq130}) similarly to {equation} (\ref{eq105}). The deviation between numerical and analytical results in the extreme classes is attributable to the low numbers of realizations falling in these classes. Parameter values: 
$\Omega/\kappa = 1/10$, $\gamma_m/\kappa = 1/10$, $\gamma_f/\kappa = 1/100$, $\Gamma_{\rm ex}/\kappa=1/1000$. 
(c) In the limit $ \Gamma_{\rm ex} \to 0 $ at $ \Gamma_{\rm ex}/2 \gamma_f $ fixed of the three-mode model, conditional mean and variance of $ P_a $ (\ref{eq132}) as functions of reduced time $ \gamma_f t $, for different values of the ratio $ r = 2 \Gamma_{\rm ex}/\gamma_f $ (increasing curves: mean, decreasing curves: variance). 
} 
\label{fig9} 
\end{figure}

We now come to the mean and the variance of $ P_\alpha $ conditioned on the value $ \mathcal{S} $ of the time-integrated {homodyning} signal $ \sigma $ (\ref{eq977}). Remarkably, we find that the conditional mean is always proportional to the signal, with a time-dependent {proportionality coefficient}, and that the conditional variance depends on time but not on the signal: 
\be
\label{eq101}
\boxed{\langle P_\alpha\rangle_{\sigma=\mathcal{S}}  = m(t) \frac{\mathcal{S}}{\sqrt{\Gamma_{\rm ex}}} \quad\mbox{where}\quad {m(t)={\frac{\Gamma_{\rm ex}t}{1+\Gamma_{\rm ex}t}}} \quad ; \quad \mbox{Var}_{\sigma=\mathcal{S}}(P_\alpha) = \mathcal{V}(t) \quad\mbox{where}\quad {\mathcal{V}(t)={\frac{1}{4(1+\Gamma_{\rm ex}t)}}}}
\ee
These expressions denote $ \Gamma_{\rm ex} $ as the nuclear spin {homodyne} squeezing rate in the one-mode model {that is at weak Faraday coupling:} 
\be
\label{eq701}
{\boxed{\Gamma_{\rm sq} \underset{\Omega\to 0}{\sim} \Gamma_{\rm ex}=\frac{\Omega^2\gamma_f}{\kappa(\gamma_m+\gamma_f)}}}
\ee
In figure \ref{fig9}a, we {plot $m(t)$ and $\mathcal{V}(t)$} as functions of the reduced time $ \tau = \Gamma_{\rm ex} t $. Just as the quantum variance in a single realization $ \mbox{Var}_\phi P_\alpha $, with which it actually coincides, the conditional variance tends asymptotically towards zero as the inverse of time. In the conditional average, the coefficient {$ m (t) $} tends towards 1 at long times. To understand this, let's relate the integrated signal (\ref{eq977}) to $ \bar{P}_\alpha $ using adiabatic expressions (\ref{eq043}) in the truncated state vector (\ref{eq039}): 
\be
\sigma(t)=\frac{1}{t} \int_0^t \dd t' \left[\sqrt{\Gamma_{\rm ex}} \bar{P}_\alpha(t') +{\frac{1}{2}\frac{\dd\zeta_s(t')}{\dd t'}}\right]
\label{eq102}
\ee
As $ \bar{P}_\alpha (t) $ stabilizes asymptotically on a single realization, and the time average of the noise $ \dd \zeta_s $ tends to zero like $ 1/t^{1/2} $, $ \sigma (+\infty) $ directly gives the value of {$ \bar{P}_\alpha(+\infty)$} up to a constant factor $ \sqrt{\Gamma_{\rm ex}} $. 

To derive the results (\ref{eq101}), we first relate the conditional variance of the operator $ P_\alpha $ to that of its quantum average in a realization $ \bar{P}_\alpha $ as follows: 
\bea
\mbox{Var}_{\sigma=\mathcal{S}}(P_\alpha) &\!\!\!\equiv\!\!\!& \langle\,\langle\phi|P_\alpha^2|\phi\rangle\,\rangle_{\sigma=\mathcal{S}} -\langle\,\langle\phi|{P}_\alpha|\phi\rangle\,\rangle_{\sigma=\mathcal{S}}^2=
\langle\,\langle\phi|P_\alpha^2|\phi\rangle-\langle\phi|P_\alpha|\phi\rangle^2\,\rangle_{\sigma=\mathcal{S}}
+ \langle \bar{P}_\alpha^2\rangle_{\sigma=\mathcal{S}}-\langle\bar{P}_\alpha\rangle^2_{\sigma=\mathcal{S}} \nonumber \\
&\!\!\!=\!\!\!& \langle \mbox{Var}_\phi P_\alpha\rangle_{\sigma=\mathcal{S}} + \mbox{Var}_{\sigma=\mathcal{S}}(\bar{P}_\alpha)
= {\frac{1}{4} \frac{1}{1+\Gamma_{\rm ex}t}} + \mbox{Var}_{\sigma=\mathcal{S}}(\bar{P}_\alpha)
\label{eq103}
\eea
where we have used {expression} (\ref{eq081}) for the quantum variance of $ P_\alpha $ in the state $ | \phi \rangle $. It therefore remains to determine the conditional probability distribution of $ \bar{P}_\alpha $ knowing that $ \sigma = \mathcal{S} $, 
\be
\label{eq104}
P(\bar{P}_\alpha=p_\alpha|\sigma=\mathcal{S}) \equiv \frac{P(\bar{P}_\alpha=p_\alpha,\sigma=\mathcal{S})}{P(\sigma=\mathcal{S})}
\ee
The random variable $ \bar{P}_\alpha (t) $, resulting from Brownian motion (\ref{eq082}), has a Gaussian probability distribution; the same applies to the temporal integral of $ \bar{P}_\alpha $ and to the noise $ \dd \zeta_s $, therefore to the signal $ \sigma $ (\ref{eq102}) which is their sum. As the variables $ \bar{P}_\alpha $ and $ \sigma $ have zero means, their joint probability distribution is characterized by their covariance matrix, or more directly by its inverse matrix, so that 
\bea
P(\bar{P}_\alpha=p_\alpha|\sigma=\mathcal{S})&\!\!\!=\!\!\!&  \frac{\frac{1}{2\pi\sqrt{\langle\bar{P}_\alpha^2\rangle_{\rm stoch}\langle\sigma^2\rangle_{\rm stoch}-\langle \sigma \bar{P}_\alpha\rangle_{\rm stoch}^2}}\exp\left(-\frac{1}{2} \frac{p_\alpha^2\langle\sigma^2\rangle_{\rm stoch} + \mathcal{S}^2 \langle \bar{P}_\alpha^2\rangle_{\rm stoch} -2 p_\alpha\mathcal{S} \langle \sigma\bar{P}_\alpha\rangle_{\rm stoch}}{\langle\bar{P}_\alpha^2\rangle_{\rm stoch}\langle\sigma^2\rangle_{\rm stoch}-\langle \sigma \bar{P}_\alpha\rangle_{\rm stoch}^2}\right)}
{\frac{1}{\sqrt{2\pi\langle\sigma^2\rangle_{\rm stoch}}} \exp\left(-\frac{\mathcal{S}^2}{2\langle\sigma^2\rangle_{\rm stoch}}\right)} \nonumber \\
&\!\!\!=\!\!\!& \frac{1}
{\sqrt{2\pi\left[\langle\bar{P}_\alpha^2\rangle_{\rm stoch} -  \langle\sigma \bar{P}_\alpha\rangle_{\rm stoch}^2/\langle\sigma^2\rangle_{\rm stoch}\right]}}
{\exp\left[-\frac{1}{2} \frac{\left(p_\alpha - \mathcal{S} \langle \sigma \bar{P}_\alpha\rangle_{\rm stoch}/\langle\sigma^2\rangle_{\rm stoch}\right)^2}{\langle\bar{P}_\alpha^2\rangle_{\rm stoch} -  \langle\sigma \bar{P}_\alpha\rangle_{\rm stoch}^2/\langle\sigma^2\rangle_{\rm stoch}}\right]}
\label{eq105}
\eea
where $ \langle \ldots \rangle_{\rm stoch} $ at time $ t $ is the average taken over all the realizations of the stochastic process $ \dd \zeta_s (t ') $ in the time interval $ [0, t] $. We deduce that, in {equations} (\ref{eq101}), 
\be
{m(t)}= \sqrt{\Gamma_{\rm ex}} \frac{\langle\sigma(t)\bar{P}_\alpha(t)\rangle_{\rm stoch}}{\langle\sigma^2(t)\rangle_{\rm stoch}}\quad\mbox{and}\quad {\mathcal{V}(t) =\frac{1}{4(1+\Gamma_{\rm ex}t)}}+\langle\bar{P}_\alpha^2(t)\rangle_{\rm stoch}-\frac{\langle\sigma(t)\bar{P}_\alpha(t)\rangle_{\rm stoch}^2}{\langle\sigma^2(t)\rangle_{\rm stoch}}
\label{eq106}
\ee
In order to determine their variances and covariance, we write $ \sigma (t) $ and $ \bar{P}_\alpha (t) $ as linear functionals of the stochastic process $ \dd \zeta_s $ and we use the fact that the Langevin forces $ \dd \zeta_s (t)/\dd t $ and $ \dd \zeta_s (t ')/\dd t' $ have a Dirac correlation function $ \delta (t-t ') $. Let us give the example of the first contribution to $ \sigma (t) $: 
\be
\int_0^t \dd t'' \bar{P}_\alpha(t'') =\int_0^t\dd t'' \int_0^{t''} \dd t' [2D(t')]^{1/2}{\frac{\dd\zeta_s(t')}{\dd t'}} =\int_0^t\dd t' \int_{t'}^{t} \dd t'' [2D(t')]^{1/2}{\frac{\dd\zeta_s(t')}{\dd t'}}
= \int_0^t \dd t' (t-t') [2D(t')]^{1/2}{\frac{\dd\zeta_s(t')}{\dd t'}}
\label{eq107}
\ee
where we changed the order of integration on $ t '$ and $ t''$ then explicitly integrated on $ t'' $. We end up with the expressions we are looking for (\ref{eq101}), the simplicity of which follows from the fact that, in one realization of the experiment, we always have 
\be
\label{eq197}
\boxed{\sigma(t) = \sqrt{\Gamma_{\rm ex}} {\frac{1+\Gamma_{\rm ex}t}{\Gamma_{\rm ex}t}} \bar{P}_\alpha(t)}
\ee
{Remarkably, the knowledge of the integrated signal $\sigma(t)$ in a realization of the experiment of duration $t$ is sufficient to prepare the nuclear spin in a well-defined pure Gaussian state (\ref{eq080}), with a parameter $u$ given by equation (\ref{eq081}) and a mean quadrature $\bar{P}_\alpha$ related to the signal by equation (\ref{eq197})}

Finally, let us return to the quadrature $ P_a $ of the unhybridized nuclear spin, which is truly usable in the experiment once the discharge is {switched-off} in the cell, {as shown by expression (\ref{eq017}) of the metrological gain in a precession measurement}. By inversion of transformation (\ref{eq034}) and by limiting {equation} (\ref{eq039}) to its first term (to the dominant order in $ \Omega $), it comes 
\be
\langle P_a\rangle_{\sigma=\mathcal{S}} = \left(\frac{\gamma_m}{\gamma_f+\gamma_m}\right)^{1/2} \, \langle P_{\alpha}\rangle_{\sigma=\mathcal{S}} \quad\mbox{and}\quad\mbox{Var}_{\sigma=\mathcal{S}}(P_a)= \frac{\gamma_f}{4(\gamma_f+\gamma_m)} + {\frac{\gamma_m}{\gamma_f+\gamma_m}}\mbox{Var}_{\sigma=\mathcal{S}}(P_\alpha)
\label{eq108}
\ee
The conditional variance of $ P_a $ at long times tends towards a nonzero value, although low in practice: this is the intrinsic limit of this nuclear spin {squeezing} scheme, which uses the metastable state of ${}^3$He as an intermediate state. 

\subsubsection{Solution of the three-mode model} 
\label{sec:sdmatm} 

The study of spin {squeezing} in the one-mode model is limited to the regime (\ref{eq033}) where the {squeezing rate} {$ \Gamma_{\rm sq}\sim\Gamma_{\rm ex} $} is the longest timescale in the system. However, it is crucial for applications to see how far we can speed up the {squeezing} process by increasing $ \Gamma_{\rm ex} $ {through}, for example, the Faraday coupling $ \Omega $ of metastable atoms to the cavity field. To this end, we obtain the analytical solution of the three-mode model by using the Gaussian character of the state vector which results, as for the one-mode model, from the initial state considered (the vacuum), from the linearity of the jump operators $ C_m $ and the quadraticity of the Hamiltonian $ H $ in the quadratures of the modes. The stochastic equation (\ref{eq076}) therefore admits as an exact solution the Gaussian ansatz generalizing that of {equation} (\ref{eq080}), 
\be
\langle p_\alpha,p_\beta,x_c|\phi(t)\rangle=\phi(\qq,t) = [8\pi \det \underline{\underline{u}}(t)]^{1/4} 
\exp\left\{-[\qq-\bar{\qq}(t)]\cdot \underline{\underline{u}}(t)\,[\qq-\bar{\qq}(t)]\right\}\equiv \eee^{-S}
\label{eq120}
\ee
where $ \underline{\underline{u}} $ is a real symmetric 3$ \times 3 $ matrix, $ \bar{\qq} $ is a real three-component vector, the coordinates $ q_\alpha = p_\alpha $ and $ q_\beta = p_\beta $ are in Fourier space ({eigenbasis} of the quadrature $ P $) and the coordinate $ q_c = x_c $ is in the ``position" space ({eigenbasis} of the quadrature $ X $). The only trick here was to choose as the metastability exchange jump operator $ C_\beta = \sqrt{\gamma_\beta} \, \ii \beta $; this choice of phase, which of course does not change the {quantum master equation} (\ref{eq030t}), remains legitimate for the evolution conditioned on the homodyne detection of the field because the {metastability jumps} are not measured. In the mixed representation of the wave function (\ref{eq120}), the Hamiltonian $ H $ is then {purely} imaginary and the jump operators are real, hence the real ansatz (\ref{eq120}). \footnote{For example, $ \ii \beta = \ii (X_\beta+\ii P_\beta) $ is represented in {momentum} by the real operator $-\partial_{p_\beta}/2-p_\beta $, and $ \beta^\dagger \beta $ by $-\partial_{p_\beta}^2/4+p_\beta^2-1/2 $.} 

To get the equations of motion on $ \underline{\underline{u}} $ and $ \bar{\qq} $, we calculate in two different ways the relative variation $ \dd \phi (\qq, t)/\phi (\qq, t) $ of the wave function, on the one hand by connecting it to the variation $ \dd S $ of the quantity $ S $ in (\ref{eq120}), separated into a deterministic part $ \dd S_{\! d} $ and {a noisy part} $ \dd S_{\! b} $, on the other hand by {inserting ansatz} (\ref{eq120}) in the stochastic equation (\ref{eq076}). By identifying the deterministic parts and the noisy parts of the two resulting forms, we obtain 
\bea
\label{eq121a}
-\dd S_{\!b} &\!\!\!=\!\!\!& \gamma_\beta^{1/2} \left(\frac{1}{2}\partial_{q_\beta} S -q_\beta+\bar{q}_\beta\right)\dd\zeta_\beta 
-\kappa^{1/2} \left(\frac{1}{2}\partial_{q_c} S -q_c +\bar{q}_c\right) \dd\zeta_c \\
-\dd S_{\!d} +\frac{1}{2} (\dd S_{\!b})^2 &\!\!\!=\!\!\!& (\Omega_\alpha q_\alpha +\Omega_\beta q_\beta) \frac{\dd t}{2} \partial_{q_c} S 
- \frac{\gamma_\beta \dd t}{2} \left\{ q_\beta^2-\frac{1}{2} +\frac{1}{4} \left[ \partial_{q_\beta}^2 S -\left(\partial_{q_\beta} S\right)^2\right] +\bar{q}_\beta \left(\partial_{q_\beta} S -2 q_\beta\right) +\bar{q}_\beta^2\right\} \nonumber \\
&\!\!\! \!\!\!&  -\frac{\kappa\dd t}{2} \left\{ q_c^2-\frac{1}{2} +\frac{1}{4} \left[ \partial_{q_c}^2 S -\left(\partial_{q_c} S\right)^2\right] +\bar{q}_c \left(\partial_{q_c} S -2 q_c\right) +\bar{q}_c^2\right\}
\label{eq121b}
\eea
It remains {to insert in} (\ref{eq121b}) the expression of $ \dd S_{\! b} $ taken from (\ref{eq121a}), by applying Ito's rule of replacing the squares of the noises by their mean, then identifying the terms of degree 2 in{ $ \qq-\bar{\qq} $} to obtain the {purely deterministic equation} linear on $ \underline{\underline{u}} $: \footnote{We notice that the quadratic terms in $ \underline{\underline{u}} $ {in the right-hand side} of (\ref{eq121b}) cancel with those of $ (\dd S_{\! b})^2/2 $ {in the left-hand side}.} 
\be
\begin{array}{lll}
\dd u_{\alpha\alpha} = -\Omega_\alpha \dd t\, u_{\alpha c} & \displaystyle\dd u_{\alpha\beta} = -\frac{\dd t}{2} (\gamma_\beta u_{\alpha\beta} + \Omega_\beta u_{\alpha c} +\Omega_\alpha u_{\beta c}) & \displaystyle\dd u_{\alpha c} = -\frac{\dd t}{2} (\kappa u_{\alpha c} + \Omega_\alpha u_{cc}) \\ 
&&\\
\dd u_{\beta\beta}= -\Omega_\beta \dd t\, u_{\beta c} + \gamma_\beta \dd t (1-u_{\beta\beta}) & \displaystyle\dd u_{\beta c} = -\frac{\dd t}{2} [(\gamma_\beta+\kappa)u_{\beta c} +\Omega_\beta u_{cc}] & \dd u_{cc}=\kappa \dd t (1- u_{cc})
\end{array}
\label{eq122}
\ee
and the terms of degree 1 in{ $ \qq-\bar{\qq} $} to obtain the stochastic linear equation {for} $ \bar{\qq} $: 
\be
\dd\bar{\qq} = \frac{1}{2} \begin{pmatrix} 0 & 0 & 0 \\
0 & -\gamma_\beta & 0 \\
\Omega_\alpha & \Omega_\beta & -\kappa 
\end{pmatrix}
\dd t \, \bar{\qq}  + 
\frac{1}{2}[\mbox{Id}-\underline{\underline{c}}(t)] \begin{pmatrix}
0 \\
\gamma_\beta^{1/2} \dd\zeta_\beta(t) \\
-\kappa^{1/2} \dd\zeta_c(t)
\end{pmatrix}
\label{eq123}
\ee
Needless to say, $ \bar{\qq} $ is the vector of the {quantum averages} of the variables $ q $ in state vector (\ref{eq120}); \footnote{We can therefore recover {equation} (\ref{eq123}) from the stochastic equation deduced from (\ref{eq076}) on the expectation value of an observable $ O $, 
$\dd\langle O\rangle=(\dd t/\ii\hbar) \langle [O,H]\rangle +(\dd t/2) \sum_m \langle C_m^\dagger [O,C_m]+\mbox{h.c.}\rangle + \sum_m [\langle O C_m+\mbox{h.c.}\rangle -\langle C_m+C_m^\dagger\rangle \langle O\rangle]\dd\zeta_m$,
{where $\langle\ldots\rangle$ is taken in state $|\phi(t)\rangle$,} by specializing it to the cases $ O = P_\alpha $, $ O = P_\beta $ and $ O = X_c $.} in addition, we have introduced the notation $ \underline{\underline{c}} $ for the inverse matrix of $ \underline{\underline{u}} $, which is none other than the quantum covariance matrix of $ q $ up to a numerical factor. We therefore have: 
\be
\langle\phi(t)|q_i|\phi(t)\rangle =\bar{q}_i(t)\quad\mbox{and}\quad\langle\phi(t)|q_i q_j|\phi(t)\rangle = \bar{q}_i(t)\bar{q}_j(t)+\frac{1}{4} c_{ij}(t) \quad \forall i,j\in\{\alpha,\beta,c\}\quad\mbox{with}\quad \underline{\underline{c}}(t)=[\underline{\underline{u}}(t)]^{-1}
\label{eq124}
\ee
The differential system (\ref{eq122}) is easily integrated for the initial condition $ \underline{\underline{u}} (0) = \mathrm{Id} $: 
\bea
\label{eq125a}
u_{\alpha\alpha}(t)& = &1+\frac{\Omega_\alpha ^2 t}{\kappa } -\frac{2 \Omega_\alpha ^2}{\kappa ^2} \left(1-\eee^{-\kappa  t/2}\right) \\
\label{eq125b}
u_{\alpha\beta}(t) &=& \frac{\Omega_\alpha\Omega_\beta}{\gamma_\beta} \left(\frac{1}{\gamma_\beta+\kappa }+\frac{1}{\kappa }\right) \left(1-\eee^{-\gamma_\beta t/2}\right)+\frac{\Omega_\alpha\Omega_\beta}{\kappa  (\kappa -\gamma_\beta)} \left(\eee^{-\kappa  t/2}-\eee^{-\gamma_\beta t/2}\right)+\frac{\Omega_\alpha\Omega_\beta}{\kappa  (\gamma_\beta +\kappa )} \left(\eee^{-(\gamma_\beta +\kappa )t/2}-\eee^{-\gamma_\beta t/2}\right)  \\
\label{eq125c}
u_{\alpha c}(t) &=& -\frac{\Omega_\alpha}{\kappa }  \left(1-\eee^{-\kappa  t/2}\right) \\
\label{eq125d}
u_{\beta\beta}(t) &=& 1 + \frac{\Omega_\beta ^2}{\gamma_\beta  (\gamma_\beta +\kappa )} \left(1-\eee^{-\gamma_\beta  t}\right) -\frac{2 \Omega_\beta ^2}{\kappa ^2-\gamma_\beta^2} \left(\eee^{-\gamma_\beta t}-\eee^{-(\gamma_\beta +\kappa )t/2}\right) \\
\label{eq125e}
u_{\beta c}(t) &=&  -\frac{ \Omega_\beta}{\gamma_\beta +\kappa }  \left(1-\eee^{-(\gamma_\beta +\kappa )t/2}\right) \\
\label{eq125f}
u_{cc}(t) &=& 1
\eea
{It would have been different if we had taken as unknown the covariance matrix $\underline{\underline{c}}(t)$, which obeys a Riccati nonlinear differential system \cite{RiccatiMabuchi}.}
Since $ \bar{\qq} $ describes a Brownian motion (partially damped because the friction matrix in (\ref{eq123}) has eigenvalues $ 0 $, $ \gamma_\beta/2 $ and $ \kappa/2 $), and since the homodyne signal averaged over the time interval $ [0, t] $ $ \sigma $ is deduced by integration, these random variables have a Gaussian statistic and we can reproduce the reasoning of {section} \ref{sec:3modesTo1Homo}. We find for the conditional mean and variance of the quadrature $ P_a $ of the nuclear spin knowing that $ \sigma = \mathcal{S} $ {(this variance determines the metrological gain (\ref{eq017})):}
\bea
\label{eq127a}
\!\!\!\!\langle P_a\rangle_{\sigma=\mathcal{S}} &\!\!\!\!=\!\!\!\!& \frac{\langle \sigma(t)\bar{P}_a(t)\rangle_{\rm stoch}}{\langle\sigma^2(t)\rangle_{\rm stoch}} \mathcal{S} \\
\label{eq127b}
\!\!\!\!\mbox{Var}_{\sigma=\mathcal{S}}(P_a) &\!\!\!\!=\!\!\!\!& \frac{1}{4}\left[\frac{\Omega_\beta^2}{\Omega^2}c_{\alpha\alpha}(t)+\frac{\Omega_\alpha^2}{\Omega^2}c_{\beta\beta}(t)-2\frac{\Omega_\alpha\Omega_\beta}{\Omega^2} c_{\alpha\beta}(t)\right]+\langle \bar{P}_a^2(t)\rangle_{\rm stoch} - \frac{\langle \sigma(t)\bar{P}_a(t)\rangle_{\rm stoch}^2}{\langle\sigma^2(t)\rangle_{\rm stoch}} = \frac{1}{4} - \frac{\langle \sigma(t)\bar{P}_a(t)\rangle_{\rm stoch}^2}{\langle\sigma^2(t)\rangle_{\rm stoch}}
\eea
The expression in brackets in {equation} (\ref{eq127b}) is the matrix element of $ \underline{\underline{c}} (t) $ in the coordinate vector $ (\Omega_\beta/\Omega,-\Omega_\alpha/\Omega, 0) $ of direction $ a $ in the rotated basis. The first term {in the middle-hand side} is therefore, {up to a factor $4$,} the quantum variance of $ P_a $ in the stochastic state $ \phi (t) $, depending on time but, let us recall, independent of the particular realization of $ \phi (t) $. The simplified expression {in the right-hand side} follows from the property (\ref{eq032}) on the unconditional mean $ \langle P_a^2 \rangle (t) = 1/4 $ and from the chain of equalities 
\be
\label{eq128}
\langle P_a^2\rangle(t)= \langle\ \langle\phi(t)|P_a^2|\phi(t)\rangle\ \rangle_{\rm stoch} = \langle\ \langle\phi(t)|P_a^2|\phi(t)\rangle -\langle\phi(t)|P_a|\phi(t)\rangle^2 + \langle\phi(t)|P_a|\phi(t)\rangle^2\ \rangle_{\rm stoch}= \langle\mbox{Var}_{\phi(t)} P_a\rangle_{\rm stoch} + \langle \bar{P}_a^2(t)\rangle_{\rm stoch}
\ee
To determine the variance and covariance of the random variables $ \bar{P}_a (t) $ and $ \sigma (t) $, it remains to calculate their amplitudes on the stochastic processes $ \dd \zeta_\beta (t ') $ and $ \dd \zeta_c (t ') $, formally integrating {equation} (\ref{eq123}) by the {method of variation of constants} for $ \bar{P}_a $ and $ \bar{X}_c $, and proceeding as in {equation} (\ref{eq107}) for $ \sigma $: 
\bea
\label{eq129a}
\hspace{-1.3cm}&&p_\beta(t,t') =-\frac{1}{2} \gamma_\beta^{1/2} \left\{\frac{\Omega_\beta}{\Omega}c_{\alpha\beta}(t')+\frac{\Omega_\alpha}{\Omega}[1-c_{\beta\beta}(t')]\eee^{-\gamma_\beta (t-t')/2}\right\} \\
\hspace{-1.3cm}&&p_c(t,t') = \frac{1}{2}\kappa^{1/2} \left\{\frac{\Omega_\beta}{\Omega} c_{\alpha c}(t')-\frac{\Omega_\alpha}{\Omega} c_{\beta c}(t') \eee^{-\gamma_\beta (t-t')/2}\right\} \\
\hspace{-1.3cm}&&\sigma_\beta(t,t') = \frac{(\gamma_\beta\kappa)^{1/2}}{2 t}\left\{ -c_{\alpha\beta}(t')[t-t'-f_\kappa(t-t')]\frac{\Omega_\alpha}{\kappa}+
[1-c_{\beta\beta}(t')][f_{\gamma_\beta}(t-t')-f_\kappa(t-t')]\frac{\Omega_\beta}{\kappa-\gamma_\beta} -c_{\beta c}(t') f_\kappa(t-t')\right\} \\
\hspace{-1.3cm}&& \sigma_c(t,t')=\frac{1}{2 t}-\frac{\kappa}{2t}\left\{-c_{\alpha c}(t')[t-t'-f_\kappa(t-t')] \frac{\Omega_\alpha}{\kappa}-c_{\beta c}(t')[f_{\gamma_\beta}(t-t')-f_\kappa(t-t')]\frac{\Omega_\beta}{\kappa-\gamma_\beta} +[1-c_{cc}(t')] f_\kappa(t-t')\right\}
\label{eq129d}
\eea
where $f_\lambda(\tau)\equiv [1-\exp(-\lambda\tau/2)]/(\lambda/2)$.
We obtain: 
\bea
\langle \sigma(t)\bar{P}_a(t)\rangle_{\rm stoch} &\!\!\!=\!\!\!& \int_0^t \dd t'\, [p_\beta(t,t') \sigma_\beta(t,t')+p_c(t,t')\sigma_c(t,t')] \quad ; \quad
\langle \sigma^2(t)\rangle_{\rm stoch} = \int_0^t \dd t'\, [\sigma_\beta^2(t,t')+\sigma_c^2(t,t')] \quad ;\nonumber \\
\langle \bar{P}_a^2(t)\rangle_{\rm stoch} &\!\!\!=\!\!\!& \int_0^t \dd t'\, [p_\beta^2(t,t')+p_c^2(t,t')]
\label{eq130}
\eea
We deduce from these results the {long time limits} \footnote{Let us give some results and {intermediate considerations}. (i) While $ c_{\beta \beta} (t ') $, $ c_{\beta c} (t') $ and $ c_{cc} (t ') $ have a finite limit when $ t' \to+\infty $ [we will need 
$c_{\beta\beta}(+\infty)=(1+\rho)^{-1}$, $c_{\beta c}(+\infty)=\Omega_\beta/((\gamma_\beta+\kappa)(1+\rho))$ with $\rho=\Omega_\beta^2\kappa/(\gamma_\beta (\gamma_\beta+\kappa)^2)$
], $ c_{\alpha \alpha} (t ') $, $ c_{\alpha \beta} (t') $ and $ c_{\alpha c} (t ') $ tend to zero as $ 1/t' $. (ii) In an integral over $ t '$ containing the exponential factor $ \exp [-\gamma_\beta (t-t')/2] $ or its square, we can replace the function which multiplies it by its limit in $ t '=+\infty $. (iii) For any uniformly bounded function $ w (t, t ') $, we can show for $ \nu \in \{\beta, c \} $ that 
$\int_0^t \dd t' [(t-t') c_{\alpha\nu}(t')+w(t,t')]^2/t^2\to \int_0^{+\infty} \dd t' c_{\alpha\nu}^2(t')$.
(iv) We then obtain the asymptotic limits 
$\langle P_a^2(t)\rangle_{\rm stoch}\to (\Omega_{\beta}/2\Omega)^2\mathcal{I}+(\Omega_\alpha/2\Omega)^2\rho/(1+\rho)$, $\langle\sigma^2(t)\rangle_{\rm stoch}\to (\Omega_\alpha^2/4\kappa)\mathcal{I}$, $\langle\sigma(t)\bar{P}_a(t)\rangle_{\rm stoch}\to (\Omega_\alpha\Omega_\beta/4\Omega\kappa^{1/2}) \mathcal{I}$ where $\mathcal{I}\equiv \int_0^{+\infty} \dd t' [\gamma_\beta c_{\alpha\beta}^2(t')+\kappa c_{\alpha c}^2(t')]$.
We thus deduce (\ref{eq131}) from (\ref{eq127a}) and from the first equality in (\ref{eq127b}), without needing to know the value of $ \mathcal{I} $. We derive from the second equality in (\ref{eq127b}) the result $ \mathcal{I} = 1 $, which we can also deduce from the equation of motion 
$\dd c_{\alpha\alpha}/\dd t=-\gamma_\beta c_{\alpha\beta}^2-\kappa c_{\alpha c}^2$
integrated between $ t = 0 $ and $ t =+\infty $. 
} 
\be
\label{eq131}
\langle{P}_a\rangle_{\sigma=\mathcal{S}} \underset{t\to+\infty}{\to}  \left(\frac{\gamma_m}{\gamma_f+\gamma_m}\right)^{1/2}\frac{\mathcal{S}}{\Gamma_{\rm ex}^{1/2}} \quad ; \quad \mbox{Var}_{\sigma=\mathcal{S}}({P}_a) \underset{t\to+\infty}{\to} \frac{1}{4} \frac{\gamma_f}{\gamma_f+\gamma_m}
\ee
with which the predictions (\ref{eq108}) of the one-mode model, however obtained within the weak coupling limit (\ref{eq033}), are in perfect agreement. 

As an application of our analytical solution of the three-mode model, let the rate $ \Gamma_{\rm ex} $ tend to zero at fixed reduced time $ \tau = \Gamma_{\rm ex} t $ while maintaining (unlike the {one-mode model}) the ratio $ \Gamma_{\rm ex}/\gamma_f $ to a non-infinitesimal constant value. The physical motivation is clear: in the {planned experiments} \cite{letter}, $ \gamma_f $ and $ \Gamma_{\rm ex} $ are of the same order of magnitude but are really much smaller than $ \gamma_m $ and $ \kappa $ (by factors $ \approx 10^{-6} $ and $ 10^{-9} $). We find in this limit: \footnote{\label{note133} In practice, it suffices to make $ \Omega_\alpha $ tend to zero at $ \tau = \Gamma_{\rm ex} t> 0 $, $ \Omega_\beta $, $ \gamma_\beta $ and $ \kappa $ fixed. In particular, this makes all exponential transients disappear {in equations (\ref{eq125a})-(\ref{eq125f})}. To simplify the calculations, it is useful to introduce the quantity 
$\rho=\Omega^2\kappa/[2\gamma_m (\kappa+2\gamma_m)^2]$
so that 
{$\rho=(\Gamma_{\rm ex}/2\gamma_f)(1+2\gamma_m/\kappa)^{-2}$}
in the limit $ \gamma_f \to 0 $. 
} 
\be
{{\boxed{\langle{P}_a\rangle_{\sigma=\mathcal{S}} \sim \frac{\Gamma_{\rm sq}t}{1+\Gamma_{\rm sq}t}\frac{\mathcal{S}}{\Gamma_{\rm ex}^{1/2}} \quad \mbox{and}\quad \mbox{Var}_{\sigma=\mathcal{S}}({P}_a) \sim \frac{1}{4} \frac{1}{1+\Gamma_{\rm sq}t}}}}
\label{eq132}
\ee
where we have introduced the true or generalized {squeezing} rate 
\be
{\boxed{\Gamma_{\rm sq}\equiv \left(\frac{1}{\Gamma_{\rm ex}}+\frac{2}{\gamma_f}\right)^{-1}}}
\label{eq135}
\ee
We find the natural scaling of the signal by $ \Gamma_{\rm ex}^{1/2} $ already observed in the one-mode model and the same functional forms in time, but we lose all relation of proportionality of type (\ref{eq197}) {between integrated signal and quadrature average in a given realization}, the conditional variance of $ \bar{P}_a $ now being $ \nequiv 0 $. \footnote{We have indeed $ \mbox{Var}_{\sigma = \mathcal{S}} (\bar{P}_a) \sim \Gamma_{\rm ex} t/[4 (1+\Gamma_{\rm ex} t)]-\Gamma_{\rm sq} t/[4 (1+\Gamma_{\rm sq} t)] $.} We represent in figure \ref{fig9}c the variation with reduced time $ \gamma_f t $ of the conditional mean and variance (\ref{eq132}) for different values of the ratio $ r = 2 \Gamma_{\rm ex}/\gamma_f $. We notice that the {squeezing} process is all the faster as $ r $ is {larger}, and that it saturates to a limiting behavior. This was predictable, because $ \Gamma_{\rm sq}$ is an increasing function of $ r $ with finite limit; at a fixed time, the conditional mean (in units of $ \mathcal{S}/\Gamma_{\rm ex}^{1/2} $) is therefore an increasing function and the conditional variance a decreasing function of $ r $, as seen in figure \ref{fig9}c. More precisely, in the weak coupling limit $ \Omega \to 0 $, where $ r \to 0 $, the generalized {squeezing} rate is equivalent to the rate $ \Gamma_{\rm ex} $ {of creation of excitations}, in agreement with the {one-mode model}, and within the limit $ r \to+\infty $, it saturates to the value $ \gamma_f/2 $. We cannot therefore {squeeze} faster than at the rate $ \gamma_f $, which is not surprising: we cannot hope to reduce the fluctuations in nuclear spin before each atom in the ground state has undergone on average {at least one} metastability exchange collision, 
{the effective rate $\gamma_f$ being in practice of the same order of magnitude as the individual rate $1/T$ in equation (\ref{eq006}) except in the case of extreme polarization, see figure \ref{fig3}a.}

\subsubsection{Effect of decoherence} 
\label{sec:edld} 
To be complete, we take into account, in the homodyne {squeezing} scheme, the finite lifetime $ (2 \gamma_0)^{-1} $ of the {metastable atoms}, which de-excite when they reach the cell walls after diffusive {motion} in the {gas}. To this end, we add a jump operator $ \sqrt{2 \gamma_0} b $ to the three-mode {quantum master equation} (\ref{eq030}). As the part other than Hermitian Hamiltonian remains quadratic in the quadratures of the modes, it can be put in reduced form by an appropriate rotation of the atomic modes, as we had already done in {section} \ref{sec:ana_omm}: one simply has to expand $ (a, b) $ in the orthonormal {eigenbasis} of the rate matrix 
\be
\underline{\underline{\Gamma}}=\begin{pmatrix}
2\gamma_f & -2\sqrt{\gamma_f\gamma_m} \\
-2\sqrt{\gamma_f\gamma_m} & 2(\gamma_0+\gamma_m)
\end{pmatrix}
\label{eq150}
\ee
with operator-valued coefficients $ \alpha $ and $ \beta $. The direction $ \beta $ remains that of the maximum eigenvalue $ \gamma_\beta $ of $ \underline{\underline{\Gamma}} $, and $ \alpha $ that of the minimum eigenvalue $ \gamma_\alpha $, now {nonzero}. This leads to the {quantum master equation} 
\be
\boxed{\frac{\dd\rho}{\dd t} = \frac{1}{\ii\hbar} \left[\hbar(\Omega_\alpha P_\alpha + \Omega_\beta P_\beta) P_c ,\rho \right] + \kappa \left( c\rho c^\dagger -\frac{1}{2}\{ c^\dagger c,\rho\}\right) + \gamma_\alpha \left( \alpha \rho \alpha^\dagger -\frac{1}{2}\{ \alpha^\dagger \alpha,\rho\}\right) + \gamma_\beta \left( \beta \rho \beta^\dagger -\frac{1}{2}\{ \beta^\dagger \beta,\rho\}\right)}
\label{eq151}
\ee
The new expression for {Faraday frequencies} $ \Omega_\alpha, \Omega_\beta $ and rates $ \gamma_\alpha, \gamma_\beta $ can be found {in} \ref{app:tmd}, which also gives the analytical expression of the mean and of the variance of the quadrature $ P_a $ of the nuclear spin conditioned on the integrated homodyne signal, in all generality. We restrict ourselves here to the physically useful limit $ \gamma_0 \ll \gamma_m $ (we still have $ \gamma_f <\gamma_m $). To lowest order in $ \gamma_0 $, the coefficients $ \Omega_\alpha $, $ \Omega_\beta $ and $ \gamma_\beta $ remain unchanged, and we have 
\be
\gamma_\alpha \simeq \frac{2\gamma_0 \gamma_f}{\gamma_m+\gamma_f}
\label{eq152}
\ee
which is the reduced rate of decoherence in the hybridized nuclear spin. Moreover, we place ourselves in the limit (\ref{eq033}), with $ \gamma_\alpha = O (\Gamma_{\rm ex}) $, which allows to evaluate the effect of decoherence using the one-mode model, {which is obtained in the same way as} in section \ref{sec:ana_omm}. The stochastic equation (\ref{eq078}) is completed as follows, 
\begin{equation}
\dd|\phi(t) \rangle = -\frac{\Gamma_{\rm ex}\dd t}{2} (P_\alpha-\bar{P}_\alpha)^2  |\phi(t) \rangle + \sqrt{\Gamma_{\rm ex}} \dd\zeta_s(t) (P_\alpha-\bar{P}_\alpha) |\phi(t) \rangle -\frac{\gamma_\alpha \dd t}{2} (\alpha^\dagger\alpha+2 \ii \bar{P}_\alpha \alpha +\bar{P}_\alpha^2) |\phi(t)\rangle + \sqrt{\gamma_\alpha} \dd\zeta_\alpha(t) ( \ii \alpha +\bar{P}_\alpha) |\phi(t)\rangle
\label{eq153}
\end{equation}
We have taken care to choose $ \gamma_\alpha^{1/2} \ii \alpha $ as the jump operator of the {effective decoherence} (the justification is the same as in section \ref{sec:sdmatm}, decoherence jumps are not measured), which allows the equation to be solved by the same real Gaussian ansatz (\ref{eq080}). This time we find 
\footnote{{In the $\epsilon\ll 1$ regime, the long-time limit of the variance of $P_\alpha$ in a single realization depends strongly on the choice of phase in the effective-decoherence jump operator, which emphasizes the unphysical character of this variance (see note \ref{note977}): if we take $\gamma_\alpha^{1/2} \alpha$ as the jump operator, we find that $\mbox{Var}_\phi P_\alpha\to \epsilon^{1/2}/4$ \cite{letter} instead of $\mbox{Var}_\phi P_\alpha\to 1/4\epsilon$ as in equation (\ref{eq154a}). More generally, the choice $\gamma_\alpha^{1/2}\exp(\ii\theta)\,\alpha$, 
$-\pi/2<\theta<\pi/2$, leads to the Riccati equation for the parameter $u$ of the (now complex) Gaussian ansatz
$\dd u = \Gamma_{\rm ex}\dd t +\gamma_\alpha \dd t \{\exp(2\ii\theta) u +[1-\exp(2\ii\theta)]/2- u^2[1+\exp(2\ii\theta)]/2\}$
so that $\mbox{Var}_\phi P_\alpha\to (1/4)\epsilon^{1/2}\sqrt{\cos\theta}\cos(\theta/2)$ in a time $\tau\sim 1/\sqrt{2\epsilon [1+\exp(2\ii\theta)]}$; the power law in $\epsilon$ obtained for $\theta=0$ is thus the rule, that obtained for $\theta=\pi/2$ is the exception.}}
\bea
\label{eq154a}
\dd u &\!\!\!=\!\!\!& [\Gamma_{\rm ex} +\gamma_\alpha (1-u)]\dd t \quad\Longrightarrow\quad {u(\tau)}= 1 +\frac{1-\exp(-\epsilon\tau)}{\epsilon} \\
\dd\bar{P}_\alpha &\!\!\!=\!\!\!& -\frac{1}{2} \gamma_\alpha \bar{P}_\alpha \dd t +\frac{\sqrt{\Gamma_{\rm ex}}\dd \zeta_s+\sqrt{\gamma_\alpha}(u-1)\dd\zeta_\alpha}{2u}
\label{eq154b}
\eea
where we have set $ \tau = \Gamma_{\rm ex} t $ and $ \epsilon = \gamma_\alpha/\Gamma_{\rm ex} $. The same Gaussianity arguments as {in} \ref{sec:3modesTo1Homo} section lead to the same dependencies in the signal $ \mathcal{S} $ of the conditional mean and variance, \footnote{We have simplified {expression} (\ref{eq155b}) using the identity $ [4 u (\tau)]^{-1}+\langle \bar{P}_\alpha^2 \rangle_{\rm stoch} = 1/4 $, which results as in equation (\ref{eq128}) from the fact that the unconditional mean $ \langle P_\alpha^2 \rangle = 1/4 $, even in the presence of decoherence.} 
\bea
\label{eq155a}
\langle P_\alpha\rangle_{\sigma=\mathcal{S}} &\!\!\!=\!\!\!& m(\tau) \frac{\mathcal{S}}{\sqrt{\Gamma_{\rm ex}}} \quad\mbox{with}\quad m(\tau)= \sqrt{\Gamma_{\rm ex}} \frac{\langle \sigma(t) \bar{P}_\alpha(t)\rangle_{\rm stoch}}{\langle \sigma^2(t)\rangle_{\rm stoch}} \\
\label{eq155b}
\mbox{Var}_{\sigma=\mathcal{S}}(P_\alpha) &\!\!\!=\!\!\!& \mathcal{V}(\tau)\quad\mbox{with}\quad \mathcal{V}(\tau) = \frac{1}{4} - \frac{\langle\sigma(t)\bar{P}_\alpha(t)\rangle_{\rm stoch}^2}{\langle\sigma^2(t)\rangle_{\rm stoch}}
\eea
and the variance and {covariance taken over} the stochastic processes {$\dd\zeta_s\equiv\dd\zeta_c$} and $ \dd \zeta_\alpha $, 
\bea
\label{eq156a}
\frac{\langle\sigma^2\rangle_{\rm stoch}}{\Gamma_{\rm ex}} &\!\!\!=\!\!\!& \int_0^\tau \frac{\dd\tau'}{\tau^2} \left\{
\left[{\frac{1}{2}}+\frac{1-\eee^{\epsilon(\tau'-\tau)/2}}{\epsilon u(\tau')}\right]^2 +
\frac{\left[u(\tau')-1\right]^2}{u^2(\tau')}  \frac{\left[1-\eee^{\epsilon(\tau'-\tau)/2}\right]^2}{\epsilon}
\right\}= {\frac{\epsilon\tau-2(1-\eee^{-\epsilon\tau/2})}{\epsilon^2\tau^2} + \frac{1}{4\tau}} \\
\label{eq156b}
\frac{\langle\sigma\bar{P}_\alpha\rangle_{\rm stoch}}{\sqrt{\Gamma_{\rm ex}}} &\!\!\!=\!\!\!& \int_0^\tau \frac{\dd\tau'}{\tau} \frac{\eee^{\epsilon(\tau'-\tau)/2}}{2 u(\tau')} \left\{{\frac{1}{2}}+\frac{1-\eee^{\epsilon(\tau'-\tau)/2}}{\epsilon u(\tau')} + \frac{\left[u(\tau')-1\right]^2}{u(\tau')}\left[1-\eee^{\epsilon(\tau'-\tau)/2}\right]\right\}= {\frac{1-\eee^{-\epsilon\tau/2}}{2\epsilon\tau}}
\eea
These expressions allow to easily evaluate the effect of decoherence on spin {squeezing} {through the metrological gain (\ref{eq017})}, see the dashed lines in {figure} \ref{fig9}a. For the practical case of a weak decoherence $ \epsilon \ll 1 $ and a time short compared to $ 1/\gamma_\alpha $, they can be expanded {to first order} in $ \epsilon $: 
\be
{\boxed{m(\tau)=\frac{\tau}{1+\tau} -\epsilon \frac{(\tau+3)\tau^2}{12(\tau+1)^2}+O(\epsilon^2\tau^2)\quad ;\quad \mathcal{V}(\tau)=\frac{1}{4(\tau+1)}+\epsilon \frac{(\tau+3/2)\tau^2}{12(\tau+1)^2}+O(\epsilon^2\tau^2)}}
\label{eq157}
\ee
We then deduces that the optimal {squeezing} on $ P_\alpha $ is obtained at a time $ t_{\rm opt} \sim (3/\Gamma_{\rm ex} \gamma_\alpha)^{1/2} $ and corresponds to a conditional variance $ \mathcal{V}_{\rm opt} \sim (\gamma_\alpha/12 \Gamma_{\rm ex})^{1/2} $.{ Note that in studies of spin {squeezing} {of alkaline gases in cavities}, we often introduce the cooperativity  $ C $ of the coupled atom-field system, defined as the square of the coupling {frequency} divided by the decay rates of the coupled states \cite{Vuletic}. In this sense, the cooperativity of the hybridized nuclear spin-field system is 
equal to
\be
\label{eq158}
C\equiv\frac{\Omega_\alpha^2}{\kappa\gamma_\alpha} = \frac{\Gamma_{\rm ex}}{\gamma_\alpha}\simeq \frac{\Omega^2}{2\gamma_0\kappa}
\ee
so that we recover the scaling law of power $-1/2 $, usual in alkalis, relating the optimal spin variance to $ C $ \cite{Vuletic}}. More generally, the decoherence has a weak effect on the nuclear spin {squeezing} as long as we stay at short times in front of $ t_{\rm opt} $. The reader will find at the end of \ref{app:tmd} an extension of these scaling laws beyond the one-mode model, i.e. for an arbitrary, not infinitesimal ratio $ \Gamma_{\rm ex}/\gamma_f $; this was retained in the abstract of the article. {The link between $ \mathcal{V}_{\rm opt} $ and cooperativity (\ref{eq158}) is then broken.} 

{\noindent {\bf Acknowledgements~:} Alice Sinatra thanks Franck Lalo\"e for helpful discussions. All authors except Yvan Castin are funded by the Horizon 2020 European research and innovation project macQsimal number 820393. Matteo Fadel thanks the Research Fund of the University of Basel for Excellent Junior Researchers.}

\appendix 
\section{{Semi-classical treatment} and reduction to three coupled spins} 
\label{app:semiclassical} 
Here we give the nonlinear equations that describe the dynamics of the system in semi-classical theory, and we linearize them for small fluctuations around a partially polarized stationary solution. 

\paragraph{Nonlinear semi-classical equations} 
Starting from the considerations and notations of {section} \ref{sec:view}, we take the average of the Heisenberg equations of motion in the quantum state of the system and perform the {decorrelation approximation} (called semi-classical in quantum optics) 
$\langle AB \rangle \simeq \langle A \rangle  \langle B \rangle $ 
where $ A $ and $ B $ are two operators, to obtain the following nonlinear evolution equations {for} the expectation values of $ \Vec{S} $ the Stokes spin of the {cavity field}, $ \Vec{I} $ the collective nuclear spin in the ground state, $ \Vec{J} $ and $ \Vec{K} $ the collective spins associated with the manifolds $ F = 3/2 $ and $ F = 1/2 $ in the metastable state, and $ \Vec{\Vec{Q}} $ the collective alignment tensor in $ F = 3/2 $, of Cartesian components $ Q_{\alpha \beta} $ : 
\begin{align}
\frac{\dd \langle S_x \rangle}{\dd t}&=  -\frac{\kappa}{2} \left(\langle S_x \rangle - \frac{n_{\rm ph}}{2}\right)- \chi \langle K_z \rangle \langle S_y \rangle  && \frac{\dd \langle S_y\rangle}{\dd t} =  -\frac{\kappa}{2} \langle S_y\rangle + \chi \langle K_z \rangle\langle S_x \rangle && \frac{\dd \langle S_z\rangle }{\dd t} =   -\frac{\kappa}{2} \langle S_z\rangle \label{DerivativesNLTotSx} \\ 
\frac{\dd \langle K_x \rangle}{\dd t}&=  \left. \frac{\dd \langle K_x \rangle}{\dd t} \right |_{\rm ME}  - \chi \langle K_y\rangle \langle S_z \rangle && \frac{\dd \langle K_y \rangle}{\dd t} =  \left. \frac{\dd \langle K_y \rangle}{\dd t} \right |_{\rm ME} + \chi\langle K_x\rangle \langle S_z \rangle &&  \frac{\dd \langle K_z \rangle}{\dd t} =  \left. \frac{\dd \langle K_z \rangle}{\dd t} \right |_{\rm ME} \\
\frac{\dd \langle \vec{J} \rangle}{\dd t}& = \left. \frac{\dd \langle \vec{J} \rangle}{\dd t} \right |_{\rm ME} && \frac{\dd \langle Q_{\alpha \beta}  \rangle}{\dd t} =  \left. \frac{\dd\langle Q_{\alpha \beta}  \rangle}{\dd t} \right |_{\rm ME} && \frac{\dd\langle \Vec{I}\rangle}{\dd t} = \left.\frac{\dd\langle \Vec{I}\rangle}{\dd t}\right |_{\rm ME} \label{eq202c}
\end{align}
The terms proportional to the loss rate $ \kappa $ of the {cavity output mirror} make $ \langle S_x \rangle $ relax towards its stationary value 
$\langle S_x \rangle_s=n_{\rm ph}/2$ 
{driven by} the laser field {polarized along} $ {x} $ injected into the cavity, and the transverse means $ \langle S_y \rangle $ and $ \langle S_z \rangle $ towards zero. The terms proportional to the Faraday coupling $ \chi $ between the cavity mode and the spin $ \Vec{K} $ derive from the Hamiltonian (\ref{eq002}). The contribution of metastability exchange collisions (ME) between {ground-state and metastable} atoms is deduced directly from the {quantum master equation} {for} the {one-atom density operator} of {references} \cite{DupontRoc, LaloeDupontLeduc} by simple multiplication or division by the total number of {ground-state atoms} $ N_{\rm cell} $ or {metastable atoms} $ n_{\rm cell} $ in the cell: \footnote{The collective expectation values are in fact related as follows to the {one-atom expectation values} $ \langle \: \: \rangle_{\rm at} $: 
$    \langle\Vec{I}\rangle = N_{\rm cell} \langle \Vec{I}\rangle_{\rm at} $, $    \langle\Vec{J}\rangle = n_{\rm cell}  \langle \Vec{J}\rangle_{\rm at} $, $ \langle\Vec{K}\rangle = n_{\rm cell}  \langle \Vec{K}\rangle_{\rm at} $, $    \langle\Vec{\Vec{Q}}\rangle = n_{\rm cell}  \langle \Vec{\Vec{Q}}\rangle_{\rm at} $, $ \langle\Vec{\Sigma}\rangle = n_{\rm cell}  \langle \Vec{\Sigma} \rangle_{\rm at} $.
} 
\begin{align}
      \left. \frac{\dd \langle\Vec{K}\rangle}{\dd t} \right |_{\rm ME}&= - \frac{7}{9\tau}\langle\Vec{K}\rangle+ \frac{1}{9\tau}\langle\Vec{J}\rangle  - \frac{1}{9\tau}\frac{n_{\rm cell} }{N_{\rm cell} }\langle\Vec{I}\rangle - \frac{4}{3\tau}\frac{1}{N_{\rm cell} }\langle\Vec{\Vec{Q}}\rangle \cdot \langle\Vec{I}\rangle \label{eq204a}\\
     \left.  \frac{\dd \langle\Vec{J}\rangle}{\dd t} \right |_{\rm ME}&= - \frac{4}{9\tau}\langle\Vec{J}\rangle + \frac{10}{9\tau}\langle\Vec{K}\rangle+ \frac{10}{9\tau}\frac{n_{\rm cell} }{N_{\rm cell} }\langle\Vec{I}\rangle + \frac{4}{3\tau}\frac{1}{N_{\rm cell} }\langle\Vec{\Vec{Q}}\rangle \cdot \langle\Vec{I}\rangle \label{eq204b}\\
     \left.  \frac{\dd \langle Q_{\alpha\beta}\rangle}{\dd t}\right |_{\rm ME} &= - \frac{2}{3\tau}\langle Q_{\alpha\beta}\rangle + \frac{1}{9\tau}\frac{1}{N_{\rm cell} }\left(3\frac{\langle I_\alpha\rangle \langle \Sigma_\beta\rangle + \langle I_\beta\rangle \langle \Sigma_\alpha\rangle}{2} - \delta_{\alpha\beta} \langle \Vec{I}\rangle \cdot \langle \Vec{\Sigma}\rangle\right)\label{eq204c}\\
        \left. \frac{\dd \langle\Vec{I}\rangle}{\dd t}\right |_{\rm ME} &= - \frac{1}{T}\langle\Vec{I}\rangle + \frac{1}{3T}\frac{N_{\rm cell} }{n_{\rm cell} }{(\langle\Vec{J}\rangle - \langle\Vec{K}\rangle)} \label{eq204d}
\end{align}
where 
$\langle \Vec{\Sigma}\rangle = \frac{2}{3} \left[\langle\Vec{J}\rangle + 2\langle\Vec{K}\rangle \right]$ 
is the expectation value of the electron spin in the metastable state. See equations (1.37b), (1.37a), (1.39) and (1.25) of {reference} \cite{LaloeDupontLeduc} (taking into account a difference of a factor $ 6 $ in the definition of the {alignment tensor}), or to equations (VIII.30), (VIII.29), (VIII.32) and (VIII.15) {of reference \cite{DupontRoc}} (by adding a Kronecker factor $ \delta_{\alpha \beta} $ omitted in (VIII.32)). Here $ 1/\tau $ and $ 1/T $, the individual metastability exchange collision rates for an atom in the metastable state and in the ground state, are in the ratio 
$T/\tau=N_{\rm cell}/ n_{\rm cell}$
since, in one unit of time, an equal number of {ground-state and metastable} atoms have undergone an exchange collision \cite{DupontRoc, LaloeDupontLeduc}. 

\paragraph{Partially polarized stationary solution} 
In a polarized stationary state of nuclear polarization $ \eta \in [0,1] $, 
\begin{align}
\langle I_x \rangle_s &= \eta \frac{N_{\rm cell} }{2} \quad\:; \quad \langle I_y \rangle_s=\langle I_z \rangle_s=0 \quad\:; \quad
\langle S_x \rangle_s =\frac{n_{\rm ph}}{2}\quad\:; \quad  \langle S_y \rangle_s = \langle S_z \rangle_s = 0  \label{eq207}
\end{align}
rotational invariance around $ {x} $ axis constrains the mean spins {in the metastable state} to be aligned {along} $ {x} $, and the mean alignment tensor to be diagonal in the Cartesian basis, with equal eigenvalues in $ {y} $ and $ {z} $ directions. The system (\ref{DerivativesNLTotSx})-(\ref{eq202c}) thus admits a stationary solution where the only nonzero expectation values in the metastable state are: 
\begin{align}
\langle K_x \rangle_s &= \frac{\eta}{2}\frac{1 - \eta^2}{3 + \eta^2}n_{\rm cell} \ \ ;\  \  \langle J_x \rangle_s =  \eta \frac{5+\eta^2}{3+\eta^2} n_{\rm cell} \  \ ;\  \  \langle \Sigma_x \rangle_s = \frac{4\eta}{3+\eta^2} n_{\rm cell} \  \ ;\  \  \langle Q_{yy} \rangle_{s} =  \langle Q_{zz} \rangle_{s} = -\frac{1}{2} \langle Q_{xx} \rangle_s = - \frac{\eta}{12}\langle \Sigma_x \rangle_s \label{eq208}
\end{align}

\paragraph{Linearized semi-classical equations} 
We now linearize {equations} (\ref{DerivativesNLTotSx})-(\ref{eq202c}) for classical fluctuations around the stationary solution (\ref{eq207})-(\ref{eq208}) by performing the substitution 
$\langle A \rangle \rightarrow \langle A\rangle_s  +\delta A$
and treating $\delta A $ to first order. By limiting ourselves to the subspace of transverse fluctuations, that is to say to the directions $ \alpha = y, z $ orthogonal to the mean spins, we obtain a closed system: 
\begin{align}
        \frac{\dd }{\dd t}\delta S_\alpha   & = -\frac{\kappa}{2} \delta S_\alpha + \chi\delta_{\alpha y}\langle S_x \rangle_s  \delta K_z  \\
\label{eq210b}
        \frac{\dd }{\dd t}\delta K_\alpha &= - \frac{7}{9\tau}\delta K_\alpha + \frac{1}{9\tau}\delta J_\alpha -\frac{2\eta}{3\tau} \delta Q_{\alpha x} - \frac{1}{9T}\left( 1 + \frac{12}{n_{\rm cell} }\langle Q_{\alpha\alpha}\rangle_s \right)\delta I_\alpha + \chi \delta_{\alpha y} \langle K_x \rangle_s \delta S_z\\
\label{eq210e}
    \frac{\dd }{\dd t}\delta J_{\alpha}   &= - \frac{4}{9\tau}\delta J_{\alpha} + \frac{10}{9\tau}\delta K_{\alpha} +\frac{2\eta}{3\tau} \delta Q_{\alpha x} + \frac{10}{9T}\left( 1 + \frac{6}{5n_{\rm cell} }\langle Q_{\alpha\alpha}\rangle \right)\delta I_{\alpha} \\
\label{eq210f}
        \frac{\dd}{\dd t}\delta Q_{\alpha x}   &=   - \frac{2}{3\tau} \delta Q_{\alpha x} + \frac{\eta}{12\tau}\delta \Sigma_\alpha + \frac{1}{6 T n_{\rm cell}} \langle \Sigma_x \rangle_s \delta I_\alpha  \\
    \frac{\dd}{\dd t}\delta I_{\alpha} &=  - \frac{1}{T}\delta I_{\alpha} + \frac{1}{3\tau}{(\delta J_{\alpha} - \delta K_{\alpha})} \label{eq210} 
\end{align}

\paragraph{Reduction to three coupled collective spins} 
By setting 
$\frac{\dd}{\dd t}\delta J_{\alpha} = 0$
in {equation} (\ref{eq210e}) and 
$\frac{\dd}{\dd t}\delta Q_{\alpha x} = 0$ 
in {equation} (\ref{eq210f}), we adiabatically eliminate the fluctuations of the collective spin $ \vec{J} $ and of the collective alignment tensor whose evolutions are governed by the {metastability exchange} only: 
\begin{equation}
\delta J_\alpha^{\rm adiab} = 2\frac{10+\eta^2}{8-\eta^2} \delta K_\alpha +\frac{12\tau}{T} \frac{5+2\eta^2}{(3+\eta^2)(8-\eta^2)} \delta I_\alpha \quad ;\quad \delta Q_{\alpha x}^{\rm adiab} = \frac{3\eta}{8-\eta^2} \delta K_\alpha + \frac{\tau}{T} \frac{\eta (13+\eta^2)}{(3+\eta^2)(8-\eta^2)} \delta I_\alpha \label{eq212}
\end{equation}
The transfer of adiabatic expressions (\ref{eq212}) in {equations} (\ref{eq210b}) and (\ref{eq210}) on $ \delta K_\alpha $ and $ \delta I_{\alpha} $ leads in the body of the article to the reduced system (\ref{eq:redS})-(\ref{eq005}) coupling the fluctuations of the three spins (\ref{eq003}), where $ \gamma_f $ and $ \gamma_m $, the effective {metastability exchange rates} between the nuclear spin and the spin $ F = 1/2 $ of the metastable, are given by equation (\ref{eq006}). 

\section{Solution of the three-mode model with decoherence for homodyne detection} 
\label{app:tmd} 

Here we give the analytical solution of the three-mode model in the presence of decoherence, see the {quantum master equation} (\ref{eq151}), for an evolution of the system conditioned on a continuous homodyne measurement of the {component polarized along $y$ of the} field {leaking out of the cavity}. The value of the coefficients $ \gamma_\alpha $, $ \gamma_\beta $, $ \Omega_\alpha $ and $ \Omega_\beta $, as well as the annihilation operators $ \alpha $ and $ \beta $, are deduced from a diagonalization of the rate matrix (\ref{eq150}). The rates $ \gamma_\alpha $ and $ \gamma_\beta $ are the eigenvalues in ascending order: 
\be
\label{eq220}
\gamma_{\alpha,\beta} = \gamma_m+\gamma_f+\gamma_0 \mp [(\gamma_m+\gamma_f+\gamma_0)^2-4\gamma_f\gamma_0]^{1/2}
\ee
In terms of the {Faraday frequencies} $ \Omega_\alpha $ and $ \Omega_\beta $, the {corresponding normalized eigenvectors} are {written as} $ (\Omega_\beta/\Omega, \Omega_\alpha/\Omega) $ and $ (-\Omega_\alpha/\Omega, \Omega_\beta/\Omega) $, so that $ \alpha = (\Omega_\beta a+\Omega_\alpha b)/\Omega $ and $ \beta = (\Omega_\beta b-\Omega_\alpha a)/\Omega $ with 
\be
\label{eq221}
\Omega_\alpha = \frac{\Omega(\gamma_f-\gamma_\alpha/2)}{[\gamma_m\gamma_f +(\gamma_f-\gamma_\alpha/2)^2]^{1/2}} \quad ; \quad
\Omega_\beta = \frac{\Omega\sqrt{\gamma_m\gamma_f}}{[\gamma_m\gamma_f +(\gamma_f-\gamma_\alpha/2)^2]^{1/2}}
\ee
with a choice of sign ensuring that $ \alpha \to a $ and $ \beta \to b $ when $ \gamma_f \to 0 $ and reproducing (\ref{eq036}) when $ \gamma_0 \to 0 $. Since the jump operator{ $ C_\alpha \propto \alpha $} describes unmeasured processes, we can, { as we did for $ C_\beta $,} take it of the form $ \sqrt{\gamma_\alpha} \ii \alpha $ and reuse the real Gaussian ansatz (\ref{eq120}) in order to solve the stochastic equation (\ref{eq076}) on the state vector. In the evolution equation for matrix $ \underline{\underline{u}} $ {that appears in the ansatz}, the indices $ \alpha $ and $ \beta $ now play symmetrical roles and we obtain 
\be
\begin{array}{lll}
\dd u_{\alpha\alpha} = - \Omega_\alpha \dd t\,  u_{\alpha c}+ \gamma_\alpha \dd t (1-u_{\alpha\alpha})  & \displaystyle \dd u_{\alpha\beta} = -\frac{\dd t}{2} [(\gamma_\alpha + \gamma_\beta) u_{\alpha\beta} + \Omega_\beta u_{\alpha c} + \Omega_\alpha u_{\beta c}] & \displaystyle \dd u_{\alpha c} = -\frac{\dd t}{2} [(\gamma_\alpha + \kappa) u_{\alpha c} + \Omega_\alpha u_{cc}] \\
&&\\
\dd u_{\beta\beta} = - \Omega_\beta \dd t\, u_{\beta c} + \gamma_\beta \dd t (1-u_{\beta\beta}) & \displaystyle \dd u_{\beta c} = -\frac{\dd t}{2} [(\gamma_\beta + \kappa) u_{\beta c} + \Omega_\beta u_{cc}] & \dd u_{cc} = \kappa \dd t (1- u_{cc}) 
\end{array}
\ee
whose solution for the initial condition $ \underline{\underline{u}} (0) = \mathrm{Id} $ is written 
\begin{align}
\label{eq223a}
    u_{\alpha\alpha}(t) = & 1 + \frac{\Omega_\alpha^2}{\gamma_\alpha(\kappa + \gamma_\alpha)}\left(1-\eee^{-\gamma_\alpha t}\right) - \frac{2\Omega_\alpha^2}{\kappa^2 - \gamma_\alpha^2}\left(\eee^{-\gamma_\alpha t}-\eee^{-(\kappa + \gamma_\alpha)t/2}\right) \\
    u_{\alpha\beta}(t) = & {\frac{\Omega_\alpha\Omega_\beta}{\gamma_\alpha + \gamma_\beta}\left(\frac{1}{\kappa + \gamma_\alpha} + \frac{1}{\kappa+\gamma_\beta}\right) \left(1 - \eee^{-(\gamma_\alpha + \gamma_\beta)t/2} \right) + \frac{\Omega_\alpha\Omega_\beta}{(\kappa-\gamma_\beta)(\kappa+\gamma_\alpha)} \left(\eee^{-(\kappa+\gamma_\alpha) t/2} - \eee^{-(\gamma_\alpha + \gamma_\beta)t/2} \right)} \nonumber \\
     & {+ \frac{\Omega_\alpha\Omega_\beta}{(\kappa-\gamma_\alpha)(\kappa+\gamma_\beta)} \left(\eee^{-(\kappa + \gamma_\beta)t/2} - \eee^{-(\gamma_\alpha + \gamma_\beta)t/2} \right)} \\
    u_{\alpha c}(t) = & -\frac{\Omega_\alpha}{\kappa + \gamma_\alpha}\left(1-\eee^{-(\kappa + \gamma_\alpha)t/2}\right)\\
    u_{\beta\beta}(t) = & 1 + \frac{\Omega_\beta^2}{\gamma_\beta(\kappa + \gamma_\beta)}\left(1-\eee^{-\gamma_\beta t}\right) - \frac{2\Omega_\beta^2}{\kappa^2 - \gamma_\beta^2}\left(\eee^{-\gamma_\beta t}-\eee^{-(\kappa + \gamma_\beta)t/2}\right) \\
\label{eq223e}
    u_{\beta c}(t) = & -\frac{\Omega_\beta}{\kappa + \gamma_\beta}\left(1-\eee^{-(\kappa + \gamma_\beta)t/2}\right) \\
    u_{cc}(t) = & \quad 1
\end{align}
The {vector of coordinate averages} $ \bar{\qq} $ {appearing in ansatz (\ref{eq120})} obeys the stochastic equation 
\be
\dd\bar{\qq} = \frac{1}{2} \begin{pmatrix} -\gamma_\alpha & 0 & 0 \\
0 & -\gamma_\beta & 0 \\
\Omega_\alpha & \Omega_\beta & -\kappa 
\end{pmatrix}
\dd t \, \bar{\qq}  + 
\frac{1}{2}[\mbox{Id}-\underline{\underline{c}}(t)] \begin{pmatrix}
\gamma_\alpha^{1/2} \dd\zeta_\alpha(t) \\
\gamma_\beta^{1/2} \dd\zeta_\beta(t) \\
-\kappa^{1/2} \dd\zeta_c(t)
\end{pmatrix}
\label{eq224}
\ee
The {unconditional expectation value} $ \langle P_a^2 \rangle $ always being {equal to} $ 1/4 $, the mean and the variance of $ P_a $ conditioned on the integrated homodyne signal {are still given} by {equations} (\ref{eq127a}) and (\ref{eq127b}), by generalizing the expressions (\ref{eq130}) of the variances and covariance of the random variables $ \bar{P}_a (t) $ and $ \sigma (t) $ in the case of three independent stochastic processes $ \dd \zeta_\alpha (t ') $, $ \dd \zeta_\beta (t') $ and $ \dd \zeta_c (t ') $ as follows: 
\begin{align}
\langle \sigma(t)\bar{P}_a(t)\rangle_{\rm stoch} &= \int_0^t \dd t'\, \sum_{\nu\in\{\alpha,\beta,c\}} p_\nu(t,t') \sigma_\nu(t,t') \quad ; \quad 
\langle\sigma^2(t)\rangle_{\rm stoch} = \int_0^t \dd t'\, \sum_{\nu\in\{\alpha,\beta,c\}} \sigma_\nu^2(t,t')\quad ; \nonumber\\
\langle \bar{P}_a^2(t)\rangle_{\rm stoch} &= \int_0^t \dd t'\, \sum_{\nu\in\{\alpha,\beta,c\}} p_\nu^2(t,t')
\label{eq225}
\end{align}
with the compact expressions of the corresponding amplitudes 
\begin{align}
\label{eq226a}
p_{\nu}(t,t') &= (-1)^{\delta_{\nu c}} \frac{\sqrt{\gamma_\nu}}{2\Omega} \left\{ \Omega_\beta \eee^{-\gamma_\alpha(t-t')/2} [\delta_{\alpha\nu} -c_{\alpha\nu}(t')]-\Omega_\alpha \eee^{-\gamma_\beta(t-t')/2} [\delta_{\beta\nu} -c_{\beta\nu}(t')]\right\} \\
\label{eq226b}
\sigma_{\nu}(t,t') &= \frac{\delta_{\nu c}}{2t} + (-1)^{\delta_{\nu c}} \frac{\sqrt{\kappa\gamma_\nu}}{2 t} \left\{
[\delta_{c\nu}-c_{c\nu}(t')]f_\kappa(t-t')+\sum_{\mu\in\{\alpha,\beta\}} \frac{\Omega_\mu}{\kappa-\gamma_\mu}[\delta_{\mu\nu}-c_{\mu\nu}(t')] [f_{\gamma_\mu}(t-t')-f_{\kappa}(t-t')]\right\} 
\end{align}
The index $ \nu $ runs on the three values $ \alpha $, $ \beta $, $ c $ and we set $ \gamma_c = \kappa $. The $ \delta $ function is that of Kronecker, and the $ f_\lambda $ function is the same as in {equations} (\ref{eq129a})-(\ref{eq129d}). 

The general solution that we have just presented includes the five rates $ \gamma_\alpha, \Gamma_{\rm ex} = \Omega_\alpha^2/\kappa, \gamma_f $ on the one hand, $ \gamma_\beta, \kappa $ on the other hand. The experimentally relevant regime is one where the last two are \og infinitely \fg \, {larger} than the first three and only contribute through unobservable transient regimes. Mathematically, we reach this limit by making $ \gamma_f $ tend to zero with $ \kappa, \gamma_m, \gamma_0 $ and $ \Omega $ fixed and with $ \tau = \Gamma_{\rm ex} t> 0 $ fixed. Then the first three rates jointly tend towards zero, that is with finite-limit ratios
$\Gamma_{\rm ex}/\gamma_f\to \Omega^2\gamma_m/[\kappa(\gamma_0+\gamma_m)^2]$ and $ \gamma_\alpha/\gamma_f\to 2 \gamma_0/(\gamma_0+\gamma_m)$,
the rate $ \gamma_\beta $ reduces to $ \gamma \equiv 2 (\gamma_0+\gamma_m) $ and the Faraday coupling $ \Omega_\beta $ to $ \Omega $. All exponential transients disappear in the matrix elements (\ref{eq223a})-(\ref{eq223e}) of $ \underline{\underline{u}} $ except those relaxing at the rate $ \gamma_\alpha $.{ The amplitudes (\ref{eq226a}) and (\ref{eq226b}) on stochastic processes reduce to 
\begin{align}
\label{eq927}
\frac{p_\alpha(t,t')}{\sqrt{\Gamma_{\rm ex}}} &= \frac{u(\tau')-1}{2 u(\tau')} \sqrt{\epsilon}\,\eee^{-\epsilon (\tau-\tau')/2}  && \frac{\sigma_\alpha(t,t')}{\Gamma_{\rm ex}} = \frac{u(\tau')-1}{\tau \, u(\tau')}\sqrt{\epsilon}\, \frac{1-\eee^{-\epsilon(\tau-\tau')/2}}{\epsilon} \\
\frac{p_\beta(t,t')}{\sqrt{\Gamma_{\rm ex}}} &= \frac{\sqrt{\rho}}{(1+\rho)u(\tau')} \eee^{-\epsilon (\tau-\tau')/2} && \frac{\sigma_\beta(t,t')}{\Gamma_{\rm ex}} =\frac{\sqrt{\rho}}{(1+\rho)\tau}\left[\frac{2}{u(\tau')} \frac{1-\eee^{-\epsilon(\tau-\tau')/2}}{\epsilon} +\rho+\frac{\gamma}{\kappa} (\rho-1)\right] \\
\frac{p_c(t,t')}{\sqrt{\Gamma_{\rm ex}}} &= \frac{(1-\rho)}{2(1+\rho)u(\tau')} \eee^{-\epsilon (\tau-\tau')/2} && \frac{\sigma_c(t,t')}{\Gamma_{\rm ex}} = \frac{1}{\tau} \left[\frac{1}{2} + \frac{1-\rho}{1+\rho} \frac{1}{u(\tau')} \frac{1-\eee^{-\epsilon(\tau-\tau')/2}}{\epsilon} + \frac{\rho}{1+\rho} \left(1+\frac{2\gamma}{\kappa}\right)\right]
\end{align}
where $ \epsilon = \gamma_\alpha/\Gamma_{\rm ex} $ as in {section} \ref{sec:edld}, the function $ u (\tau) $ is given by {equation} (\ref{eq154a}) and the notation 
$\rho=\Omega^2\kappa/[\gamma(\kappa+\gamma)^2]$
generalizes the one of {footnote} \ref{note133}.} Relations (\ref{eq127a}) and (\ref{eq127b}) remain valid, with the new expressions for the variance and covariance 
\be
\label{eq228}
\boxed{\frac{\langle \sigma^2 \rangle_{\rm stoch}}{\Gamma_{\rm ex}}= {\frac{\epsilon\tau-2(1-\eee^{-\epsilon\tau/2})}{\epsilon^2\tau^2} +\frac{\Gamma_{\rm ex}}{4\tau \Gamma_{\rm sq}^{\rm gen}}} \quad\quad\mbox{and}\quad\quad
\frac{\langle \sigma \bar{P}_a \rangle_{\rm stoch}}{\sqrt{\Gamma_{\rm ex}}} = {\frac{1-\eee^{-\epsilon\tau/2}}{2\epsilon\tau}}}
\ee
and the {generalized} squeezing rate 
\be
\label{eq229}
\boxed{\Gamma_{\rm sq}^{\rm gen} = \left[\frac{1}{\Gamma_{\rm ex}}+\frac{2(\gamma_0+\gamma_m)}{\gamma_f\gamma_m}\right]^{-1}}
\ee
which reproduce the variance and covariance (\ref{eq156a}) and (\ref{eq156b}) of the one-mode model with decoherence when $ \Gamma_{\rm ex}/\gamma_f \to 0 $ and the {spin squeezing rate} (\ref{eq135}) of the three-mode model without decoherence when $ \gamma_0 \to 0 $. The new results can be simplified within the useful limit of weak {effective decoherence} $ \gamma_\alpha/\Gamma_{\rm ex} \to 0 $ by a {order-one expansion} in $ \epsilon $, which allows to generalize (\ref{eq157}) as follows on the conditional mean and variance at a non-infinitesimal value of $ \Gamma_{\rm ex}/\gamma_f $: 
\begin{align}
\label{eq230a}
m(t) &= \frac{\Gamma_{\rm sq}^{\rm gen} t}{1+\Gamma_{\rm sq}^{\rm gen} t} - \frac{\gamma_\alpha}{\Gamma_{\rm sq}^{\rm gen}} \frac{(3+\Gamma_{\rm sq}^{\rm gen} t)(\Gamma_{\rm sq}^{\rm gen} t)^2}{12(1+\Gamma_{\rm sq}^{\rm gen} t)^2}+O[(\gamma_\alpha t)^2]\\
\label{eq230b}
\mathcal{V}(t)&=\frac{1}{4(1+\Gamma_{\rm sq}^{\rm gen} t)} + \frac{\gamma_\alpha}{\Gamma_{\rm sq}^{\rm gen}} \frac{(\Gamma_{\rm sq}^{\rm gen}t+3/2)(\Gamma_{\rm sq}^{\rm gen} t)^2}{12(1+\Gamma_{\rm sq}^{\rm gen} t)^2}+O[(\gamma_\alpha t)^2]
\end{align}
{ This generalization simply amounts to replace $ \tau $ by $ \Gamma_{\rm sq}^{\rm gen} t $ and $ \epsilon $ by $ \gamma_\alpha/\Gamma^{\rm gen}_{\rm sq} $ {in the right-hand sides} of (\ref{eq157}). \footnote{ It is in fact valid for all orders in $ \epsilon $ since the proposed replacement does not change $ \epsilon \tau $ (always equal to $ \gamma_\alpha t $) and {transforms equations} (\ref{eq156a}) and (\ref{eq156b}) into equation (\ref{eq228}).}} The optimal {squeezing} on $ P_a $ is then obtained at a time $ t_{\rm opt} \sim (3/\Gamma_{\rm sq}^{\rm gen} \gamma_\alpha)^{1/2} $ and corresponds to a conditional variance $ \mbox{Var}_{\sigma = \mathcal{S}}^{\rm opt} (P_a) \sim (\gamma_\alpha/12 \Gamma_{\rm sq}^{\rm gen})^{1/2} $; {the optimal metrological gain is deduced from this by  equation (\ref{eq017}).}

\def\bysame{\leavevmode ---------\thinspace}
\def\cdrandname{\&}
\providecommand\cdrnumero{no.~}
\providecommand{\cdredsname}{eds.}
\providecommand{\cdredname}{ed.}
\providecommand{\cdrchapname}{chap.}
\providecommand{\cdrmastersthesisname}{Memoir}
\providecommand{\cdrphdthesisname}{PhD Thesis}

\end{document}


\selectlanguage{french}
\begin{frontmatter}



\title{Étude théorique de la compression de spin nucléaire par mesure quantique non destructive en continu -- {\sl Nuclear spin squeezing by continuous quantum non-demolition measurement: a theoretical study}}


\author[ad1]{Alan Serafin}
\author[ad1]{Yvan Castin}
\author[ad2]{Matteo Fadel}
\author[ad2]{Philipp Treutlein}
\author[ad1]{Alice Sinatra}

\address[ad1]{Laboratoire Kastler Brossel, ENS-Universit\'e PSL, CNRS, Universit\'e de la Sorbonne et Coll\`ege de France, 24 rue Lhomond, 75231 Paris, France}
\address[ad2]{Département de physique, Université de Bâle, Klingelbergstrasse 82, 4056 Bâle, Suisse}

\begin{abstract}
Nous proposons de tirer parti du très faible couplage du spin nucléaire de l'hélium 3 fondamental à son environnement pour produire des états quantiques macroscopiques à très longue durée de vie, ici des états comprimés du spin nucléaire, dans un gaz en cellule à température ordinaire. Pour effectuer une mesure quantique non destructive sur une composante transverse du spin nucléaire collectif préalablement polarisé, on allume temporairement une décharge dans le gaz, ce qui peuple l'état métastable de l'hélium 3. Le spin collectif correspondant au niveau $F=1/2$ des métastables s'hybride alors légèrement avec celui des fondamentaux par collision d'échange de métastabilité. Il reste à mesurer en continu le champ lumineux sortant d'une cavité optique, où il a interagi de manière dispersive avec le spin collectif du métastable, pour accéder aux fluctuations du spin nucléaire.  Dans un modèle de trois spins collectifs couplés (nucléaire, du métastable et de Stokes de la lumière) dans l'approximation de Primakoff, et pour deux schémas de mesure, nous calculons les moments de la composante comprimée $I_z$ du spin nucléaire collectif conditionnés au signal optique moyenné sur le temps d'observation $t$. Dans le schéma de comptage de photons, nous retrouvons que l'observable comprimée est $I_z^2$ plutôt que $I_z$. Dans le schéma de détection homodyne, nous résolvons analytiquement l'équation stochastique sur l'état du système conditionné à la mesure ; la moyenne conditionnelle de $I_z$ dépend linéairement du signal et la variance conditionnelle de $I_z$ n'en dépend pas.  La variance conditionnelle décroît comme {$(\Gamma_{\rm sq}t)^{-1}$}, où le taux de compression {$\Gamma_{\rm sq}$}, que nous calculons explicitement, est linéaire en l'intensité lumineuse dans la cavité à faible couplage atome-champ et sature à fort couplage au taux effectif d'échange de métastabilité dans l'état fondamental, proportionnel à la densité d'atomes métastables.  Enfin, nous tenons compte de la désexcitation des métastables sur les parois, qui induit une décohérence du spin nucléaire avec un taux ramené $\gamma_\alpha$. Elle impose une limite {$\propto (\gamma_\alpha/\Gamma_{\rm sq})^{1/2}$} sur la variance conditionnelle atteinte en un temps {$\propto (\gamma_\alpha\Gamma_{\rm sq})^{-1/2}$}. Une version multilingue est disponible sur l'archive ouverte HAL à l'adresse \url{https://hal.archives-ouvertes.fr/hal-03083577}.
\\
\noindent{\small{\it Mots-cl\'es~:} compression de spin ; hélium 3 ; spin nucléaire ; métrologie quantique ; fonctions d'onde stochastiques}
\\
\selectlanguage{english}
\vskip 0.05\baselineskip
\noindent{\bf Abstract}
\vskip 0.5\baselineskip
\noindent
We propose to take advantage of the very weak coupling of the ground-state helium-3 nuclear spin to its environment to produce very long-lived macroscopic quantum states, here nuclear spin squeezed states, in a gas cell at room temperature. To perform a quantum non-demolition measurement of a transverse component of the previously polarized collective nuclear spin, a discharge is temporarily switched on in the gas, which populates helium-3 metastable state. The collective spin corresponding to the $F=1/2$ metastable level then hybridizes slightly with the one in the ground state by metastability exchange collisions. To access the nuclear spin fluctuations, one continuously measures the light field leaking out of an optical cavity, where it has interacted dispersively with the metastable state collective spin. In a model of three coupled collective spins (nuclear, metastable and Stokes for light) in the Primakoff approximation, and for two measurement schemes, we calculate the moments of the collective nuclear spin squeezed component $ I_z $ conditioned on the optical signal averaged over the observation time $ t $. In the photon counting scheme, we find that the squeezed observable is $ I_z^2 $ rather than $ I_z $. In the homodyne detection scheme, we analytically solve the stochastic equation for the state of the system conditioned to the measurement; the conditional expectation value of $ I_z $ depends linearly on the signal and the conditional variance of $ I_z $ does not depend on it. The conditional variance decreases as $ (\Gamma_{\rm sq}t)^{-1} $, where the squeezing rate $ \Gamma_{\rm sq}$, which we calculate explicitly, depends linearly on the light intensity in the cavity at weak atom-field coupling and saturates at strong coupling to the ground state metastability exchange effective rate, proportional to the metastable atom density. Finally, we take into account the de-excitation of metastable atoms at the walls, which induces nuclear spin decoherence with an effective rate $ \gamma_\alpha$. It imposes a limit $ \propto (\gamma_\alpha/\Gamma_{\rm sq})^{1/2} $ on the conditional variance reached in a time $ \propto(\gamma_\alpha\Gamma_{\rm sq})^{-1/2}$. A multilingual version is available on the open archive HAL at \url{https://hal.archives-ouvertes.fr/hal-03083577}.
\\
\noindent{\small {\it Keywords:} spin squeezing ; helium 3 ; nuclear spin ; quantum metrology ; stochastic wave functions}

\noindent 
\vskip 0.5\baselineskip
\selectlanguage{french}
\end{abstract} 
\end{frontmatter}


\section{Introduction}
L'hélium 3 dans son état fondamental jouit de la propriété remarquable d'avoir un spin {$1/2$} purement nucléaire donc parfaitement isolé du monde extérieur, même dans un environnement aussi hostile aux cohérences quantiques que celui d'un gaz d'hélium dans une cellule centimétrique à température ordinaire et à une pression de l'ordre du millibar. Par des techniques de polarisation nucléaire bien maîtrisées, atteignant un taux de 90 $ \% $, on peut préparer alors de manière routinière (par exemple à des fins d'imagerie pulmonaire par résonance magnétique nucléaire \cite{MacFall1996}) un spin nucléaire collectif géant avec une durée de vie extrêmement longue. Ainsi, un temps de cohérence $T_2$ supérieur à 60 heures a été mesuré dans des dispositifs de magnétométrie ultra-précise \cite{Heil2010}, et semble limité seulement par le temps de décroissance longitudinal $ T_1 $ dû aux collisions avec les parois. \footnote{Des temps $T_1$ de plusieurs centaines d'heures peuvent même être obtenus \cite{Nacher2017}.} Ces valeurs font du spin nucléaire macroscopique dans un gaz à l'ambiante un système {\sl idéal} pour la production, l'étude et l'utilisation d’états intriqués, et donc un compétiteur des gaz d'atomes froids et des condensats de Bose-Einstein en métrologie et traitement quantique de l'information \cite{PezzeRMP}. Déjà en 2005, nous avions pressenti que les spins nucléaires de l'hélium 3 pourraient donner naissance à des mémoires quantiques \cite{DantanReinaudi2005} ou à des états quantiques non locaux \cite{Reinaudi2007} de très longue durée de vie. Depuis, des percées expérimentales ont été accomplies dans le domaine de la compression de spin, notamment au moyen de mesures quantiques non destructives (QND) dans des gaz atomiques d'alcalins interagissant avec un mode du champ électromagnétique {\cite{PezzeRMP,stroboscopic,Kasevich,Xiao}}, qui ont permis d'obtenir récemment un état comprimé de spin d'une durée de vie d'une seconde dans l'état hyperfin fondamental du rubidium dans des conditions métrologiques \cite{Myles2020}. {On rappelle que, par analogie avec les états comprimés de la lumière en optique quantique, un spin (ici collectif) $\vec{I}$ est dit être dans un état comprimé si, dans une direction $Oz$ orthogonale à celle $Ox$ du spin moyen, il admet des fluctuations réduites par rapport à la limite quantique standard \footnote{\label{note0}{Le bruit quantique standard correspond à des écarts-types $\Delta I_y^{\rm st}=\Delta I_z^{\rm st}=(|\langle I_x\rangle|/2)^{1/2}$, c'est-à-dire au cas d'égalité avec symétrie de révolution des fluctuations autour du spin moyen dans l'inégalité de Heisenberg $\Delta I_y \Delta I_z \geq |\langle I_x\rangle|/2$, nos spins étant ici sans dimension. Une discussion plus approfondie montre que, dans la géométrie considérée, le rapport bruit sur signal dans une mesure de fréquence de précession est en fait proportionnel à $\xi=(2I)^{1/2}\Delta I_z/|\langle I_x\rangle|$ (ceci fait apparaître naturellement l'angle des fluctuations de la direction du spin dans le plan $xOz$) plutôt qu'au rapport $\Delta I_z/\Delta I_z^{\rm st}$ naïvement attendu \cite{Wineland}. Dans le schéma que nous proposons, nous dégradons $\xi^2$ d'un facteur $\approx 2$ en utilisant un état de spin nucléaire partiellement polarisé $\langle I_x\rangle/I<1$, voir la figure \ref{fig3}, mais nous ambitionnons de gagner beaucoup plus avec la compression de spin.}} ; ceci entraîne une précision accrue dans le pointage angulaire du spin moyen dans cette direction, donc de sa fréquence de précession autour de l'axe $Oy$, dans tout dispositif métrologique de type horloge atomique ou magnétomètre \cite{Wineland,Ueda} qui s'est suffisamment affranchi des bruits techniques pour atteindre la limite quantique standard \cite{Santarelli}.}

Transposer la technique de compression par mesure QND des spins hyperfins des alcalins au spin nucléaire de l'hélium 3 représente cependant un réel défi, en raison même de la particularité dudit spin, son faible couplage à l'environnement. L'état fondamental singulet de l'hélium 3, séparé en énergie d'environ 20 eV {du premier état excité}, n'est pas directement accessible par laser. Cependant, au moyen d'une décharge oscillante, une petite fraction des atomes du gaz, de l'ordre de $ 10^{-6} $, peut être portée dans l'état triplet métastable, un excellent point de départ pour des transitions optiques dans le proche infrarouge. L'orientation des spins nucléaires s'obtient alors au travers d'un processus indirect,  le pompage optique par échange de métastabilité \cite{Nacher2017}. Dans un premier temps, le moment cinétique est transféré par interaction laser-matière des photons aux atomes métastables, a priori à leur spin électronique (le seul à être fortement couplé au champ {laser}) mais a posteriori aussi à leur spin nucléaire grâce au couplage hyperfin. Dans un deuxième temps, on tire parti des collisions d'échange de métastabilité entre atomes métastables et atomes dans l’état fondamental pour orienter les spins nucléaires dans l’état fondamental, avec une échelle de temps limitée par la faible densité des atomes dans l'état métastable et de l'ordre de la seconde. Même si la collision d'échange de métastabilité peut transférer des corrélations quantiques (voir les références \cite{DantanReinaudi2005,Reinaudi2007}), on ne peut s'attendre à ce qu'une seule mesure sur une petite fraction des atomes ($ 10^{-6} $) projette l'ensemble du système dans un état comprimé. La solution que nous proposons est d'effectuer une mesure QND  en continu démultipliée par une cavité optique résonnante {dans laquelle est placée la cellule}. En effet, bien que les atomes métastables aient individuellement une durée de vie relativement courte (ils perdent leurs corrélations quantiques et retombent dans l'état fondamental à chaque collision {avec} les parois de la cellule), faire une mesure  destructive en continu de la lumière qui sort de la cavité après interaction avec les atomes métastables revient à effectuer une mesure QND  en continu sur le spin nucléaire collectif dans l'état fondamental, ce qui le {prépare dans} l'état comprimé souhaité sans affecter sa durée de vie.\footnote{{Ce type de mesure diffère donc de la mesure instantanée des traités de mécanique quantique, qui projette l'état du système dans un sous-espace propre de l'observable mesurée selon le postulat de von Neumann.}} {Comme nous le verrons, notre procédure de réduction des fluctuations de spin se résume plus précisément par le diagramme:}
{\small
\begin{center}
\begin{tabular}{ccccc}
\!\!\!\fbox{\parbox{2.8cm}{spin collectif $\vec{I}$ dans l'état fondamental}} & \parbox{2.7cm}{collisions d'échange $\leftarrow\!\!\!\xrightarrow{\hspace{2.2cm}}$ \hspace*{2mm} de métastabilité} & \fbox{\parbox{3cm}{spin collectif dans l'état métastable}} & \parbox{1.8cm}{effet Faraday $\leftarrow\!\!\!\xrightarrow{\hspace{1.3cm}}$} & \fbox{\parbox{2.5cm}{champ en cavité polarisé selon $Ox$} } \\
$\Big\uparrow$ & & & & $\Big\downarrow$  \\
\!\!\!\fbox{\parbox{2.8cm}{réduction des fluctuations de $I_z^2$,  $I_z$}} & $\xleftarrow{\hspace{2cm}}$ & \fbox{\parbox{3cm}{mesure en continu non destructive de $I_z^2$, $I_z$}} & \parbox{2.9cm}{comptage de photons, $\xleftarrow{\hspace{2.3cm}}$ détection homodyne} &  \fbox{\parbox{2.5cm}{mesure en continu du champ sortant polarisé selon $Oy$}}
\end{tabular}
\end{center}
}

Ce travail donne une présentation théorique détaillée du mécanisme de compression et de ses limites; une étude de faisabilité plus pointue prenant en compte les valeurs expérimentalement accessibles des paramètres est effectuée dans la référence \cite{letter}. Très récemment, des idées similaires ont été mises en avant dans un système physique différent, le mélange alcalin-gaz rare \cite{Firstenberg2020, katz2019quantum}. Nous sommes confiants que la manipulation quantique des spins nucléaires à longue durée de vie est promise à un développement rapide, ouvrant de nouvelles perspectives pour la recherche fondamentale et les applications, {en particulier en magnétométrie \cite{magnetth1,magnetth2,magnetexp1,magnetth3,Xiao}}.

\section{Vue d'ensemble et description semi-classique} 
\label{sec:vue}

\begin{figure}[t] 
\centerline{\includegraphics [scale = 0.44]{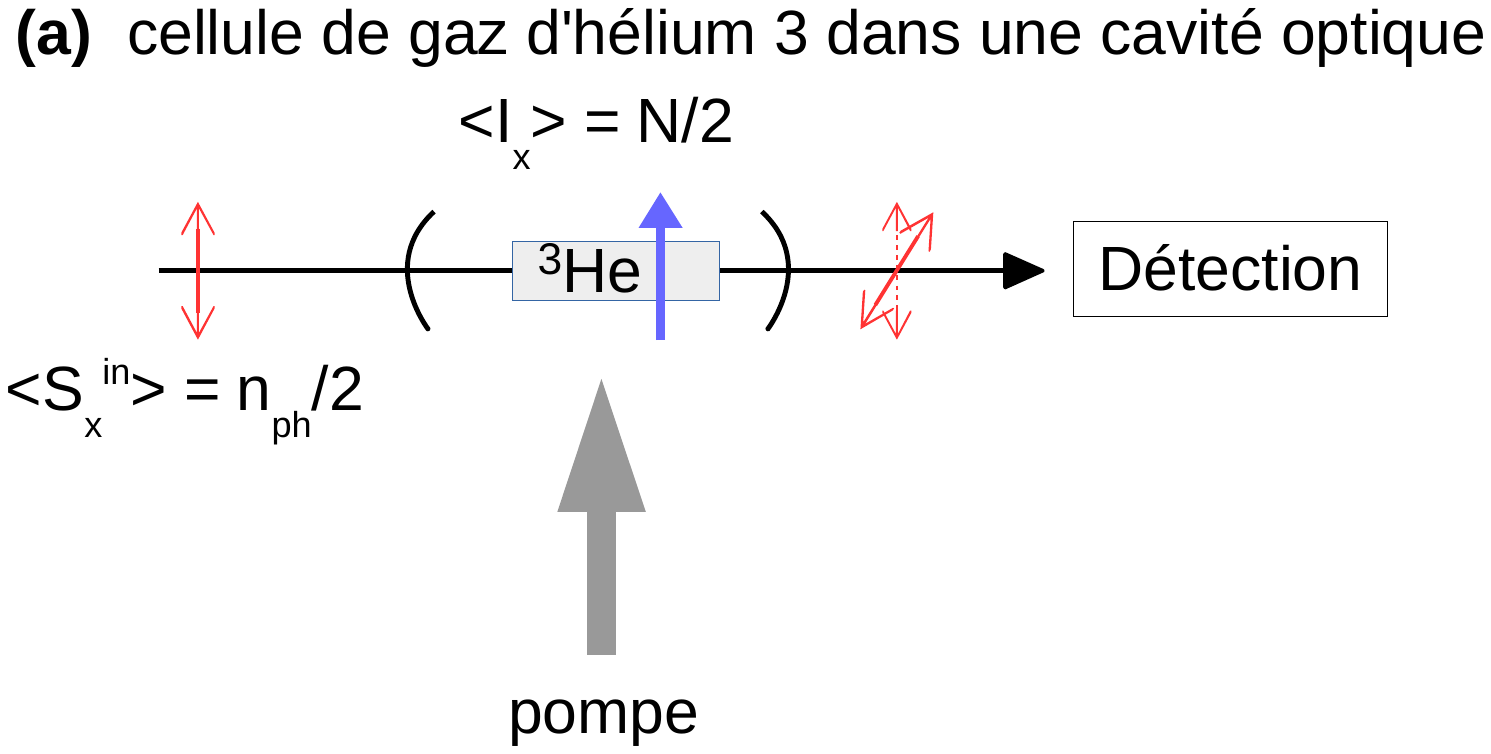} 
\includegraphics [scale = 0.42]{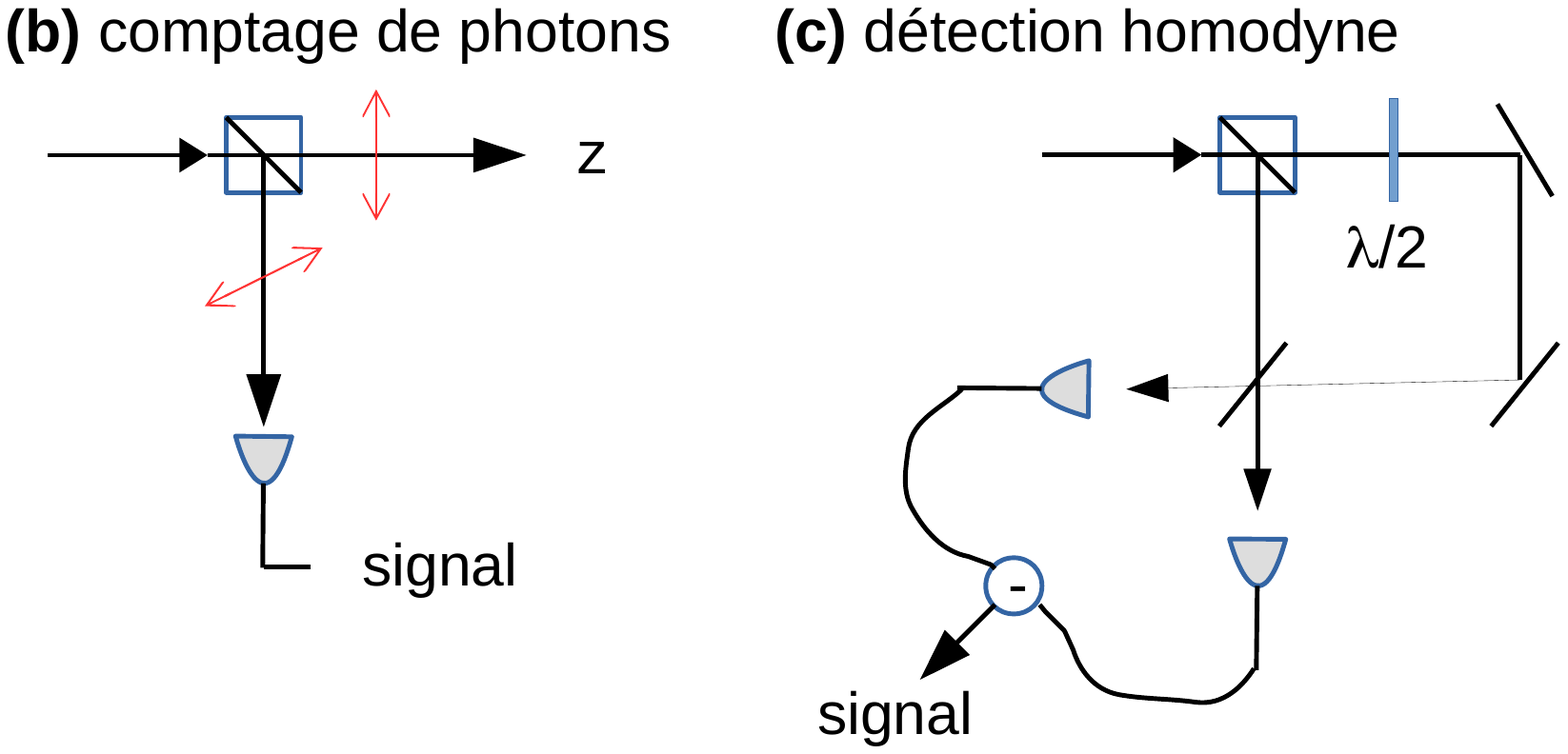}} 
\caption{Vue d'ensemble du dispositif. (a) Cellule en verre centimétrique remplie d'un gaz d'hélium 3 à température ambiante et placée dans une cavité optique d'axe $ Oz $ (axe horizontal sur la figure). Le spin de Stokes de la lumière et les spins atomiques (nucléaires et des métastables) sont polarisés linéairement selon $Ox$ (axe vertical sur la figure). Le mode du champ en cavité polarisé selon $Oy$, initialement vide, se peuple par effet Faraday sous l'action des fluctuations quantiques du spin des métastables selon $Oz$ lors de sa propagation dans le gaz. {On le mesure en continu à l'extérieur de la cavité par l'une des deux méthodes suivantes:} (b) les photons sortant de la cavité polarisés selon $Oy$ sont séparés de ceux polarisés selon $Ox$ par un cube polariseur puis détectés dans le régime de comptage de photons; (c) on effectue une détection homodyne d'une quadrature du champ sortant polarisé selon $Oy$, en utilisant comme oscillateur local le champ sortant polarisé selon $Ox$ (dont on a fait tourner au préalable la polarisation avec une lame demi-onde pour l'amener selon $Oy$).}
\label{fig1} 
\end{figure} 

Le système physique considéré est représenté sur la figure~\ref{fig1}. Une cellule remplie d'un gaz partiellement polarisé de quelques mbars d'atomes d'hélium 3 pur est placée à l'intérieur d'une cavité optique. Alors que la majorité des atomes restent dans l'état singulet fondamental $ 1^{1} S $ de l'hélium, une faible décharge porte une infime fraction des atomes, en général $ \simeq 10^{-6} $, dans l'état triplet métastable $ 2^{3} S$. D'une part, on injecte dans la cavité un faisceau laser se propageant selon l'axe de la cavité $Oz$ et polarisé linéairement selon la direction $Ox$, qui est également la direction de polarisation de l'échantillon atomique, pour exciter la transition $ 2^{3} S-2^{3} P $ avec un grand désaccord en fréquence. D'autre part, les atomes dans l'état métastable $ 2^{3} S$ (de spin hyperfin électronique et nucléaire) sont couplés aux atomes dans l'état fondamental (de spin purement nucléaire) par des collisions d'échange de métastabilité ; {de façon remarquable, bien que chaque collision d'échange soit individuellement incohérente, ceci conduit à un couplage macroscopique bien défini entre les spins collectifs correspondants \cite{DupontRoc,LaloeDupontLeduc}.} Comme l'interaction de Faraday avec les atomes métastables fait légèrement tourner autour de l'axe $Oz$ la polarisation de la lumière initialement dirigée selon $Ox$, {d'un angle proportionnel} à la composante du spin collectif des métastables selon $Oz$ comme nous le verrons, une mesure destructive  en continu de la composante de polarisation selon $Oy$ du champ sortant de la cavité (i) par comptage de photons comme indiqué sur la figure~\ref{fig1}b ou (ii) par détection homodyne comme sur la figure~\ref{fig1}c, réalise in fine une mesure quantique non destructive  en continu du spin nucléaire collectif selon $Oz$ des atomes d'hélium 3 dans l’état fondamental. 

Dans le reste de cette section, par un traitement semi-classique des fluctuations des spins autour de {leurs valeurs moyennes dans} l'état stationnaire, nous réduisons notre système physique complexe à celui plus simple de trois spins collectifs couplés, dont la section \ref{sec:qt} donnera une description quantique.

\begin{figure}[t] 
\centerline{\includegraphics [scale = 0.30]{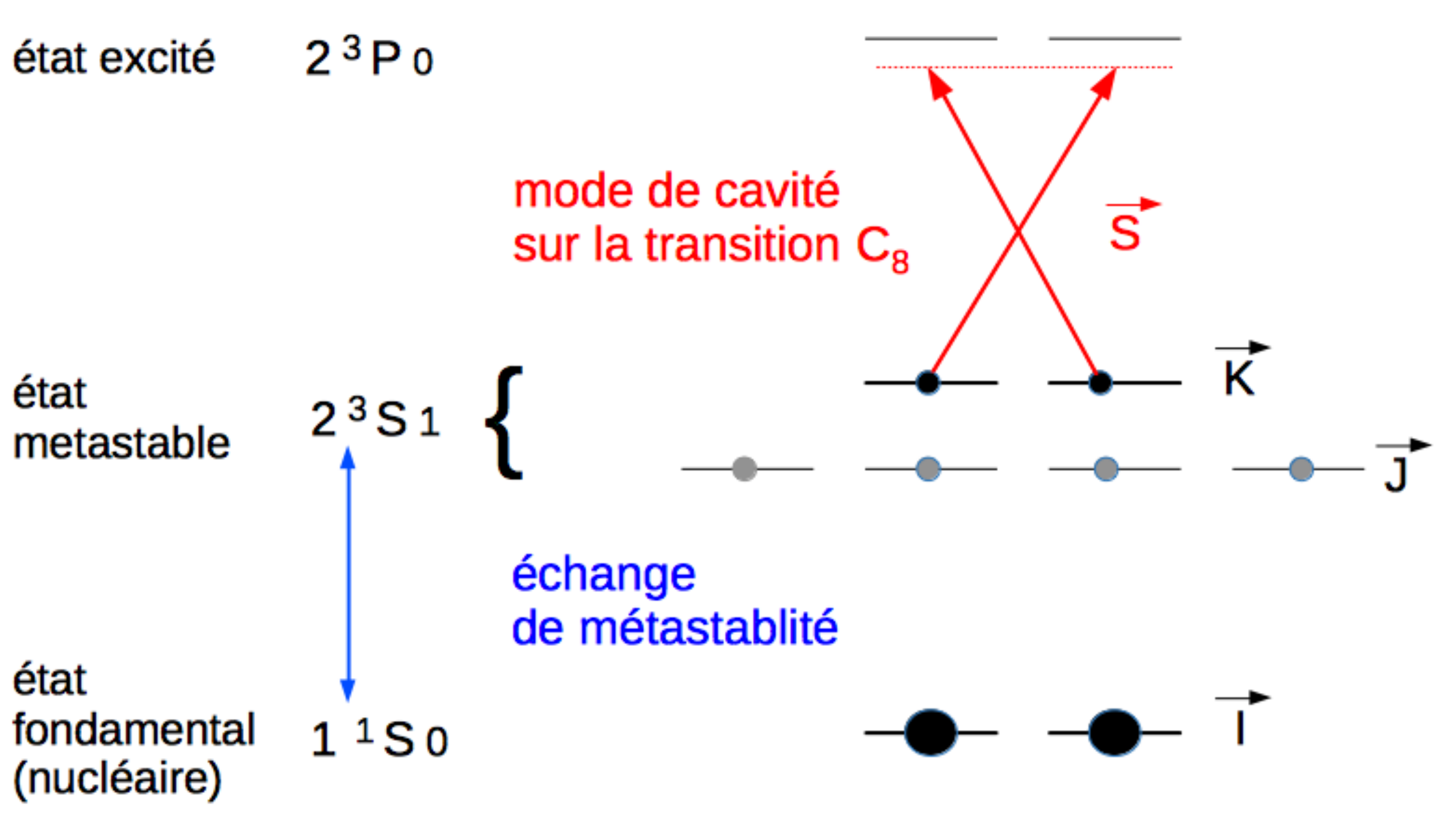}} 
\caption{Niveaux d'énergie utiles de l'atome ${}^3$He (les sous-niveaux Zeeman correspondent au choix de $Oz$ comme axe de quantification, les atomes étant polarisés selon $Ox$). Le mode du champ en cavité polarisé selon $Ox$ excite la transition $ C_8 $ entre le niveau $F=1/2$ de l'état métastable $2^{3}S_1$ et le plus haut niveau d'énergie $F=1/2$ de l'état excité $2{}^3P$, avec un désaccord en fréquence  négatif beaucoup plus grand en valeur absolue que la mi-largeur Doppler de l'état excité (de l'ordre de 1 GHz),  afin que la classe de vitesse résonnante avec le laser soit presque vide, mais beaucoup plus faible que le clivage hyperfin de 6,74 GHz dans l'état métastable (et a fortiori que le clivage fin $2{}^3P_1-2{}^3P_0$ de 29,6 GHz dans l'état excité),  afin que le niveau métastable $F=3/2$ soit très peu affecté par le laser.  (Note : l'espacement en fréquence ne permet pas de satisfaire largement à ces deux contraintes, et l'on ne peut exclure que le couplage de $F=3/2$ au champ n'ait un petit effet sur la dynamique de compression; nous le négligeons ici mais on pourrait en tenir compte avec un hamiltonien plus complet que notre modèle minimal (\ref{eq002}), comme celui de la référence \cite{Dantan2007}. {Nous avons par ailleurs vérifié, au moyen de cette référence, qu'il n'existe pas de fréquence laser \og magique\fg\, loin de résonance permettant d'annuler la contribution du tenseur d'alignement du niveau $F=3/2$ à l'opérateur de déplacement lumineux.})  Les six sous-niveaux de l'état métastable $2^{3}S_1$ sont couplés aux deux sous-niveaux (purement nucléaires) de l'état fondamental $1^{1}S_0$ par les collisions d'échange de métastabilité.}
\label{fig2} 
\end{figure}
\begin{figure}[t] 
\centerline{\includegraphics [width = 0.40 \textwidth]{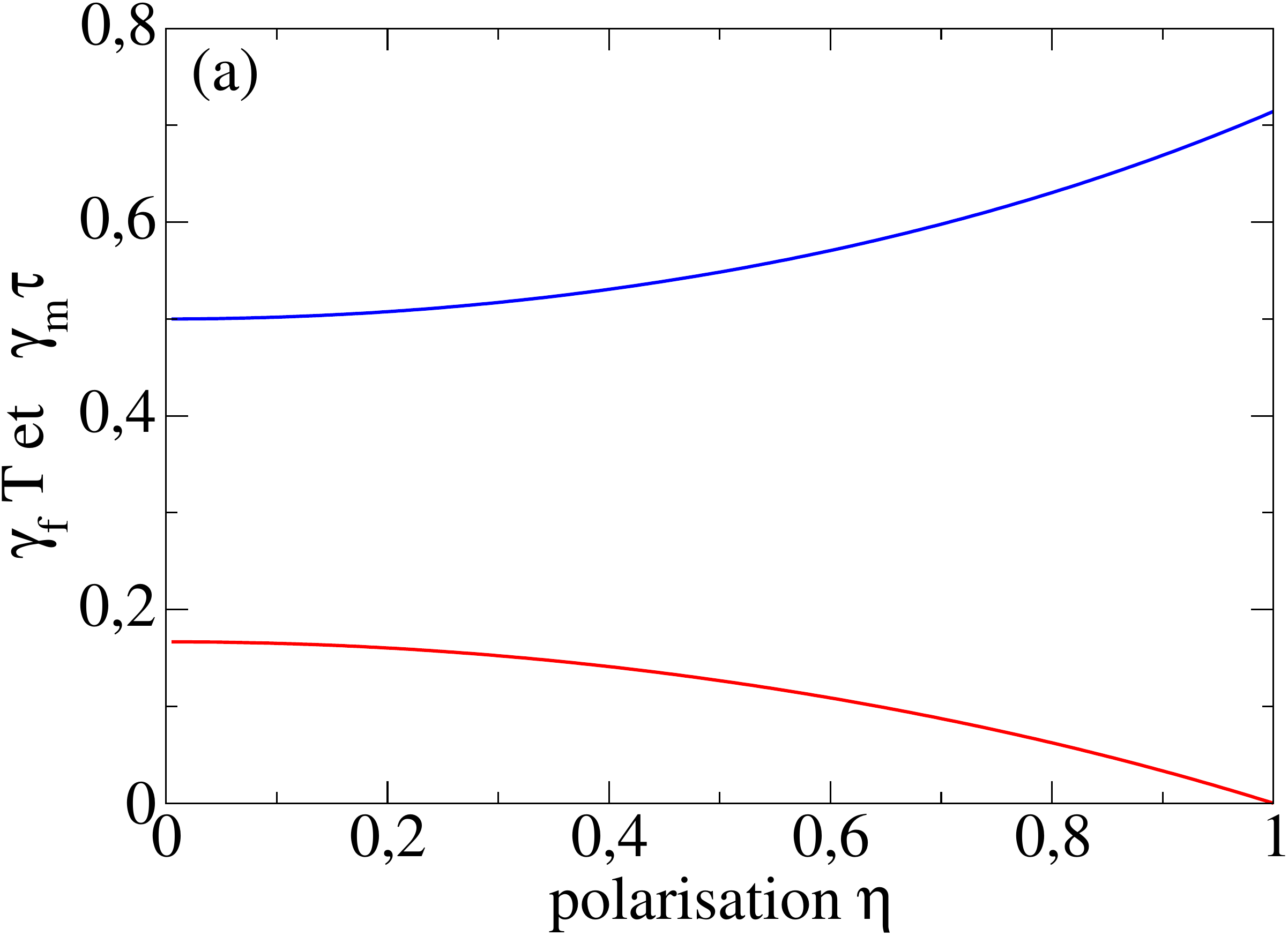} \quad\quad \includegraphics [width = 0.40 \textwidth]{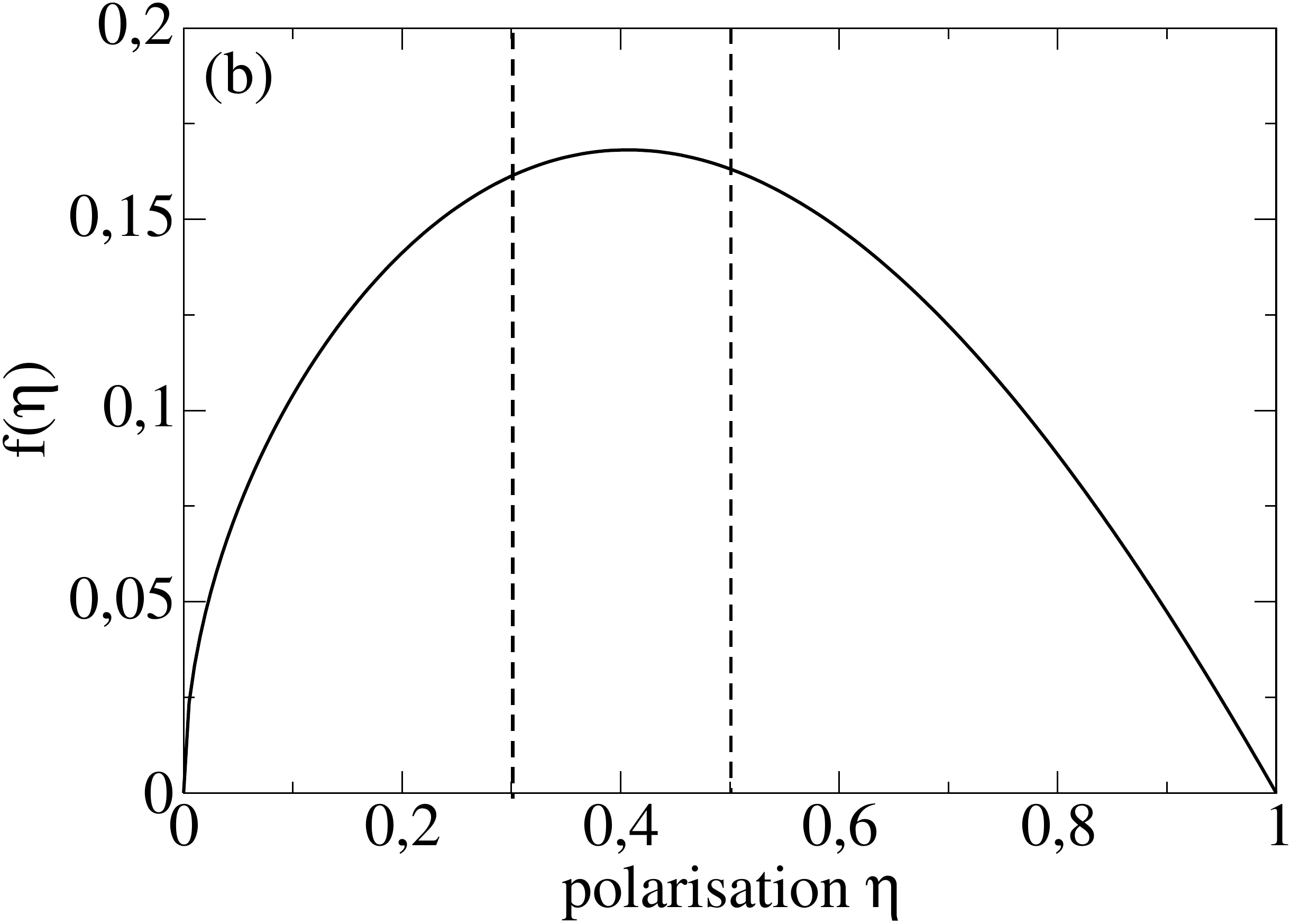}} 
\caption{(a) Taux effectifs d'échange de métastabilité $ \gamma_f $ (courbe du bas, en rouge) et $ \gamma_m $ (courbe du haut, en bleu) d'après l'équation (\ref{eq006}) {dans l'état fondamental et métastable} en fonction de la polarisation nucléaire $ \eta $, normalisés par les taux des collisions d'échange de métastabilité $1/T$ et $1/\tau$ subies par {un atome fondamental et métastable} dans le gaz. (b) Dépendance en polarisation nucléaire de la  pulsation de Faraday $\Omega_\alpha$ {entrant dans le taux de création d'excitations par le couplage de Faraday dans le mode bosonique nucléaire hybridé (\ref{eq054}) et dans les taux de compression de spin (\ref{eq868}) et (\ref{eq135}),} dans la limite $\gamma_f\ll\gamma_m$ ; plus précisément, on représente le facteur $f(\eta)=\sqrt{\eta}\frac{1-\eta^2}{3 + \eta^2}$ tel que 
$\Omega_\alpha \simeq \Omega (\gamma_f/\gamma_m)^{1/2} = \chi \sqrt{n_{\rm ph}n_{\rm cell}}\sqrt{\frac{n_{\rm cell}}{N_{\rm cell}}}f(\eta)$.
Lorsque la polarisation varie entre 0,3 et 0,5 (lignes tiretées verticales), $ f (\eta) $ s'écarte de 4 \% de son maximum $\simeq 0,17 $ atteint en $ \eta = 0,42 $. {Il est donc avantageux de se placer près de cette valeur de $\eta$ pour réduire la dérive temporelle de $\Omega_\alpha$ due à un léger amortissement de la polarisation nucléaire pendant la compression de spin (en effet, le processus de pompage optique est alors interrompu et la durée de vie de la polarisation est réduite par la présence de la décharge, elle devient de l'ordre de $\gamma_\alpha^{-1}$, où $\gamma_\alpha$ est le taux de décohérence ramené de la section \ref{sec:edld}).}}
\label{fig3} 
\end{figure}

 
La structure atomique pertinente de l'atome ${}^3$He et les transitions excitées par le champ en cavité sont représentées sur la figure \ref{fig2}. On appelle $ \Vec{I} $ le spin nucléaire collectif dans l'état fondamental, $ \Vec{J} $ et $ \Vec{K} $ les spins collectifs associés aux multiplicités hyperfines $F=3/2$ et $F=1/2$ dans l'état métastable. Pour la lumière se propageant selon $Oz$, nous introduisons le spin de Stokes \cite{Dantan2007} construit à partir des opérateurs de création et d'annihilation d'un photon dans les modes {de cavité} polarisés linéairement selon $Ox$ et $Oy $ : \footnote{\label{note2} De manière équivalente, on peut construire le spin de Stokes $ \vec{S} $ en utilisant les opérateurs d'annihilation dans les modes polarisés circulairement, $ c_1 = \frac{1}{\sqrt{2}} (c_x-\ii c_y) $, $ c_2 = \frac{1}{\sqrt{2}} (c_x+\ii c_y) $ \cite{Mandel1999}, auquel cas $ S_z = \frac{1}{2} \left (c_1^\dagger c_1-c_2^\dagger c_2 \right) $.} 
\begin{equation}
S_x = \frac{1}{2}\left( c_x^\dagger c_x -  c_y^\dagger c_y \right) \quad ; \quad 
S_y = \frac{1}{2}\left( c_x^\dagger c_y +  c_y^\dagger c_x \right) \quad ; \quad 
S_z =  \frac{1}{2\ii}\left( c_x^\dagger c_y -  c_y^\dagger c_x \right) \label{eq001} \,.
\end{equation}
Nous supposons pour simplifier que la cellule est éclairée uniformément par le mode de cavité. Dans la limite d'un grand désaccord et d'une faible saturation de la transition atomique par le champ, l'état excité $2^3 P$ peut être éliminé adiabatiquement et l'interaction hamiltonienne entre le spin du métastable $ \vec{K} $ et le spin de Stokes $ \vec{S} $ prend la forme de Faraday \cite{Dantan2007}: 
\begin{equation}
H = \hbar \chi K_z S_z 
\label{eq002}
\end{equation}
qui n'est autre que l'opérateur de déplacement lumineux des sous-niveaux Zeeman dans {le niveau métastable $F=1/2$}, comme on le voit bien sur la forme de $S_z$ dans la note \ref{note2}. Les équations non linéaires couplées décrivant l'évolution des spins moyens sont données dans l'\ref{app:semiclassical}, voir les équations~(\ref{DerivativesNLTotSx})-(\ref{eq202c}). Outre l'évolution due à l'hamiltonien de Faraday (\ref{eq002}) et aux collisions d'échange de métastabilité, elles incluent la contribution des termes liouvilliens habituels dans l'équation pilote décrivant l'injection d'un champ cohérent polarisé selon $Ox $ dans la cavité et les pertes dues au miroir de sortie, dont l'effet combiné conduit à $\langle S_x \rangle = n_{\rm ph}/2 $ dans l'état stationnaire en l'absence d'atomes, $n_{\rm ph}$ étant le nombre moyen de photons dans le mode polarisé selon $Ox$. Ces équations sont ensuite linéarisées autour d'une solution stationnaire partiellement polarisée (\ref{eq207})-(\ref{eq208}), et les fluctuations du spin $ \vec{J} $ et du tenseur d'alignement collectif dans $F=3/2$ sont éliminées adiabatiquement \footnote{Nous pensons que cette approximation non mathématiquement contrôlée est raisonnable pour l'expérience proposée, car le spin $\vec{J}$ n'est pas directement couplé à la lumière donc n'est pas directement affecté par la mesure du champ  en continu. En revanche, si l'on éliminait de même les fluctuations du spin $\vec{K}$, directement couplé au champ, on commettrait une erreur non négligeable sur la dynamique de compression du spin dans le cas de la détection par comptage de photons (revenant à omettre le saut double $C_d$ dans l'équation pilote (\ref{eq030oneMode}) et le taux $\Gamma_0$ dans le nombre moyen de photons comptés (\ref{eq064}) donc à sous-estimer fortement le nombre de photodétections requises pour atteindre un niveau de compression donné), mais une erreur négligeable dans le cas de la détection homodyne, comme nous l'avons vérifié sur le modèle à un mode de la section \ref{sec:ana_omm}.} pour obtenir des équations couplées sur les fluctuations des trois spins collectifs $\vec{I}$, $\vec{K}$ et $\vec{S}$, dont les valeurs moyennes stationnaires sont données par: 
\begin{equation}
\langle \vec{I}\rangle_s= \frac{N}{2} \, \vec{u}_x 
\quad ; \quad
\langle \vec{K} \rangle_s= \frac{n}{2} \, \vec{u}_x
\quad ; \quad
\langle \vec{S} \rangle_s=\frac{n_{\rm ph}}{2} \, \vec{u}_x \label{eq003}
\end{equation}
Ici $ \vec{u}_x $ est le vecteur unitaire selon $Ox$, $ N $ et $ n $ sont les nombres effectifs d'atomes fondamentaux et métastables participant à la dynamique des spins collectifs. Comme nous le montrons dans l'\ref{app:semiclassical}, ces nombres effectifs sont renormalisés par rapport aux nombres totaux vrais $ N_{\rm cell} $ et $ n_{\rm cell} $ dans la cellule, par des facteurs dépendant de la polarisation: 
\begin{equation}
N=\eta \, N_{\rm cell} 
\quad ; \quad
 n=\left( \frac{1-\eta^2}{3+\eta^2} \right) \eta \, n_{\rm cell} \label{eq004} \\
\end{equation}
où $ \eta \in [0,1] $ est la polarisation nucléaire,\footnote{Notons que $ n = 0 $ dans le cas entièrement polarisé $\eta=1$. En effet, toute la population de l'état métastable se trouve alors dans le sous-niveau Zeeman extrême $ m_x = 3/2 $ de l'état hyperfin $F=3/2$ et la multiplicité $F=1/2$ est vide.} et les équations semi-classiques sur les fluctuations des trois spins collectifs s'écrivent: 
\begin{align}    
        \frac{\dd}{\dd t}\delta S_z &=  -\frac{\kappa}{2} \delta S_z \quad \quad \quad \quad \quad \quad \quad  \quad \quad \:\:
        \frac{\dd}{\dd t}\delta S_y   =   -\frac{\kappa}{2} \delta S_y + \chi\langle S_x \rangle_s  \delta K_z \label{eq:redS} \\
                \frac{\dd}{\dd t}\delta I_z &=  - \gamma_f \delta I_z + \gamma_m \delta K_z \quad \quad \quad \quad \quad \quad 
                  \frac{\dd}{\dd t}\delta I_y =  - \gamma_f \delta I_y + \gamma_m \delta K_y \label{eq:redI} \\
    \frac{\dd}{\dd t}\delta K_z &= - \gamma_m \delta K_z + \gamma_f \delta I_z \quad \quad \quad \quad  \quad \quad
        \frac{\dd}{\dd t}\delta K_y =    - \gamma_m \delta K_y + \gamma_f \delta I_y+ \chi \langle K_x \rangle_s \delta S_z \label{eq005}
\end{align}
Ici, $\kappa$ est le taux de perte de la cavité, $ \gamma_m $ et $ \gamma_f $ sont les taux effectifs d'échange de métastabilité dans l'état métastable et dans l'état fondamental. Ces derniers dépendent de la polarisation nucléaire comme ci-dessous et sur la figure~\ref{fig3}a, et sont dans le même rapport que les nombres d'atomes effectifs $ N $ et $ n $ (\ref{eq004}) constituant les spins collectifs: 
\begin{equation}
{\gamma_f = \frac{4 + \eta^2}{8 - \eta^2} \, \frac{1-\eta^2}{3+\eta^2} \, \frac{1}{T} \quad ;\quad \gamma_m =  \frac{4 + \eta^2}{8 - \eta^2}  \, \frac{1}{\tau} \quad ;}\quad  \frac{\gamma_m}{\gamma_f}= \frac{N}{n} \gg 1
\label{eq006}
\end{equation}
les taux individuels des collisions d'échange de métastabilité $1/T$ et $1/\tau$ subies par un atome dans l'état fondamental et dans l'état {métastable} étant proportionnels à $n_{\rm cell}$ et $N_{\rm cell}$ {avec la même constante de proportionnalité}.  Sur la figure \ref{fig3}b, nous montrons également la dépendance en polarisation nucléaire du couplage de Faraday effectif $\Omega_\alpha$ (\ref{eq036}) entre la lumière et le spin nucléaire hybridé par le métastable, qui contrôle le taux de compression de spin {comme nous le verrons.}

\section{Description quantique} 
\label{sec:qt} 
Dans la section \ref{sec:vue}, nous avons vu que l'on peut modéliser notre système physique complexe sous la forme de trois spins collectifs couplés (\ref{eq003}): le spin nucléaire $ \vec{I} $ dans l'état fondamental, le spin $ \vec{K} $ dans le niveau hyperfin $ F = 1/2 $ de l'état métastable et le spin de Stokes $ \vec{S} $ du champ lumineux en cavité. Dans cette section, nous présentons le traitement quantique complet de ce modèle. Après avoir introduit l'approximation de Primakoff, nous passons à la description quantique de l'échange de métastabilité qui couple les spins nucléaire et métastable. 

\subsection{Approximation de Primakoff {et gain métrologique dû à la compression}} 
\label{sub: primakov} 
Initialement, le spin nucléaire collectif $ \vec{I} $, le spin collectif du métastable $ \vec{K} $ et le spin de Stokes $ \vec{S} $ de la lumière sont polarisés selon $Ox $, et le resteront pendant toute la procédure expérimentale. Dans l'approximation de Holstein-Primakoff, qui assimile les composantes de spin macroscopiques selon $Ox $ à des variables classiques, les composantes $Oy $ et $Oz $ restantes, orthogonales aux spins moyens, se comportent comme les opérateurs de quadratures (parties hermitienne et antihermitienne d'opérateurs d'annihilation donc canoniquement conjuguées, $[X,P]=\ii/2$) de trois modes bosoniques $a$, $b$, $c$ : \footnote{Si nous considérons un grand spin $ \vec{S} $ polarisé selon $Ox $, nous pouvons approximer la composante de spin dans cette direction par une variable classique, en posant
$\hat{S}_x \simeq \langle \hat{S}_x \rangle$ si bien que $[ \hat{S}_y/\!\sqrt{2\langle{\hat{S}_x\rangle}}\, , \, \hat{S}_z/\!\sqrt{2\langle{\hat{S}_x\rangle}}] \simeq \ii /2$.
} 
\begin{align}
\frac{I_y}{\sqrt{N}} \stackrel{\mbox{\scriptsize{Primakoff}}}{\simeq}  X_a = \frac{a+a^\dagger}{2} \quad &;  \quad
\frac{K_y}{\sqrt{n}} \stackrel{\mbox{\scriptsize{Primakoff}}}{\simeq}  X_b = \frac{b+b^\dagger}{2} \quad &;  \quad
\frac{S_y}{\sqrt{n_{\rm ph}}} \stackrel{\mbox{\scriptsize{Primakoff}}}{\simeq}  X_c = \frac{c+c^\dagger}{2} 
\label{eq014a} \\
\frac{I_z}{\sqrt{N}} \stackrel{\mbox{\scriptsize{Primakoff}}}{\simeq}  P_a = \frac{a-a^\dagger}{2\ii} \quad &; \quad 
\frac{K_z}{\sqrt{n}} \stackrel{\mbox{\scriptsize{Primakoff}}}{\simeq}  P_b = \frac{b-b^\dagger}{2\ii} \quad &; \quad
\frac{S_z}{\sqrt{n_{\rm ph}}} \stackrel{\mbox{\scriptsize{Primakoff}}}{\simeq}  P_c = \frac{c-c^\dagger}{2\ii}
\label{eq014b} 
\end{align}
{Nous avons tenu compte des valeurs moyennes (\ref{eq003}) dans la normalisation.} Faisons le lien avec la représentation bosonique exacte (\ref{eq001}) {du spin de Stokes} en écrivant:
\begin{align}
\frac{S_y}{\sqrt{n_{\rm ph}}} - \ii \frac{S_z}{\sqrt{n_{\rm ph}}} =\frac{1}{\sqrt{n_{\rm ph}}} c^\dagger_y c_x \stackrel{\mbox{\scriptsize{Primakoff}}}{\simeq} c_y^\dagger \quad \quad ; \quad \quad \frac{S_y}{\sqrt{n_{\rm ph}}} + \ii\frac{S_z}{\sqrt{n_{\rm ph}}} =\frac{1}{\sqrt{n_{\rm ph}}}  c^\dagger_x c_y\stackrel{\mbox{\scriptsize{Primakoff}}}{\simeq} c_y
\end{align}
Ceci montre que l'opérateur de création $ c^\dagger $ dans (\ref{eq014a})-(\ref{eq014b}), identifié avec $ c_y^\dagger $ dans l'approximation de Primakoff, transfère un photon du mode de cavité fortement peuplé par un état cohérent polarisé selon $Ox $ dans le mode de cavité initialement vide polarisé selon $Oy $. Dans l'approximation de Primakoff, l'hamiltonien du couplage de Faraday atome-champ (\ref{eq002}) s'écrit: 
\begin{align}
H = \hbar \Omega P_b P_c \quad \quad \mbox{avec}  \quad \quad \Omega = \chi \sqrt{nn_{\rm ph}} \label{eq016} \,.
\end{align}
Comme $\chi$ ne dépend pas de l'intensité du champ dans la cavité, $\Omega^2$ est proportionnel à cette intensité. {Pour finir, écrivons en termes des variables de Primakoff le paramètre $\xi$ de la référence \cite{Wineland} quantifiant le niveau de compression de spin utilisable dans un magnétomètre (le gain métrologique est d'autant plus élevé que $\xi$ est plus faible, voir notre note \ref{note0}), $Oz$ étant la direction transverse au spin moyen de plus forte compression {et le spin nucléaire collectif étant de nombre quantique $I=N_{\rm cell}/2$:}
\begin{equation}
\label{eq017}
\boxed{\xi^2 \equiv \frac{2I\,\mbox{Var}_{\rm signal}(I_z)}{\langle I_x\rangle^2} = \frac{4\,\mbox{Var}_{\rm signal}(P_a)}{\eta}}
\end{equation}
La polarisation nucléaire $\eta$ étant fixée, et une mesure quantique non destructive étant effectuée en continu sur le spin nucléaire, il faut chercher à minimiser la variance de $P_a$ conditionnée au signal de mesure à définir, en la faisant descendre le plus en dessous possible de sa valeur initiale $1/4$.}

\subsection{Équation pilote pour l'échange de métastabilité} 
\label{subsec:MEC} 
Considérons dans cette sous-section l'évolution du système due au seul échange de métastabilité ($\chi=0$).  Dans un traitement quantique, les équations classiques (\ref{eq:redI})-(\ref{eq005}) deviennent des équations stochastiques incluant les fluctuations quantiques. Dans l'approximation de Primakoff, cela donne pour les quadratures $ X $ dans l'état métastable et fondamental: 
\begin{equation}
\dd  X_a = -\gamma_f X_a  \dd t  + \sqrt{\gamma_m \gamma_f} X_b  \dd t + \dd X_a^{\rm stoch}
\quad ; \quad 
\dd X_b = -\gamma_m X_b  \dd t + \sqrt{\gamma_m \gamma_f} X_a  \dd t + \dd X_b^{\rm stoch} 
\label{eq018}
\end{equation} 
où l'on a utilisé la troisième égalité de l'équation (\ref{eq006}). Les bruits de Langevin $ \dd X_i^{\rm stoch} $, avec $ i \in \{a,b\} $, ont une moyenne nulle, sont des variables aléatoires indépendantes à des temps différents et ont des variances et des covariances à temps égaux calculées dans la référence \cite{DantanReinaudi2005}: 
\begin{equation}
\langle  \dd X_i^{\rm stoch} \dd X_j^{\rm stoch} \rangle = D_{ij} \dd t \quad \mbox{avec} \quad  
{\underline{\underline{D}}}=\frac{1}{2} \begin{pmatrix} \gamma_f & -\sqrt{\gamma_m \gamma_f} \\  -\sqrt{\gamma_m \gamma_f} & \gamma_m \end{pmatrix}
\label{eq019}
\end{equation}
On a des équations de même forme que (\ref{eq018}) pour les quadratures $ P_i $,  avec d'autres bruits de Langevin $\dd P_i^{\rm stoch}$, de même matrice de covariance que l'équation~(\ref{eq019}) entre eux mais de matrice de covariance avec les bruits $ \dd X_i^{\rm stoch} $ donnée par 
\begin{equation}
\langle  \dd X_i^{\rm stoch} \dd P_j^{\rm stoch} \rangle = {\cal D}_{ij} \dd t \quad \mbox{avec} \quad  
{\underline{\underline{\cal D}}}= \ii {\underline{\underline{D}}}
\end{equation}
Pour le calcul des valeurs moyennes et des variances des observables atomiques, cette formulation stochastique équivaut à une équation pilote sur l'opérateur densité atomique $ \rho_{\rm at} $ des deux modes bosoniques  $a$ et $b$:
\begin{equation}
\frac{\dd \rho_{\rm at}}{\dd t} =  C\rho_{\rm at} C^\dagger -\frac{1}{2}\{ C^\dagger C,\rho_{\rm at} \} \quad \mbox{avec} \quad  C=\sqrt{2\gamma_f}a-\sqrt{2\gamma_m}b
\label{eq021}
\end{equation}
En effet, la représentation stochastique de Langevin de l'équation pilote (\ref{eq021}) pour un opérateur quelconque $ A $ s'écrit 
\begin{equation}
\dd A = \frac{\dd t}{2} \left\{ C^\dagger [A,C ] - [A,C^\dagger] C \right\} + \dd A^{\rm stoch} \quad \mbox{où} \quad  
 \dd A^{\rm stoch}  = [C^\dagger,A] \dd B + \dd B^\dagger [A,C] 
\end{equation}
et $ \dd B $ est un opérateur stochastique markovien de moyenne nulle, de matrice de covariance à temps égaux 
\begin{equation}
\langle \dd B \, \dd B^\dagger \rangle = \dd t \quad;\quad \langle \dd B \, \dd B \rangle =  \langle \dd B^\dagger \dd B^\dagger \rangle= \langle \dd B^\dagger \dd B \rangle =0
\end{equation}
Pour être complets, esquissons un autre raisonnement, qui fait l'économie des bruits de Langevin quantiques. Il suffit d'admettre que les équations d'évolution sur les moyennes $\langle X_i\rangle$ et $\langle P_i\rangle$ tirées de (\ref{eq:redI})-(\ref{eq005}) dérivent d'une équation pilote de la forme de Lindblad (\ref{eq030Arbitrary}). Comme ces équations sont linéaires, les opérateurs de saut $C_m$ encadrant $\rho_{\rm at}$ dans l'équation pilote sont des combinaisons linéaires de $a$ et $b$. Ceci redonne (\ref{eq021}).

\subsection{Équation pilote à trois modes} 
L'évolution complète, comprenant l'interaction {atome-champ d'hamiltonien hermitien $H$} (\ref{eq016}), l'échange de métastabilité et les pertes de la cavité, est décrite par l'équation pilote \footnote{Nous négligeons ici l'évolution interne des modes atomiques (précession de spin) en supposant que les sous-niveaux Zeeman sont dégénérés dans l'état fondamental et dans {le niveau} métastable $ F = 1/2 $, {donc} que le champ magnétique extérieur est nul, $ \vec{B}=\vec{0}$. {Cette hypothèse simplificatrice appelle les commentaires suivants. (i) Dans l'expérience, on prévoit d'imposer un champ directeur selon $Ox$ de l'ordre du $\mu$T pour éviter une précession \og sauvage \fg\, du spin moyen autour d'un champ magnétique résiduel de direction inconnue \cite{letter}. D'un point de vue théorique, on se place alors dans le référentiel tournant autour de $Ox$ à la pulsation de Larmor correspondante $\omega_L^{(x)}$ du spin nucléaire pour éliminer ce champ directeur [il faudrait en toute rigueur compenser l'écart entre les pulsations de Larmor métastable et fondamentale, par exemple au moyen d'un champ magnétique fictif créé par déplacement lumineux, mais cette précaution semble superflue car la pulsation de Larmor métastable reste petite devant le taux effectif d'échange de métastabilité $\gamma_m$]; il faut en principe faire tourner aussi la cavité à la vitesse angulaire $\omega_L^{(x)}$ pour qu'elle soit immobile dans le référentiel tournant ; si la cavité reste fixe dans le référentiel du laboratoire, on peut établir le couplage de Faraday et effectuer une mesure du champ sortant de la cavité non plus en continu mais de manière stroboscopique (à chaque fois que la direction à comprimer en spin se confond avec l'axe optique) \cite{stroboscopic}. (ii) S'il existe pendant la phase de compression un petit champ magnétique statique parasite (en plus du champ directeur) dans le plan $yOz$, d'angle $\phi$ avec $Oy$, ce champ tourne à la pulsation $-\omega_{\rm L}^{(x)}$ dans le référentiel tournant ; dans l'équation stochastique du modèle à un mode (\ref{eq078}) pour la détection homodyne, ceci ajoute une contribution d'hamiltonien hermitien
$H_\alpha = \sqrt{N} \hbar\omega_L^{(y)} (\Omega_\beta/\Omega) [\cos(\phi-\omega_L^{(x)}t) X_\alpha+\sin(\phi-\omega_L^{(x)}t) P_\alpha]$ 
où $\omega_L^{(y)}\ll\omega_L^{(x)}$ est la pulsation de Larmor du spin nucléaire dans le champ parasite et $\Omega_\beta/\Omega$ est donné par l'équation (\ref{eq036}).  Le terme en $P_\alpha$ s'absorbe dans un changement de phase de l'ansatz (\ref{eq080}). Le terme en $X_\alpha$ ne change pas le paramètre $u$ de l'ansatz (\ref{eq080}) donc la variance de $P_\alpha$ dans une réalisation mais ajoute une partie déterministe oscillante à $\bar{P}_\alpha$, à savoir 
$(\sqrt{N}\omega_L^{(y)}/2\omega_L^{(x)})(\Omega_\beta/\Omega)[\sin(\phi-\omega_L^{(x)}t)-\sin\phi]$
et, d'après (\ref{eq102}), une partie déterministe 
$\sqrt{\Gamma_{\rm ex}} (\Omega_\beta/\Omega) (\sqrt{N} \omega_L^{(y)}/2\omega_L^{(x)})\{[\cos(\phi-\omega_L^{(x)}t)-\cos\phi]/(\omega_L^{(x)}t)-\sin\phi\}$
au signal intégré $\sigma(t)$. Le terme en $\sin\phi$ dans le signal peut être annulé en jouant sur l'instant initial \og $t=0$\fg\, de la phase de compression ; les autres contributions s'annulent pour une durée $t$ de l'expérience multiple entier de $2\pi/\omega_L^{(x)}$. (iii) Si l'on veut utiliser l'état comprimé de spin pour mesurer un champ magnétique, on éteint la décharge et le champ directeur et on s'arrange pour que le champ à mesurer $\vec{B}_{\rm mes}$ soit orienté selon $Oy$. Le spin nucléaire collectif précesse alors dans le plan $xOz$ d'un angle qu'il faut mesurer pour remonter à $B_{\rm mes}$, et la direction de compression initiale $Oz$ est justement celle qu'il faut pour réduire l'incertitude de pointage angulaire sur le spin. Une autre stratégie consiste à partir d'un état polarisé ordinaire, non comprimé, du spin nucléaire et à effectuer, décharge allumée mais champ directeur éteint, la mesure en continu du champ lumineux sortant de la cavité {\sl en présence} de $B_{\rm mes} \vec{u}_y$ ; dans le modèle à un mode avec détection homodyne de la section \ref{sec:3modesTo1Homo}, on retombe alors sur les propositions de magnétométrie des références \cite{magnetth1,magnetth2}. }}
\begin{equation}
\frac{\dd \rho}{\dd t} = \frac{1}{\ii\hbar} \left[H ,\rho \right] + \kappa \left( c\rho c^\dagger -\frac{1}{2}\{ c^\dagger c,\rho\}\right) + C\rho C^\dagger -\frac{1}{2}\{ C^\dagger C,\rho \}
\label{eq030}
\end{equation}
où $ C $ est l'opérateur de saut pour l'échange de métastabilité (\ref{eq021}), $ \kappa $ est le taux de perte de la cavité, $ \gamma_m $ et $ \gamma_f $ sont les taux {effectifs} d'échange de métastabilité {dans l'état} métastable et fondamental. 

Initialement, les trois modes sont dans l'état vide correspondant à un état polarisé pour les trois spins.  Pour cet état initial, les premiers moments des quadratures restent nuls, et l'on peut obtenir un système fermé d'équations sur les seconds moments.  On trouve que les quadratures $ P $ restent  de variances constantes et de covariances nulles dans les trois modes,
\begin{equation}
\langle P_a^2 \rangle(t) =  \langle P_b^2 \rangle(t) =  \langle P_c^2 \rangle(t) =  \frac{1}{4}\quad ; \quad \langle P_a P_b\rangle(t)=\langle P_a P_c\rangle(t)=\langle P_b P_c\rangle(t)=0
\label{eq032} 
\end{equation}
que la variance $\langle X_c^2\rangle$ reste bornée et que les covariances $\langle X_a X_c\rangle$ et $\langle X_b X_c\rangle$ restent nulles, tandis que les variances  et la covariance des quadratures  $X_a$ et $X_b$, et donc le nombre d'excitations dans les modes atomiques,\footnote{Pour l'état initial considéré, on a à tout temps $\langle X_a\rangle=0$ et $ \langle X_a^2 \rangle-\frac{1}{4} = \langle a^\dagger a \rangle $, où $ \langle a^\dagger a \rangle $ est le nombre moyen d'excitations dans le mode de spin nucléaire, si bien que Var $ X_a = \langle a^\dagger a \rangle+\frac{1}{4} $; {en effet, on a $a^\dagger a+1/2=X_a^2+P_a^2$}. Les mêmes relations valent pour les deux autres modes.}  divergent linéairement en temps, du moins tant que l'approximation de Primakoff est applicable.  Nous donnons ici explicitement seulement les comportements aux temps longs :
\be
\begin{array}{ll}
\displaystyle\langle X_a^2 \rangle (t)\underset{t \to +\infty}{=} \frac{\gamma_m \gamma_f}{(\gamma_m+\gamma_f)^2}\frac{\Omega^2 t}{4\kappa} +O(1) & \quad\displaystyle \langle X_b^2 \rangle (t) \underset{t \to +\infty}{=}  \frac{\gamma_f^2}{(\gamma_m+\gamma_f)^2}\frac{\Omega^2 t}{4\kappa} +O(1)\\
&\\
\displaystyle \langle X_a X_b\rangle(t) \underset{t \to +\infty}{=} \frac{\gamma_m^{1/2}\gamma_f^{3/2}}{(\gamma_m+\gamma_f)^2}\frac{\Omega^2 t}{4\kappa}+O(1)  & \quad\displaystyle \langle X_c^2\rangle (t) -  \frac{1}{4} \underset{t \to +\infty}{\to} \left(\frac{\Omega}{2\kappa}\right)^2 \left( 1 - \frac{2\gamma_m}{\kappa + 2(\gamma_m+\gamma_f)} \right)
\end{array}
\label{eq031}
\ee

\subsection{Modèle à un mode} 
\label{sec:ana_omm} 

Dans cette sous-section, nous établissons une équation pilote à un mode décrivant l'évolution lente du spin nucléaire dans la limite 
\begin{equation}
\Gamma_{\rm ex} \ll \gamma_f < \gamma_m \quad\mbox{et}\quad \Gamma_{\rm ex} \ll \kappa
\label{eq033}
\end{equation}
{où $\Gamma_{\rm ex}$ est un taux de création d'excitations dans le mode bosonique nucléaire hybridé $\alpha$} défini plus {bas sous l'effet du couplage de Faraday} (il suffit de savoir ici que $\Gamma_{\rm ex}\propto \Omega^2$ si bien que (\ref{eq033}) est une limite de couplage de Faraday faible $\Omega\to 0$). Il convient à cette fin d'introduire  les opérateurs bosoniques d'annihilation dans une base astucieusement tournée, au moyen des combinaisons linéaires suivantes des opérateurs $ a $ et $ b $: 
\begin{equation}
\alpha=\sqrt{\frac{\gamma_m}{\gamma_m+\gamma_f}}a+\sqrt{\frac{\gamma_f}{\gamma_m+\gamma_f}}b 
\quad \quad ; \quad \quad
\beta=\sqrt{\frac{\gamma_m}{\gamma_m+\gamma_f}}b-\sqrt{\frac{\gamma_f}{\gamma_m+\gamma_f}}a
\label{eq034}
\end{equation}
$ \alpha $ et $ \beta $ correspondent en effet aux modes propres de la partie d'échange de métastabilité de l'équation pilote à trois modes (\ref{eq030}) (en pratique, on a $ \gamma_m \gg \gamma_f $, voir l'équation (\ref{eq006}), si bien que le mode $ \beta $ correspond au spin du métastable légèrement hybridé avec le spin de l'état fondamental, et $ \alpha $ au spin nucléaire légèrement hybridé avec le spin du métastable). Tandis que le mode $ \alpha $ subit une divergence en temps de son nombre moyen d'excitations {(d'où la possibilité de définir un taux $\Gamma_{\rm ex}$)}, le mode $ \beta $ est fortement amorti et tend vers une valeur stationnaire  (voir les résultats (\ref{eq032}) et (\ref{eq031}), qui montrent que {$\langle P_\beta^2\rangle=1/4$ et} $\langle X_\beta^2\rangle = O(1)$ où $X_\beta=(\beta+\beta^\dagger)/2$), ce qui permettra de l'éliminer adiabatiquement, tout comme le champ en cavité. Dans cette nouvelle base, l'équation pilote à trois modes (\ref{eq030}) prend la forme 
\begin{equation}
\boxed{
\frac{\dd \rho}{\dd t} = \frac{1}{\ii\hbar} \left[ H,\rho \right] + \kappa \left( c\rho c^\dagger -\frac{1}{2}\{ c^\dagger c,\rho\}\right) 
+ \gamma_\beta\left( \beta \rho \beta^\dagger -\frac{1}{2}\{ \beta^\dagger \beta,\rho\}\right)}
\label{eq030t}
\end{equation}
où  $\gamma_\beta\equiv 2(\gamma_m+\gamma_f)$ et, en notant  $P_\alpha=(\alpha-\alpha^\dagger)/2\ii$ et $ P_\beta=(\beta-\beta^\dagger)/2\ii$ les quadratures $P$ des nouveaux modes,
\begin{equation}
H=\hbar(\Omega_\alpha P_\alpha + \Omega_\beta P_\beta) P_c \quad \mbox{avec}\quad \Omega_\alpha\equiv \Omega \sqrt{\frac{\gamma_f}{\gamma_m+\gamma_f}} \quad\mbox{et}\quad \Omega_\beta{\equiv}\Omega  \sqrt{\frac{\gamma_m}{\gamma_m+\gamma_f}}
\label{eq036}
\end{equation}
{La référence \cite{Sorensen} explique en toute généralité comment effectuer une élimination adiabatique au niveau de l'équation pilote. Ici, nous préférons l'effectuer,} comme dans la référence \cite{CastinMolmer_adel}, en couplage de Faraday faible $\Omega\to 0$ dans le formalisme des fonctions d'onde Monte-Carlo \cite{JOSAB, CastinDalibard}, où l'opérateur densité solution de l'équation pilote (\ref{eq030t}) s'obtient par moyenne de cas purs sur des réalisations stochastiques indépendantes, chaque réalisation correspondant à l'évolution déterministe d'un vecteur d'état  non normalisé $|\psi(t)\rangle$ sous l'action de l'hamiltonien effectif non hermitien 
\begin{equation}
H_{\rm eff}=H-\frac{\ii\hbar}{2} \left( \kappa c^\dagger c + \gamma_\beta \beta^\dagger \beta \right)
\label{eq010_MC}
\end{equation}
interrompue aléatoirement par des sauts quantiques (évolutions discontinues $|\psi\rangle\to C |\psi\rangle$) d'opérateurs de saut
\begin{equation}
C_c=\sqrt{\kappa}c \quad \mbox{et} \quad C_\beta=\sqrt{\gamma_\beta}\beta \,.
\label{eq038}
\end{equation}
En l'absence du couplage cohérent $ \Omega $ dans (\ref{eq036}) le mode métastable hybridé et le mode de cavité restent dans l'état vide initial. Au premier ordre en $ \Omega $, cet état est couplé à des états à une excitation dans la cavité (par l'action de $ P_c $) et à zéro ou une excitation dans le mode du métastable hybridé (par l'action de $ P_\alpha $ ou de $P_\beta $). Nous pouvons alors tronquer le vecteur d'état Monte-Carlo $|\psi\rangle$ dans la base de Fock $ \{| n_\alpha \rangle_{\rm fond} | n_\beta \rangle_{\scriptsize\mbox{méta}} | n_c \rangle_{\rm cav} \} $ comme suit,
\begin{equation}
|\psi\rangle = |\psi_\alpha^{00} \rangle |0\rangle |0\rangle +
|\psi_\alpha^{01} \rangle |0\rangle |1\rangle +  |\psi_\alpha^{11} \rangle |1\rangle |1\rangle
\label{eq039}
\end{equation}
en commettant une erreur de norme $O(\Omega^2)$. Sous l'effet de l'hamiltonien effectif (\ref{eq010_MC}), les composantes rapides $ | \psi_\alpha^{01} \rangle $ et $ | \psi_\alpha^{11} \rangle $ rejoignent exponentiellement un régime de suivi adiabatique de la composante lente $ | \psi_\alpha^{00} \rangle $ avec des taux $ \kappa/2 $ ou $ (\kappa+\gamma_\beta)/2$. D'où leur élimination adiabatique dans la limite (\ref{eq033})\footnote{Dans le suivi adiabatique, les probabilités d'occupation des composantes excitées sont
$\langle\psi_\alpha^{11}|\psi_\alpha^{11}\rangle_{\rm adiab}/\langle\psi|\psi\rangle = [\Omega_\beta^2/4(\kappa+\gamma_\beta)^2] \langle\psi_\alpha^{00}|\psi_\alpha^{00}\rangle/\langle\psi|\psi\rangle $
et
$ \langle \psi_\alpha^{01}|\psi_\alpha^{01}\rangle_{\rm adiab}/\langle\psi|\psi\rangle =(\Gamma_{\rm ex}/\kappa) \langle \psi_\alpha^{00}|P_\alpha^2|\psi_\alpha^{00}\rangle/\langle\psi|\psi\rangle$
où l'on a utilisé (\ref{eq054}). Dans la limite (\ref{eq033}), on vérifie aisément qu'elles sont $\ll 1$, si bien que presque toute la population est dans la composante $ | \psi_\alpha^{00} \rangle | 0 \rangle | 0 \rangle $ comme il se doit, ce qui nous permettra dans la suite de remplacer $\langle\psi|\psi\rangle$ par $\langle\psi_\alpha^{00}|\psi_\alpha^{00}\rangle$. On vérifie également qu'une autre condition de validité de l'élimination adiabatique, à savoir la lenteur de l'évolution du spin nucléaire hybridé $\alpha$ par rapport aux variables rapides, qui s'écrit ici $\Gamma_{\rm ex},\Gamma_0 \ll \kappa, \kappa+\gamma_\beta$, est satisfaite. Ces considérations ne permettent cependant pas de montrer que la condition $\Gamma_{\rm ex}\ll\gamma_f$ est nécessaire (sauf si $\kappa\ll\gamma_\beta$). Pour le voir en toute généralité, nous poussons à l'ordre $\Omega^4$ le calcul de l'hamiltonien effectif
$H_{\rm eff}^{00}=P H_{\rm eff}P + P H Q (zQ-Q H_{\rm eff} Q)^{-1} QHP$
dans le sous-espace $n_\beta=n_c=0$ sur lequel $P$ projette (ici $Q=\mathbb{1}-P$ et $z=O(\Omega^2)$). Qualitativement, à cet ordre, par action de $H_\alpha$ puis $H_\beta$ sur $|\psi_\alpha^{00}\rangle |0\rangle |0\rangle$ (avec la notation évidente $H=H_\alpha+H_\beta$), on crée virtuellement une excitation $\beta$ seule, relaxant au taux $\gamma_\beta/2$, d'où la condition d'adiabaticité supplémentaire $\Gamma_0\ll \gamma_\beta$ ; jointe à $\Gamma_0\ll\kappa$ et $\gamma_f<\gamma_m$, elle implique $\Gamma_{\rm ex}\ll\gamma_f$ puisque 
$\Gamma_{\rm ex}/\gamma_f = (\Gamma_0/\kappa+\Gamma_0/\gamma_\beta)(4\gamma_\beta/\gamma_m) < 16 (\Gamma_0/\kappa+\Gamma_0/\gamma_\beta)$.
Quantitativement, nous trouvons une correction au coefficient de $P_\alpha^2$ dans $H_{\rm eff}^{\rm 00}$ de type $H_\alpha G_0 H_\beta G_0 H_\beta G_0 H_\alpha$ ($G_0$ est la résolvante de $H_{\rm eff}$ pour $\Omega=0$) de la forme $\hbar \Gamma_{\rm ex} \Omega_\beta^2/\gamma_\beta\kappa$, qui doit être négligeable, ce qui impose $\Omega_\beta^2/\gamma_\beta\kappa\ll 1$, c'est-à-dire $\Gamma_{\rm ex}\ll \gamma_f$ compte tenu de $\gamma_f<\gamma_m$. Les corrections au terme scalaire sont négligeables dès que $\Gamma_0\ll \gamma_\beta,\kappa$, et le nouveau terme en $P_\alpha^4$ qui apparaît est négligeable devant $\hbar \Gamma_{\rm ex} P_\alpha^2$ pour $P_\alpha=O(1)$ si $\Gamma_{\rm ex}\ll \kappa$.}
\begin{equation}
|\psi_\alpha^{11}\rangle_{\rm adiab} \simeq  \frac{\ii\Omega_\beta}{2(\kappa+\gamma_\beta)} |\psi_\alpha^{00}\rangle \quad \mbox{et} \quad
|\psi_\alpha^{01}\rangle_{\rm adiab} \simeq \frac{\Omega_\alpha}{\kappa} \,P_\alpha |\psi_\alpha^{00}\rangle \label{eq043}
\end{equation}
On reporte les expressions de 
$|\psi_\alpha^{11}\rangle_{\rm adiab}$, $|\psi_\alpha^{01}\rangle_{\rm adiab}$ 
dans l'équation d'évolution  hamiltonienne de $ | \psi_\alpha^{00} \rangle $ pour obtenir 
\begin{equation}
\ii\hbar \frac{\dd }{\dd t} |\psi_\alpha^{00}\rangle = -\frac{\ii\hbar}{2} \left( \Gamma_{\rm ex} P_\alpha^2  + \Gamma_0 \right) |\psi_\alpha^{00}\rangle \equiv H_{\rm eff}^{00} \: |\psi_\alpha^{00}\rangle 
\label{eq045}
\end{equation} 
où l'on a introduit les taux
\begin{equation}
\boxed{
\Gamma_{\rm ex}=\frac{\Omega_\alpha^2}{\kappa} \quad\mbox{et}\quad \Gamma_0=\frac{\Omega_\beta^2}{4(\kappa+\gamma_\beta)}}
\label{eq054}
\end{equation}
{En} étudiant l'effet de l'opérateur de saut de cavité  $C_c$ et de saut d'échange de métastabilité $ C_\beta $ sur le vecteur d'état (\ref{eq039}), nous pouvons interpréter l'hamiltonien effectif de l'équation~(\ref{eq045}). (i) Considérons d'abord l'effet d'un saut de cavité, qui se produit à l'instant $ t $ avec un taux
$\kappa (\langle \psi_\alpha^{11}|\psi_\alpha^{11}\rangle+\langle \psi_\alpha^{01}|\psi_\alpha^{01}\rangle)_{\rm adiab}/\langle\psi_\alpha^{00}|\psi_\alpha^{00}\rangle$.
Juste après le saut, le vecteur d'état, initialement en régime de suivi adiabatique, devient 
\begin{align}
|\psi(t^+)\rangle= C_c|\psi(t^-)\rangle_{\rm adiab}  \propto  |\psi_\alpha^{01}(t^-)\rangle_{\rm adiab} |0\rangle|0\rangle + |\psi_\alpha^{11}(t^-)\rangle_{\rm adiab} |1\rangle|0\rangle
\end{align}
 C'est la superposition d'une composante instable $|1\rangle|0\rangle$ et d'une composante stable $|0\rangle|0\rangle$.
Avec une probabilité 
$\langle\psi_\alpha^{11}|\psi_\alpha^{11}\rangle_{\rm adiab}/(\langle\psi_\alpha^{01}|\psi_\alpha^{01}\rangle+\langle\psi_\alpha^{11}|\psi_\alpha^{11}\rangle)_{\rm adiab}$
le saut de cavité est alors suivi d'un saut d'échange de métastabilité avant que le {vecteur d'état du} système n'ait le temps de rejoindre sa valeur adiabatique. On a dans ce cas un \og saut double \fg, qui en définitive n'affecte pas la composante $ | \psi_\alpha^{00} (t^-) \rangle $ puisque 
\begin{equation}
C_\beta C_c |\psi(t^-)\rangle_{\rm adiab} \propto  |\psi_\alpha^{00}(t^-)\rangle |0\rangle |0\rangle
\end{equation}
Ce processus  contribue au terme scalaire (proportionnel à l'identité) dans l'hamiltonien effectif de l'équation~(\ref{eq045}). Avec la probabilité  complémentaire
$\langle\psi_\alpha^{01}|\psi_\alpha^{01}\rangle_{\rm adiab}/(\langle\psi_\alpha^{01}|\psi_\alpha^{01}\rangle+\langle\psi_\alpha^{11}|\psi_\alpha^{11}\rangle)_{\rm adiab}$
le {vecteur d'état} rejoint sa valeur adiabatique avant que d'autres sauts ne se produisent, et est asservi à $ {|\psi_\alpha^{(01)}(t^-)\rangle_{\rm adiab}\propto} P_\alpha | \psi_\alpha^{00} (t^-) \rangle $, c'est-à-dire que la composante lente $ | \psi_\alpha^{00} (t^-) \rangle $ a subi  de manière effective un saut quantique simple avec un opérateur de saut proportionnel à $ P_\alpha $. Ce processus correspond au premier terme, proportionnel à $ P_\alpha^2 $, dans l'hamiltonien effectif de l'équation~(\ref{eq045}). 
(ii) Supposons ensuite que le saut  à l'instant $t$ est un saut d'échange de métastabilité, ce qui se produit avec un taux
$\gamma_\beta \langle \psi_\alpha^{11}|\psi_\alpha^{11}\rangle_{\rm adiab}/\langle\psi_\alpha^{00}|\psi_\alpha^{00}\rangle$.
On vérifie dans ce cas que  le vecteur d'état après le saut, $C_\beta |\psi(t^-)\rangle$, est entièrement instable et subit presque immédiatement un second saut, un saut de cavité. L'effet total correspond là encore à un saut double et à l'action d'un opérateur scalaire sur la composante lente. Nous tirons de cette discussion les taux de saut simple et de saut double suivants :
\bea
\Gamma_s &=& \frac{\kappa (\langle \psi_\alpha^{11}|\psi_\alpha^{11}\rangle+\langle \psi_\alpha^{01}|\psi_\alpha^{01}\rangle)_{\rm adiab}}{\langle\psi_\alpha^{00}|\psi_\alpha^{00}\rangle} \frac{\langle\psi_\alpha^{01}|\psi_\alpha^{01}\rangle_{\rm adiab}}{(\langle\psi_\alpha^{01}|\psi_\alpha^{01}\rangle+\langle\psi_\alpha^{11}|\psi_\alpha^{11}\rangle)_{\rm adiab}}=\Gamma_{\rm ex} \frac{\langle\psi_\alpha^{00}|P_\alpha^2|\psi_\alpha^{00}\rangle}{\langle\psi_\alpha^{00}|\psi_\alpha^{00}\rangle}\equiv\Gamma_{\rm ex} \langle P_\alpha^2\rangle \\
\Gamma_d &=& \frac{\kappa (\langle \psi_\alpha^{11}|\psi_\alpha^{11}\rangle+\langle \psi_\alpha^{01}|\psi_\alpha^{01}\rangle)_{\rm adiab}}{\langle\psi_\alpha^{00}|\psi_\alpha^{00}\rangle} \frac{\langle\psi_\alpha^{11}|\psi_\alpha^{11}\rangle_{\rm adiab}}{(\langle\psi_\alpha^{01}|\psi_\alpha^{01}\rangle+\langle\psi_\alpha^{11}|\psi_\alpha^{11}\rangle)_{\rm adiab}} + \frac{\gamma_\beta \langle \psi_\alpha^{11}|\psi_\alpha^{11}\rangle_{\rm adiab}}{\langle\psi_\alpha^{00}|\psi_\alpha^{00}\rangle} = \Gamma_0
\eea
On obtient finalement l'équation pilote à un mode décrivant l'évolution lente de l'opérateur densité $\rho_\alpha$ du mode bosonique $ \alpha $  (hybridé mais presque purement de spin nucléaire):
\begin{equation}
\boxed{
\frac{\dd \rho_{\alpha}}{\dd t} =  C_s\rho_{\alpha} C_s^\dagger -\frac{1}{2}\{ C_s^\dagger C_s,\rho_{\alpha}\} +  C_d\rho_{\alpha} C_d^\dagger -\frac{1}{2}\{ C_d^\dagger C_d,\rho_{\alpha}\}}
\label{eq030oneMode}
\end{equation}
en termes de deux sauts quantiques, le saut  simple (uniquement de cavité) $ C_s $ et le saut double  (de cavité et d'échange de métastabilité dans cet ordre ou dans l'autre) $ C_d $ :
\begin{equation}
C_s= \sqrt{\Gamma_{\rm ex}} P_\alpha \quad ; \quad C_d=\sqrt{\Gamma_0} \mathbb{1}
\end{equation}
De l'équation (\ref{eq030oneMode})  intégrée pour l'état initial vide de $\alpha$, on tire :
\begin{equation}
\langle X_\alpha^2 \rangle = \frac{1}{4} (1+\Gamma_{\rm ex}t) \quad ; \quad \langle P_\alpha^2 \rangle = \frac{1}{4} 
\end{equation}
{ce qui désigne effectivement $\Gamma_{\rm ex}$ comme un taux de création d'excitations dans le mode $\alpha$.} En revenant à la base atomique initiale (non tournée) et en limitant le vecteur d'état (\ref{eq039}) à son premier terme, on retrouve l'équation (\ref{eq032}) et les trois premiers résultats de l'équation (\ref{eq031}) du modèle à trois modes, {pourtant valables à couplage de Faraday $\Omega$ quelconque, pas nécessairement infinitésimal}. Enfin, le nombre moyen de photons polarisés selon $Oy$ sortant de la cavité par unité de temps, donné dans le modèle à un mode par $\Gamma_0+\Gamma_{\rm ex}/4$ comme le montrera l'équation (\ref{eq064}), est en accord avec la valeur exacte $\kappa \langle c^\dagger c\rangle_{s}$ où le nombre moyen stationnaire de photons polarisés selon $Oy$ dans la cavité $\langle c^\dagger c\rangle_{s}=\langle X_c^2\rangle_{s}-1/4$ est le dernier résultat de (\ref{eq031}).\footnote{En revanche, la valeur de $\langle c^\dagger c\rangle_{\rm adiab}$ dans la forme adiabatique (\ref{eq043}) du vecteur d'état ne représente pas ce nombre. La solution du paradoxe tient à l'existence de la voie de désexcitation (ii), celle de l'annihilation en premier saut de l'excitation $n_\beta=1$ dans le mode du métastable immédiatement suivie par la perte d'un photon en cavité. Le vrai taux de sortie de photons polarisés selon $Oy$ est donc $\kappa \langle c^\dagger c\rangle_{\rm adiab}+\gamma_\beta \langle \beta^\dagger\beta\rangle_{\rm adiab}$.}

\section{Mesure quantique non destructive du spin nucléaire en continu} 
\label{sec:QND} 

Les moyennes quantiques calculées dans la section \ref{sec:qt} correspondent seulement aux moyennes d'ensemble  sur un nombre infini de {réalisations de l'expérience}. Dans cette section, nous étudions  ce qui nous intéresse vraiment, l'évolution du système dans une ou plusieurs réalisations  données, conditionnée aux résultats d'une mesure  en continu sur la lumière polarisée selon $Oy $ sortant de  la cavité. Pour cela, nous revenons à la formulation en termes de fonctions d'onde Monte-Carlo, comme dans la section \ref{sec:qt}, où des trajectoires  stochastiques $|\psi(t)\rangle$ correspondant à une succession particulière de sauts quantiques reconstruisent l'opérateur densité  du système conditionné à  des résultats de mesure \cite{CastinDalibard}. La forme  précise des opérateurs de saut Monte-Carlo, qui n'est pas unique dans la reformulation stochastique d'une équation pilote,  est alors déterminée par les mesures particulières effectuées.

\subsection{Étude de la compression par comptage de photons : résultats analytiques à un mode et numériques à trois modes}
\label{sec:dpcdp}
Supposons que l'on compte {en continu} et directement  (par photodétection) le nombre de photons polarisés selon $Oy $ sortant de la cavité (voir la figure~\ref{fig1}b), comme l'a proposé la référence \cite{Milburn1993}. L'opérateur de saut associé à cette mesure est $\sqrt{\kappa}c$, de sorte que l'équation pilote à trois modes (\ref{eq030}) est déjà sous la bonne forme pour analyser  l'évolution du vecteur d'état $|\psi(t)\rangle$ conditionnée à la mesure. 

 Il en va de même dans la limite d'un faible couplage de Faraday, $\Omega\to 0$, qui conduit au modèle à un mode {de la section \ref{sec:ana_omm}}. Comme les opérateurs de saut $ C_d $ et $ C_s $ de son équation pilote (\ref{eq030oneMode}) correspondent tous deux à la perte en cavité d'un photon polarisé selon $Oy$ (rappelons-le, $ C_d $ résulte d'un saut de cavité immédiatement suivi ou précédé d'un saut d'échange de métastabilité, et $ C_s $ d'un saut simple de cavité), la mesure ne peut faire la distinction entre les deux, et l'opérateur densité conditionné à un nombre donné  $n$ de photons détectés est obtenu en moyennant sur des réalisations ayant ce même nombre {\it total} $n$ de sauts.  Un  vecteur d'état Monte-Carlo non normalisé ayant subi ces $n$ sauts pendant la durée $t$ s'écrit 
\begin{equation}
|\psi(t)\rangle = \eee^{-\frac{\ii}{\hbar}{H}^{00}_{\rm eff}(t-t_n)} \, C_{\epsilon_{n}} \, \eee^{-\frac{\ii}{\hbar}{H}^{00}_{\rm eff}(t_n-t_{n-1})}
\, C_{\epsilon_{n-1}} \ldots \, C_{\epsilon_{1}} \,  \, \eee^{-\frac{\ii}{\hbar}{H}_{\rm eff}^{00} t_1} |\psi(0)\rangle \label{eq057}
\end{equation}
où  $ \epsilon_{k}\in\{s,d\} $ et  $t_k$ sont le type et l'instant du $ k $ème saut, $H_{\rm eff}^{00}$ est l'hamiltonien effectif (\ref{eq045}) {et nous notons le vecteur d'état du modèle à un mode $|\psi\rangle$ plutôt que $|\psi_\alpha^{00}\rangle$ pour alléger}. La moyenne quantique d'une observable $ O $ s'obtient en moyennant sur toutes les trajectoires possibles, donc en sommant sur le nombre et le type des sauts et en intégrant sur leurs instants: 
\begin{equation}
\langle O \rangle(t) = \sum_n \int_{0<t_1<t_2 \ldots <t_n<t} \dd t_1 \, \dd t_2 \ldots \dd t_n \, \sum_{(\epsilon_k)_{1 \leq k \leq n} \in \{ s,d \}^n }
\langle \psi(t) |O| \psi(t) \rangle
\label{eq058}
\end{equation}
où la norme au carré de chaque  vecteur d'état non normalisé $ | \psi (t) \rangle $ donne automatiquement sa densité de probabilité \cite{LiYun}.  En prenant $ O = \mathbb{1} $, nous en déduisons la probabilité que $ n $ sauts se soient produits dans l'intervalle de temps {$[0, t]$:}
\begin{equation}
\Pi_n(t) =  \int_{0<t_1<t_2 \ldots <t_n<t} \dd t_1 \, \dd t_2 \ldots \dd t_n \, \sum_{(\epsilon_k)_{1 \leq k \leq n} \in \{ s,d \}^n }
\langle \psi(t) | \psi(t) \rangle
\label{eq059}
\end{equation}
Pour évaluer (\ref{eq059}), nous tirons parti du fait que tous les opérateurs de saut dans (\ref{eq057}) et leurs conjugués hermitiens commutent entre eux et avec $H_{\rm eff}^{00}$.\footnote{{Pour cette raison, garder l'information sur les instants des sauts ne permet pas, par post-sélection, d'augmenter l'efficacité de la compression de spin. En effet, l'opérateur densité $\rho_\alpha(t)|_{t_1,\ldots, t_n}$ sachant que $n$ sauts se sont produits aux instants $t_1,\ldots,t_n$ conduit à la même distribution de probabilité de $P_\alpha$ que l'opérateur densité $\rho_\alpha(t)|_n$ sachant seulement qu'il y a eu $n$ sauts pendant $[0,t]$.}}  En utilisant les identités
\begin{equation}
 \sum_{\epsilon_n=s,d}\ldots \sum_{\epsilon_1=s,d} \left( C_{\epsilon_{n}}^\dagger C_{\epsilon_{n}} \ldots   C_{\epsilon_{1}}^\dagger C_{\epsilon_{1}} \right) =
 \left( \sum_{\epsilon_n=s,d} C_{\epsilon_{n}}^\dagger C_{\epsilon_{n}} \right) \ldots  \left( \sum_{\epsilon_n=s,d}  C_{\epsilon_{1}}^\dagger C_{\epsilon_{1}} \right) = \left( \Gamma_{\rm ex} P_\alpha^2 + \Gamma_0\mathbb{1} \right)^n
\end{equation}
et en injectant une relation de fermeture dans la base propre de $ P_\alpha $ telle que
$P_\alpha |p_\alpha\rangle = p_\alpha |p_\alpha\rangle$, 
après avoir intégré sur les temps $t_k$ comme le permet le produit télescopique des opérateurs d'évolution, nous obtenons 
\begin{equation}
\Pi_n(t)=\frac{t^n}{n!} \int_{-\infty}^{+\infty} \, \dd p_\alpha \, \left( \Gamma_{\rm ex} p_\alpha^2 +  \Gamma_0 \right)^n 
\eee^{-\Gamma_{\rm ex}p_\alpha^2t} \eee^{-\Gamma_0t} \Pi (p_\alpha,0)=
\binom{2n}{n}\frac{(\Gamma_{\rm ex}t/8)^n\, \eee^{-\Gamma_0 t}}{(1+\Gamma_{\rm ex}t/2)^{n+1/2}}  \,
\Phi \left(-n , \frac{1}{2} -n ; {\Gamma_0} t+  \frac{2\Gamma_0}{\Gamma_{\rm ex}}\right)
\label{eq062}
\end{equation}
où $ \Pi (p_\alpha, 0) $ est la distribution de probabilité initiale de $ p_\alpha $ (une gaussienne de moyenne nulle et de variance $ 1/4 $) et $ \Phi $ est la fonction hypergéométrique confluente de Kummer ${}_1F_1 $. On remarque que (\ref{eq062}) est en fait une moyenne gaussienne sur $ p_\alpha $ d'une loi de Poisson de paramètre 
$\lambda=( \Gamma_{\rm ex} p_\alpha^2 + \Gamma_0)t$.
On en déduit la moyenne et la variance du nombre de photodétections pendant la durée $ t $: 
\begin{equation}
\langle n \rangle = \left(\Gamma_0 +\frac{1}{4} \Gamma_{\rm ex}\right) t \quad ; \quad 
\mbox{Var} \, n =  \langle n \rangle + \frac{(\Gamma_{\rm ex}t)^2 }{8} \label{eq064}
\end{equation}
Toujours en utilisant l'équation~(\ref{eq062}), nous accédons à la distribution de probabilité de $ p_\alpha $ sachant que $ n $ photons ont été détectés dans l'intervalle de temps $ [0, t] $:
\begin{equation}
\Pi_t(p_\alpha|n)=\frac{1}{\Pi_n(t)} \frac{t^n}{n!}  \left( \Gamma_{\rm ex} p_\alpha^2 +\Gamma_0\right)^n 
\eee^{-\Gamma_{\rm ex}p_\alpha^2t} \eee^{-\Gamma_0 t} \Pi (p_\alpha,0)
\label{eq065}
\end{equation} 
{Comme on pouvait s'y attendre, c'est une fonction paire de $p_\alpha$, la photodétection donnant accès seulement à l'intensité du champ sortant polarisé selon $Oy$ et ne pouvant distinguer entre des valeurs opposées $\pm p_\alpha$ de la quadrature $P_\alpha$ du spin nucléaire hybridé selon $Oz$. Il en résulte donc une compression des fluctuations de $P_\alpha^2$ plutôt que de $P_\alpha$, ce que nous caractérisons par} la moyenne et la variance conditionnelles de $ P_\alpha^2 $ sachant que $ n $ photons ont été détectés pendant $ t $, {déduites de (\ref{eq065}):}
\begin{equation}
\langle P_\alpha^2 \rangle_n = \frac{(n+1)}{\Gamma_{\rm ex} t} \frac{\Pi_{n+1}(t)}{\Pi_n(t)} - \frac{\Gamma_0}{\Gamma_{\rm ex}} \quad ; \quad {\mbox{Var}_n (P_\alpha^2)\equiv \langle P_\alpha^4\rangle_n -\langle P_\alpha^2\rangle_n^2 = \frac{(n+1)^2}{(\Gamma_{\rm ex} t)^2} \left[ \frac{(n+2)\Pi_{n+2}(t)}{(n+1)\Pi_n(t)} -\frac{\Pi^2_{n+1}(t)}{\Pi_n^2(t)}\right]}
\label{eq066}
\end{equation}
Enfin, au moyen de l'équation~(\ref{eq065}), nous trouvons que pour $ \Gamma_{\rm ex} t\to +\infty$, la distribution de probabilité de $ p_\alpha^2 $ conditionnée au nombre $ n $ de photodétections est piquée autour {d'une valeur  $ p_0^2 $ avec une variance conditionnelle tendant vers zéro} \footnote{D'après l'équation~(\ref{eq064}), le second membre de {la première équation dans} (\ref{eq068}) est asymptotiquement de l'ordre de l'unité pour une séquence de photodétection typique. {Cette équation} n'a en fait de sens que pour $p_0^2$ positif donc $n> \Gamma_0 t$ ; alors, les équivalents {(\ref{eq068})} s'appliquent lorsque l'écart entre les deux pics dans $\Pi_t(p_\alpha|n)$ est beaucoup plus grand que leur largeur,  ce qui impose 
$2 p_0^2\gg n^{1/2}/\Gamma_{\rm ex }t = (\Gamma_0+ \Gamma_{\rm ex} p_0^2)^{1/2}/\Gamma_{\rm ex} t^{1/2}$.
{Pour les obtenir, on pose $n=\gamma t$ avec $\gamma>\Gamma_0$, puis on écrit (\ref{eq065}) sous la forme $\exp[-t S(p_\alpha)]\Pi(p_\alpha,0)$ et on quadratise $S(p_\alpha)$ autour de ses minima.}}:
{
\begin{equation}
\boxed{
p_0^2-\frac{1}{4} = \frac{n-\langle n \rangle}{\Gamma_{\rm ex} t} \quad \mbox{d'où}\quad 
\langle P_\alpha^2\rangle_n \underset{\Gamma_{\rm ex}t\to +\infty}{\sim} p_0^2} \quad ; \quad
\boxed{\mbox{Var}_n (P_\alpha^2)\underset{\Gamma_{\rm ex}t {\to+\infty}}{ \sim } \frac{n}{(\Gamma_{\rm ex}t)^2} \to 0 \label{eq068}}
\end{equation}
}
{En remplaçant dans cette expression $n$ par sa valeur moyenne et en tenant compte de la valeur $1/8$ de la variance de $P_\alpha^2$ dans l'état initial, nous aboutissons au taux de compression du spin nucléaire par comptage de photons:}
\be
\label{eq868}
{\boxed{\Gamma_{\rm sq} = \frac{\Gamma_{\rm ex}^2}{8(\Gamma_0+\frac{1}{4}\Gamma_{\rm ex})}}}
\ee
De même, la distribution de probabilité conditionnelle de $ p_\alpha $ présente deux pics à $\pm p_0$ comme on peut le voir sur la fonction de Wigner de la figure~\ref{fig7}b, obtenue par simulation numérique de l'évolution conditionnelle du système aux temps longs dans le modèle à un mode (\ref{eq030oneMode}).\footnote{\label{note600}{L'absence de franges montre qu'on a préparé un mélange statistique plutôt qu'une superposition cohérente de deux états comprimés en la quadrature $P_\alpha$.  On trouve en effet à partir de l'équation (\ref{eq058}) que
$\langle p_0|\rho_n(t)|-p_0\rangle/\langle p_0|\rho_n(t)|p_0\rangle=[(\Gamma_0-\Gamma_{\rm ex} p_0^2)/(\Gamma_0+\Gamma_{\rm ex} p_0^2)]^n\simeq \exp(-2\Gamma_{\rm ex} t p_0^2)\lll 1$
sur la figure \ref{fig7}b, $\rho_n(t)$ étant l'opérateur densité conditionnel. La méthode de Laplace donne aux temps longs la distribution de Wigner conditionnelle 
$W_t(x_\alpha,p_\alpha|n)\sim\pi\Pi_t(p_\alpha|n)\exp[-2 x_\alpha^2/\Gamma_{\rm ex}t s(p_\alpha)]/\sqrt{\pi\Gamma_{\rm ex}t s(p_\alpha)/2}$ avec $s(p_\alpha)=[1+(\Gamma_0+\Gamma_{\rm ex}p_0^2)/(\Gamma_0+\Gamma_{\rm ex}p_\alpha^2)]/2$.
On remarque que $s(\pm p_0)=1$ et $s(p_\alpha)\simeq 1$ pour $\Gamma_0\gg\Gamma_{\rm ex}$.  Alors que $W_t(0,0|n)$ est écrasé exponentiellement en temps (dans la limite $\Gamma_{\rm ex}\ll\Gamma_0$, il vient
$W_t(0,0|n)\sim(\Gamma_{\rm ex}/\Gamma_0)^{1/2}p_0\exp(2 p_0^2)\exp(-4 p_0^4 \Gamma_{\rm sq}t)$),
on a aussi $W_t(0,0|n)=\langle 2\varepsilon\rangle_n$ où $\varepsilon=\pm 1$ est la parité de la fonction d'onde Monte-Carlo $\psi(p_\alpha,t)$. Dans une simulation numérique, on a donc seulement une décroissance lente $\langle\varepsilon\rangle_n\approx 1/\sqrt{\mathcal{N}_n}$ où $\mathcal{N}_n$ est le nombre de trajectoires ayant subi $n$ sauts pendant $t$ ; ceci conduit à des franges non physiques à valeurs négatives dans la distribution de Wigner près de l'axe $O x_\alpha$. Pour minimiser cet effet et le rendre imperceptible à une résolution en $p_\alpha$ pas trop haute ($\dd p_\alpha=0,044$ sur la figure \ref{fig7}b), on arrête la simulation Monte-Carlo à un stade où il y a exactement le même nombre $\mathcal{N}_n/2$ de fonctions d'onde paires et impaires. {Pour obtenir une superposition cohérente d'états comprimés, il faudrait effectuer une post-sélection supplémentaire, en se limitant aux réalisations Monte-Carlo de fonction d'onde $\psi(p_\alpha,t)$ de parité fixée $\varepsilon$ (ayant subi un nombre pair de sauts simples si $\varepsilon=1$, un nombre impair sinon, l'opérateur de saut $C_s$ changeant la parité).  Dans l'opérateur densité conditionnel correspondant, on a alors
$\langle p_0| \rho_{n,\varepsilon}(t)|-p_0\rangle/\langle p_0|\rho_{n,\varepsilon}(t)|p_0\rangle = \overline{\psi(-p_0,t)\psi(p_0,t)}^{n,\varepsilon}/\overline{\psi(p_0,t)\psi(p_0,t)}^{n,\varepsilon}=\varepsilon$
sans que la structure à deux pics de la distribution de $P_\alpha$ ne soit affectée aux temps longs car
$\Pi_t(p_\alpha|n,\varepsilon)/\Pi_t(p_\alpha|n) \propto 1+\varepsilon[(\Gamma_0-\Gamma_{\rm ex} p_\alpha^2)/(\Gamma_0+\Gamma_{\rm ex}p_\alpha^2)]^n \to 1$
lorsque $n\to +\infty$ à $p_\alpha$ non nul fixé. La distribution de Wigner $W_t(x_\alpha,p_\alpha|n,\varepsilon)$ présente désormais des franges positives et négatives d'amplitude maximale $2$ sur l'axe $p_\alpha=0$. Cette technique de filtrage vient à bout aussi des mécanismes de décohérence de la section \ref{sec:edld} car les fonctions d'onde Monte-Carlo restent de parité bien définie après action de l'opérateur de saut correspondant $\gamma_\alpha^{1/2}\alpha$. Cette idée de contrôler la décohérence au travers de mesures de parité est bien connue en électrodynamique quantique en cavité \cite{Raimond}.}}} {Ceci montre bien que,} lors d'une réalisation  donnée de l'expérience,  {la photodétection en continu} des photons  polarisés selon $Oy$ sortant de la cavité rend de plus en plus certaine la valeur de $ P_\alpha^2 $, et donc {dans une large mesure} de $ I_z^2 $, le carré de la composante selon $Oz $ du spin nucléaire collectif, {comme on le voit en reliant} dans la limite $\Omega\to 0$ les moments conditionnels de $P_a^2$, c'est-à-dire de $I_z^2$, à ceux de $P_\alpha^2$: 
\begin{equation}
\label{eq968}
\langle P_a^2\rangle_n =  \frac{\gamma_m}{\gamma_f+\gamma_m} \langle P_\alpha^2\rangle_n + \frac{\gamma_f/4}{\gamma_f+\gamma_m}\quad ;\quad \mbox{Var}_n( P_a^2)= \frac{\gamma_m^2}{(\gamma_f+\gamma_m)^2}\mbox{Var}_n(P_\alpha^2) + 
\frac{\gamma_f \gamma_m}{(\gamma_f+\gamma_m)^2} \langle P_\alpha^2\rangle_n + \frac{\gamma_f^2/8}{(\gamma_f+\gamma_m)^2}
\end{equation}
{Puisque la compression porte sur $P_a^2$ plutôt que sur $P_a$, la distribution angulaire conditionnelle du spin nucléaire collectif est bimodale (elle présente comme celle de $P_\alpha$ deux pics bien séparés pourvu que $\gamma_f/\gamma_m\ll (2 p_0)^2$); ces structures plus étroites que la limite quantique standard permettent quand même un pointage angulaire plus précis qu'avec un état non comprimé. {Nous redéfinissons donc le gain métrologique (\ref{eq017}) en remplaçant au troisième membre de cette équation la variance conditionnelle de $P_a$ par le carré de la mi-largeur $\delta P_a$ des pics centrés en $\pm P_{a,0}$ de la distribution de probabilité conditionnelle de $P_a$, puis en assimilant le centre $P_{a,0}^2$ et la largeur $\delta(P_a^2)$ de la distribution de $P_a^2$ à la moyenne et à l'écart-type conditionnels de $P_a^2$:
\be
\label{eq067}
\xi^2 = \frac{4(\delta P_a)^2}{\eta} = \frac{(2 P_{a,0} \delta P_a)^2}{\eta P_{a,0}^2} = \frac{[\delta(P_a^2)]^2}{\eta P_{a,0}^2}=\frac{\mbox{Var}_n(P_a^2)}{\eta \langle P_a^2\rangle_n}
\ee
Remarquons que cette expression ne se déduit pas de la méthode des moments exposée dans la section II.B.6 de la référence \cite{PezzeRMP} en prenant $P_a^2$ comme estimateur, car l'effet de la transformation unitaire $\exp(2 \ii\theta X_a)$ (en pratique, une précession du spin nucléaire autour d'un champ magnétique selon $Oy$) n'est pas de déplacer le pic dans la distribution de $P_a^2$ mais de le cliver en deux pics centrés en $(P_{a,0}\pm\theta)^2$.
}}

\begin{figure}[t]
\centerline{\includegraphics [width = 0.4 \textwidth,height=0.23\textwidth]{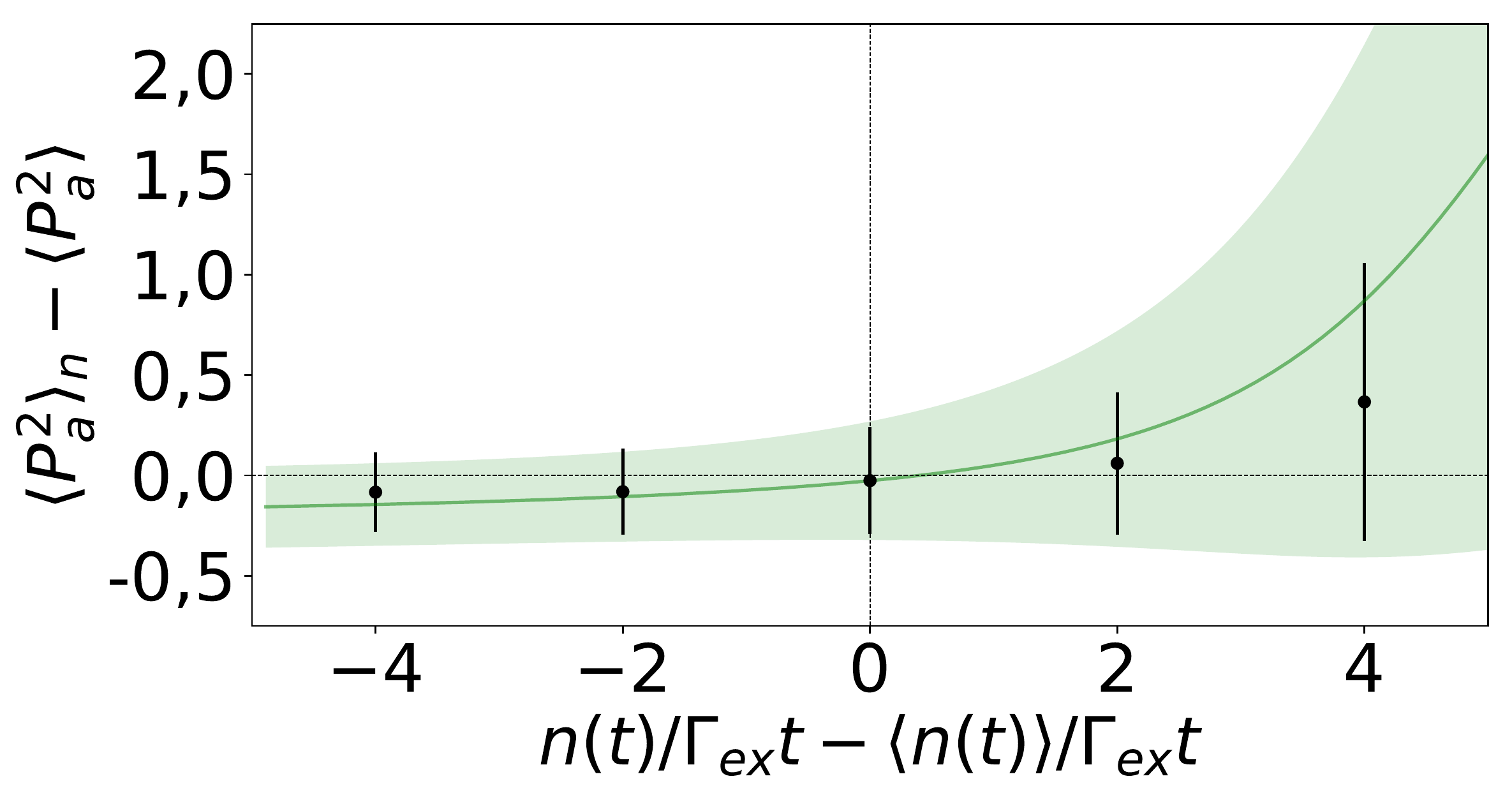} (a) \quad \includegraphics [width=0.4\textwidth,height=0.23\textwidth]{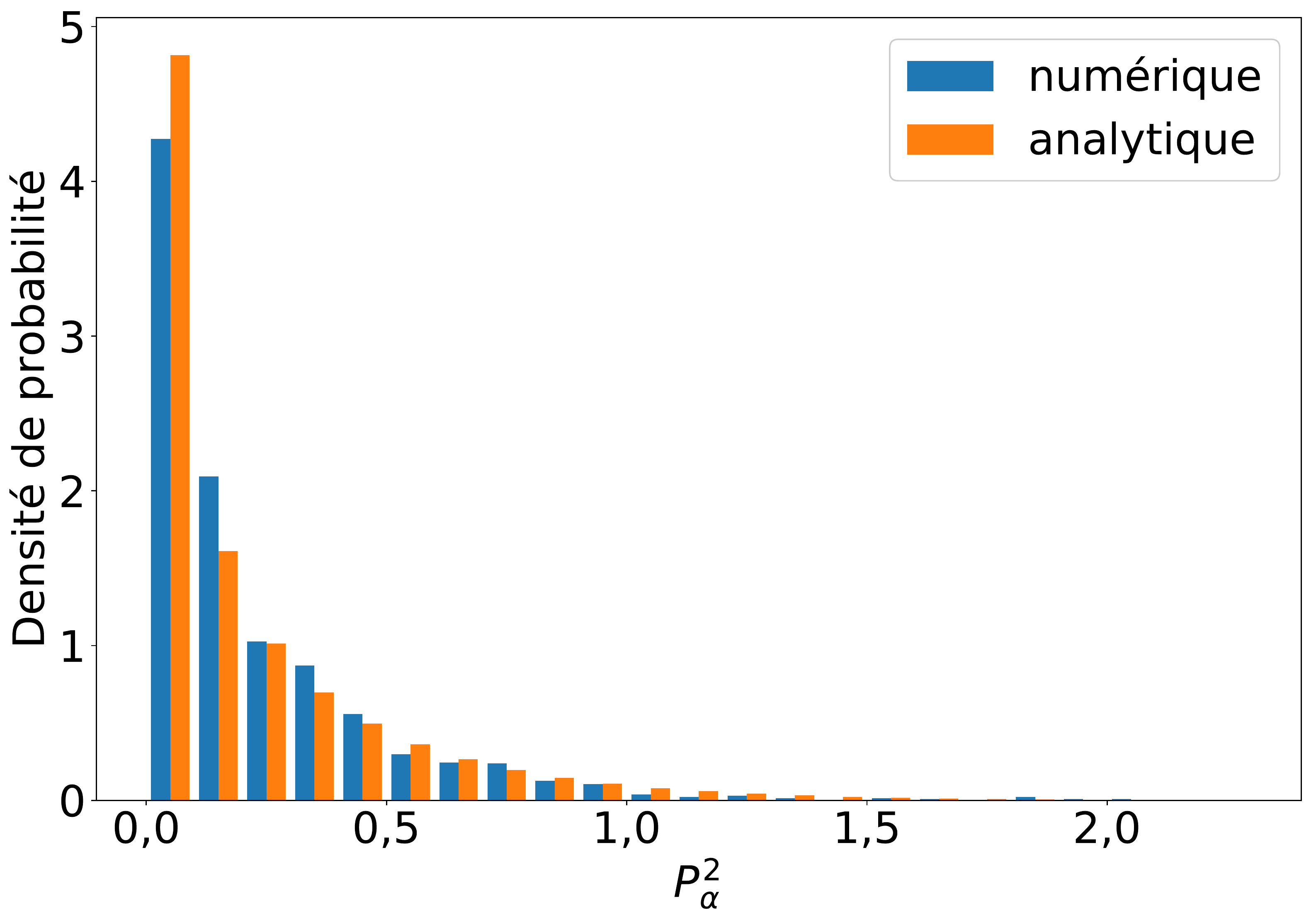} (b)} 
\caption{Compression de $P_a^2$ par comptage de photons aux temps courts, $ \Gamma_{\rm ex} t = 15 $, {où $\Gamma_{\rm ex}$ est le taux (\ref{eq054}) de création d'excitations dans le mode bosonique nucléaire hybridé $\alpha$}. (a) Moyenne et écart-type conditionnels de la quadrature du spin nucléaire au carré $P_a^2$ sachant que $ n $ photodétections ont eu lieu dans l'intervalle de temps $ [0, t] $,  en fonction de ce nombre $ n $. L'écart-type est représenté sous la forme d'un intervalle de confiance. La moyenne inconditionnelle $\langle P_a^2\rangle=1/4$ est indépendante du temps, voir l'équation (\ref{eq032}). Points noirs et barres d'erreur~: simulation numérique du modèle à 3 modes avec 3000 réalisations ; ligne verte et zone colorée : prédictions analytiques tirées des équations (\ref{eq062}), (\ref{eq066}) et (\ref{eq968}) du modèle à un mode. En pratique, les points noirs sont obtenus après moyenne sur des classes de valeurs de $n$ centrées sur ces points (dans une classe donnée, les trajectoires ont des nombres de photodétections proches mais des histoires indépendantes pour les sauts d'échange de métastabilité auxquelles l'expérimentateur n'a pas accès). Paramètres du modèle à 3 modes:
$\Omega/\kappa = 1/3$, $\gamma_m/\kappa = 1/10$,  $\gamma_f/\kappa = 1/1000$ (si bien que $\Gamma_{\rm ex}/\kappa=1/909$), $n_a^{\rm max} = 64, n_b^{\rm max} =n_c^{\rm max}=8$.  
Ceci correspond à
$\Gamma_0/\Gamma_{\rm ex}=12\,500/601\simeq 20,8$ {et $\Gamma_{\rm sq}/\Gamma_{\rm ex}=601/101\, 202\simeq 1/168$}
{où $\Gamma_{\rm sq}$ est le taux de compression (\ref{eq868}).} (b) Pour la classe centrée sur $n=\langle n(t)\rangle$, histogramme des valeurs conditionnelles de $P_\alpha^2$. Barres bleues : simulation numérique du modèle à trois modes ; barres orange : prédictions analytiques tirées de l'équation (\ref{eq065}) du modèle à un mode.
}
\label{fig6} 
\end{figure} 

Effectuons pour terminer une vérification numérique de ces prédictions analytiques dans le modèle à trois modes.  Sur la figure {\ref{fig6}a}, nous représentons la moyenne conditionnelle du carré $ P_a^2 $ de la quadrature de spin nucléaire sachant que $ n $ photodétections se sont produites dans l'intervalle de temps $ [0, t] $, avec $ \Gamma_{\rm ex} t = 15 $ (points noirs), en fonction de ce nombre $ n $. L'ensemble des réalisations est divisé en 5 classes correspondant à un nombre de photodétections tombant dans un intervalle donné, et les points noirs sont obtenus en moyennant sur les réalisations dans une même classe. Les résultats numériques sont proches des prédictions analytiques tirées de (\ref{eq066}) et (\ref{eq968}) et représentées en vert, sauf dans les classes extrêmes qui comportent un nombre trop faible de réalisations. En revanche, les prédictions analytiques asymptotiques {(\ref{eq068})}, non représentées, seraient en désaccord avec les simulations des deux modèles car le temps {$t=15/\Gamma_{\rm ex}$ n'est pas assez long, il est très inférieur au temps de compression $1/\Gamma_{\rm sq}$}. {Sur la figure \ref{fig6}b, nous représentons la distribution de probabilité conditionnelle de $P_\alpha^2$ correspondant à la classe centrale de la figure \ref{fig6}a ; il y a là aussi bon accord entre analytique à un mode et numérique à trois modes.}
\begin{figure}[t]
\begin{center}
\subfloat [] 
{\includegraphics [width = 0.45 \textwidth,height=50mm]{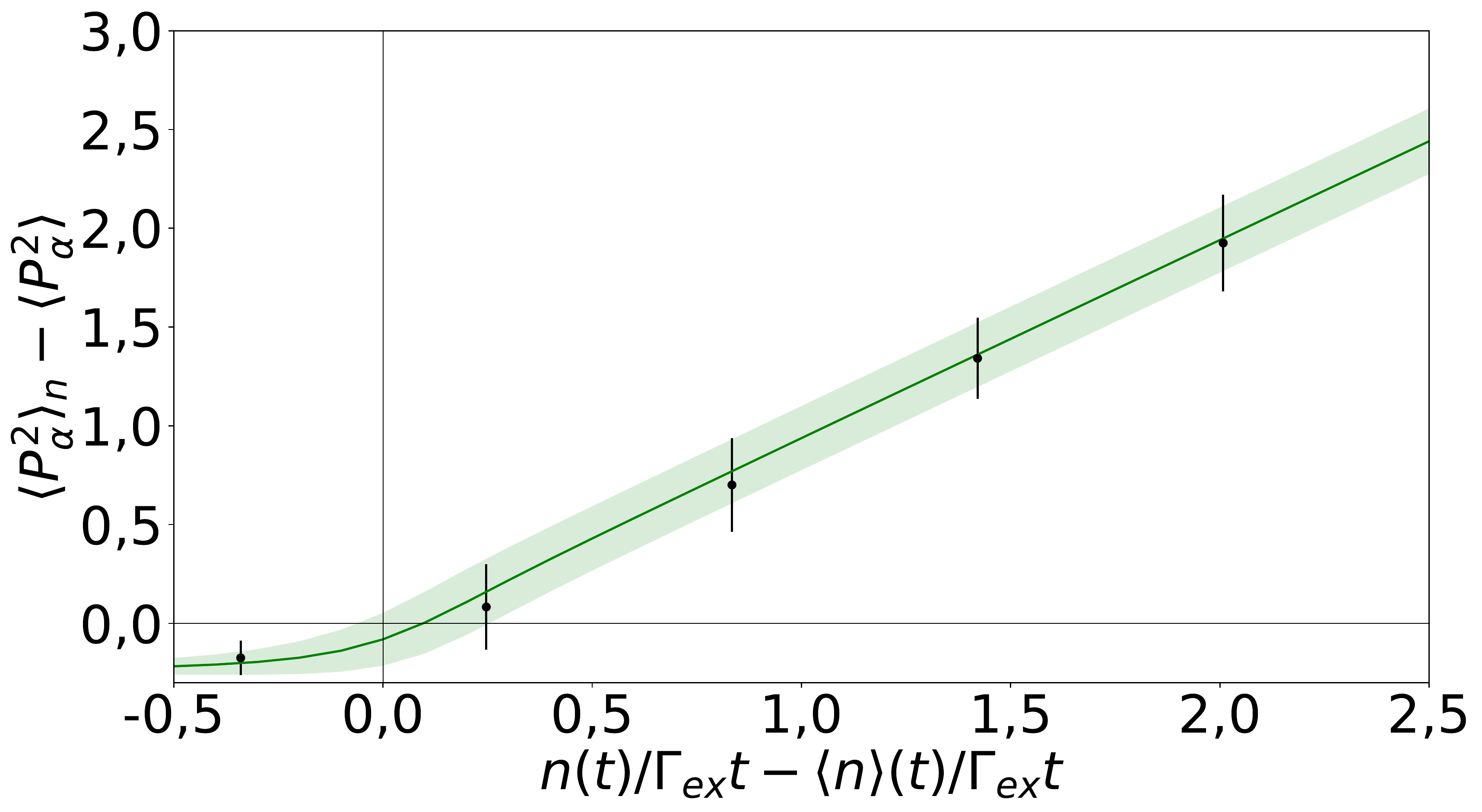}} \quad\quad
\subfloat [] 
{\includegraphics [width = 0.4 \textwidth,height=55mm]{fig7c.pdf}} 
\end{center}
\caption{Compression de $P_\alpha^2$ par comptage de photons aux temps longs dans le modèle à un mode (\ref{eq030oneMode}).  (a) Moyenne et écart-type conditionnels de $P_\alpha^2$ sachant que le nombre de photodétections $n$ tombe dans une classe de valeurs donnée, de manière similaire à la figure \ref{fig6}a mais pour $ \Gamma_{\rm ex} t = 1000$ et 2000 réalisations (ce temps long {rendrait} difficile une simulation dans le modèle à {trois} modes).  (b) Distribution de Wigner du mode bosonique nucléaire  hybridé dans l'espace des quadratures $ (X_\alpha, P_\alpha) $ à $ \Gamma_{\rm ex} t = 1000 $, obtenue par moyennage des dyades $|\psi(t)\rangle\langle\psi(t)|$ {dont le nombre de photodétections tombe dans} la 3ème classe de (a) {(302 trajectoires sur 5000 réalisations)}. Elle présente deux lignes de {crête mais pas de franges d'interférence (voir la note \ref{note600}).}}
\label{fig7} 
\end{figure} 
Sur la figure \ref{fig7}, nous explorons justement les temps longs dans le modèle à un mode, avec $ \Gamma_{\rm ex} t = 1000 $ {c'est-à-dire $\Gamma_{\rm sq}t\simeq 5,94$}. La figure~\ref{fig7}a, qui est l'équivalent de la figure~\ref{fig6}a, montre que $ \langle P_\alpha^2 \rangle_n $ est alors relié au nombre de photodétections $ n $ comme dans la prédiction analytique {(\ref{eq068})}, c'est-à-dire selon la première bissectrice dans les unités de la figure, avec un écart-type conditionnel dans (\ref{eq068}) à peu près constant {$\simeq (\Gamma_0 t)^{1/2}/\Gamma_{\rm ex}t\simeq 1/(\Gamma_{\rm sq} t)^{1/2}$} car $\Gamma_0$ est ici $\gg \Gamma_{\rm ex}$.

\subsection{Étude de la compression par détection homodyne : solution analytique à un mode et à trois modes}
Nous supposons maintenant que les photons sortant de la cavité polarisés selon $Oy$ sont mesurés en continu par détection homodyne \cite{Wiseman2002}, comme sur la figure \ref{fig1}c. Il nous faut d'abord trouver les bonnes équations stochastiques donnant l'évolution du vecteur d'état du système conditionnée à la détection homodyne, puisque les opérateurs de saut apparaissant naturellement dans l'écriture  (\ref{eq030t}) ou (\ref{eq030oneMode}) de l'équation pilote à trois modes ou à un mode sont inadaptés. Nous présentons ensuite quelques résultats analytiques obtenus dans le modèle à un mode puis dans le modèle à trois modes, avant de discuter brièvement l'effet du temps de cohérence fini des atomes métastables. {Afin d'obtenir ces résultats, nous avons utilisé le fait que, pour l'état initial vide considéré ici, le vecteur d'état conditionnel est donné exactement à tout temps par un ansatz gaussien \cite{KlausGauss}, quel que soit le nombre de modes du modèle, en présence ou en absence de décohérence.}

\subsubsection{Formulation stochastique adaptée de l'équation pilote} 
Une équation pilote générale de la forme de Lindblad {\cite{Lindblad}}
\begin{equation}
\frac{\dd \rho}{\dd t} = \frac{1}{\ii\hbar} \left[H,\rho \right] + \sum_m C_m\rho C_m^\dagger - \frac{1}{2} \{ C_m^\dagger C_m, \rho \}
\label{eq030Arbitrary}
\end{equation}
avec $ H $ la partie  hermitienne de l'hamiltonien et  $C_m$ les opérateurs de saut, peut être réécrite de manière équivalente en ajoutant une constante arbitraire aux opérateurs de saut et/ou en les mélangeant par combinaison linéaire unitaire quelconque. Afin de tenir compte d'une détection homodyne  sur le champ sortant, on forme, à partir d'un opérateur de saut $ C_m $  correspondant à une photodétection, les deux opérateurs de saut \og homodynes \fg\, $ D_{m, \pm} $ \cite{CastinDalibard} 
\begin{equation}
D_{m,+}=\frac{\mu \mathbb{1} + C_m}{\sqrt{2}} \quad;\quad  D_{m,-}=\frac{\mu \mathbb{1} - C_m}{\sqrt{2}}
\label{eq072}
\end{equation}
où $ \mu^2 $ a les dimensions d'une  pulsation. La mesure de la différence  des taux de saut $ D^\dagger_+D_+-D^\dagger_-D_-$ donne alors accès à une quadrature de $ C_m $.  Ainsi, pour $ \mu $ réel et $C_m$ correspondant à l'opérateur saut de cavité $C_c$, voir l'équation (\ref{eq038}), la différence entre les nombres de photons $ N_\pm $ détectés pendant le court intervalle de temps $ \Delta t $ dans les deux voies de sortie de la figure \ref{fig1}c, qui constitue par définition le signal homodyne,
\begin{equation}
N_+=(D_{c,+}^\dagger D_{c,+}^{\phantom{\dagger}}) \, \Delta t \quad ;\quad N_-=(D_{c,-}^\dagger D_{c,-}^{\phantom{\dagger}}) \, \Delta t  \quad ;\quad 
\frac{N_+-N_-}{2\, \mu}= \frac{c+c^\dagger}{2} \sqrt{\kappa} \Delta t
\label{eq075}
\end{equation}
donne accès à $X_c$~; c'est bien la quadrature du champ conjuguée à $P_c$ donc translatée d'une quantité proportionnelle à $P_b$ et au temps sous l'action de l'hamiltonien $H$ (\ref{eq016}), ce qui renseigne sur $P_a$ au travers des collisions d'échange de métastabilité.  Dans le cas de l'équation pilote à 3 modes (\ref{eq030t}), on est obligé d'appliquer la procédure de dédoublement (\ref{eq072}) {\it a priori} seulement à l'opérateur de saut de cavité.  En pratique, nous {l'appliquerons} également à l'opérateur de saut $C_\beta$, c'est-à-dire que nous dédoublerons par homodynage {\it tous} les opérateurs de saut $C_m$, afin d'éviter l'inconfort d'une représentation mixte mêlant sauts quantiques {discrets} et évolution stochastique continue, voir l'équation (\ref{eq076}) à venir. Dans le cas de l'équation pilote à un mode (\ref{eq030oneMode}), il faut de toute façon \og homodyner\fg\, les deux opérateurs de saut $C_s$ et $C_d$, puisque chacun s'accompagne de la perte d'un photon en cavité, comme l'explique la section \ref{sec:ana_omm}. 

Dans la limite d'un oscillateur local de grande amplitude $ \mu $,  on peut faire comme si $\Delta t$ était infinitésimal \footnote{Cette approximation est valable pour une résolution en temps, c'est-à-dire un pas temporel $ \Delta t $, telle que $ \mu^{-2} \ll \Delta t \ll \kappa^{-1} $, où $ \kappa $ est en pratique le taux d'évolution le plus rapide du système dans l'expérience.} et représenter l'évolution de la fonction d'onde Monte-Carlo, {cette fois normalisée à l'unité}, par une équation stochastique non linéaire continue sans sauts quantiques \cite{CastinDalibard,Helvetica,Percival} en point de vue d'Ito :
\be
\fbox{\parbox{0.85\textwidth}{
\vspace{-3mm}
\begin{align}
\dd|\phi(t)\rangle & = -\frac{\ii}{\hbar}H|\phi(t)\rangle \dd t 
    - \frac{1}{2} \sum_m \left(C_m^\dagger C_m - \langle \phi(t) | C_m + C_m^\dagger |\phi(t) \rangle C_m + \frac{1}{4}\langle \phi(t) | C_m + C_m^\dagger|\phi(t) \rangle^2\right)|\phi(t)\rangle \dd t\nonumber \\
   & + \sum_m \left( C_m -\frac{1}{2}\langle \phi(t) | C_m + C_m^\dagger |\phi(t) \rangle \right)|\phi(t)\rangle\,\dd\zeta_m(t) \nonumber
\end{align}
\vspace{-3mm}
}}
\label{eq076}
\ee
où, à chaque opérateur de saut $ C_m $ dans l'écriture initiale de l'équation pilote, on associe un processus stochastique en temps continu $ \dd \zeta_m (t)$, à valeurs réelles, gaussien, de moyenne nulle, de variance $ \dd t $, statistiquement indépendant des autres processus et sans mémoire.  Au même niveau d'approximation,  l'opérateur de signal homodyne (\ref{eq075}) est remplacé par la somme de sa moyenne et d'un bruit classique représentant ses fluctuations, qui n'est autre que le $\dd\zeta_m$ correspondant \cite{CastinDalibard}: 
\begin{equation}
\frac{N_+-N_-}{2 \mu } = \frac{\sqrt{\kappa}\langle \phi | c+c^\dagger |\phi\rangle}{2} \dd t + {\frac{1}{2}}\dd \zeta_c \label{eq077}
\end{equation}
 En pratique, plus que l'historique d'homodynage, c'est-à-dire la dépendance en temps détaillée du signal de détection homodyne, c'est sa moyenne temporelle sur un intervalle de temps $[0,t]$ qui est facilement accessible dans une expérience {et permet une post-sélection des états (c'est-à-dire des réalisations expérimentales) sur un ensemble de poids statistique non négligeable}. Nous introduisons donc le signal intégré ayant la dimension de la racine d'une pulsation,
\be
\boxed{
\sigma(t) \equiv \frac{N_+^{\rm tot}-N_-^{\rm tot}}{2\mu t}= \frac{1}{t} \int_0^t  \dd t' \left[\sqrt{\kappa} \langle\phi(t')|X_c |\phi(t')\rangle + {\frac{1}{2}\frac{\dd\zeta_c(t')}{\dd t'}}\right]}
\label{eq977}
\ee
et nous calculerons dans la suite la moyenne et la variance de la quadrature $P_a$ du spin nucléaire conditionnées à $\sigma$.\footnote{{\label{note977} Rappelons que la moyenne et la variance d'une observable (ici $P_a$) dans une seule réalisation $|\phi(t)\rangle$ de l'équation stochastique, donc de l'expérience, n'ont en général pas de sens physique, car il n'y a pas de mesure possible de la valeur moyenne d'une observable dans une seule réalisation, il faut au contraire moyenner sur un grand nombre de réalisations de l'expérience ayant évolué pendant $t$ à partir du même cas pur ou opérateur densité initial. Un contre-exemple correspond au cas où l'on peut, par des mesures en continu, remonter à la dépendance en temps de {\sl tous} les processus stochastiques $\dd\zeta_m$ ; c'est le cas du modèle à un mode soumis à une détection homodyne, la dépendance en temps du signal (\ref{eq077}) fixant celle de l'unique processus stochastique $\dd\zeta_c$. Ceci permettrait alors, en principe, de sélectionner, parmi un grand nombre de réalisations de l'expérience, celles conduisant à l'état $|\phi(t)\rangle$ choisi, et d'en déduire les moyennes d'observables dans $|\phi(t)\rangle$ ; en pratique, ce serait irréaliste, compte tenu du poids statistique infinitésimal des réalisations à garder.}}

\subsubsection{Résultats analytiques dans le modèle à un mode} \label{sec:3modesTo1Homo} 
\begin{figure}[t]
\begin{center}
{\includegraphics [width = 0.4 \textwidth]{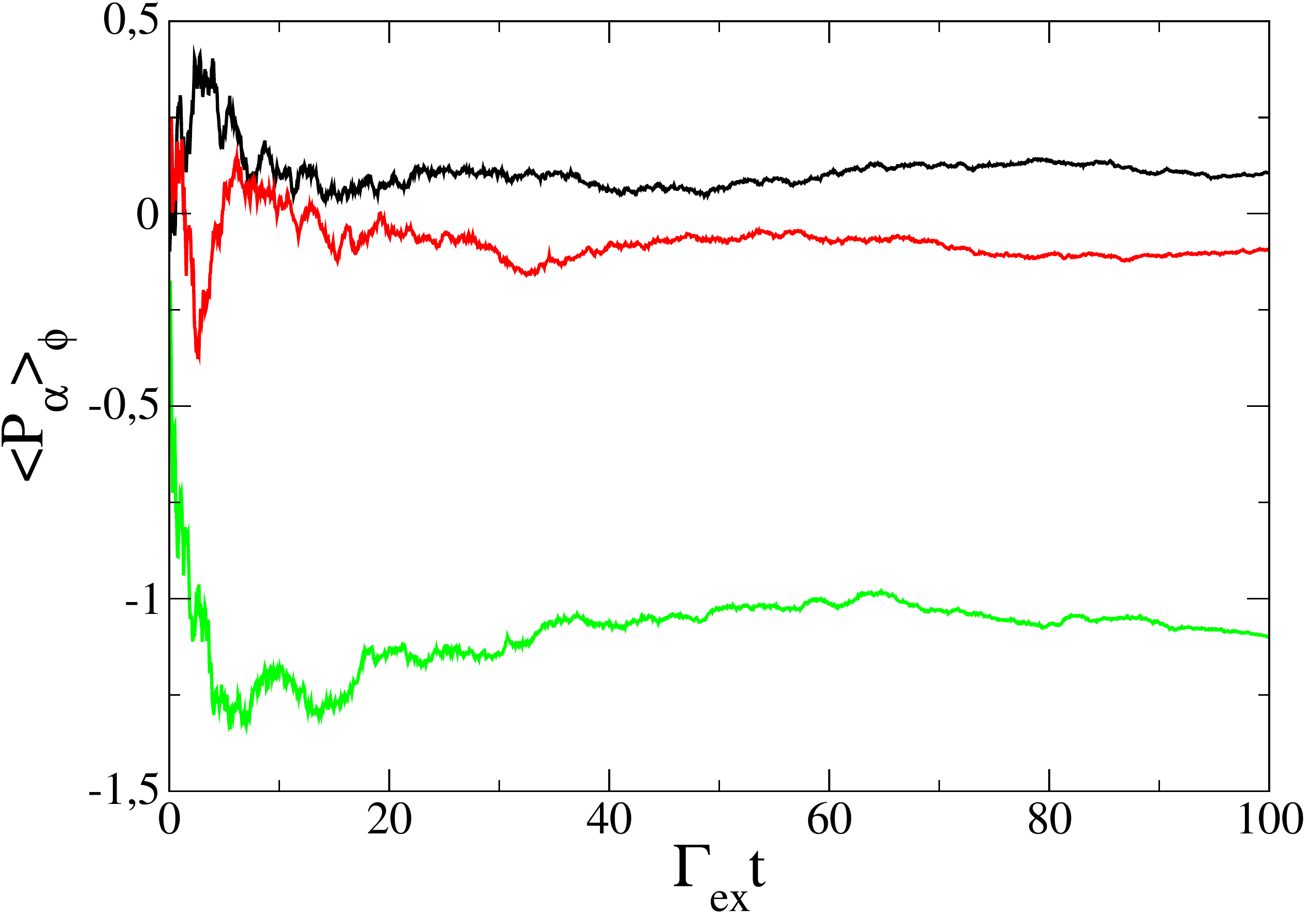}(a)}\quad
{\includegraphics [width = 0.4 \textwidth]{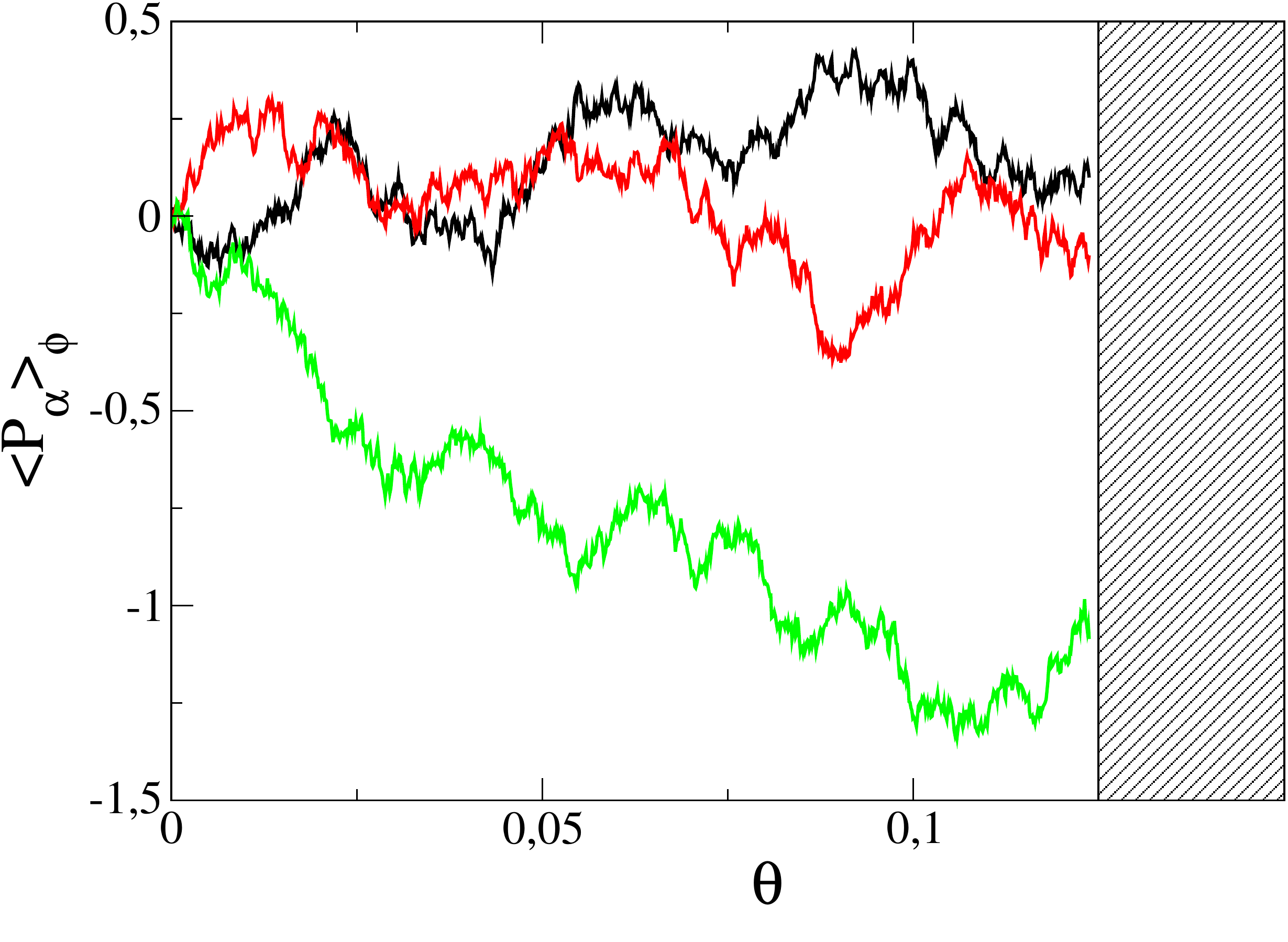}(b)} 
\end{center}
\caption{Dans le cas de la compression par détection homodyne  en continu, marche au hasard (\ref{eq082}) effectuée dans le modèle à un mode par la valeur moyenne quantique de la quadrature $P_\alpha$ du spin nucléaire dans une réalisation donnée de l'expérience. (a) Valeur moyenne en fonction du temps vrai $t$ {adimensionné par le taux (\ref{eq054}) de création d'excitations} pour trois réalisations de l'expérience ; il s'agit d'un mouvement brownien étiré convergeant aux temps longs vers une valeur fixe mais imprédictible. (b) Idem en fonction du temps renormalisé compact $\theta$ (\ref{eq083}) ; il s'agit cette fois d'un mouvement brownien ordinaire mais limité à $\theta\leq 1/8$.}
\label{fig8} 
\end{figure} 
Écrivons explicitement l'équation stochastique (\ref{eq076}) pour le modèle à un mode (\ref{eq030oneMode}) :
\begin{equation}
\dd|\phi(t) \rangle = -\frac{\dd t}{2} \Gamma_{\rm ex} [P_\alpha-\bar{P}_\alpha(t)]^2  |\phi(t) \rangle + \sqrt{\Gamma_{\rm ex}} \dd\zeta_s(t) [P_\alpha-\bar{P}_\alpha(t)] |\phi(t) \rangle 
\label{eq078}
\end{equation}
avec 
$\bar{P}_\alpha(t)\equiv \langle \phi(t) | P_\alpha |\phi(t)  \rangle$.
Le fait marquant est que les sauts associés à l'opérateur $C_d$ proportionnel à l'identité, qui ajoutaient du bruit dans la détection par comptage de photons de la section \ref{sec:dpcdp}, ne donnent pas de contribution, l'opérateur $C_d$ disparaissant de l'équation d'évolution conditionnelle dans le cas homodyne. En effet, les photons émis lors de ces sauts proviennent de la composante $|1\rangle |1\rangle$ du vecteur d'état (\ref{eq039}) contenant une excitation $\beta$, ce qui les rend optiquement incohérents avec le champ lumineux injecté en cavité, c'est-à-dire avec la composante $|0\rangle |0\rangle$ de (\ref{eq039}), au sens où $|1\rangle |1\rangle$ contribue à $\langle c^\dagger c\rangle$ mais pas à $\langle c+c^\dagger\rangle$. Il ne reste donc que le processus stochastique $\dd\zeta_s$ associé à l'opérateur de saut $C_s$.  Ce processus se confond avec celui $\dd\zeta_c$ apparaissant dans le signal de détection homodyne (\ref{eq077}), $\dd\zeta_s\equiv \dd\zeta_c$, fait admis ici mais qui sera établi en revenant au modèle à trois modes dans la section \ref{sec:sdmatm}.

 L'équation stochastique (\ref{eq078}) présente un terme de bruit linéaire et un terme déterministe quadratique en l'opérateur $P_\alpha$, réels dans l'espace de Fourier.  Pour l'état initial considéré ici, elle est donc résolue exactement par un ansatz gaussien sur la fonction d'onde en représentation impulsion, réel et correctement normalisé pour la relation de commutation $[X_\alpha,P_\alpha]=\ii/2$:
\begin{equation}
\langle p_\alpha | \phi(t)\rangle = [2\pi u(t)]^{1/4} \exp\{-u(t)[p_\alpha-\bar{P}_\alpha(t)]^2\}
\label{eq080}
\end{equation}
En revanche, la gaussianité est perdue dans la compression par photodétection de la section \ref{sec:dpcdp}.
En utilisant le calcul d'Ito,\footnote{On ne garde que les termes linéaires en $\dd t$ ou en le bruit, et on remplace systématiquement les termes quadratiques $ \dd \zeta_s^2 $ par leur moyenne $ \dd t $.} on trouve que $u$ suit une équation d'évolution déterministe, à intégrer avec la condition initiale $u(0)=1$:
\begin{equation}
\dd u(t)=\Gamma_{\rm ex}\dd t \quad \mbox{donc}\quad u(t)=1+\Gamma_{\rm ex} t\quad\mbox{et}\quad\mbox{Var}_{\phi} P_\alpha(t)\equiv \frac{1}{4 u(t)} = \frac{1}{4} \frac{1}{1+\Gamma_{\rm ex}t}
\label{eq081}
\end{equation}
où nous avons donné aussi la variance de $P_\alpha$ dans l'état $|\phi\rangle$. En revanche, l'équation sur la valeur moyenne de $P_\alpha$ dans $|\phi\rangle$ est purement stochastique, avec un coefficient de diffusion $D(t)$ dépendant du temps et la condition initale $\bar{P}_\alpha(0)=0$:
\begin{equation}
\dd\bar{P}_\alpha(t)=[2 D(t)]^{1/2} \dd\zeta_s(t) \quad\mbox{avec}\quad D(t)= \frac{\Gamma_{\rm ex}}{8 u(t)^2} = \frac{\Gamma_{\rm ex}}{8(1+\Gamma_{\rm ex}t)^2}
\label{eq082}
\end{equation}
Comme $D(t)$ est d'intégrale finie, $\bar{P}_\alpha(t)$ se stabilise asymptotiquement (aux temps longs) à une valeur fixe sur une seule réalisation, comme on le voit sur la figure \ref{fig8}, avec une variance dans l'état quantique $\mbox{Var}_\phi P_\alpha$ tendant vers $0$. Ce phénomène de \og convergence stochastique\fg\, vers un état propre de l'observable mesurée (en l'occurrence $P_\alpha$) est attendu dans la description d'une mesure quantique par une équation de diffusion du vecteur d'état \cite{Helvetica,Percival,Gisin}. Pour le montrer ici, on introduit un temps $\theta$ renormalisé en termes duquel $\bar{P}_\alpha$ effectue un mouvement brownien ordinaire avec un coefficient de diffusion unité, et on remarque que ce temps est borné:
\begin{equation}
\theta=\int_0^t\dd t' D(t') = \frac{{\Gamma_{\rm ex}} t}{8(1+\Gamma_{\rm ex} t)}\underset{t\to +\infty}{\to} \theta_\infty=\frac{1}{8}
\label{eq083}
\end{equation}
À l'instant renormalisé $\theta_\infty$, $\bar{P}_\alpha$ suit une loi gaussienne de moyenne nulle et de variance $1/4$~: $\bar{P}_\alpha$ a donc la même distribution de probabilité asymptotique ($t\to +\infty$) que celle de l'observable $P_\alpha$ dans l'état quantique initial du spin nucléaire.

\begin{figure}[t]
\centerline{\includegraphics [width = 0.33 \textwidth]{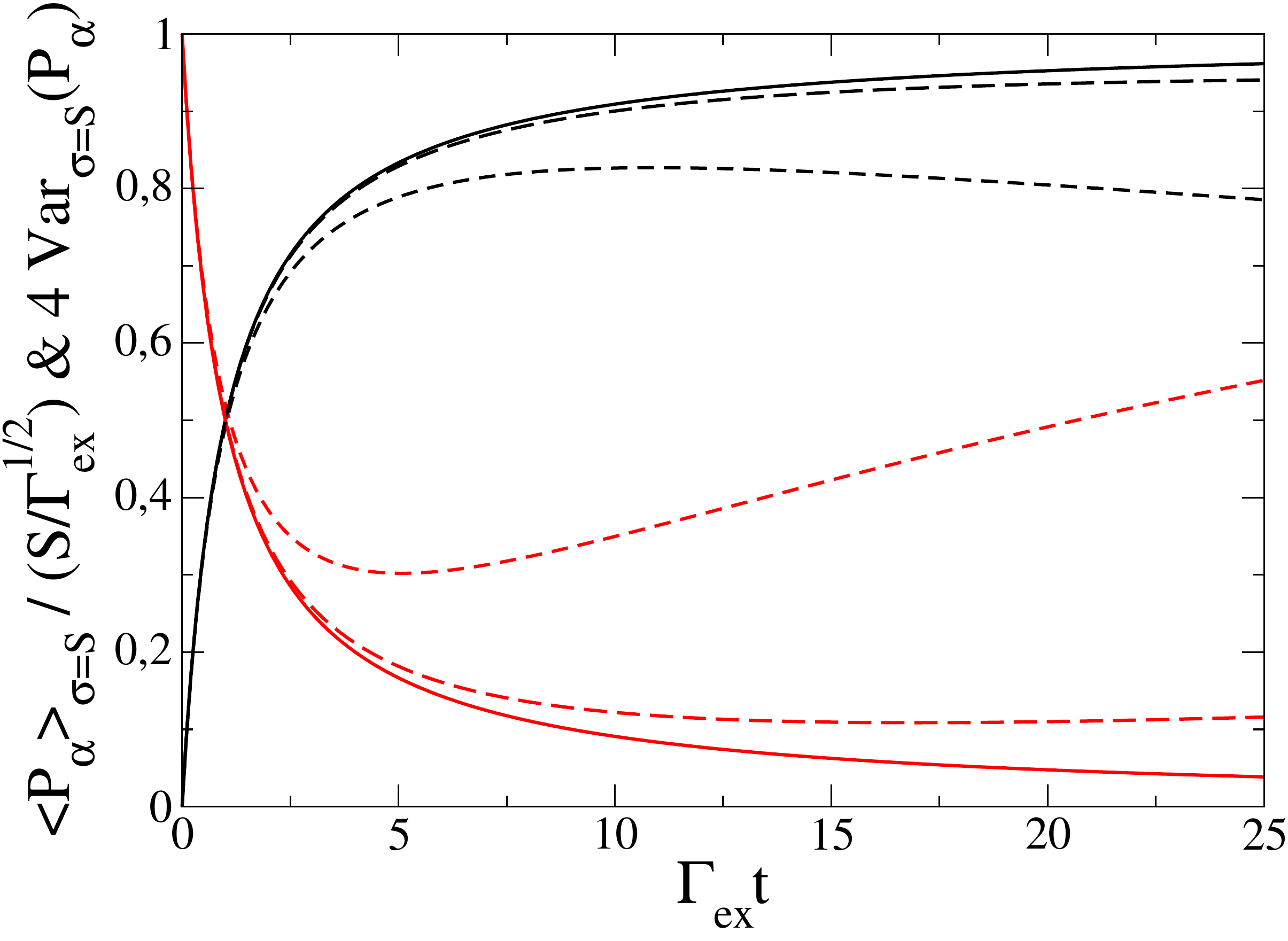}(a) \includegraphics [width = 0.33 \textwidth,height=0.23\textwidth]{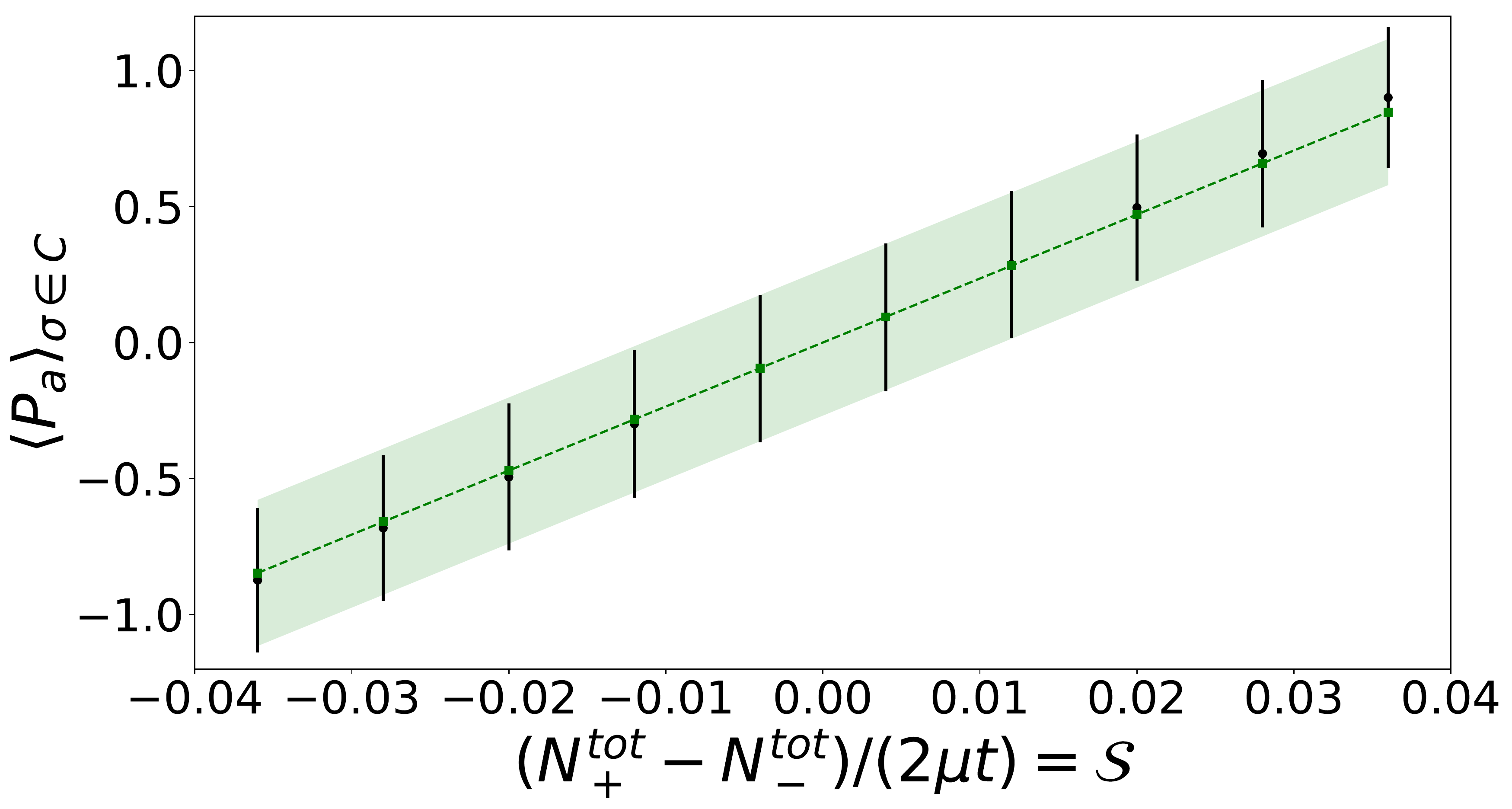}(b) \includegraphics [width = 0.33 \textwidth]{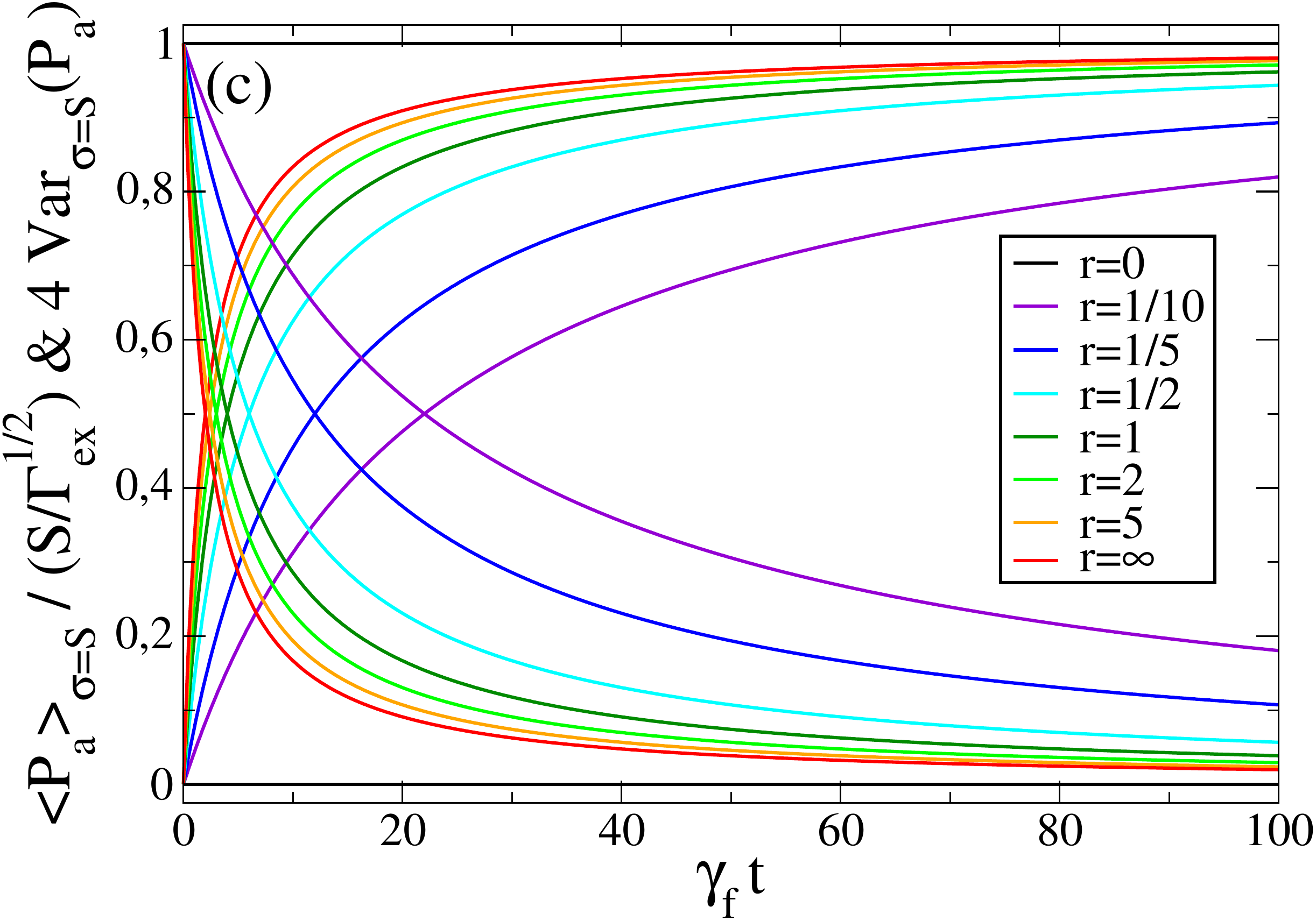}(c)} 
\caption{Compression {du spin nucléaire} par mesure homodyne en continu. (a) Dans le modèle à un mode, moyenne (en noir) et variance (multipliée par 4, en rouge) de la quadrature $P_\alpha$ du spin nucléaire hybridé conditionnées au signal d'homodynage intégré $\sigma$, en fonction du temps d'intégration $t$. Traits pleins : expressions analytiques (\ref{eq101}). Tiretés : expressions (\ref{eq155a}) et (\ref{eq155b}) en présence de décohérence {par désexcitation des métastables sur les parois de la cellule} (tireté long : $\epsilon=1/100$, tireté court : $\epsilon=1/10$, avec $\epsilon=\gamma_\alpha/\Gamma_{\rm ex}$ et $\gamma_\alpha$ le taux de décohérence ramené (\ref{eq152})). (b) Dans le modèle à trois modes, pour $\Gamma_{\rm ex}t=5$, moyenne et écart-type de la quadrature $P_a$ du spin nucléaire conditionnées à l'appartenance du signal $\sigma$ à une classe $C$, l'intervalle de valeurs de $\sigma$ ayant été partagé en 10 classes de même largeur {(les valeurs $\mathcal{S}$ de $\sigma$ sont en unités de $\Gamma_{\rm ex}^{1/2}$ sur l'axe des abscisses)}. L'écart-type est représenté sous la forme d'un intervalle de confiance. En noir : simulation numérique de l'équation stochastique (\ref{eq076}) avec $1079$ réalisations. Tireté vert et zone colorée : résultats exacts tirés des relations (\ref{eq127a}) et (\ref{eq127b}), et de l'expression analytique de la distribution de probabilité conditionnelle de $\bar{P}_a$ en termes des variances et covariance (\ref{eq130}) sur le modèle de l'équation (\ref{eq105}). L'écart entre numérique et analytique dans les classes extrêmes est imputable aux faibles nombres de réalisations tombant dans ces classes. Valeurs des paramètres :
$\Omega/\kappa = 1/10$, $\gamma_m/\kappa = 1/10$, $\gamma_f/\kappa = 1/100$, $\Gamma_{\rm ex}/\kappa=1/1000$. 
(c) Dans la limite $\Gamma_{\rm ex}\to 0$ à $\Gamma_{\rm ex}/2\gamma_f$ fixé du modèle à trois modes, moyenne et variance conditionnelles de $P_a$ (\ref{eq132}) en fonction du temps réduit $\gamma_f t$, pour différentes valeurs du rapport $r=2\Gamma_{\rm ex}/\gamma_f$ (courbes croissantes : moyenne, courbes décroissantes : variance). 
}
\label{fig9} 
\end{figure}

Venons-en maintenant à la moyenne et à la variance de $P_\alpha$ conditionnées à la valeur $\mathcal{S}$ du signal d'homodynage intégré en temps $\sigma$ (\ref{eq977}). De façon remarquable, nous trouvons que la moyenne conditionnelle est toujours proportionnelle au signal, avec un coefficient de proportionnalité dépendant du temps, et que la variance conditionnelle dépend du temps mais pas du signal:
\be
\label{eq101}
\boxed{\langle P_\alpha\rangle_{\sigma=\mathcal{S}}  = m(t) \frac{\mathcal{S}}{\sqrt{\Gamma_{\rm ex}}} \quad\mbox{où}\quad {m(t)={\frac{\Gamma_{\rm ex}t}{1+\Gamma_{\rm ex}t}}} \quad ; \quad \mbox{Var}_{\sigma=\mathcal{S}}(P_\alpha) = \mathcal{V}(t) \quad\mbox{où}\quad {\mathcal{V}(t)={\frac{1}{4(1+\Gamma_{\rm ex}t)}}}}
\ee
Ces expressions désignent $\Gamma_{\rm ex}$ comme le taux de compression du spin nucléaire {par homodynage} dans le modèle à un mode {donc à faible couplage de Faraday:}
\be
\label{eq701}
{\boxed{\Gamma_{\rm sq} \underset{\Omega\to 0}{\sim} \Gamma_{\rm ex}=\frac{\Omega^2\gamma_f}{\kappa(\gamma_m+\gamma_f)}}}
\ee
Sur la figure \ref{fig9}a, nous représentons {$m(t)$} et {$\mathcal{V}(t)$} en fonction du temps réduit {$\Gamma_{\rm ex}t$}. De même que la variance quantique sur une réalisation $\mbox{Var}_\phi P_\alpha$, avec laquelle elle coïncide en fait, la variance conditionnelle tend asymptotiquement vers zéro comme l'inverse du temps. Dans la moyenne conditionnelle, le coefficient {$m(t)$} tend vers 1 aux temps longs. Pour le comprendre, relions le signal intégré (\ref{eq977}) à $\bar{P}_\alpha$ en utilisant les expressions adiabatiques (\ref{eq043}) dans le vecteur d'état tronqué (\ref{eq039}):
\be
\sigma(t)=\frac{1}{t} \int_0^t \dd t' \left[\sqrt{\Gamma_{\rm ex}} \bar{P}_\alpha(t') +{\frac{1}{2}\frac{\dd\zeta_s(t')}{\dd t'}}\right]
\label{eq102}
\ee
Comme $\bar{P}_\alpha(t)$ se stabilise asymptotiquement sur une seule réalisation, et que la moyenne temporelle du bruit $\dd\zeta_s$ tend vers zéro comme $1/t^{1/2}$ presque sûrement, $\sigma(+\infty)$ donne directement la valeur de {$\bar{P}_\alpha(+\infty)$} à un facteur constant $\sqrt{\Gamma_{\rm ex}}$ près.

Pour établir les résultats (\ref{eq101}), relions d'abord la variance conditionnelle de l'opérateur $P_\alpha$ à celle de sa moyenne quantique sur une réalisation $\bar{P}_\alpha$ comme suit:
\bea
\mbox{Var}_{\sigma=\mathcal{S}}(P_\alpha) &\!\!\!\equiv\!\!\!& \langle\,\langle\phi|P_\alpha^2|\phi\rangle\,\rangle_{\sigma=\mathcal{S}} -\langle\,\langle\phi|{P}_\alpha|\phi\rangle\,\rangle_{\sigma=\mathcal{S}}^2=
\langle\,\langle\phi|P_\alpha^2|\phi\rangle-\langle\phi|P_\alpha|\phi\rangle^2\,\rangle_{\sigma=\mathcal{S}}
+ \langle \bar{P}_\alpha^2\rangle_{\sigma=\mathcal{S}}-\langle\bar{P}_\alpha\rangle^2_{\sigma=\mathcal{S}} \nonumber \\
&\!\!\!=\!\!\!& \langle \mbox{Var}_\phi P_\alpha\rangle_{\sigma=\mathcal{S}} + \mbox{Var}_{\sigma=\mathcal{S}}(\bar{P}_\alpha)
= {\frac{1}{4} \frac{1}{1+\Gamma_{\rm ex}t}} + \mbox{Var}_{\sigma=\mathcal{S}}(\bar{P}_\alpha)
\label{eq103}
\eea
où nous avons utilisé l'expression (\ref{eq081}) de la variance quantique de $P_\alpha$ dans l'état $|\phi\rangle$. Il reste donc à déterminer la distribution de probabilité conditionnelle de $\bar{P}_\alpha$ sachant que $\sigma=\mathcal{S}$, 
\be
\label{eq104}
P(\bar{P}_\alpha=p_\alpha|\sigma=\mathcal{S}) \equiv \frac{P(\bar{P}_\alpha=p_\alpha,\sigma=\mathcal{S})}{P(\sigma=\mathcal{S})}
\ee
Or, la variable aléatoire $\bar{P}_\alpha(t)$, résultant d'un mouvement brownien (\ref{eq082}), a une distribution de probabilité gaussienne ; il en va de même pour l'intégrale temporelle de $\bar{P}_\alpha$ et du bruit $\dd\zeta_s$, donc du signal $\sigma$ (\ref{eq102}) qui en est la somme. Comme les variables $\bar{P}_\alpha$ et $\sigma$ sont de moyennes nulles, leur distribution de probabilité conjointe est caractérisée par leur matrice de covariance, ou plus directement par sa matrice inverse, si bien que
\bea
P(\bar{P}_\alpha=p_\alpha|\sigma=\mathcal{S})&\!\!\!=\!\!\!&  \frac{\frac{1}{2\pi\sqrt{\langle\bar{P}_\alpha^2\rangle_{\rm stoch}\langle\sigma^2\rangle_{\rm stoch}-\langle \sigma \bar{P}_\alpha\rangle_{\rm stoch}^2}}\exp\left(-\frac{1}{2} \frac{p_\alpha^2\langle\sigma^2\rangle_{\rm stoch} + \mathcal{S}^2 \langle \bar{P}_\alpha^2\rangle_{\rm stoch} -2 p_\alpha\mathcal{S} \langle \sigma\bar{P}_\alpha\rangle_{\rm stoch}}{\langle\bar{P}_\alpha^2\rangle_{\rm stoch}\langle\sigma^2\rangle_{\rm stoch}-\langle \sigma \bar{P}_\alpha\rangle_{\rm stoch}^2}\right)}
{\frac{1}{\sqrt{2\pi\langle\sigma^2\rangle_{\rm stoch}}} \exp\left(-\frac{\mathcal{S}^2}{2\langle\sigma^2\rangle_{\rm stoch}}\right)} \nonumber \\
&\!\!\!=\!\!\!& \frac{1}
{\sqrt{2\pi\left[\langle\bar{P}_\alpha^2\rangle_{\rm stoch} -  \langle\sigma \bar{P}_\alpha\rangle_{\rm stoch}^2/\langle\sigma^2\rangle_{\rm stoch}\right]}}
{\exp\left[-\frac{1}{2} \frac{\left(p_\alpha - \mathcal{S} \langle \sigma \bar{P}_\alpha\rangle_{\rm stoch}/\langle\sigma^2\rangle_{\rm stoch}\right)^2}{\langle\bar{P}_\alpha^2\rangle_{\rm stoch} -  \langle\sigma \bar{P}_\alpha\rangle_{\rm stoch}^2/\langle\sigma^2\rangle_{\rm stoch}}\right]}
\label{eq105}
\eea
où $\langle\ldots\rangle_{\rm stoch}$ à l'instant $t$ est la moyenne prise sur toutes les réalisations du processus stochastique $\dd\zeta_s(t')$ sur l'intervalle de temps $[0,t]$. On en déduit que, dans les équations (\ref{eq101}),
\be
{m(t)}= \sqrt{\Gamma_{\rm ex}} \frac{\langle\sigma(t)\bar{P}_\alpha(t)\rangle_{\rm stoch}}{\langle\sigma^2(t)\rangle_{\rm stoch}}\quad\mbox{et}\quad {\mathcal{V}(t) =\frac{1}{4(1+\Gamma_{\rm ex}t)}}+\langle\bar{P}_\alpha^2(t)\rangle_{\rm stoch}-\frac{\langle\sigma(t)\bar{P}_\alpha(t)\rangle_{\rm stoch}^2}{\langle\sigma^2(t)\rangle_{\rm stoch}}
\label{eq106}
\ee
Afin de déterminer leurs variances et covariance, on écrit $\sigma(t)$ et $\bar{P}_\alpha(t)$ comme des fonctionnelles linéaires du processus stochastique $\dd\zeta_s$ et on utilise le fait que les forces de Langevin $\dd\zeta_s(t)/\dd t$ et $\dd\zeta_s(t')/\dd t'$ ont une fonction de corrélation de Dirac $\delta(t-t')$. Donnons l'exemple de la première contribution à $\sigma(t)$:
\be
\int_0^t \dd t'' \bar{P}_\alpha(t'') =\int_0^t\dd t'' \int_0^{t''} \dd t' [2D(t')]^{1/2}{\frac{\dd\zeta_s(t')}{\dd t'}} =\int_0^t\dd t' \int_{t'}^{t} \dd t'' [2D(t')]^{1/2}{\frac{\dd\zeta_s(t')}{\dd t'}}
= \int_0^t \dd t' (t-t') [2D(t')]^{1/2}{\frac{\dd\zeta_s(t')}{\dd t'}}
\label{eq107}
\ee
où l'on a changé l'ordre d'intégration sur $t'$ et $t''$ puis intégré explicitement sur $t''$. On aboutit aux expressions cherchées (\ref{eq101}), dont la simplicité découle du fait que, sur une réalisation de l'expérience, on a toujours
\be
\label{eq197}
\boxed{\sigma(t) = \sqrt{\Gamma_{\rm ex}} {\frac{1+\Gamma_{\rm ex}t}{\Gamma_{\rm ex}t}} \bar{P}_\alpha(t)}
\ee
{De manière remarquable, la connaissance du seul signal intégré $\sigma(t)$ dans une réalisation de l'expérience de durée $t$ suffit à préparer le spin nucléaire dans un cas pur gaussien bien défini (\ref{eq080}), avec un paramètre $u$ donné par l'équation (\ref{eq081}) et une quadrature moyenne $\bar{P}_\alpha$ reliée au signal par l'équation (\ref{eq197}).}

Pour terminer, revenons à la quadrature $P_a$ du spin nucléaire non hybridé, qui est celle véritablement utilisable dans l'expérience une fois la décharge éteinte dans la cellule, {comme le montre l'expression (\ref{eq017}) du gain métrologique dans une mesure de précession}. En quittant la base tournée par inversion de la transformation (\ref{eq034}) et en limitant l'équation (\ref{eq039}) à son premier terme (à l'ordre dominant en $\Omega$), il vient
\be
\langle P_a\rangle_{\sigma=\mathcal{S}} = \left(\frac{\gamma_m}{\gamma_f+\gamma_m}\right)^{1/2} \, \langle P_{\alpha}\rangle_{\sigma=\mathcal{S}} \quad\mbox{et}\quad\mbox{Var}_{\sigma=\mathcal{S}}(P_a)= \frac{\gamma_f}{4(\gamma_f+\gamma_m)} + {\frac{\gamma_m}{\gamma_f+\gamma_m}}\mbox{Var}_{\sigma=\mathcal{S}}(P_\alpha)
\label{eq108}
\ee
La variance  conditionnelle de $ P_a $ aux temps longs tend vers une valeur non nulle, bien que faible en pratique~: c'est la limite intrinsèque de ce schéma de compression de spin nucléaire, qui utilise l'état métastable de ${}^3$He comme intermédiaire.

\subsubsection{Solution du modèle à trois modes}
\label{sec:sdmatm}

L'étude de la compression de spin dans le cadre du modèle à un mode est limitée au régime (\ref{eq033}) où le taux de compression {$\Gamma_{\rm sq}\sim\Gamma_{\rm ex}$} correspond à l'échelle de temps la plus longue du système. Il est cependant crucial pour les applications de voir jusqu'où on peut accélérer le processus de compression en augmentant $\Gamma_{\rm ex}$ donc, par exemple, le couplage de Faraday $\Omega$ des atomes métastables au champ en cavité. À cette fin, nous obtenons la solution analytique du modèle à trois modes en utilisant le caractère gaussien du vecteur d'état qui résulte, comme pour le modèle à un mode, de l'état initial considéré (le vide), de la linéarité des opérateurs de saut $C_m$ et de la quadraticité de l'hamiltonien $H$ en les quadratures des modes.  L'équation stochastique (\ref{eq076}) admet donc comme solution exacte l'ansatz gaussien généralisant celui de l'équation (\ref{eq080}),
\be
\langle p_\alpha,p_\beta,x_c|\phi(t)\rangle=\phi(\qq,t) = [8\pi \det \underline{\underline{u}}(t)]^{1/4} 
\exp\left\{-[\qq-\bar{\qq}(t)]\cdot \underline{\underline{u}}(t)\,[\qq-\bar{\qq}(t)]\right\}\equiv \eee^{-S}
\label{eq120}
\ee
où $\underline{\underline{u}}$ est une matrice $3\times 3$ symétrique réelle, $\bar{\qq}$ est un vecteur à trois composantes réelles, les coordonnées $q_\alpha=p_\alpha$ et $q_\beta=p_\beta$ sont dans l'espace de Fourier (base propre de la quadrature $P$) et la coordonnée $q_c=x_c$ est dans l'espace des {\og positions \fg} (base propre de la quadrature $X$).  La seule astuce ici était de choisir comme opérateur de saut d'échange de métastabilité $C_\beta=\sqrt{\gamma_\beta}\, \ii \beta$ ; ce choix de phase, qui ne change bien entendu pas l'équation pilote (\ref{eq030t}), reste légitime pour l'évolution conditionnée à la détection homodyne du champ car les sauts de métastabilité ne sont pas mesurés. Dans la représentation mixte de la fonction d'onde (\ref{eq120}), l'hamiltonien $H$ est alors imaginaire pur et les opérateurs de saut sont réels, d'où l'ansatz réel (\ref{eq120}).\footnote{Par exemple, $\ii\beta=\ii(X_\beta+\ii P_\beta)$ est représenté en impulsion par l'opérateur réel $-\partial_{p_\beta}/2-p_\beta$, et $\beta^\dagger\beta$ par $-\partial_{p_\beta}^2/4+p_\beta^2-1/2$.}

Pour obtenir les équations du mouvement sur $\underline{\underline{u}}$ et $\bar{\qq}$, nous calculons de deux manières différentes la variation relative $\dd\phi(\qq,t)/\phi(\qq,t)$ de la fonction d'onde, d'une part en la reliant à la variation $\dd S$ de la quantité $S$ dans (\ref{eq120}), séparée en une partie déterministe $\dd S_{\!d}$ et une partie de bruit $\dd S_{\!b}$, d'autre part en reportant l'ansatz (\ref{eq120}) dans l'équation stochastique (\ref{eq076}). En identifiant les parties déterministes et les parties bruitées des deux formes qui en résultent, nous obtenons
\bea
\label{eq121a}
-\dd S_{\!b} &\!\!\!=\!\!\!& \gamma_\beta^{1/2} \left(\frac{1}{2}\partial_{q_\beta} S -q_\beta+\bar{q}_\beta\right)\dd\zeta_\beta 
-\kappa^{1/2} \left(\frac{1}{2}\partial_{q_c} S -q_c +\bar{q}_c\right) \dd\zeta_c \\
-\dd S_{\!d} +\frac{1}{2} (\dd S_{\!b})^2 &\!\!\!=\!\!\!& (\Omega_\alpha q_\alpha +\Omega_\beta q_\beta) \frac{\dd t}{2} \partial_{q_c} S 
- \frac{\gamma_\beta \dd t}{2} \left\{ q_\beta^2-\frac{1}{2} +\frac{1}{4} \left[ \partial_{q_\beta}^2 S -\left(\partial_{q_\beta} S\right)^2\right] +\bar{q}_\beta \left(\partial_{q_\beta} S -2 q_\beta\right) +\bar{q}_\beta^2\right\} \nonumber \\
&\!\!\! \!\!\!&  -\frac{\kappa\dd t}{2} \left\{ q_c^2-\frac{1}{2} +\frac{1}{4} \left[ \partial_{q_c}^2 S -\left(\partial_{q_c} S\right)^2\right] +\bar{q}_c \left(\partial_{q_c} S -2 q_c\right) +\bar{q}_c^2\right\}
\label{eq121b}
\eea
Il reste à reporter dans (\ref{eq121b}) l'expression de $\dd S_{\!b}$ tirée de (\ref{eq121a}), en appliquant la règle d'Ito de remplacement des carrés des bruits par leur moyenne, puis à identifier les termes de degré 2 en $\qq-\bar{\qq}$ pour obtenir l'équation linéaire purement déterministe sur $\underline{\underline{u}}$:\footnote{On remarque que les termes quadratiques en $\underline{\underline{u}}$ au second membre de (\ref{eq121b}) se compensent avec ceux de $(\dd S_{\!b})^2/2$ au premier membre.}
\be
\begin{array}{lll}
\dd u_{\alpha\alpha} = -\Omega_\alpha \dd t\, u_{\alpha c} & \displaystyle\dd u_{\alpha\beta} = -\frac{\dd t}{2} (\gamma_\beta u_{\alpha\beta} + \Omega_\beta u_{\alpha c} +\Omega_\alpha u_{\beta c}) & \displaystyle\dd u_{\alpha c} = -\frac{\dd t}{2} (\kappa u_{\alpha c} + \Omega_\alpha u_{cc}) \\ 
&&\\
\dd u_{\beta\beta}= -\Omega_\beta \dd t\, u_{\beta c} + \gamma_\beta \dd t (1-u_{\beta\beta}) & \displaystyle\dd u_{\beta c} = -\frac{\dd t}{2} [(\gamma_\beta+\kappa)u_{\beta c} +\Omega_\beta u_{cc}] & \dd u_{cc}=\kappa \dd t (1- u_{cc})
\end{array}
\label{eq122}
\ee
et les termes de degré 1 en $\qq-\bar{\qq}$ pour obtenir l'équation linéaire stochastique sur $\bar{\qq}$ :
\be
\dd\bar{\qq} = \frac{1}{2} \begin{pmatrix} 0 & 0 & 0 \\
0 & -\gamma_\beta & 0 \\
\Omega_\alpha & \Omega_\beta & -\kappa 
\end{pmatrix}
\dd t \, \bar{\qq}  + 
\frac{1}{2}[\mbox{Id}-\underline{\underline{c}}(t)] \begin{pmatrix}
0 \\
\gamma_\beta^{1/2} \dd\zeta_\beta(t) \\
-\kappa^{1/2} \dd\zeta_c(t)
\end{pmatrix}
\label{eq123}
\ee
Faut-il le préciser, $\bar{\qq}$ est le vecteur des moyennes quantiques des variables $q$ dans le vecteur d'état (\ref{eq120});\footnote{On peut donc retrouver l'équation (\ref{eq123}) à partir de l'équation stochastique déduite de (\ref{eq076}) sur la moyenne d'une observable $O$, 
$\dd\langle O\rangle=(\dd t/\ii\hbar) \langle [O,H]\rangle +(\dd t/2) \sum_m \langle C_m^\dagger [O,C_m]+\mbox{h.c.}\rangle + \sum_m [\langle O C_m+\mbox{h.c.}\rangle -\langle C_m+C_m^\dagger\rangle \langle O\rangle]\dd\zeta_m$,
où $\langle\ldots\rangle$ est prise dans l'état $|\phi(t)\rangle$, en la spécialisant aux cas $O=P_\alpha$, $O=P_\beta$ et $O=X_c$.} par ailleurs, on a introduit la notation $\underline{\underline{c}}$ pour la matrice inverse de $\underline{\underline{u}}$, qui n'est autre que la matrice de covariance quantique des $q$ à un facteur numérique près. On a donc:
\be
\langle\phi(t)|q_i|\phi(t)\rangle =\bar{q}_i(t)\quad\mbox{et}\quad\langle\phi(t)|q_i q_j|\phi(t)\rangle = \bar{q}_i(t)\bar{q}_j(t)+\frac{1}{4} c_{ij}(t) \quad \forall i,j\in\{\alpha,\beta,c\}\quad\mbox{avec}\quad \underline{\underline{c}}(t)=[\underline{\underline{u}}(t)]^{-1}
\label{eq124}
\ee
Le système différentiel (\ref{eq122}) s'intègre aisément pour la condition initale $\underline{\underline{u}}(0)=\mathrm{Id}$:
\bea
\label{eq125a}
u_{\alpha\alpha}(t)& = &1+\frac{\Omega_\alpha ^2 t}{\kappa } -\frac{2 \Omega_\alpha ^2}{\kappa ^2} \left(1-\eee^{-\kappa  t/2}\right) \\
\label{eq125b}
u_{\alpha\beta}(t) &=& \frac{\Omega_\alpha\Omega_\beta}{\gamma_\beta} \left(\frac{1}{\gamma_\beta+\kappa }+\frac{1}{\kappa }\right) \left(1-\eee^{-\gamma_\beta t/2}\right)+\frac{\Omega_\alpha\Omega_\beta}{\kappa  (\kappa -\gamma_\beta)} \left(\eee^{-\kappa  t/2}-\eee^{-\gamma_\beta t/2}\right)+\frac{\Omega_\alpha\Omega_\beta}{\kappa  (\gamma_\beta +\kappa )} \left(\eee^{-(\gamma_\beta +\kappa )t/2}-\eee^{-\gamma_\beta t/2}\right)  \\
\label{eq125c}
u_{\alpha c}(t) &=& -\frac{\Omega_\alpha}{\kappa }  \left(1-\eee^{-\kappa  t/2}\right) \\
\label{eq125d}
u_{\beta\beta}(t) &=& 1 + \frac{\Omega_\beta ^2}{\gamma_\beta  (\gamma_\beta +\kappa )} \left(1-\eee^{-\gamma_\beta  t}\right) -\frac{2 \Omega_\beta ^2}{\kappa ^2-\gamma_\beta^2} \left(\eee^{-\gamma_\beta t}-\eee^{-(\gamma_\beta +\kappa )t/2}\right) \\
\label{eq125e}
u_{\beta c}(t) &=&  -\frac{ \Omega_\beta}{\gamma_\beta +\kappa }  \left(1-\eee^{-(\gamma_\beta +\kappa )t/2}\right) \\
\label{eq125f}
u_{cc}(t) &=& 1
\eea
{Il en serait allé autrement si nous avions pris comme inconnue la matrice de covariance $\underline{\underline{c}}(t)$, qui obéit à un système différentiel non linéaire de Riccati \cite{RiccatiMabuchi}.} Comme $\bar{\qq}$ décrit un mouvement brownien (partiellement amorti car la matrice de frottement dans (\ref{eq123}) est de valeurs propres $0$, $\gamma_\beta/2$ et $\kappa/2$), et comme le signal homodyne moyenné sur l'intervalle de temps $[0,t]$ $\sigma$ s'en déduit par intégration, ces variables aléatoires ont une statistique gaussienne et nous pouvons reproduire le raisonnement de la section \ref{sec:3modesTo1Homo}. Nous trouvons pour la moyenne et la variance conditionnelles de la quadrature $P_a$ du spin nucléaire sachant que $\sigma=\mathcal{S}$, {cette variance déterminant le gain métrologique (\ref{eq017}):}
\bea
\label{eq127a}
\!\!\!\!\langle P_a\rangle_{\sigma=\mathcal{S}} &\!\!\!\!=\!\!\!\!& \frac{\langle \sigma(t)\bar{P}_a(t)\rangle_{\rm stoch}}{\langle\sigma^2(t)\rangle_{\rm stoch}} \mathcal{S} \\
\label{eq127b}
\!\!\!\!\mbox{Var}_{\sigma=\mathcal{S}}(P_a) &\!\!\!\!=\!\!\!\!& \frac{1}{4}\left[\frac{\Omega_\beta^2}{\Omega^2}c_{\alpha\alpha}(t)+\frac{\Omega_\alpha^2}{\Omega^2}c_{\beta\beta}(t)-2\frac{\Omega_\alpha\Omega_\beta}{\Omega^2} c_{\alpha\beta}(t)\right]+\langle \bar{P}_a^2(t)\rangle_{\rm stoch} - \frac{\langle \sigma(t)\bar{P}_a(t)\rangle_{\rm stoch}^2}{\langle\sigma^2(t)\rangle_{\rm stoch}} = \frac{1}{4} - \frac{\langle \sigma(t)\bar{P}_a(t)\rangle_{\rm stoch}^2}{\langle\sigma^2(t)\rangle_{\rm stoch}}
\eea
L'expression entre crochets dans l'équation (\ref{eq127b}) est l'élément de matrice de $\underline{\underline{c}}(t)$ dans le vecteur $(\Omega_\beta/\Omega,-\Omega_\alpha/\Omega,0)$ des coordonnées de la direction $a$ dans la base tournée, {donc, à un facteur $4$ près,} la variance quantique de $P_a$ dans l'état stochastique $\phi(t)$, dépendant du temps mais, rappelons-le, indépendante de la réalisation particulière de $\phi(t)$. L'expression simplifiée au troisième membre découle de la propriété (\ref{eq032}) sur la moyenne non conditionnelle $\langle P_a^2\rangle(t)=1/4$ et de la chaîne d'égalités
\be
\label{eq128}
\langle P_a^2\rangle(t)= \langle\ \langle\phi(t)|P_a^2|\phi(t)\rangle\ \rangle_{\rm stoch} = \langle\ \langle\phi(t)|P_a^2|\phi(t)\rangle -\langle\phi(t)|P_a|\phi(t)\rangle^2 + \langle\phi(t)|P_a|\phi(t)\rangle^2\ \rangle_{\rm stoch}= \langle\mbox{Var}_{\phi(t)} P_a\rangle_{\rm stoch} + \langle \bar{P}_a^2(t)\rangle_{\rm stoch}
\ee
Il reste, pour déterminer les variance et covariance des variables aléatoires $\bar{P}_a(t)$ et $\sigma(t)$, à calculer leurs amplitudes sur les processus stochastiques $\dd\zeta_\beta(t')$ et $\dd\zeta_c(t')$, en intégrant formellement l'équation (\ref{eq123}) par la méthode de variation de la constante  pour $\bar{P}_a$ et $\bar{X}_c$, et en procédant comme dans l'équation (\ref{eq107}) pour $\sigma$:
\bea
\label{eq129a}
\hspace{-1.3cm}&&p_\beta(t,t') =-\frac{1}{2} \gamma_\beta^{1/2} \left\{\frac{\Omega_\beta}{\Omega}c_{\alpha\beta}(t')+\frac{\Omega_\alpha}{\Omega}[1-c_{\beta\beta}(t')]\eee^{-\gamma_\beta (t-t')/2}\right\} \\
\hspace{-1.3cm}&&p_c(t,t') = \frac{1}{2}\kappa^{1/2} \left\{\frac{\Omega_\beta}{\Omega} c_{\alpha c}(t')-\frac{\Omega_\alpha}{\Omega} c_{\beta c}(t') \eee^{-\gamma_\beta (t-t')/2}\right\} \\
\hspace{-1.3cm}&&\sigma_\beta(t,t') = \frac{(\gamma_\beta\kappa)^{1/2}}{2 t}\left\{ -c_{\alpha\beta}(t')[t-t'-f_\kappa(t-t')]\frac{\Omega_\alpha}{\kappa}+
[1-c_{\beta\beta}(t')][f_{\gamma_\beta}(t-t')-f_\kappa(t-t')]\frac{\Omega_\beta}{\kappa-\gamma_\beta} -c_{\beta c}(t') f_\kappa(t-t')\right\} \\
\hspace{-1.3cm}&& \sigma_c(t,t')=\frac{1}{2 t}-\frac{\kappa}{2t}\left\{-c_{\alpha c}(t')[t-t'-f_\kappa(t-t')] \frac{\Omega_\alpha}{\kappa}-c_{\beta c}(t')[f_{\gamma_\beta}(t-t')-f_\kappa(t-t')]\frac{\Omega_\beta}{\kappa-\gamma_\beta} +[1-c_{cc}(t')] f_\kappa(t-t')\right\}
\label{eq129d}
\eea
où $f_\lambda(\tau)\equiv [1-\exp(-\lambda\tau/2)]/(\lambda/2)$.
On obtient:
\bea
\langle \sigma(t)\bar{P}_a(t)\rangle_{\rm stoch} &\!\!\!=\!\!\!& \int_0^t \dd t'\, [p_\beta(t,t') \sigma_\beta(t,t')+p_c(t,t')\sigma_c(t,t')] \quad ; \quad
\langle \sigma^2(t)\rangle_{\rm stoch} = \int_0^t \dd t'\, [\sigma_\beta^2(t,t')+\sigma_c^2(t,t')] \quad ;\nonumber \\
\langle \bar{P}_a^2(t)\rangle_{\rm stoch} &\!\!\!=\!\!\!& \int_0^t \dd t'\, [p_\beta^2(t,t')+p_c^2(t,t')]
\label{eq130}
\eea
On déduit de ces résultats les limites aux temps longs \footnote{Donnons quelques résultats et considérations intermédiaires. (i) Alors que $c_{\beta\beta}(t')$, $c_{\beta c}(t')$ et $c_{cc}(t')$ ont une limite finie lorsque $t'\to +\infty$ [on aura besoin de 
$c_{\beta\beta}(+\infty)=(1+\rho)^{-1}$, $c_{\beta c}(+\infty)=\Omega_\beta/((\gamma_\beta+\kappa)(1+\rho))$ avec $\rho=\Omega_\beta^2\kappa/(\gamma_\beta (\gamma_\beta+\kappa)^2)$
], $c_{\alpha\alpha}(t')$, $c_{\alpha\beta}(t')$ et $c_{\alpha c}(t')$ tendent vers zéro comme $1/t'$. (ii) Dans une intégrale sur $t'$ contenant le facteur exponentiel $\exp[-\gamma_\beta(t-t')/2]$ ou son carré, on peut remplacer la fonction qui le multiplie par sa limite en $t'=+\infty$. (iii) Pour toute fonction uniformément bornée $w(t,t')$, on peut montrer pour $\nu\in\{\beta,c\}$ que 
$\int_0^t \dd t' [(t-t') c_{\alpha\nu}(t')+w(t,t')]^2/t^2\to \int_0^{+\infty} \dd t' c_{\alpha\nu}^2(t')$.
(iv) On obtient alors les limites asymptotiques 
$\langle P_a^2(t)\rangle_{\rm stoch}\to (\Omega_{\beta}/2\Omega)^2\mathcal{I}+(\Omega_\alpha/2\Omega)^2\rho/(1+\rho)$, $\langle\sigma^2(t)\rangle_{\rm stoch}\to (\Omega_\alpha^2/4\kappa)\mathcal{I}$, $\langle\sigma(t)\bar{P}_a(t)\rangle_{\rm stoch}\to (\Omega_\alpha\Omega_\beta/4\Omega\kappa^{1/2}) \mathcal{I}$ où $\mathcal{I}\equiv \int_0^{+\infty} \dd t' [\gamma_\beta c_{\alpha\beta}^2(t')+\kappa c_{\alpha c}^2(t')]$.
Nous déduisons ainsi (\ref{eq131}) de (\ref{eq127a}) et de la première égalité dans (\ref{eq127b}), sans avoir besoin de connaître la valeur de $\mathcal{I}$. On tire de la seconde égalité dans (\ref{eq127b}) le résultat $\mathcal{I}=1$, que l'on peut déduire aussi de l'équation du mouvement 
$\dd c_{\alpha\alpha}/\dd t=-\gamma_\beta c_{\alpha\beta}^2-\kappa c_{\alpha c}^2$
intégrée entre $t=0$ et $t=+\infty$.
}
\be
\label{eq131}
\langle{P}_a\rangle_{\sigma=\mathcal{S}} \underset{t\to+\infty}{\to}  \left(\frac{\gamma_m}{\gamma_f+\gamma_m}\right)^{1/2}\frac{\mathcal{S}}{\Gamma_{\rm ex}^{1/2}} \quad ; \quad \mbox{Var}_{\sigma=\mathcal{S}}({P}_a) \underset{t\to+\infty}{\to} \frac{1}{4} \frac{\gamma_f}{\gamma_f+\gamma_m}
\ee
avec lesquelles les prédictions (\ref{eq108}) du modèle à un mode, pourtant obtenues dans la limite de couplage faible (\ref{eq033}), sont en accord parfait.

En application de notre solution analytique du modèle à trois modes, faisons tendre le taux $\Gamma_{\rm ex}$ vers zéro à temps réduit $\tau = \Gamma_{\rm ex} t$ fixé en maintenant (contrairement au modèle à un mode) le rapport $\Gamma_{\rm ex}/\gamma_f$ à une valeur constante non infinitésimale. La motivation physique est claire : dans les expériences projetées \cite{letter}, $\gamma_f$ et $\Gamma_{\rm ex}$ sont du même ordre de grandeur mais sont vraiment beaucoup plus petits que $\gamma_m$ et $\kappa$ (par des facteurs $\approx 10^{-6}$ et $10^{-9}$). Nous trouvons dans cette limite:\footnote{\label{note133} En pratique, il suffit de faire tendre $\Omega_\alpha$ vers zéro à $\tau=\Gamma_{\rm ex} t>0$, $\Omega_\beta$, $\gamma_\beta$ et $\kappa$ fixés. En particulier, ceci fait disparaître tous les transitoires exponentiels {dans les équations (\ref{eq125a})-(\ref{eq125f})}. Pour simplifier les calculs, il est utile d'introduire la quantité 
$\rho=\Omega^2\kappa/[2\gamma_m (\kappa+2\gamma_m)^2]$
si bien que
{$\rho=(\Gamma_{\rm ex}/2\gamma_f)(1+2\gamma_m/\kappa)^{-2}$}
dans la limite $\gamma_f\to 0$.
}
\be
{{\boxed{\langle{P}_a\rangle_{\sigma=\mathcal{S}} \sim \frac{\Gamma_{\rm sq}t}{1+\Gamma_{\rm sq}t}\frac{\mathcal{S}}{\Gamma_{\rm ex}^{1/2}} \quad \mbox{et}\quad \mbox{Var}_{\sigma=\mathcal{S}}({P}_a) \sim \frac{1}{4} \frac{1}{1+\Gamma_{\rm sq}t}}}}
\label{eq132}
\ee
où l'on a introduit le taux de compression {\rm du spin nucléaire}
\be
{\boxed{\Gamma_{\rm sq}\equiv \left(\frac{1}{\Gamma_{\rm ex}}+\frac{2}{\gamma_f}\right)^{-1}}}
\label{eq135}
\ee
On retrouve l'adimensionnement naturel du signal par $\Gamma_{\rm ex}^{1/2}$ déjà constaté dans le modèle à un mode et les mêmes formes fonctionnelles en temps, mais on perd toute relation de proportionnalité de type (\ref{eq197}) {entre signal intégré et moyenne de quadrature sur une réalisation}, la variance conditionnelle de $\bar{P}_a$ étant désormais $\nequiv 0$. \footnote{On a en effet $\mbox{Var}_{\sigma=\mathcal{S}}(\bar{P}_a)\sim \Gamma_{\rm ex}t/[4(1+\Gamma_{\rm ex}t)]-{\Gamma_{\rm sq}t/[4(1+\Gamma_{\rm sq}t)]}$.} Nous représentons sur la figure \ref{fig9}c la dépendance en le temps adimensionné $\gamma_f t$ de la moyenne et de la variance conditionnelles (\ref{eq132}) pour différentes valeurs du rapport $r=2\Gamma_{\rm ex}/\gamma_f$. On remarque que le processus de compression est d'autant plus rapide que $r$ est plus grand, et qu'il sature à un comportement limite. C'était prévisible, car {$\Gamma_{\rm sq}$} est une fonction croissante de $r$ de limite finie ; à temps fixé, la moyenne conditionnelle (en unités de $\mathcal{S}/\Gamma_{\rm ex}^{1/2}$) est donc une fonction croissante et la variance conditionnelle une fonction décroissante de $r$, comme on le voit sur la figure \ref{fig9}c. Plus précisément, dans la limite de couplage faible $\Omega\to 0$, où $r\to 0$, le taux de compression est équivalent au taux {de création d'excitations} $\Gamma_{\rm ex}$, en accord avec le modèle à un mode, et dans la limite $r\to +\infty$, il sature à la valeur $\gamma_f/2$. On ne peut donc comprimer plus rapidement qu'au taux $\gamma_f$, ce qui n'est pas surprenant : on ne peut espérer réduire les fluctuations du spin nucléaire avant que chaque atome dans l'état fondamental n'ait subi en moyenne au moins une collision d'échange de métastabilité, {le taux effectif $\gamma_f$ étant en pratique du même ordre de grandeur que le taux individuel $1/T$ dans l'équation (\ref{eq006}) sauf en cas d'extrême polarisation, voir la figure \ref{fig3}a.}

\subsubsection{Effet de la décohérence} 
\label{sec:edld}
Pour être complets, nous tenons compte, dans le schéma de compression homodyne, de la durée de vie finie $(2\gamma_0)^{-1}$ des atomes métastables, qui se désexcitent lorsqu'ils atteignent les parois de la cellule après un mouvement diffusif dans le gaz. À cette fin, nous ajoutons un opérateur de saut $ \sqrt{2\gamma_0} b $ à l'équation pilote à trois modes (\ref{eq030}). Comme la partie autre qu'hamiltonienne hermitienne reste quadratique en les quadratures des modes, elle peut être mise sous forme réduite par une rotation appropriée des modes atomiques, comme nous l'avions fait déjà dans la section \ref{sec:ana_omm} : il faut décomposer {\sl le vecteur} $(a,b)$ dans la base propre orthonormale de la matrice des taux
\be
\underline{\underline{\Gamma}}=\begin{pmatrix}
2\gamma_f & -2\sqrt{\gamma_f\gamma_m} \\
-2\sqrt{\gamma_f\gamma_m} & 2(\gamma_0+\gamma_m)
\end{pmatrix}
\label{eq150}
\ee
avec des coefficients à valeur opérateur $\alpha$ et $\beta$. La direction $\beta$ reste celle de la valeur propre maximale $\gamma_\beta$ de $\underline{\underline{\Gamma}}$, et $\alpha$ celle de la valeur propre minimale $\gamma_\alpha$, désormais non nulle. Ceci conduit à l'équation pilote
\be
\boxed{\frac{\dd\rho}{\dd t} = \frac{1}{\ii\hbar} \left[\hbar(\Omega_\alpha P_\alpha + \Omega_\beta P_\beta) P_c ,\rho \right] + \kappa \left( c\rho c^\dagger -\frac{1}{2}\{ c^\dagger c,\rho\}\right) + \gamma_\alpha \left( \alpha \rho \alpha^\dagger -\frac{1}{2}\{ \alpha^\dagger \alpha,\rho\}\right) + \gamma_\beta \left( \beta \rho \beta^\dagger -\frac{1}{2}\{ \beta^\dagger \beta,\rho\}\right)}
\label{eq151}
\ee
La nouvelle expression des pulsations de Faraday $\Omega_\alpha,\Omega_\beta$ et des taux $\gamma_\alpha,\gamma_\beta$ se trouve dans l'\ref{app:tmd}, qui donne aussi l'expression analytique de la moyenne et de la variance de la quadrature $P_a$ du spin nucléaire conditionnées au signal homodyne intégré, en toute généralité. Nous nous restreignons ici à la limite physiquement utile $\gamma_0 \ll \gamma_m$ (on a toujours $\gamma_f<\gamma_m$). À l'ordre le plus bas en $\gamma_0$, les coefficients $\Omega_\alpha$, $\Omega_\beta$ et $\gamma_\beta$ restent inchangés, et l'on a
\be
\gamma_\alpha \simeq \frac{2\gamma_0 \gamma_f}{\gamma_m+\gamma_f}
\label{eq152}
\ee
ce qui n'est autre que le taux de décohérence ramené dans le spin nucléaire hybridé. De plus, nous nous plaçons dans la limite (\ref{eq033}), avec $\gamma_\alpha = O (\Gamma_{\rm ex})$, ce qui permet d'évaluer l'effet de la décohérence en utilisant le modèle à un mode, dont l'obtention reste la même que dans la section \ref{sec:ana_omm}. L'équation stochastique (\ref{eq078}) est complétée comme suit,
\begin{equation}
\dd|\phi(t) \rangle = -\frac{\Gamma_{\rm ex}\dd t}{2} (P_\alpha-\bar{P}_\alpha)^2  |\phi(t) \rangle + \sqrt{\Gamma_{\rm ex}} \dd\zeta_s(t) (P_\alpha-\bar{P}_\alpha) |\phi(t) \rangle -\frac{\gamma_\alpha \dd t}{2} (\alpha^\dagger\alpha+2 \ii \bar{P}_\alpha \alpha +\bar{P}_\alpha^2) |\phi(t)\rangle + \sqrt{\gamma_\alpha} \dd\zeta_\alpha(t) ( \ii \alpha +\bar{P}_\alpha) |\phi(t)\rangle
\label{eq153}
\end{equation}
Nous avons pris soin de choisir $\gamma_\alpha^{1/2}\ii\alpha$ comme opérateur de saut de la décohérence ramenée (la justification est la même que dans la section \ref{sec:sdmatm}, les sauts de décohérence n'étant pas mesurés), ce qui permet de résoudre l'équation par le même ansatz gaussien réel (\ref{eq080}). Nous trouvons cette fois\footnote{{Dans le régime $\epsilon\ll 1$, la limite aux temps longs de la variance de $P_\alpha$ sur une seule réalisation dépend fortement du choix de phase dans l'opérateur de saut de la décohérence ramenée, ce qui souligne le caractère non physique de cette variance (voir la note \ref{note977}): si l'on prend $\gamma_\alpha^{1/2} \alpha$ comme opérateur de saut, on trouve que $\mbox{Var}_\phi P_\alpha\to \epsilon^{1/2}/4$ \cite{letter} au lieu de $\mbox{Var}_\phi P_\alpha\to 1/4\epsilon$ comme dans l'équation (\ref{eq154a}). Plus généralement, le choix $\gamma_\alpha^{1/2}\exp(\ii\theta)\,\alpha$, $-\pi/2<\theta<\pi/2$, conduit à l'équation de Riccati sur le paramètre $u$ de l'ansatz gaussien (désormais complexe) 
$\dd u = \Gamma_{\rm ex}\dd t +\gamma_\alpha \dd t \{\exp(2\ii\theta) u +[1-\exp(2\ii\theta)]/2- u^2[1+\exp(2\ii\theta)]/2\}$
si bien que $\mbox{Var}_\phi P_\alpha\to (1/4)\epsilon^{1/2}\sqrt{\cos\theta}\cos(\theta/2)$ en un temps $\tau\sim 1/\sqrt{2\epsilon [1+\exp(2\ii\theta)]}$; la loi de puissance en $\epsilon$ obtenue pour $\theta=0$ est donc la règle, celle obtenue pour $\theta=\pi/2$ est l'exception.}}
\bea
\label{eq154a}
\dd u &\!\!\!=\!\!\!& [\Gamma_{\rm ex} +\gamma_\alpha (1-u)]\dd t \quad\Longrightarrow\quad {u(\tau)}= 1 +\frac{1-\exp(-\epsilon\tau)}{\epsilon} \\
\dd\bar{P}_\alpha &\!\!\!=\!\!\!& -\frac{1}{2} \gamma_\alpha \bar{P}_\alpha \dd t +\frac{\sqrt{\Gamma_{\rm ex}}\dd \zeta_s+\sqrt{\gamma_\alpha}(u-1)\dd\zeta_\alpha}{2u}
\label{eq154b}
\eea
où nous avons posé $\tau=\Gamma_{\rm ex}t$ et $\epsilon=\gamma_\alpha/\Gamma_{\rm ex}$. Les mêmes arguments de gaussianité que dans la section \ref{sec:3modesTo1Homo} conduisent aux mêmes dépendances en le signal $\mathcal{S}$ de la moyenne et de la variance conditionnelles,\footnote{Nous avons simplifié l'expression (\ref{eq155b}) à l'aide de l'identité $[4 u(\tau)]^{-1}+\langle\bar{P}_\alpha^2\rangle_{\rm stoch} =1/4$, qui résulte comme dans l'équation (\ref{eq128}) du fait que la moyenne inconditionnelle $\langle P_\alpha^2\rangle=1/4$, même en présence de décohérence.}
\bea
\label{eq155a}
\langle P_\alpha\rangle_{\sigma=\mathcal{S}} &\!\!\!=\!\!\!& m(\tau) \frac{\mathcal{S}}{\sqrt{\Gamma_{\rm ex}}} \quad\mbox{avec}\quad m(\tau)= \sqrt{\Gamma_{\rm ex}} \frac{\langle \sigma(t) \bar{P}_\alpha(t)\rangle_{\rm stoch}}{\langle \sigma^2(t)\rangle_{\rm stoch}} \\
\label{eq155b}
\mbox{Var}_{\sigma=\mathcal{S}}(P_\alpha) &\!\!\!=\!\!\!& \mathcal{V}(\tau)\quad\mbox{avec}\quad \mathcal{V}(\tau) = \frac{1}{4} - \frac{\langle\sigma(t)\bar{P}_\alpha(t)\rangle_{\rm stoch}^2}{\langle\sigma^2(t)\rangle_{\rm stoch}}
\eea
et les variance et covariance prises sur les processus stochastiques {$\dd\zeta_s\equiv\dd\zeta_c$} et $\dd\zeta_\alpha$,
\bea
\label{eq156a}
\frac{\langle\sigma^2\rangle_{\rm stoch}}{\Gamma_{\rm ex}} &\!\!\!=\!\!\!& \int_0^\tau \frac{\dd\tau'}{\tau^2} \left\{
\left[{\frac{1}{2}}+\frac{1-\eee^{\epsilon(\tau'-\tau)/2}}{\epsilon u(\tau')}\right]^2 +
\frac{\left[u(\tau')-1\right]^2}{u^2(\tau')}  \frac{\left[1-\eee^{\epsilon(\tau'-\tau)/2}\right]^2}{\epsilon}
\right\}= {\frac{\epsilon\tau-2(1-\eee^{-\epsilon\tau/2})}{\epsilon^2\tau^2} + \frac{1}{4\tau}} \\
\label{eq156b}
\frac{\langle\sigma\bar{P}_\alpha\rangle_{\rm stoch}}{\sqrt{\Gamma_{\rm ex}}} &\!\!\!=\!\!\!& \int_0^\tau \frac{\dd\tau'}{\tau} \frac{\eee^{\epsilon(\tau'-\tau)/2}}{2 u(\tau')} \left\{{\frac{1}{2}}+\frac{1-\eee^{\epsilon(\tau'-\tau)/2}}{\epsilon u(\tau')} + \frac{\left[u(\tau')-1\right]^2}{u(\tau')}\left[1-\eee^{\epsilon(\tau'-\tau)/2}\right]\right\}= {\frac{1-\eee^{-\epsilon\tau/2}}{2\epsilon\tau}}
\eea
Ces expressions permettent d'évaluer facilement l'effet de la décohérence sur la compression de spin {au travers du gain métrologique (\ref{eq017})}, voir les tiretés sur la figure \ref{fig9}a. Pour le cas utile en pratique d'une faible décohérence $\epsilon\ll 1$ et d'un temps court devant $1/\gamma_\alpha$, elles peuvent être développées au premier ordre en $\epsilon$:
\be
{\boxed{m(\tau)=\frac{\tau}{1+\tau} -\epsilon \frac{(\tau+3)\tau^2}{12(\tau+1)^2}+O(\epsilon^2\tau^2)\quad ;\quad \mathcal{V}(\tau)=\frac{1}{4(\tau+1)}+\epsilon \frac{(\tau+3/2)\tau^2}{12(\tau+1)^2}+O(\epsilon^2\tau^2)}}
\label{eq157}
\ee
On en déduit que la compression optimale sur $P_\alpha$ est obtenue à un temps $t_{\rm opt}\sim(3/\Gamma_{\rm ex}\gamma_\alpha)^{1/2}$ et correspond à une variance conditionnelle $\mathcal{V}_{\rm opt}\sim(\gamma_\alpha/12\Gamma_{\rm ex})^{1/2}$. Remarquons qu'on introduit souvent, dans les études de compression de spin dans les gaz d'atomes alcalins en cavité, la coopérativité $C$ du système couplé atome-champ, définie comme le carré de la pulsation de couplage divisé par les taux de décroissance des états couplés \cite{Vuletic}. En ce sens, la coopérativité du système spin nucléaire hybridé-champ vaut
\be
\label{eq158}
C\equiv\frac{\Omega_\alpha^2}{\kappa\gamma_\alpha} = \frac{\Gamma_{\rm ex}}{\gamma_\alpha}\simeq \frac{\Omega^2}{2\gamma_0\kappa}
\ee
si bien que nous retrouvons la loi d'échelle d'exposant $-1/2$, habituelle dans les alcalins, reliant la variance de spin optimale à $C$ \cite{Vuletic}.  Plus généralement, la décohérence a un effet faible sur la compression du spin nucléaire tant qu'on reste à des temps courts devant $t_{\rm opt}$. Le lecteur trouvera en fin d'\ref{app:tmd} une extension de ces lois d'échelle au-delà du modèle à un mode, c'est-à-dire pour un rapport $\Gamma_{\rm ex}/\gamma_f$ quelconque, non infinitésimal ; c'est elle qui a été retenue dans le résumé de l'article. Le lien entre $\mathcal{V}_{\rm opt}$ et la coopérativité (\ref{eq158}) est alors rompu.

{\noindent {\bf Remerciements~:} Alice Sinatra remercie Franck Lalo\"e pour d'utiles discussions. Tous les auteurs à l'exception d'Yvan Castin bénéficient d'un financement par le projet européen de recherche et d'innovation Horizon 2020 macQsimal numéro 820393. Matteo Fadel remercie le Fonds de recherche de l'Université de Bâle pour les jeunes chercheurs d'excellence.}

\appendix 
\section{Traitement semi-classique et réduction à trois spins couplés} 
\label{app:semiclassical} 
Nous donnons ici les équations non linéaires qui décrivent la dynamique du système dans la théorie semi-classique, et nous les linéarisons pour de faibles fluctuations autour d'une solution stationnaire partiellement polarisée. 

\paragraph{Équations semi-classiques non linéaires} 
En partant des considérations et des notations de la section \ref{sec:vue}, nous prenons la moyenne des équations du mouvement de Heisenberg dans l'état quantique du système et effectuons l'approximation de décorrélation (dite semi-classique en optique quantique)
$\langle AB \rangle \simeq \langle A \rangle  \langle B \rangle $ 
où $A$ et $B$ sont deux opérateurs, pour obtenir les équations d'évolution non linéaires suivantes sur les moyennes de $\Vec{S}$ spin de Stokes du champ lumineux en cavité, $\Vec{I}$ spin nucléaire collectif dans l'état fondamental, $\Vec{J}$ et $\Vec{K}$ spins collectifs associés aux multiplicités  $F=3/2$ et $F=1/2$ dans l'état métastable, et $\Vec{\Vec{Q}}$ tenseur d'alignement collectif dans $F=3/2$, de composantes cartésiennes  $Q_{\alpha\beta}$: 
\begin{align}
\frac{\dd \langle S_x \rangle}{\dd t}&=  -\frac{\kappa}{2} \left(\langle S_x \rangle - \frac{n_{\rm ph}}{2}\right)- \chi \langle K_z \rangle \langle S_y \rangle  && \frac{\dd \langle S_y\rangle}{\dd t} =  -\frac{\kappa}{2} \langle S_y\rangle + \chi \langle K_z \rangle\langle S_x \rangle && \frac{\dd \langle S_z\rangle }{\dd t} =   -\frac{\kappa}{2} \langle S_z\rangle \label{DerivativesNLTotSx} \\ 
\frac{\dd \langle K_x \rangle}{\dd t}&=  \left. \frac{\dd \langle K_x \rangle}{\dd t} \right |_{\rm ECH}  - \chi \langle K_y\rangle \langle S_z \rangle && \frac{\dd \langle K_y \rangle}{\dd t} =  \left. \frac{\dd \langle K_y \rangle}{\dd t} \right |_{\rm ECH} + \chi\langle K_x\rangle \langle S_z \rangle &&  \frac{\dd \langle K_z \rangle}{\dd t} =  \left. \frac{\dd \langle K_z \rangle}{\dd t} \right |_{\rm ECH} \\
\frac{\dd \langle \vec{J} \rangle}{\dd t}& = \left. \frac{\dd \langle \vec{J} \rangle}{\dd t} \right |_{\rm ECH} && \frac{\dd \langle Q_{\alpha \beta}  \rangle}{\dd t} =  \left. \frac{\dd\langle Q_{\alpha \beta}  \rangle}{\dd t} \right |_{\rm ECH} && \frac{\dd\langle \Vec{I}\rangle}{\dd t} = \left.\frac{\dd\langle \Vec{I}\rangle}{\dd t}\right |_{\rm ECH} \label{eq202c}
\end{align}
Les termes proportionnels au taux de perte $\kappa$ {du miroir de sortie} de la cavité font relaxer $\langle S_x\rangle$ vers sa valeur stationnaire 
$\langle S_x \rangle_s=n_{\rm ph}/2$ 
forcée par le champ laser polarisé selon $Ox$ injecté dans la cavité, et les moyennes transverses $\langle S_y\rangle$ et $\langle S_z\rangle$ vers zéro. Les termes proportionnels au couplage de Faraday $\chi$ entre le mode de cavité et le spin $\Vec{K}$ dérivent de l'hamiltonien (\ref{eq002}). La contribution des collisions d'échange de métastabilité (ECH) entre atomes fondamentaux et métastables se déduit directement de l'équation pilote sur l'opérateur densité à un atome des références \cite{DupontRoc,LaloeDupontLeduc} par simple multiplication ou division par le nombre total de fondamentaux $N_{\rm cell}$ ou métastables $n_{\rm cell}$ dans la cellule:\footnote{Les moyennes collectives sont en effet reliées comme suit aux moyennes à un atome $ \langle \: \: \rangle_{\rm at} $:
$    \langle\Vec{I}\rangle = N_{\rm cell} \langle \Vec{I}\rangle_{\rm at} $, $    \langle\Vec{J}\rangle = n_{\rm cell}  \langle \Vec{J}\rangle_{\rm at} $, $ \langle\Vec{K}\rangle = n_{\rm cell}  \langle \Vec{K}\rangle_{\rm at} $, $    \langle\Vec{\Vec{Q}}\rangle = n_{\rm cell}  \langle \Vec{\Vec{Q}}\rangle_{\rm at} $, $ \langle\Vec{\Sigma}\rangle = n_{\rm cell}  \langle \Vec{\Sigma} \rangle_{\rm at} $.
} 
\begin{align}
      \left. \frac{\dd \langle\Vec{K}\rangle}{\dd t} \right |_{\rm ECH}&= - \frac{7}{9\tau}\langle\Vec{K}\rangle+ \frac{1}{9\tau}\langle\Vec{J}\rangle  - \frac{1}{9\tau}\frac{n_{\rm cell} }{N_{\rm cell} }\langle\Vec{I}\rangle - \frac{4}{3\tau}\frac{1}{N_{\rm cell} }\langle\Vec{\Vec{Q}}\rangle \cdot \langle\Vec{I}\rangle \label{eq204a}\\
     \left.  \frac{\dd \langle\Vec{J}\rangle}{\dd t} \right |_{\rm ECH}&= - \frac{4}{9\tau}\langle\Vec{J}\rangle + \frac{10}{9\tau}\langle\Vec{K}\rangle+ \frac{10}{9\tau}\frac{n_{\rm cell} }{N_{\rm cell} }\langle\Vec{I}\rangle + \frac{4}{3\tau}\frac{1}{N_{\rm cell} }\langle\Vec{\Vec{Q}}\rangle \cdot \langle\Vec{I}\rangle \label{eq204b}\\
     \left.  \frac{\dd \langle Q_{\alpha\beta}\rangle}{\dd t}\right |_{\rm ECH} &= - \frac{2}{3\tau}\langle Q_{\alpha\beta}\rangle + \frac{1}{9\tau}\frac{1}{N_{\rm cell} }\left(3\frac{\langle I_\alpha\rangle \langle \Sigma_\beta\rangle + \langle I_\beta\rangle \langle \Sigma_\alpha\rangle}{2} - \delta_{\alpha\beta} \langle \Vec{I}\rangle \cdot \langle \Vec{\Sigma}\rangle\right)\label{eq204c}\\
        \left. \frac{\dd \langle\Vec{I}\rangle}{\dd t}\right |_{\rm ECH} &= - \frac{1}{T}\langle\Vec{I}\rangle + \frac{1}{3T}\frac{N_{\rm cell} }{n_{\rm cell} }{(\langle\Vec{J}\rangle - \langle\Vec{K}\rangle)} \label{eq204d}
\end{align}
où 
$\langle \Vec{\Sigma}\rangle = \frac{2}{3} \left[\langle\Vec{J}\rangle + 2\langle\Vec{K}\rangle \right]$ 
est la valeur moyenne du spin électronique dans l'état métastable. On se reportera aux équations (1.37b), (1.37a), (1.39) et (1.25) de la référence \cite{LaloeDupontLeduc} (en tenant compte d'un écart d'un facteur $6$ sur la définition du tenseur d'alignement), ou aux équations (VIII.30), (VIII.29), (VIII.32) et (VIII.15) {de la référence \cite{DupontRoc}} (en ajoutant un facteur de Kronecker $\delta_{\alpha\beta}$ omis dans (VIII.32)). Ici $ 1/\tau $ et $ 1/T $, les taux individuels des collisions d'échange de métastabilité subies par un atome dans l'état métastable et dans l'état fondamental, sont dans le rapport 
$T/\tau=N_{\rm cell}/ n_{\rm cell}$
puisque, dans l'unité de temps, un nombre égal d'atomes fondamentaux et d'atomes métastables ont subi une collision d'échange \cite{DupontRoc,LaloeDupontLeduc}. 

\paragraph{Solution stationnaire partiellement polarisée} 
Dans un état stationnaire polarisé de polarisation nucléaire $ \eta \in [0,1] $,
\begin{align}
\langle I_x \rangle_s &= \eta \frac{N_{\rm cell} }{2} \quad\:; \quad \langle I_y \rangle_s=\langle I_z \rangle_s=0 \quad\:; \quad
\langle S_x \rangle_s =\frac{n_{\rm ph}}{2}\quad\:; \quad  \langle S_y \rangle_s = \langle S_z \rangle_s = 0  \label{eq207}
\end{align}
l'invariance par rotation d'axe $Ox$ contraint les spins moyens {dans l'état métastable} à être alignés selon $Ox$, et le tenseur d'alignement moyen à être diagonal dans la base cartésienne, avec des valeurs propres égales selon les directions $Oy$ et $Oz$. Le système (\ref{DerivativesNLTotSx})-(\ref{eq202c}) admet ainsi une solution stationnaire avec comme seules moyennes non nulles dans l'état métastable: 
\begin{align}
\langle K_x \rangle_s &= \frac{\eta}{2}\frac{1 - \eta^2}{3 + \eta^2}n_{\rm cell} \ \ ;\  \  \langle J_x \rangle_s =  \eta \frac{5+\eta^2}{3+\eta^2} n_{\rm cell} \  \ ;\  \  \langle \Sigma_x \rangle_s = \frac{4\eta}{3+\eta^2} n_{\rm cell} \  \ ;\  \  \langle Q_{yy} \rangle_{s} =  \langle Q_{zz} \rangle_{s} = -\frac{1}{2} \langle Q_{xx} \rangle_s = - \frac{\eta}{12}\langle \Sigma_x \rangle_s \label{eq208}
\end{align}

\paragraph{Équations semi-classiques linéarisées} 
On linéarise maintenant les équations (\ref{DerivativesNLTotSx})-(\ref{eq202c}) en les fluctuations classiques autour de la solution stationnaire (\ref{eq207})-(\ref{eq208}) en effectuant la substitution 
$\langle A \rangle \rightarrow \langle A\rangle_s  +\delta A$
et en traitant $\delta A$ au premier ordre. En nous limitant au sous-espace des fluctuations transverses, c'est-à-dire aux directions $ \alpha = y, z $ orthogonales aux spins moyens, nous obtenons un système fermé:
\begin{align}
        \frac{\dd }{\dd t}\delta S_\alpha   & = -\frac{\kappa}{2} \delta S_\alpha + \chi\delta_{\alpha y}\langle S_x \rangle_s  \delta K_z  \\
\label{eq210b}
        \frac{\dd }{\dd t}\delta K_\alpha &= - \frac{7}{9\tau}\delta K_\alpha + \frac{1}{9\tau}\delta J_\alpha -\frac{2\eta}{3\tau} \delta Q_{\alpha x} - \frac{1}{9T}\left( 1 + \frac{12}{n_{\rm cell} }\langle Q_{\alpha\alpha}\rangle_s \right)\delta I_\alpha + \chi \delta_{\alpha y} \langle K_x \rangle_s \delta S_z\\
\label{eq210e}
    \frac{\dd }{\dd t}\delta J_{\alpha}   &= - \frac{4}{9\tau}\delta J_{\alpha} + \frac{10}{9\tau}\delta K_{\alpha} +\frac{2\eta}{3\tau} \delta Q_{\alpha x} + \frac{10}{9T}\left( 1 + \frac{6}{5n_{\rm cell} }\langle Q_{\alpha\alpha}\rangle \right)\delta I_{\alpha} \\
\label{eq210f}
        \frac{\dd}{\dd t}\delta Q_{\alpha x}   &=   - \frac{2}{3\tau} \delta Q_{\alpha x} + \frac{\eta}{12\tau}\delta \Sigma_\alpha + \frac{1}{6 T n_{\rm cell}} \langle \Sigma_x \rangle_s \delta I_\alpha  \\
    \frac{\dd}{\dd t}\delta I_{\alpha} &=  - \frac{1}{T}\delta I_{\alpha} + \frac{1}{3\tau}{(\delta J_{\alpha} - \delta K_{\alpha})} \label{eq210} 
\end{align}

\paragraph{Réduction à trois spins collectifs couplés} 
En posant
$\frac{\dd}{\dd t}\delta J_{\alpha} = 0$
dans l'équation (\ref{eq210e}) et
$\frac{\dd}{\dd t}\delta Q_{\alpha x} = 0$ 
dans l'équation (\ref{eq210f}), nous éliminons adiabatiquement les fluctuations du spin collectif $ \vec{J} $ et du tenseur d'alignement collectif dont les évolutions sont régies par l'échange de métastabilité uniquement: 
\begin{equation}
\delta J_\alpha^{\rm adiab} = 2\frac{10+\eta^2}{8-\eta^2} \delta K_\alpha +\frac{12\tau}{T} \frac{5+2\eta^2}{(3+\eta^2)(8-\eta^2)} \delta I_\alpha \quad ;\quad \delta Q_{\alpha x}^{\rm adiab} = \frac{3\eta}{8-\eta^2} \delta K_\alpha + \frac{\tau}{T} \frac{\eta (13+\eta^2)}{(3+\eta^2)(8-\eta^2)} \delta I_\alpha \label{eq212}
\end{equation}
Le report des expressions adiabatiques (\ref{eq212}) dans les équations (\ref{eq210b}) et (\ref{eq210}) sur $ \delta K_\alpha $ et $ \delta I_{\alpha}$ conduit dans le corps de l'article au système réduit (\ref{eq:redS})-(\ref{eq005}) couplant les fluctuations des trois spins (\ref{eq003}), où $ \gamma_f $ et $ \gamma_m $, les taux effectifs d'échange de métastabilité entre le spin nucléaire et le spin $ F = 1/2 $ du métastable,  sont donnés par l'équation (\ref{eq006}).

\section{Solution du modèle à trois modes avec décohérence pour la détection homodyne}
\label{app:tmd}

Nous donnons ici la solution analytique du modèle à trois modes en présence de décohérence, voir l'équation pilote (\ref{eq151}), pour une évolution du système conditionnée à une mesure homodyne en continu {de la composante polarisée selon $Oy$} du champ sortant de la cavité. La valeur des coefficients $\gamma_\alpha$, $\gamma_\beta$, $\Omega_\alpha$ et $\Omega_\beta$, ainsi que des opérateurs d'annihilation $\alpha$ et $\beta$, se déduit d'une diagonalisation de la matrice des taux (\ref{eq150}). Les taux $\gamma_\alpha$ et $\gamma_\beta$ en sont les valeurs propres rangées par ordre croissant:
\be
\label{eq220}
\gamma_{\alpha,\beta} = \gamma_m+\gamma_f+\gamma_0 \mp [(\gamma_m+\gamma_f+\gamma_0)^2-4\gamma_f\gamma_0]^{1/2}
\ee
En termes des pulsations de Faraday $\Omega_\alpha$ et $\Omega_\beta$, les vecteurs propres normalisés correspondants s'écrivent $(\Omega_\beta/\Omega,\Omega_\alpha/\Omega)$ et $(-\Omega_\alpha/\Omega,\Omega_\beta/\Omega)$, si bien que $\alpha=(\Omega_\beta a + \Omega_\alpha b)/\Omega$ et $\beta=(\Omega_\beta b - \Omega_\alpha a)/\Omega$ avec
\be
\label{eq221}
\Omega_\alpha = \frac{\Omega(\gamma_f-\gamma_\alpha/2)}{[\gamma_m\gamma_f +(\gamma_f-\gamma_\alpha/2)^2]^{1/2}} \quad ; \quad
\Omega_\beta = \frac{\Omega\sqrt{\gamma_m\gamma_f}}{[\gamma_m\gamma_f +(\gamma_f-\gamma_\alpha/2)^2]^{1/2}}
\ee
dans un choix de signe assurant que $\alpha\to a$ et $\beta\to b$ lorsque $\gamma_f\to 0$ et reproduisant (\ref{eq036}) lorsque $\gamma_0\to 0$. Puisque l'opérateur de saut $C_\alpha\propto\alpha$ décrit des processus non mesurés, nous pouvons, comme nous l'avons fait pour $C_\beta$, le prendre de la forme $\sqrt{\gamma_\alpha}\ii\alpha$ et réutiliser l'ansatz gaussien réel (\ref{eq120}) afin de résoudre l'équation stochastique (\ref{eq076}) sur le vecteur d'état. Dans l'équation d'évolution {de} la matrice $\underline{\underline{u}}$ {apparaissant dans l'ansatz}, les indices $\alpha$ et $\beta$ jouent désormais des rôles symétriques et l'on obtient
\be
\begin{array}{lll}
\dd u_{\alpha\alpha} = - \Omega_\alpha \dd t\,  u_{\alpha c}+ \gamma_\alpha \dd t (1-u_{\alpha\alpha})  & \displaystyle \dd u_{\alpha\beta} = -\frac{\dd t}{2} [(\gamma_\alpha + \gamma_\beta) u_{\alpha\beta} + \Omega_\beta u_{\alpha c} + \Omega_\alpha u_{\beta c}] & \displaystyle \dd u_{\alpha c} = -\frac{\dd t}{2} [(\gamma_\alpha + \kappa) u_{\alpha c} + \Omega_\alpha u_{cc}] \\
&&\\
\dd u_{\beta\beta} = - \Omega_\beta \dd t\, u_{\beta c} + \gamma_\beta \dd t (1-u_{\beta\beta}) & \displaystyle \dd u_{\beta c} = -\frac{\dd t}{2} [(\gamma_\beta + \kappa) u_{\beta c} + \Omega_\beta u_{cc}] & \dd u_{cc} = \kappa \dd t (1- u_{cc}) 
\end{array}
\ee
dont la solution pour la condition initiale $\underline{\underline{u}}(0)=\mathrm{Id}$ s'écrit
\begin{align}
\label{eq223a}
    u_{\alpha\alpha}(t) = & 1 + \frac{\Omega_\alpha^2}{\gamma_\alpha(\kappa + \gamma_\alpha)}\left(1-\eee^{-\gamma_\alpha t}\right) - \frac{2\Omega_\alpha^2}{\kappa^2 - \gamma_\alpha^2}\left(\eee^{-\gamma_\alpha t}-\eee^{-(\kappa + \gamma_\alpha)t/2}\right) \\
    u_{\alpha\beta}(t) = & {\frac{\Omega_\alpha\Omega_\beta}{\gamma_\alpha + \gamma_\beta}\left(\frac{1}{\kappa + \gamma_\alpha} + \frac{1}{\kappa+\gamma_\beta}\right) \left(1 - \eee^{-(\gamma_\alpha + \gamma_\beta)t/2} \right) + \frac{\Omega_\alpha\Omega_\beta}{(\kappa-\gamma_\beta)(\kappa+\gamma_\alpha)} \left(\eee^{-(\kappa+\gamma_\alpha) t/2} - \eee^{-(\gamma_\alpha + \gamma_\beta)t/2} \right)} \nonumber \\
     & {+ \frac{\Omega_\alpha\Omega_\beta}{(\kappa-\gamma_\alpha)(\kappa+\gamma_\beta)} \left(\eee^{-(\kappa + \gamma_\beta)t/2} - \eee^{-(\gamma_\alpha + \gamma_\beta)t/2} \right)} \\
    u_{\alpha c}(t) = & -\frac{\Omega_\alpha}{\kappa + \gamma_\alpha}\left(1-\eee^{-(\kappa + \gamma_\alpha)t/2}\right)\\
    u_{\beta\beta}(t) = & 1 + \frac{\Omega_\beta^2}{\gamma_\beta(\kappa + \gamma_\beta)}\left(1-\eee^{-\gamma_\beta t}\right) - \frac{2\Omega_\beta^2}{\kappa^2 - \gamma_\beta^2}\left(\eee^{-\gamma_\beta t}-\eee^{-(\kappa + \gamma_\beta)t/2}\right) \\
\label{eq223e}
    u_{\beta c}(t) = & -\frac{\Omega_\beta}{\kappa + \gamma_\beta}\left(1-\eee^{-(\kappa + \gamma_\beta)t/2}\right) \\
    u_{cc}(t) = & \quad 1
\end{align}
Le vecteur des coordonnées moyennes $\bar{\qq}$ {apparaissant dans l'ansatz (\ref{eq120})} obéit à l'équation stochastique
\be
\dd\bar{\qq} = \frac{1}{2} \begin{pmatrix} -\gamma_\alpha & 0 & 0 \\
0 & -\gamma_\beta & 0 \\
\Omega_\alpha & \Omega_\beta & -\kappa 
\end{pmatrix}
\dd t \, \bar{\qq}  + 
\frac{1}{2}[\mbox{Id}-\underline{\underline{c}}(t)] \begin{pmatrix}
\gamma_\alpha^{1/2} \dd\zeta_\alpha(t) \\
\gamma_\beta^{1/2} \dd\zeta_\beta(t) \\
-\kappa^{1/2} \dd\zeta_c(t)
\end{pmatrix}
\label{eq224}
\ee
La moyenne non conditionnelle $\langle P_a^2\rangle$ valant toujours $1/4$, la moyenne et la variance de $P_a$ conditionnées au signal homodyne intégré restent données par les équations (\ref{eq127a}) et (\ref{eq127b}), en généralisant les expressions (\ref{eq130}) des variances et covariance des variables aléatoires $\bar{P}_a(t)$ et $\sigma(t)$ au cas de trois processus stochastiques indépendants $\dd\zeta_\alpha(t')$, $\dd\zeta_\beta(t')$ et $\dd\zeta_c(t')$ comme suit:
\begin{align}
\langle \sigma(t)\bar{P}_a(t)\rangle_{\rm stoch} &= \int_0^t \dd t'\, \sum_{\nu\in\{\alpha,\beta,c\}} p_\nu(t,t') \sigma_\nu(t,t') \quad ; \quad 
\langle\sigma^2(t)\rangle_{\rm stoch} = \int_0^t \dd t'\, \sum_{\nu\in\{\alpha,\beta,c\}} \sigma_\nu^2(t,t')\quad ; \nonumber\\
\langle \bar{P}_a^2(t)\rangle_{\rm stoch} &= \int_0^t \dd t'\, \sum_{\nu\in\{\alpha,\beta,c\}} p_\nu^2(t,t')
\label{eq225}
\end{align}
avec les expressions compactes des amplitudes correspondantes
\begin{align}
\label{eq226a}
p_{\nu}(t,t') &= (-1)^{\delta_{\nu c}} \frac{\sqrt{\gamma_\nu}}{2\Omega} \left\{ \Omega_\beta \eee^{-\gamma_\alpha(t-t')/2} [\delta_{\alpha\nu} -c_{\alpha\nu}(t')]-\Omega_\alpha \eee^{-\gamma_\beta(t-t')/2} [\delta_{\beta\nu} -c_{\beta\nu}(t')]\right\} \\
\label{eq226b}
\sigma_{\nu}(t,t') &= \frac{\delta_{\nu c}}{2t} + (-1)^{\delta_{\nu c}} \frac{\sqrt{\kappa\gamma_\nu}}{2 t} \left\{
[\delta_{c\nu}-c_{c\nu}(t')]f_\kappa(t-t')+\sum_{\mu\in\{\alpha,\beta\}} \frac{\Omega_\mu}{\kappa-\gamma_\mu}[\delta_{\mu\nu}-c_{\mu\nu}(t')] [f_{\gamma_\mu}(t-t')-f_{\kappa}(t-t')]\right\} 
\end{align}
L'indice $\nu$ court sur les trois valeurs $\alpha$, $\beta$, $c$ et l'on a posé $\gamma_c=\kappa$. La fonction $\delta$ est celle de Kronecker, et la fonction $f_\lambda$ est la même que dans les équations (\ref{eq129a})-(\ref{eq129d}).

La solution générale que nous venons d'exposer comporte les cinq taux $\gamma_\alpha,\Gamma_{\rm ex}=\Omega_\alpha^2/\kappa,\gamma_f$ d'une part, $\gamma_\beta,\kappa$ d'autre part. Le régime pertinent expérimentalement est celui où les deux derniers sont \og infiniment\fg\, plus grands que les trois premiers et ne contribuent qu'au travers de régimes transitoires inobservables. Mathématiquement, on accède à cette limite en faisant tendre $\gamma_f$ vers zéro à $\kappa,\gamma_m,\gamma_0$ et $\Omega$ fixés et à $\tau=\Gamma_{\rm ex} t>0$ fixé. Alors les trois premiers taux tendent conjointement vers zéro, c'est-à-dire dans des rapports de limite finie et non nulle 
$\Gamma_{\rm ex}/\gamma_f\to \Omega^2\gamma_m/[\kappa(\gamma_0+\gamma_m)^2]$ et $ \gamma_\alpha/\gamma_f\to 2 \gamma_0/(\gamma_0+\gamma_m)$,
le taux $\gamma_\beta$ se réduit à $\gamma\equiv 2(\gamma_0+\gamma_m)$ et le couplage de Faraday $\Omega_\beta$ à $\Omega$. Tous les transitoires exponentiels disparaissent dans les éléments de matrice (\ref{eq223a})-(\ref{eq223e}) de $\underline{\underline{u}}$ sauf ceux relaxant au taux $\gamma_\alpha$. Les amplitudes (\ref{eq226a}) et (\ref{eq226b}) sur les processus stochastiques se réduisent à
\begin{align}
\label{eq927}
\frac{p_\alpha(t,t')}{\sqrt{\Gamma_{\rm ex}}} &= \frac{u(\tau')-1}{2 u(\tau')} \sqrt{\epsilon}\,\eee^{-\epsilon (\tau-\tau')/2}  && \frac{\sigma_\alpha(t,t')}{\Gamma_{\rm ex}} = \frac{u(\tau')-1}{\tau \, u(\tau')}\sqrt{\epsilon}\, \frac{1-\eee^{-\epsilon(\tau-\tau')/2}}{\epsilon} \\
\frac{p_\beta(t,t')}{\sqrt{\Gamma_{\rm ex}}} &= \frac{\sqrt{\rho}}{(1+\rho)u(\tau')} \eee^{-\epsilon (\tau-\tau')/2} && \frac{\sigma_\beta(t,t')}{\Gamma_{\rm ex}} =\frac{\sqrt{\rho}}{(1+\rho)\tau}\left[\frac{2}{u(\tau')} \frac{1-\eee^{-\epsilon(\tau-\tau')/2}}{\epsilon} +\rho+\frac{\gamma}{\kappa} (\rho-1)\right] \\
\frac{p_c(t,t')}{\sqrt{\Gamma_{\rm ex}}} &= \frac{(1-\rho)}{2(1+\rho)u(\tau')} \eee^{-\epsilon (\tau-\tau')/2} && \frac{\sigma_c(t,t')}{\Gamma_{\rm ex}} = \frac{1}{\tau} \left[\frac{1}{2} + \frac{1-\rho}{1+\rho} \frac{1}{u(\tau')} \frac{1-\eee^{-\epsilon(\tau-\tau')/2}}{\epsilon} + \frac{\rho}{1+\rho} \left(1+\frac{2\gamma}{\kappa}\right)\right]
\end{align}
où $\epsilon=\gamma_\alpha/\Gamma_{\rm ex}$ comme dans la section \ref{sec:edld}, la fonction $u(\tau)$ est donnée par l'équation (\ref{eq154a}) et la notation 
$\rho=\Omega^2\kappa/[\gamma(\kappa+\gamma)^2]$
généralise celle de la note \ref{note133}. Les relations (\ref{eq127a}) et (\ref{eq127b}) restent valables, avec les nouvelles expressions des variance et covariance 
\be
\label{eq228}
\boxed{\frac{\langle \sigma^2 \rangle_{\rm stoch}}{\Gamma_{\rm ex}}= {\frac{\epsilon\tau-2(1-\eee^{-\epsilon\tau/2})}{\epsilon^2\tau^2} +\frac{\Gamma_{\rm ex}}{4\tau \Gamma_{\rm sq}^{\rm gen}}} \quad\quad\mbox{et}\quad\quad
\frac{\langle \sigma \bar{P}_a \rangle_{\rm stoch}}{\sqrt{\Gamma_{\rm ex}}} = {\frac{1-\eee^{-\epsilon\tau/2}}{2\epsilon\tau}}}
\ee
et du taux de compression {généralisé}
\be
\label{eq229}
\boxed{\Gamma_{\rm sq}^{\rm gen} = \left[\frac{1}{\Gamma_{\rm ex}}+\frac{2(\gamma_0+\gamma_m)}{\gamma_f\gamma_m}\right]^{-1}}
\ee
qui reproduisent les variance et covariance (\ref{eq156a}) et (\ref{eq156b}) du modèle à un mode avec décohérence lorsque $\Gamma_{\rm ex}/\gamma_f\to 0$ et le taux de compression {de spin} (\ref{eq135}) du modèle à trois modes sans décohérence lorsque $\gamma_0\to 0$. Les nouveaux résultats peuvent être simplifiés dans la limite utile de faible décohérence ramenée $\gamma_\alpha/\Gamma_{\rm ex}\to 0$ par un développement à l'ordre un en $\epsilon$, ce qui permet de généraliser comme suit les résultats (\ref{eq157}) sur la moyenne et la variance conditionnelles à une valeur de $\Gamma_{\rm ex}/\gamma_f$ non infinitésimale:
\begin{align}
\label{eq230a}
m(t) &= \frac{\Gamma_{\rm sq}^{\rm gen} t}{1+\Gamma_{\rm sq}^{\rm gen} t} - \frac{\gamma_\alpha}{\Gamma_{\rm sq}^{\rm gen}} \frac{(3+\Gamma_{\rm sq}^{\rm gen} t)(\Gamma_{\rm sq}^{\rm gen} t)^2}{12(1+\Gamma_{\rm sq}^{\rm gen} t)^2}+O[(\gamma_\alpha t)^2]\\
\label{eq230b}
\mathcal{V}(t)&=\frac{1}{4(1+\Gamma_{\rm sq}^{\rm gen} t)} + \frac{\gamma_\alpha}{\Gamma_{\rm sq}^{\rm gen}} \frac{(\Gamma_{\rm sq}^{\rm gen}t+3/2)(\Gamma_{\rm sq}^{\rm gen} t)^2}{12(1+\Gamma_{\rm sq}^{\rm gen} t)^2}+O[(\gamma_\alpha t)^2]
\end{align}
Cette généralisation revient simplement à remplacer $\tau$ par $\Gamma_{\rm sq}^{\rm gen}t$ et $\epsilon$ par $\gamma_\alpha/\Gamma^{\rm gen}_{\rm sq}$ dans les seconds membres de (\ref{eq157}).\footnote{Elle vaut en fait à tous les ordres en $\epsilon$ puisque le remplacement proposé ne change pas $\epsilon\tau$ (toujours égal à $\gamma_\alpha t$) et fait passer des équations (\ref{eq156a}) et (\ref{eq156b}) à l'équation (\ref{eq228}).} La compression optimale sur $P_a$ est alors obtenue au bout d'un temps $t_{\rm opt}\sim (3/\Gamma_{\rm sq}^{\rm gen}\gamma_\alpha)^{1/2}$ et correspond à une variance conditionnelle $\mbox{Var}_{\sigma=\mathcal{S}}^{\rm opt}(P_a)\sim (\gamma_\alpha/12\Gamma_{\rm sq}^{\rm gen})^{1/2}$; {le gain métrologique optimal s'en déduit par l'équation (\ref{eq017}).}

\def\bysame{\leavevmode ---------\thinspace}
\def\cdrandname{\&}
\providecommand\cdrnumero{no.~}
\providecommand{\cdredsname}{eds.}
\providecommand{\cdredname}{ed.}
\providecommand{\cdrchapname}{chap.}
\providecommand{\cdrmastersthesisname}{Memoir}
\providecommand{\cdrphdthesisname}{PhD Thesis}